%% file: thesis.tex
\documentclass[phd,10pt]{psuthesis}
\pdfoutput=1
\usepackage[utf8]{inputenc}
\usepackage[final]{ifdraft}

\newif\ifusepgf
\usepgffalse

\ifdraft{\usepackage[bleed]{lettermargins}}{}

\usepackage[boldvectors,braket,particle,units]{physymb}
\usepackage{hepnicenames}

\usepackage{mathtools}
\MHInternalSyntaxOn
\MH_set_boolean_T:n {outer_mult}
\MHInternalSyntaxOff

\usepackage[table]{xcolor}

\ifusepgf
 \usepackage{tikz}[2013-12-19]
 \usepackage{pgfplots}[2014-02-27]
 \usepackage{pgfplotstable}

 \ifdraft{
  \usepackage[colorinlistoftodos,textsize=small]{todonotes}
  \reversemarginpar 
 }{
  \usepackage[disable]{todonotes}
 }
\else
 \newcommand{\todo}[2][]{}
\fi
\newcommand{\note}[2][]{\todo[color=yellow,#1]{#2}}
\newcommand{\iftime}[2][]{\todo[color=cyan!20,#1]{#2}}

\usepackage[mathscr]{eucal} 
\usepackage{eqlist} 
\usepackage{tabularx} 
\usepackage{booktabs}
\usepackage{slashed}
\usepackage{iterm}
\usepackage{subcaption}
\usepackage{floatrow}

\usepackage[backend=biber,style=phys,bibstyle=physplus,articletitle=false,biblabel=brackets,chaptertitle=false,pageranges=false,doi,eprint,autocite=plain,maxnames=4,minnames=1]{biblatex}
\input{stackexchangebiblatex}

\usepackage[Lenny]{fncychap}
\ChTitleVar{\Huge\sffamily\bfseries}

\input{commands}

\ifusepgf
 \input{graphicscommands}
\else
 \newcommand\setplotshift[1]{}
 \usepackage{epstopdf}
 \usepackage{tikzexternal}
 \tikzsetexternalprefix{generated-figures/}
\fi

\usepackage{hyperref}
\hypersetup{pdftitle={Probing Hadron Structure in Proton-Nucleus Collisions},pdfauthor={David Zaslavsky}}
\usepackage{bookmark}

\ifdraft{
 \usepackage[right,final]{showlabels}
}{}

\title{Probing Hadron Structure in Proton-Nucleus Collisions}
\author{David Zaslavsky}
\dept{Physics}
\degreedate{December 2014}
\copyrightyear{2014}
\documenttype{Dissertation}
\submittedto{The Graduate School}
\collegesubmittedto{Eberly College of Science}

\numberofreaders{4}
\advisor[Dissertation Advisor, Chair of Committee]{Anna M. Sta\'sto}{Associate Professor of Physics}
\readerone[]{Irina Mocioiu}{Associate Professor of Physics}
\readertwo[]{Mark Strikman}{Distinguished Professor of Physics}
\readerthree[]{James Brannick}{Professor of Mathematics}
\readerfour[Graduate Program Chair]{Richard Robinett}{Professor of Physics}

\begin{document}
\frontmatter

%

\psutitlepage

\psucommitteepage

%
\pagestyle{plain}
\chapter*{Abstract}
    \begin{singlespace}
        \input{abstract.tex}

    \end{singlespace}
\newpage

\thesistableofcontents

\thesislistoffigures

\thesislistoftables

\ifdraft{
 \todototoc
 \listoftodos
}{}

%
\chapter{Acknowledgments}
\begin{singlespace}
    \input{acknowledgements.tex}
\end{singlespace}

\vfill
\hrule
\vspace{1ex}
{\small This document is distributed under the \href{https://creativecommons.org/licenses/by/4.0/}{Creative Commons Attribution 4.0 License}.}

\thesismainmatter

\include{introduction}

\include{structure}
\include{cgc}
\include{correlation}
\include{crosssection}
\include{xsecanalysis}
\include{beyondnlo}

\bookmarksetup{startatroot}
\include{conclusion}

\appendix
\include{grouptheory}
\include{lightcone}
\include{identities}
\printbibliography[heading=bibintoc]
\backmatter

\input{vita.tex}

\end{document}

%% file: stackexchangebiblatex.tex
\newbibmacro{author+profile}{%
  \iffieldundef{nameaddon}{%
    \usebibmacro{author}%
  }{%
    \ifhyperref{%
      \href{\thefield{nameaddon}}{\usebibmacro{author}}%
    }{%
      \usebibmacro{author}%
    }%
  }%
}

\DeclareBibliographyDriver{stackexchange}{%
  \usebibmacro{bibindex}%
  \usebibmacro{begentry}%
  \usebibmacro{author+profile}%
  \setunit{\labelnamepunct}\newblock
  \usebibmacro{title}%
  \setunit{\addperiod\addspace}\newblock
  \printfield{howpublished}%
  \newunit\newblock%
  \usebibmacro{date}%
  \setunit{\addperiod\addspace}\newblock
  \usebibmacro{url+urldate}%
  \newunit\newblock
  \usebibmacro{addendum+pubstate}%
  \setunit{\bibpagerefpunct}\newblock
  \iftoggle{bbx:related}
    {\usebibmacro{related:init}%
     \usebibmacro{related}}
    {}%
  \usebibmacro{finentry}
}

\DeclareFieldFormat[stackexchange]{titlecase}{#1}

\DeclareSourcemap{
  \maps[datatype=bibtex]{
    \map{
      \step[fieldsource=howpublished, match=\regexp{.+Stack\sExchange}, final=true]
      \step[typesource=misc, typetarget=stackexchange, final=true]
    }
    \map{
      \pertype{stackexchange}
      \step[fieldsource=author, match=\regexp{.+\s*\(https?://.+\)}]
      \step[fieldset=nameaddon, origfieldval]
      \step[fieldsource=author, match=\regexp{(.+)\s*\(https?://.+\)}, replace=\regexp{$1}]
      \step[fieldsource=nameaddon, match=\regexp{.+\s*\((https?://.+)\)}, replace=\regexp{$1}]
      \step[fieldsource=note, match=\regexp{.+\(version:\s\d{4}-\d{2}-\d{2}\)}]
      \step[fieldset=date, origfieldval]
      \step[fieldset=note, null]
      \step[fieldsource=date, match=\regexp{.+\(version:\s(\d{4}-\d{2}-\d{2})\)}, replace=\regexp{$1}]
      \step[fieldset=eprint, null]
    }
  }
}

%% file: commands.tex
\newcommand\definesymbol[3]{\newcommand#1{\ensuremath{#2}}}

\DeclareMathOperator\Ci{Ci}
\DeclareMathOperator\Si{Si}

\DeclareMathOperator\pathorder{\mathcal{P}}

\definesymbol\massnumber{A}{Atomic mass number of the target nucleus}
\definesymbol\centrality{c}{centrality coefficient}
\definesymbol\scamp{\mathcal{A}}{Scattering amplitude}
\definesymbol\rapidity{Y}{Rapidity}
\definesymbol\pseudorapidity{\eta}{Pseudrapidity}
\definesymbol\Sperp{S_\perp}{Transverse area of the target nucleus}
\definesymbol\sqs{\sqrt{s}}{Center-of-mass energy}
\definesymbol\pperp{p_\perp}{Transverse momentum}
\definesymbol\pgammaperp{p_{\Pphoton\perp}}{Transverse momentum of a photon}
\definesymbol\ppionperp{p_{\Ppi\perp}}{Transverse momentum of a neutral pion}
\definesymbol\pTcut{p_{\perp\text{cut}}}{Lower cutoff on transverse momentum}
\definesymbol\kperp{k_\perp}{Transverse momentum of a gluon}
\definesymbol\kiperp{k_{1\perp}}{Transverse momentum of a gluon}
\definesymbol\kpperp{k_\perp'}{Transverse momentum conjugate to a coordinate}
\definesymbol\kgperp{k_{\Pgluon\perp}}{Transverse momentum of a gluon}
\definesymbol\gluonkperp{k_{\Pgluon\perp}}{Transverse momentum of a gluon}
\definesymbol\kgiperp{k_{\Pgluon1\perp}}{Transverse momentum of a gluon}
\definesymbol\kqperp{k_{\Pquark\perp}}{Transverse momentum of a quark}
\definesymbol\qperp{q_{\perp}}{Transverse momentum integration variable}
\definesymbol\qiperp{q_{1\perp}}{Transverse momentum integration variable}
\definesymbol\qiiperp{q_{2\perp}}{Transverse momentum integration variable}
\definesymbol\qiiiperp{q_{3\perp}}{Transverse momentum integration variable}
\definesymbol\qipperp{q_{1\perp}'}{Transverse momentum integration variable}
\definesymbol\qiipperp{q_{2\perp}'}{Transverse momentum integration variable}
\definesymbol\vphotonmass{M}{Virtual photon mass, $\sqrt{p_\Pphoton^\mu p_{\Pphoton\mu}}$}
\definesymbol\epsM{\epsilon_M}{Abbreviation for $\sqrt{1 - z}M$}
\definesymbol\dipoleampl{\mathcal{N}}{Color dipole scattering amplitude}
\definesymbol\Qs{Q_s}{saturation scale}
\definesymbol\xp{x_p}{Bj\"orken $x$ for projectile parton}
\definesymbol\xg{x_g}{Bj\"orken $x$ for target gluon}
\definesymbol\xplo{x_p^\text{LO}}{Bj\"orken $x$ for projectile parton in leading order kinematics}
\definesymbol\xglo{x_g^\text{LO}}{Bj\"orken $x$ for target gluon in leading order kinematics}
\definesymbol\xpnlo{x_p^\text{NLO}}{Bj\"orken $x$ for projectile parton in next-to-leading order kinematics}
\definesymbol\xgnlo{x_g^\text{NLO}}{Bj\"orken $x$ for target gluon in next-to-leading order kinematics}
\definesymbol\z{z}{momentum fraction for fragmentation}
\definesymbol\xisym{\xi}{kinematic variable} 
\definesymbol\inty{y}{transformed integration variable} 
\definesymbol\dipoleG{G}{unintegrated(?) momentum-space dipole gluon distribution}
\definesymbol\dipolephi{\phi}{unintegrated momentum-space dipole gluon distribution}
\definesymbol\dipoleN{\mathcal{N}}{dipole cross section}
\definesymbol\dipoleS{S_{x_g}^{(2)}}{unintegrated dipole gluon distribution}
\definesymbol\tripoleS{S_{x_g}^{(3)}}{unintegrated tripole gluon distribution}
\definesymbol\quadrupoleS{S_{x_g}^{(4)}}{unintegrated quadrupole gluon distribution}
\definesymbol\sextupoleS{S_{x_g}^{(6)}}{unintegrated sextupole gluon distribution}
\definesymbol\dipoleF{\mathcal{F}_{x_g}}{Fourier transform of unintegrated dipole gluon distribution}
\definesymbol\adjointF{\tilde{\mathcal{F}}_{x_g}}{Fourier transform of unintegrated adjoint dipole gluon distribution}
\definesymbol\quadrupoleG{\mathcal{G}_{x_g}}{Fourier transform of unintegrated quadrupole gluon distribution}
\definesymbol\hardfactor{\mathcal{H}}{hard factor}
\definesymbol\mandelstams{s}{total center-of-mass energy squared (Mandelstam variable)}
\definesymbol\mandelstamt{t}{momentum transfer squared (Mandelstam variable)} 
\definesymbol\mandelstamu{u}{crossed momentum transfer squared (Mandelstam variable)}
\definesymbol\partonmandelstams{\hat{s}}{squared center-of-mass energy of partons}
\definesymbol\partonmandelstamt{\hat{t}}{parton-level squared momentum transfer}
\definesymbol\partonmandelstamu{\hat{u}}{parton-level crossed squared momentum transfer}
\definesymbol\SUIII{\mathrm{SU}(3)}{3D special unitary group}
\definesymbol\suiii{\mathfrak{su}(3)}{Lie algebra of the 3D special unitary group}
\definesymbol\Nc{N_c}{number of colors}
\definesymbol\CF{C_F}{color factor}
\definesymbol\Nf{N_f}{number of flavors}
\definesymbol\TR{T_R}{group constant}
\definesymbol\xperp{x_\perp}{transverse position of one pole of a color dipole}
\definesymbol\yperp{y_\perp}{transverse position of the other pole of a color dipole}
\definesymbol\bperp{b_\perp}{transverse position of an emitted color dipole}
\definesymbol\xpperp{x_\perp'}{transverse position of a color dipole}
\definesymbol\ypperp{y_\perp'}{transverse position of a color dipole}
\definesymbol\bpperp{b_\perp'}{transverse position of a color dipole}
\definesymbol\Rperp{R_\perp}{impact parameter}
\definesymbol\rperp{r_\perp}{transverse extent of an original color dipole}
\definesymbol\sperp{s_\perp}{transverse extent of an emitted color dipole}
\definesymbol\tperp{t_\perp}{transverse extent of another emitted color dipole}
\definesymbol\rpperp{r_\perp'}{transverse position coordinate}
\definesymbol\spperp{s_\perp'}{transverse position coordinate}
\definesymbol\tpperp{t_\perp'}{transverse position coordinate}
\definesymbol\uperp{u_\perp}{transverse position coordinate}
\definesymbol\vperp{v_\perp}{transverse position coordinate}
\definesymbol\upperp{u_\perp'}{transverse position coordinate}
\definesymbol\vpperp{v_\perp'}{transverse position coordinate}
\definesymbol\gs{g_\text{s}}{strong coupling}
\definesymbol\alphas{\alpha_\text{s}}{strong coupling}
\definesymbol\alphaem{\alpha_\text{em}}{electromagnetic coupling}
\definesymbol\eulergamma{\gamma_E}{Euler's constant}
\definesymbol\czero{c_0}{$e^{-2\gamma_E}$}
\definesymbol\genericS{\mathcal{S}}{a generic function of the gluon distribution}
\definesymbol\structureone{F_1}{first structure function}
\definesymbol\structuretwo{F_2}{second structure function}
\definesymbol\structureF{F}{structure function}
\definesymbol\msbar{\overline{\text{MS}}}{MS-bar renormalization scheme}
\definesymbol\Ncoll{N_\text{coll}}{number of individual nucleon-nucleon collisions in a heavy ion collision}
\definesymbol\phasespace{\mathcal{PS}}{phase space}
\definesymbol\gaugevector{\eta}{vector that fixes the choice of light cone gauge}
\definesymbol\gluonfield{A}{gluon field}
\definesymbol\generator{T}{generator of a group}
\definesymbol\eformfac{G_E}{electric form factor}
\definesymbol\intpdf{f_{i/h}}{integrated parton distribution}
\definesymbol\intg{g}{integrated gluon distribution}
\renewcommand\digamma{\psi}
\newcommand\elementmatrix{E}
\newcommand\h[4]{\hardfactor^{(#1)}_{#2#3#4}}
\newcommand\hampl[4]{\expandafter\hat\h #1#2#3#4}
\newcommand\hfampl[4]{\expandafter\tilde\h #1#2#3#4}
\newcommand\F[5]{F^{(#2)}_{#1#3#4#5}}
\makeatletter
\def\Fs{\@ifstar\Fssym\Fsterm}
\newcommand\Fssym{F_s}
\newcommand\Fsterm[4]{\F{s}{#1}{#2}{#3}{#4}}
\def\Fn{\@ifstar\Fnsym\Fnterm}
\newcommand\Fnsym{F_n}
\newcommand\Fnterm[4]{\F{n}{#1}{#2}{#3}{#4}}
\def\Fd{\@ifstar\Fdsym\Fdterm}
\newcommand\Fdsym{F_d}
\newcommand\Fdterm[4]{\F{d}{#1}{#2}{#3}{#4}}
\def\xsec{\@ifstar\xsecsym\xsecterm}
\newcommand\xsecsym{\frac{\udddc\sigma}{\udc Y\uddc\vec p_\perp}}
\newcommand\xsecterm[4]{\biggl(\frac{\udddc\sigma}{\udc Y\uddc\vec p_\perp}\biggr)_{#2#3#4}^{(#1)}}
\makeatother
\newcommand\cgcavg[1]{\braket{#1}_{\xg}}

\newcommand\defn{\coloneqq} 
\newcommand\pp{\ensuremath{\mathrm{pp}}}
\newcommand\pA{\ensuremath{\mathrm{p}A}}

\newcommand\dAu{\ensuremath{\mathrm{dAu}}}
\newcommand\pPb{\ensuremath{\mathrm{pPb}}}

\definesymbol\besselJzero{J_0}{Bessel function}
\definesymbol\besselKzero{K_0}{Bessel function}
\definesymbol\besselKone{K_1}{Bessel function}
\newmuskip\pFqskip
\mathchardef\pFcomma=\mathcode`, 

\DeclareRobustCommand{\Pnucleus}{\HepGenParticle{A}{}{}\xspace}
\DeclareRobustCommand{\Panything}{\HepGenParticle{X}{}{}\xspace}
\DeclareRobustCommand{\Phadron}{\HepGenParticle{h}{}{}\xspace}
\DeclareRobustCommand{\Pdeuteron}{\HepGenParticle{d}{}{}\xspace}

\newcommand{\DYcgcp}[1][]{k_{\textmd{\Pgluon}#1}}
\newcommand{\DYcgcpperp}{\DYcgcp[\perp]}
\newcommand{\DYtotalp}[1][]{q_{#1}}
\newcommand{\DYtotalpperp}{\DYtotalp[\perp]}
\newcommand{\DYphotonp}[1][]{p_{\textmd{\Pphoton}#1}}
\newcommand{\DYphotonpperp}{\DYphotonp[\perp]}
\newcommand{\DYquarkp}[1][]{k_{\textmd{\Pquark}#1}}
\newcommand{\DYquarkpperp}{\DYquarkp[\perp]}
\newcommand{\DYpionp}[1][]{p_{\textmd{\Ppi}#1}}
\newcommand{\DYpionpperp}{\DYpionp[\perp]}
\newcommand{\DYprotonp}[1][]{p_{\textmd{\Pproton}#1}}
\newcommand{\DYnucleusp}[1][]{p_{\textmd{\Pnucleus}#1}}
\newcommand{\DYanythingp}[1][]{p_{\textmd{\Panything}#1}}
\newcommand{\DYphotonpfrac}{z_{\textmd{\Pphoton}}}
\newcommand{\DYpionpfrac}{z_{\textmd{\Ppi}}}
\newcommand{\DYfragmentationfrac}{z_2}
\newcommand{\DYphotontotalpfrac}{z}

\newcommand{\xsecquarkp}[1][]{k_{\textmd{\Pquark}#1}}
\newcommand{\xsecquarkpperp}{\xsecquarkp[\perp]}
\newcommand{\xseccgcp}[1][]{q_{#1}}
\newcommand{\xseccgcpperp}{\xseccgcp[\perp]}
\newcommand{\xsecgluonp}[1][]{k_{\textmd{\Pgluon}#1}}
\newcommand{\xsecgluonpperp}{\xsecgluonp[\perp]}
\newcommand{\xseciquarkp}[1][]{p_{\textmd{\Pquark}#1}}

\newcommand{\xsecprotonp}[1][]{p_{\textmd{\Pproton}#1}}
\newcommand{\xsecnucleusp}[1][]{p_{\textmd{\Pnucleus}#1}}
\newcommand{\xsectotalp}[1][]{k_{#1}'}
\newcommand{\xsectotalpperp}[1][]{k_{#1\perp}'}
\newcommand{\xsechadronp}[1][]{p_{#1}}
\newcommand{\xsechadronpperp}{\xsechadronp[\perp]}
\newcommand{\xsecanythingp}[1][]{p_{\textmd{\Panything}#1}}
\newcommand{\xsecquarkpfrac}{\xi}

%% file: graphicscommands.tex
\usepackage{epstopdf}
\usetikzlibrary{calc}
\usetikzlibrary{decorations.pathmorphing}
\usetikzlibrary{decorations.markings}
\usetikzlibrary{external}
\usetikzlibrary{fadings}
\usetikzlibrary{fit}
\usetikzlibrary{intersections}
\usetikzlibrary{patterns}
\usetikzlibrary{shadings}
\usetikzlibrary{shadows}
\usetikzlibrary{shapes}
\usetikzlibrary{shapes.geometric}

\usepgfplotslibrary{groupplots}
\usepgfplotslibrary{fillbetween}
\usepgfplotslibrary{units}

\tikzsetexternalprefix{generated-figures/}
\tikzset{
 /pgf/number format/sci generic={
  mantissa sep=\times,
  exponent={10^{#1}}
 }
}
\pgfdeclarelayer{background}
\pgfdeclarelayer{foreground}
\pgfsetlayers{background,main,foreground}
\usepackage{drawproton}
\colorlet{protonbg}{gray!30!white}

\colorlet{partondist}{green!60!black}
\colorlet{gluondist}{red!80!black}
\colorlet{fragmentation}{cyan!60!black}
\colorlet{kinfactor}{blue!40!red!70!black}

\tikzset{higgs/.style={
 dashed}}
\tikzset{quark/.style={
 postaction={decorate},decoration={markings,mark=at position .55 with {\arrow[
  ]{>}}}}}
\tikzset{antiquark/.style={
 postaction={decorate},decoration={markings,mark=at position .55 with {\arrow[
  ]{<}}}}}
\tikzset{lepton/.style={
 postaction={decorate},decoration={markings,mark=at position .55 with {\arrow[
  ]{>}}}}}
\tikzset{antilepton/.style={
 postaction={decorate},decoration={markings,mark=at position .55 with {\arrow[
  ]{<}}}}}
\tikzset{gluon/.style={decorate, 
 decoration={coil,amplitude=3pt,segment length=5pt}}}
\tikzset{small gluon/.style={decorate, 
 decoration={coil,amplitude=2pt,segment length=2pt}}}
\tikzset{tiny gluon/.style={decorate, 
 decoration={coil,amplitude=0.6pt,segment length=1pt}}}
\tikzset{photon/.style={decorate, 
 decoration={snake,amplitude=2pt,segment length=5pt}}}
\tikzset{weak/.style={decorate, 
 decoration={zigzag,amplitude=3pt,segment length=5pt}}}
\tikzset{specquark/.style={quark
}}
\tikzset{baryon/.style={quark,preaction={draw=blue,line width=10pt,shorten >=0.3pt,shorten <=0.3pt},preaction={draw=white,line width=9pt}}}

\newlength\baryonlinespacing
\tikzset{baryon/.style={to path={-- (\tikztotarget) \tikztonodes},
  execute at begin to={
   \draw ($(\tikztostart)!\baryonlinespacing!90:(\tikztotarget)$)--($(\tikztotarget)!\baryonlinespacing!-90:(\tikztostart)$);
   \draw[quark] (\tikztostart)--(\tikztotarget);
   \draw ($(\tikztostart)!\baryonlinespacing!-90:(\tikztotarget)$)--($(\tikztotarget)!\baryonlinespacing!90:(\tikztostart)$);
  }}
 }
\tikzset{right half baryon/.style={to path={-- (\tikztotarget) \tikztonodes},
  execute at begin to={
   \draw[quark] (\tikztostart)--(\tikztotarget);
   \draw ($(\tikztostart)!\baryonlinespacing!-90:(\tikztotarget)$)--($(\tikztotarget)!\baryonlinespacing!90:(\tikztostart)$);
  }}
 }
\tikzset{left half baryon/.style={to path={-- (\tikztotarget) \tikztonodes},
  execute at begin to={
   \draw ($(\tikztostart)!\baryonlinespacing!90:(\tikztotarget)$)--($(\tikztotarget)!\baryonlinespacing!-90:(\tikztostart)$);
   \draw[quark] (\tikztostart)--(\tikztotarget);
  }}
 }
\tikzset{nucleus/.style={to path={-- (\tikztotarget) \tikztonodes},
  execute at begin to={
   \draw ($(\tikztostart)!1.5\baryonlinespacing!90:(\tikztotarget)$)--($(\tikztotarget)!1.5\baryonlinespacing!-90:(\tikztostart)$);
   \draw ($(\tikztostart)!1.2\baryonlinespacing!90:(\tikztotarget)$)--($(\tikztotarget)!1.2\baryonlinespacing!-90:(\tikztostart)$);
   \draw ($(\tikztostart)!0.9\baryonlinespacing!90:(\tikztotarget)$)--($(\tikztotarget)!0.9\baryonlinespacing!-90:(\tikztostart)$);
   \draw ($(\tikztostart)!0.3\baryonlinespacing!90:(\tikztotarget)$)--($(\tikztotarget)!0.3\baryonlinespacing!-90:(\tikztostart)$);
   \draw[quark] (\tikztostart)--(\tikztotarget);
   \draw ($(\tikztostart)!0.3\baryonlinespacing!-90:(\tikztotarget)$)--($(\tikztotarget)!0.3\baryonlinespacing!90:(\tikztostart)$);
   \draw ($(\tikztostart)!0.9\baryonlinespacing!-90:(\tikztotarget)$)--($(\tikztotarget)!0.9\baryonlinespacing!90:(\tikztostart)$);
   \draw ($(\tikztostart)!1.2\baryonlinespacing!-90:(\tikztotarget)$)--($(\tikztotarget)!1.2\baryonlinespacing!90:(\tikztostart)$);
   \draw ($(\tikztostart)!1.5\baryonlinespacing!-90:(\tikztotarget)$)--($(\tikztotarget)!1.5\baryonlinespacing!90:(\tikztostart)$);
  }}
 }
\tikzset{blob/.style={ellipse,minimum height=0.8cm,minimum width=0.3cm}}
\tikzset{interaction/.style={ellipse,minimum height=0.3cm,minimum width=0.3cm}}

\tikzset{xq2shading/.style={rounded corners=5pt,drop shadow,preaction={fill=white,draw=black,line width=0.2pt},opacity=0.15,top color=red!70!magenta,bottom color=cyan,middle color=red!70!magenta!50!cyan!30!white,shading angle=45}}
\tikzset{plot grid/.style={help lines,opacity=0.5}}
\tikzset{axis/.style={help lines,->}}
\tikzset{axis2/.style={axis,<->}}

\pgfplotsset{compat=newest}
\pgfplotsset{
 exp yticklabels off/.style={
  yticklabel={
   \pgfmathfloatparsenumber{\tick}
   \pgfmathfloatexp{\pgfmathresult}
   \pgfmathprintnumber{\pgfmathresult}
  }
 },
 tick scale binop=\times,
 width=7cm
}
\makeatletter
\def\parsenode[#1]#2\pgf@nil{%
    \tikzset{label node/.style={#1}}
    \def\nodetext{#2}
}

\tikzset{
    add node at x/.style 2 args={
        name path global=plot line,
        /pgfplots/execute at end plot visualization/.append={
                \begingroup
                \@ifnextchar[{\parsenode}{\parsenode[]}#2\pgf@nil
            \path [name path global = position line #1-1]
                ({axis cs:#1,0}|-{rel axis cs:0,0}) --
                ({axis cs:#1,0}|-{rel axis cs:0,1});
            \path [xshift=1pt, name path global = position line #1-2]
                ({axis cs:#1,0}|-{rel axis cs:0,0}) --
                ({axis cs:#1,0}|-{rel axis cs:0,1});
            \path [
                name intersections={
                    of={plot line and position line #1-1},
                    name=left intersection
                },
                name intersections={
                    of={plot line and position line #1-2},
                    name=right intersection
                },
                label node/.append style={pos=1}
            ] (left intersection-1) -- (right intersection-1)
            node [label node]{\nodetext};
            \endgroup
        }
    },
    add node at y/.style 2 args={
        name path global=plot line,
        /pgfplots/execute at end plot visualization/.append={
                \begingroup
                \@ifnextchar[{\parsenode}{\parsenode[]}#2\pgf@nil
            \path [name path global = position line #1-1]
                ({axis cs:0,#1}-|{rel axis cs:0,0}) --
                ({axis cs:0,#1}-|{rel axis cs:1,1});
            \path [yshift=1pt, name path global = position line #1-2]
                ({axis cs:0,#1}-|{rel axis cs:0,0}) --
                ({axis cs:0,#1}-|{rel axis cs:1,1});
            \path [
                name intersections={
                    of={plot line and position line #1-1},
                    name=left intersection
                },
                name intersections={
                    of={plot line and position line #1-2},
                    name=right intersection
                },
                label node/.append style={pos=1}
            ] (left intersection-1) -- (right intersection-1)
            node [label node] {\nodetext};
            \endgroup
        }
    }
}
\makeatother
\tikzset{
 inline label at y/.style 2 args={
  add node at y={#1}{[fill=white,font=\tiny,sloped]#2}
 }
}

\pgfplotsset{
 use lhcf plot range/.style={
  ymax=1.3,
  ymin=1e-4,
  xmin=0,
  xmax=0.6
 },
 result axis/.style={
  ymin=1e-6,
  xlabel={$p_\perp [\si{GeV}]$},
  ylabel={$\frac{\udddc N}{\udc \eta\uddc p_\perp} \bigl[\si{GeV^{-2}}\bigr]$},
  legend style={cells={anchor=west}},
  legend columns=1,
  y filter/.code={\ifx\pgfmathresult\empty\def\pgfmathresult{-40}\fi}
 },
 data plot/.style={
  black,line width={0.05pt},mark=*,mark size={0.9pt},error bars/y dir=both,error bars/y explicit
 }
}

\pgfplotsset{
 result plot marks/.style={
  /tikz/mark=x,
  /tikz/mark size={0.5pt},
  /tikz/mark options={fill},
 },
 LO/.style={
  /pgfplots/result plot fill/.style={
   /tikz/draw=none,
   /tikz/preaction={
    /tikz/fill=#1,
    /tikz/fill opacity=0.2
   },
   /tikz/pattern=crosshatch,
   /tikz/pattern color=#1,
   /pgfplots/area legend
  },
  /pgfplots/result plot line/.style={
   /tikz/draw=#1,
   /tikz/mark=x,
   /tikz/mark size={0.5pt},
   /tikz/mark options={fill=#1},
   /pgfplots/forget plot
  }
 },
 NLO/.style={
  /pgfplots/result plot fill/.style={
   /tikz/draw=none,
   /tikz/fill=#1,
   /tikz/fill opacity=0.6,
   /pgfplots/area legend
  },
  /pgfplots/result plot line/.style={
   /tikz/draw=#1,
   /tikz/mark=x,
   /tikz/mark size={0.5pt},
   /tikz/mark options={fill=#1},
   /pgfplots/forget plot
  }
 },
 exact/.style={
  /pgfplots/result plot fill/.style={
   /tikz/draw=none,
   /tikz/preaction={
    /tikz/fill=#1,
    /tikz/fill opacity=0.2
   },
   /tikz/pattern=grid,
   /tikz/pattern color=#1,
   /pgfplots/area legend
  },
  /pgfplots/result plot line/.style={
   /tikz/draw=#1,
   /tikz/mark=x,
   /tikz/mark size={0.5pt},
   /tikz/mark options={fill=#1},
   /pgfplots/forget plot
  }
 },
 resummed/.style={
  /pgfplots/result plot fill/.style={
   /tikz/draw=none,
   /tikz/preaction={
    /tikz/fill=#1,
    /tikz/fill opacity=0.2
   },
   /tikz/pattern=crosshatch dots,
   /tikz/pattern color=#1,
   /pgfplots/area legend
  },
  /pgfplots/result plot line/.style={
   /tikz/draw=#1,
   /tikz/mark=x,
   /tikz/mark size={0.5pt},
   /tikz/mark options={fill=#1},
   /pgfplots/forget plot
  }
 },
 rcBK GBW LO/.style={
  LO=cyan!50!blue
 },
 rcBK GBW NLO/.style={
  NLO=red
 },
 rcBK GBW exact/.style={
  exact=green!70!black
 },
 rcBK GBW resummed 1/.style={
  resummed=green!50!black!70!purple
 },
 rcBK GBW resummed 2/.style={
  resummed=orange!80!black
 },
 rcBK GBW resummed 3/.style={
  resummed=purple!70!white
 },
 BK LO/.style={
  LO=blue
 },
 BK NLO/.style={
  NLO=magenta
 },
 MV LO/.style={
  LO=green!50!black
 },
 MV NLO/.style={
  NLO=orange
 },
 GBW LO/.style={
  LO=green
 },
 GBW NLO/.style={
  NLO=brown
 }
}

\pgfplotsset{
 xsec LO/.default=1,
 xsec LO/.style={
  table/x=pT,
  table/y expr={#1*\thisrow{lo-mean}}
 },
 xsec LO plus sigma/.default=1,
 xsec LO plus sigma/.style={
  table/x=pT,
  table/y expr={#1*(\thisrow{lo-mean}+\thisrow{lo-stddev})}
 },
 xsec LO minus sigma/.default=1,
 xsec LO minus sigma/.style={
  table/x=pT,
  table/y expr={#1*(\thisrow{lo-mean}-\thisrow{lo-stddev})}
 },
 xsec LO upper error estimate/.default=1,
 xsec LO upper error estimate/.style={
  table/x=pT,
  table/y expr={#1*(\thisrow{lo-mean}+\thisrow{lo-errbound})}
 },
 xsec LO lower error estimate/.default=1,
 xsec LO lower error estimate/.style={
  table/x=pT,
  table/y expr={#1*(\thisrow{lo-mean}-\thisrow{lo-errbound})}
 },
 xsec NLO/.default=1,
 xsec NLO/.style={
  table/x=pT,
  table/y expr={#1*(\thisrow{lo-mean}+\coupling{\thisrow{mu2}}*\thisrow{nlo-mean})}
 },
 xsec NLO plus sigma/.default=1,
 xsec NLO plus sigma/.style={
  table/x=pT,
  table/y expr={#1*(\thisrow{lo-mean}+\thisrow{lo-stddev}+\coupling{\thisrow{mu2}}*(\thisrow{nlo-mean}+\thisrow{nlo-stddev}))}
 },
 xsec NLO minus sigma/.default=1,
 xsec NLO minus sigma/.style={
  table/x=pT,
  table/y expr={#1*(\thisrow{lo-mean}-\thisrow{lo-stddev}+\coupling{\thisrow{mu2}}*(\thisrow{nlo-mean}-\thisrow{nlo-stddev}))}
 },
 xsec NLO upper error estimate/.default=1,
 xsec NLO upper error estimate/.style={
  table/x=pT,
  table/y expr={#1*(\thisrow{lo-mean}+\thisrow{lo-errbound}+\coupling{\thisrow{mu2}}*(\thisrow{nlo-mean}+\thisrow{nlo-errbound}))}
 },
 xsec NLO lower error estimate/.default=1,
 xsec NLO lower error estimate/.style={
  table/x=pT,
  table/y expr={#1*(\thisrow{lo-mean}-\thisrow{lo-errbound}+\coupling{\thisrow{mu2}}*(\thisrow{nlo-mean}-\thisrow{nlo-errbound}))}
 },
 brahms xsec LO/.style={
  xsec LO=\hadronconversionfactor
 },
 brahms xsec NLO/.style={
  xsec NLO=\hadronconversionfactor
 }
}

\pgfplotsset{
 correlation graph/.style={
  xtick={0,1.5708,3.14159,4.7123889,6.2831853},
  xticklabels={$0$,$\frac{\pi}{2}$,$\pi$,$\frac{3\pi}{2}$,$2\pi$},
  legend style={font=\small},
  scaled ticks=false,
  /pgf/number format/fixed,
  /pgf/number format/precision=3,
  max space between ticks=60,
  try min ticks=3,
  domain=0:2*pi,
  xlabel=$\Delta\phi$,
  ylabel=$C^{\text{DY}}(\Delta\phi)$,
  ymin=0,
  mark size=0.6pt,
  mark options={mark=*}
 },
 ktdist correlation graph/.style={
  correlation graph,
  mark options=,
  no markers,
  stack plots=y,
  cycle list name=ktdist colors,
  axis on top,
  every axis plot/.append style={fill}
 },
 colormap={ktdist colors}{
  color=(blue)
  color=(violet)
  color=(magenta)
  color=(red)
  color=(orange)
  color=(yellow)
  color=(green)
  color=(cyan)
  color=(blue)
 }
}
\pgfplotscreateplotcyclelist{ktdist colors}{
 {violet},
 {magenta},
 {red},
 {orange},
 {yellow},
 {green},
 {cyan},
 {blue}
}
\newcommand\pgfplotsktdistlegend[1]{%
 \begin{axis}[
  title=,
  xlabel=,ylabel=,zlabel=,
  legend entries=,
  grid=none,
  ymin=0,ymax=1,xmin=-4,xmax=4,
  plot graphics/ymin=0,plot graphics/ymax=1,plot graphics/xmin=-4,plot graphics/xmax=4,
  enlargelimits=false,
  scale only axis,
  width=0.3cm,height=3cm,
  xticklabel pos=left,
  xtick={-4,-2,...,4},
  xticklabels={$\SI{e-4}{GeV}$,$\num{e-2}$,$\vphantom{\num{e0}}1$,$\num{e2}$,$\num{e4}$},
  ytick=\empty,
  view={-90}{90},
  samples=9,
  domain=-4:4,
  samples y=2,
  domain y=0:1,
  zmin=0,zmax=1,
  anchor=west,
  xshift=0.3cm,
  at={#1}
 ]
  \addplot3[surf,mark=none,shader=flat corner] expression {x};
 \end{axis}%
}

\pgfplotscreateplotcyclelist{rainbow}{red,orange,yellow!50!black,green!80!black,cyan,blue,purple}

\newcommand\setplotshift[2]{\expandafter\newcommand\csname plotshift-#1\endcsname{#2}}
\newcommand\plotshift[1]{\csname plotshift-#1\endcsname}

\newcommand\resultplot[4]{
 { 
  \expandafter\pgfplotstablevertcat\expandafter\himudata\csname #4himu\endcsname
  \expandafter\pgfplotstablevertcat\expandafter\lomudata\csname #4lomu\endcsname
  \addplot[#1,result plot line,name path=hi mu] table[x expr=#2,y expr=#3] {\himudata};
  \addplot[#1,result plot line,forget plot,name path=lo mu] table[x expr=#2,y expr=#3] {\lomudata};
  \addplot[#1,result plot fill] fill between[of=hi mu and lo mu];
 }
}

\newcommand\fixedcoupling[1]{1}
\newcommand\runningcoupling[1]{pi/(2.25*(ln(greater(#1,1)*#1+notgreater(#1,1)*1) - ln(0.0588)))/0.2}
\newcommand\coupling[1]{\runningcoupling{#1}}

\def\readdata#1#2#3#4{
 \def\filenamepattern##1{#1}
 \expandafter\pgfplotstablesort\csname #4himu\endcsname{\filenamepattern{#3}}
 \expandafter\pgfplotstablesort\csname #4lomu\endcsname{\filenamepattern{#2}}
}

 \readdata{datafiles/brahms-dAu-mu#1-GBW.Y32.summary.dat}{10}{50}{brahmsGBW}
 \readdata{datafiles/brahms-dAu-mu#1-MV24.Y32.summary.dat}{10}{50}{brahmsMV}
 \readdata{datafiles/brahms-dAu-mu#1-BK10.Y32.summary.dat}{10}{50}{brahmsBK}
 \readdata{datafiles/brahms-dAu-mu#1-rcBKL01.Y22.summary.dat}{10}{50}{brahmsrcBKYA}
 \readdata{datafiles/brahms-dAu-mu#1-rcBKL01.Y32.summary.dat}{10}{50}{brahmsrcBKYB}
 \readdata{datafiles/brahms-dAu-mu#1-rcBKL01-e.Y22.summary.dat}{10}{50}{brahmsrcBKYC}
 \readdata{datafiles/brahms-dAu-mu#1-rcBKL01-e.Y32.summary.dat}{10}{50}{brahmsrcBKYD}
 \readdata{datafiles/star-dAu-mu#1-rcBKL01.Y4.summary.dat}{10}{50}{starrcBK}
 \readdata{datafiles/star-dAu-mu#1-rcBKL01-e.Y4.summary.dat}{10}{50}{starrcBKe}
 \readdata{datafiles/brahms-pp-mu#1-rcBKL01pp.Y22.summary.dat}{10}{50}{brahmspprcBKYA}
 \readdata{datafiles/brahms-pp-mu#1-rcBKL01pp.Y32.summary.dat}{10}{50}{brahmspprcBKYB}
 \readdata{datafiles/star-pp-mu#1-rcBKL01pp.Y4.summary.dat}{10}{50}{starpprcBK}
 \readdata{datafiles/brahms-pp-mu#1-rcBKL01.Y22.summary.dat}{10}{50}{brahmsscaledpprcBKYA}
 \readdata{datafiles/brahms-pp-mu#1-rcBKL01.Y32.summary.dat}{10}{50}{brahmsscaledpprcBKYB}
 \readdata{datafiles/star-pp-mu#1-rcBKL01.Y4.summary.dat}{10}{50}{starscaledpprcBK}
 \readdata{datafiles/brahms-dAu-mu#1-rcBKL01-r3.Y22.summary.dat}{10}{50}{brahmsrcBKRYA}
 \readdata{datafiles/brahms-dAu-mu#1-rcBKL01-r3.Y32.summary.dat}{10}{50}{brahmsrcBKRYB}
 \readdata{datafiles/brahms-dAu-mu#1-rcBKL01-r6.Y22.summary.dat}{10}{50}{brahmsrcBKRYC}
 \readdata{datafiles/brahms-dAu-mu#1-rcBKL01-r6.Y32.summary.dat}{10}{50}{brahmsrcBKRYD}
 \readdata{datafiles/brahms-dAu-mu#1-rcBKL01-r10.Y22.summary.dat}{10}{50}{brahmsrcBKRYE}
 \readdata{datafiles/brahms-dAu-mu#1-rcBKL01-r10.Y32.summary.dat}{10}{50}{brahmsrcBKRYF}
 \readdata{datafiles/brahms-dAu-mu#1-rcBKL01-e.Y22.summary.dat}{10}{50}{brahmsrcBKeYA}
 \readdata{datafiles/brahms-dAu-mu#1-rcBKL01-e.Y32.summary.dat}{10}{50}{brahmsrcBKeYB}
 \readdata{datafiles/star-dAu-mu#1-rcBKL01-e.Y4.summary.dat}{10}{50}{starrcBKe}
 \pgfplotstablecreatecol[copy column from table={\brahmsscaledpprcBKYAlomu}{lo-mean}]{pp-lo-mean}{\brahmsrcBKYAlomu}
 \pgfplotstablecreatecol[copy column from table={\brahmsscaledpprcBKYAlomu}{nlo-mean}]{pp-nlo-mean}{\brahmsrcBKYAlomu}
 \pgfplotstablecreatecol[copy column from table={\brahmsscaledpprcBKYAhimu}{lo-mean}]{pp-lo-mean}{\brahmsrcBKYAhimu}
 \pgfplotstablecreatecol[copy column from table={\brahmsscaledpprcBKYAhimu}{nlo-mean}]{pp-nlo-mean}{\brahmsrcBKYAhimu}
 \pgfplotstablecreatecol[copy column from table={\brahmsscaledpprcBKYBlomu}{lo-mean}]{pp-lo-mean}{\brahmsrcBKYBlomu}
 \pgfplotstablecreatecol[copy column from table={\brahmsscaledpprcBKYBlomu}{nlo-mean}]{pp-nlo-mean}{\brahmsrcBKYBlomu}
 \pgfplotstablecreatecol[copy column from table={\brahmsscaledpprcBKYBhimu}{lo-mean}]{pp-lo-mean}{\brahmsrcBKYBhimu}
 \pgfplotstablecreatecol[copy column from table={\brahmsscaledpprcBKYBhimu}{nlo-mean}]{pp-nlo-mean}{\brahmsrcBKYBhimu}
 \pgfplotstablecreatecol[copy column from table={\starscaledpprcBKlomu}{lo-mean}]{pp-lo-mean}{\starrcBKlomu}
 \pgfplotstablecreatecol[copy column from table={\starscaledpprcBKlomu}{nlo-mean}]{pp-nlo-mean}{\starrcBKlomu}
 \pgfplotstablecreatecol[copy column from table={\starscaledpprcBKhimu}{lo-mean}]{pp-lo-mean}{\starrcBKhimu}
 \pgfplotstablecreatecol[copy column from table={\starscaledpprcBKhimu}{nlo-mean}]{pp-nlo-mean}{\starrcBKhimu}
 \pgfplotstablecreatecol[copy column from table={\brahmspprcBKYAlomu}{lo-mean}]{nofit-pp-lo-mean}{\brahmsrcBKYAlomu}
 \pgfplotstablecreatecol[copy column from table={\brahmspprcBKYAlomu}{nlo-mean}]{nofit-pp-nlo-mean}{\brahmsrcBKYAlomu}
 \pgfplotstablecreatecol[copy column from table={\brahmspprcBKYAhimu}{lo-mean}]{nofit-pp-lo-mean}{\brahmsrcBKYAhimu}
 \pgfplotstablecreatecol[copy column from table={\brahmspprcBKYAhimu}{nlo-mean}]{nofit-pp-nlo-mean}{\brahmsrcBKYAhimu}
 \pgfplotstablecreatecol[copy column from table={\brahmspprcBKYBlomu}{lo-mean}]{nofit-pp-lo-mean}{\brahmsrcBKYBlomu}
 \pgfplotstablecreatecol[copy column from table={\brahmspprcBKYBlomu}{nlo-mean}]{nofit-pp-nlo-mean}{\brahmsrcBKYBlomu}
 \pgfplotstablecreatecol[copy column from table={\brahmspprcBKYBhimu}{lo-mean}]{nofit-pp-lo-mean}{\brahmsrcBKYBhimu}
 \pgfplotstablecreatecol[copy column from table={\brahmspprcBKYBhimu}{nlo-mean}]{nofit-pp-nlo-mean}{\brahmsrcBKYBhimu}
 \pgfplotstablecreatecol[copy column from table={\starpprcBKlomu}{lo-mean}]{nofit-pp-lo-mean}{\starrcBKlomu}
 \pgfplotstablecreatecol[copy column from table={\starpprcBKlomu}{nlo-mean}]{nofit-pp-nlo-mean}{\starrcBKlomu}
 \pgfplotstablecreatecol[copy column from table={\starpprcBKhimu}{lo-mean}]{nofit-pp-lo-mean}{\starrcBKhimu}
 \pgfplotstablecreatecol[copy column from table={\starpprcBKhimu}{nlo-mean}]{nofit-pp-nlo-mean}{\starrcBKhimu}
 \newcommand\brahmsNcoll{7.2}
 \newcommand\starNcoll{7.5}
 \pgfplotstablesort[sort cmp={float <}]\brahmslomurcBKloYsep{datafiles/brahms-dAu-mu10-rcBKL01.Y22.separated.dat}
 \input{xsecdata.tex}
 \newcommand\hadronconversionfactor{1.3}
 \pgfplotstableread{datafiles/kovr-satscale22.dat}{\kovrsatscaleA}
 \pgfplotstableread{datafiles/kovr-satscale32.dat}{\kovrsatscaleB}
 \pgfplotstableread{datafiles/brahms-dAu-mu10-GBWx100.Y32.summary.dat}{\brahmsGBWxA}
 \pgfplotstableread{datafiles/brahms-dAu-mu10-GBWx200.Y32.summary.dat}{\brahmsGBWxB}
 \pgfplotstableread{datafiles/brahms-dAu-mu10-GBWx300.Y32.summary.dat}{\brahmsGBWxC}
 \pgfplotstableread{datafiles/brahms-dAu-mu10-GBWx400.Y32.summary.dat}{\brahmsGBWxD}
 \pgfplotstableread{datafiles/brahms-dAu-mu10-GBWx500.Y32.summary.dat}{\brahmsGBWxE}
 \pgfplotstableread{datafiles/brahms-dAu-mu10-GBWx600.Y32.summary.dat}{\brahmsGBWxF}
 \pgfplotstableread{datafiles/brahms-dAu-mu10-GBWc007.Y32.summary.dat}{\brahmsGBWcA}
 \pgfplotstableread{datafiles/brahms-dAu-mu10-GBWc01.Y32.summary.dat}{\brahmsGBWcB}
 \pgfplotstableread{datafiles/brahms-dAu-mu10-GBWc03.Y32.summary.dat}{\brahmsGBWcC}
 \pgfplotstableread{datafiles/brahms-dAu-mu10-GBWc05.Y32.summary.dat}{\brahmsGBWcD}
 \pgfplotstableread{datafiles/brahms-dAu-mu10-GBWc07.Y32.summary.dat}{\brahmsGBWcE}
 \pgfplotstableread{datafiles/brahms-dAu-mu10-GBWl001.Y32.summary.dat}{\brahmsGBWlA}
 \pgfplotstableread{datafiles/brahms-dAu-mu10-GBWl005.Y32.summary.dat}{\brahmsGBWlB}
 \pgfplotstableread{datafiles/brahms-dAu-mu10-GBWl010.Y32.summary.dat}{\brahmsGBWlC}
 \pgfplotstableread{datafiles/brahms-dAu-mu10-GBWl050.Y32.summary.dat}{\brahmsGBWlD}
 \pgfplotstableread{datafiles/brahms-dAu-mu10-GBWl100.Y32.summary.dat}{\brahmsGBWlE}
 \pgfplotstableread{datafiles/brahms-dAu-mu10-GBWl500.Y32.summary.dat}{\brahmsGBWlF}
 \pgfplotstableread{datafiles/brahms-dAu-mu10-GBWl900.Y32.summary.dat}{\brahmsGBWlG}
 \readdata{datafiles/brahms-dAu-mu#1-GBWpT075.Y32.summary.dat}{10}{50}{brahmsGBWpT075}
 \readdata{datafiles/brahms-dAu-mu#1-GBWpT129.Y32.summary.dat}{10}{50}{brahmsGBWpT129}
 \readdata{datafiles/brahms-dAu-mu#1-GBWpT169.Y32.summary.dat}{10}{50}{brahmsGBWpT169}
 \readdata{datafiles/brahms-dAu-mu#1-GBWpT255.Y32.summary.dat}{10}{50}{brahmsGBWpT255}
 \pgfplotstablesort[sort cmp={float <},sort key={mu2factor}]\rhicvsmurcBK{datafiles/rhicvsmu-dAu-rcBKL01.Y4.summary.dat}
 \pgfplotstablesort[sort cmp={float <},sort key={mu2factor}]\lhcvsmurcBKhiY{datafiles/lhcvsmu-pPb-rcBKL01.Y591.summary.dat}
 \pgfplotstableread{datafiles/pdfdataQ210.csv}\pdfdatatableA
 \pgfplotstableread{datafiles/pdfdataQ2100.csv}\pdfdatatableB

%% file: xsecdata.tex
\pgfplotstablesort[row sep=\\]\brahmsdAuloY{
pt          ptlo                pthi                yield       staterr         syserr         \\
0.65        0.6                 0.7                 0.388093    0.00243152      0.0582139      \\
0.75        0.7                 0.8                 0.241047    0.00225041      0.036157       \\
0.85        0.8                 0.9                 0.165855    0.000938941     0.0248783      \\
0.95        0.9                 1                   0.112226    0.00107049      0.0168339      \\
1.09        1                   1.2                 0.0611503   0.000486684     0.00917254     \\
1.29        1.2                 1.4                 0.0299976   0.000307453     0.00449965     \\
1.49        1.4                 1.6                 0.0136606   0.000190933     0.00204909     \\
1.69        1.6                 1.8                 0.0068023   0.000125613     0.00102034     \\
1.89        1.8                 2                   0.00399323  9.32567e-05     0.000598985    \\
2.13        2.0                 2.3                 0.00205668  5.16416e-05     0.000308502    \\
2.55        2.3                 2.9                 0.000565847 9.04178e-06     8.48771e-05    \\
3.33        2.9                 4.0                 9.39018e-05 6.84736e-06     1.40853e-05    \\
}
\pgfplotstablesort[row sep=\\]\brahmsdAuhiY{
pt          ptlo                pthi                yield       staterr         syserr         \\
0.25        0.2                 0.3                 2.01147     0.0258206       0.3017205      \\
0.35        0.3                 0.4                 1.45906     0.0174999       0.218859       \\
0.45        0.4                 0.5                 0.933742    0.0114546       0.1400613      \\
0.55        0.5                 0.6                 0.466025    0.00760817      0.06990375     \\
0.65        0.6                 0.7                 0.293151    0.00555747      0.04397265     \\
0.75        0.7                 0.8                 0.142542    0.00358952      0.0213813      \\
0.85        0.8                 0.9                 0.085915    0.00174568      0.01288725     \\
0.95        0.9                 1                   0.0559974   0.00123521      0.00839961     \\
1.09        1                   1.2                 0.0233523   0.000140574     0.00467046     \\
1.29        1.2                 1.4                 0.0104938   8.29051e-05     0.00209876     \\
1.49        1.4                 1.6                 0.00506653  4.97884e-05     0.001013306    \\
1.69        1.6                 1.8                 0.00246804  3.15367e-05     0.000493608    \\
1.89        1.8                 2                   0.00120714  1.93553e-05     0.000241428    \\
2.13        2                   2.3                 0.000459623 9.66696e-06     6.89434e-05    \\
2.55        2.3                 2.9                 0.000124681 3.42328e-06     1.870215e-05   \\
3.31        2.9                 4                   1.22812e-05 7.80291e-07     1.84218e-06    \\
}
\pgfplotstablesort[row sep=\\]\brahmspploY{
pt          ptlo                pthi                yield       staterr         syserr          \\
0.35        0.3                 0.4                 0.431377    0.080198        0.06470655      \\
0.45        0.4                 0.5                 0.30279     0.0456967       0.0454185       \\
0.55        0.5                 0.6                 0.155592    0.0288541       0.0233388       \\
0.65        0.6                 0.7                 0.126558    0.00338659      0.0189837       \\
0.75        0.7                 0.8                 0.0691939   0.00158536      0.010379085     \\
0.85        0.8                 0.9                 0.0451841   0.00102526      0.006777615     \\
0.95        0.9                 1                   0.0272665   0.000758827     0.004089975     \\
1.09        1                   1.2                 0.0140419   0.00033972      0.00280838      \\
1.29        1.2                 1.4                 0.00641209  0.000223705     0.001282418     \\
1.49        1.4                 1.6                 0.00277971  0.000125871     0.000555942     \\
1.69        1.6                 1.8                 0.00130404  8.47478e-05     0.000260808     \\
1.89        1.8                 2                   0.000692191 5.98538e-05     0.0001384382    \\
2.13        2                   2.3                 0.000269266 2.4238e-05      4.048899e-05    \\
2.55        2.3                 2.9                 9.03446e-05 8.10509e-06     1.355169e-05    \\
3.33        2.9                 4                   1.04874e-05 2.56088e-06     1.57311e-06     \\
}
\pgfplotstablesort[row sep=\\]\brahmspphiY{
pt          ptlo                pthi                yield       staterr         syserr          \\
0.25        0.2                 0.3                 0.920648    0.0087473       0.1380972       \\
0.35        0.3                 0.4                 0.615452    0.00589208      0.0923178       \\
0.45        0.4                 0.5                 0.347337    0.00361243      0.05210055      \\
0.55        0.5                 0.6                 0.176466    0.00238555      0.0264699       \\
0.65        0.6                 0.7                 0.101806    0.00165747      0.0152709       \\
0.75        0.7                 0.8                 0.0490655   0.00118409      0.007359825     \\
0.85        0.8                 0.9                 0.0288363   0.000511404     0.004325445     \\
0.95        0.9                 1                   0.0174729   0.000342913     0.002620935     \\
1.09        1                   1.2                 0.00755556  4.44288e-05     0.001511112     \\
1.29        1.2                 1.4                 0.00286486  2.39533e-05     0.000572972     \\
1.49        1.4                 1.6                 0.00134034  1.42094e-05     0.000268068     \\
1.69        1.6                 1.8                 0.000540773 8.23377e-06     0.0001081546    \\
1.89        1.8                 2                   0.000256595 4.92646e-06     5.1319e-05      \\
2.13        2                   2.3                 8.6293e-05  2.34913e-06     1.294395e-05    \\
2.55        2.3                 2.9                 2.30826e-05 7.97715e-07     3.46239e-06     \\
3.31        2.9                 4.                  2.09339e-06 1.81017e-07     3.14008e-07     \\
}
\pgfplotstablesort[row sep=\\]\brahmsRdAuloY{
pt          ptlo                pthi                RdAu        staterr         syserr          \\
0.65        0.6                 0.7                 0.425928    0.0117058       0.0638892       \\
0.75        0.7                 0.8                 0.483863    0.0119712       0.0725795       \\
0.85        0.8                 0.9                 0.509839    0.0119233       0.0764759       \\
0.95        0.9                 1                   0.571681    0.0168185       0.0857521       \\
1.1         1                   1.2                 0.604868    0.0154052       0.0907301       \\
1.3         1.2                 1.4                 0.649795    0.023628        0.0974693       \\
1.5         1.4                 1.6                 0.682589    0.032348        0.102388        \\
1.7         1.6                 1.8                 0.724524    0.0489497       0.108679        \\
1.9         1.8                 2                   0.801287    0.0717699       0.120193        \\
2.15        2                   2.3                 1.0609      0.0991426       0.159135        \\
2.6         2.3                 2.9                 0.869934    0.0792727       0.13049         \\
3.45        2.9                 4                   1.24364     0.316933        0.186546        \\
}
\pgfplotstablesort[row sep=\\]\brahmsRdAuhiY{
pt          ptlo                pthi                RdAu        staterr         syserr          \\
0.25        0.2                 0.3                 0.305851    0.00488135      0.04587765      \\
0.35        0.3                 0.4                 0.324309    0.00493462      0.04864635      \\
0.45        0.4                 0.5                 0.371094    0.00594363      0.0556641       \\
0.55        0.5                 0.6                 0.372553    0.00792005      0.05588295      \\
0.65        0.6                 0.7                 0.40528     0.010171        0.0607919       \\
0.75        0.7                 0.8                 0.382655    0.0129897       0.05739825      \\
0.85        0.8                 0.9                 0.425644    0.0115142       0.0638466       \\
0.95        0.9                 1                   0.460111    0.0136285       0.06901665      \\
1.1         1                   1.2                 0.444456    0.00377164      0.0666684       \\
1.3         1.2                 1.4                 0.528807    0.00615162      0.07932105      \\
1.5         1.4                 1.6                 0.537941    0.00778982      0.08069115      \\
1.7         1.6                 1.8                 0.653997    0.0130781       0.09809955      \\
1.9         1.8                 2                   0.700217    0.0179423       0.10503255      \\
2.15        2                   2.3                 0.8092      0.0282973       0.12138         \\
2.6         2.3                 2.9                 0.796981    0.0358501       0.11954715      \\
3.45        2.9                 4                   0.824708    0.0879926       0.1237062       \\
}
\newcommand\stardAusigmainel{2.21e6} 
\pgfplotstablesort[row sep=\\]\stardAu{
 epilo          epihi           epi         pt          xsec        staterr     syserr \\
 25             30              27.1        1.03        9900        100         1500   \\
 30             35              32.2        1.21        2750        40          290    \\
 35             40              37.2        1.38        970         20          120    \\
 40             45              42.1        1.55        315         11          48     \\
 45             50              47.1        1.72        110         6           25     \\
 50             55              52.0        1.91        46          4           10     \\
}
\newcommand\starppsigmainel{42e3} 
\pgfplotstablesort[row sep=\\]\starppA{
 epilo          epihi           epi         pt          xsec        staterr     syserr  \\
 30             35              32.2        2.34        0.571       0.018       0.057   \\
 35             40              37.1        2.74        0.141       0.008       0.016   \\
 40             45              42.3        3.01        0.0422      0.0043      0.0052  \\
 45             50              46.9        3.40        0.0134      0.0022      0.0035  \\
}
\pgfplotstablesort[row sep=\\]\starppB{
 epilo          epihi           epi         pt          xsec        staterr     syserr  \\
 30             35              32.2        1.48        10.8        0.1         2.4     \\
 35             40              37.1        1.68        3.68        0.06        0.66    \\
 40             45              42.1        1.88        1.00        0.03        0.16    \\
 45             50              47.1        2.08        0.342       0.015       0.066   \\
 50             55              51.9        2.25        0.084       0.006       0.021   \\
}
\pgfplotstablesort[row sep=\\]\starppC{
 epilo          epihi           epi         pt          xsec        staterr     syserr  \\
 25             30              27.2        1.01        77.9        0.9         6.0     \\
 30             35              32.1        1.19        25.7        0.4         2.2     \\
 35             40              37.1        1.37        7.92        0.10        0.74    \\
 40             45              42.1        1.56        2.63        0.05        0.29    \\
 45             50              47.1        1.73        0.90        0.03        0.11    \\
 50             55              52.0        1.94        0.302       0.025       0.049   \\
}
\pgfplotstablesort[row sep=\\]\starRdAu{
pt              RdAu    staterr     syserr  \\
1.03            0.322   0.005       0.056   \\
1.21            0.272   0.006       0.036   \\
1.38            0.311   0.008       0.048   \\
1.55            0.303   0.012       0.057   \\
1.72            0.309   0.019       0.081   \\
1.91            0.38    0.05        0.11    \\
}
\pgfplotstablesort[row sep=\\]\lhcfpPbIV{
 ptmin  ptmax   rate    loerr       hierr   \\
 0.10   0.15    1.17e-1 -7.83e-2    8.91e-2 \\
 0.15   0.20    1.00e-1 -7.75e-2    8.79e-2 \\
 0.20   0.25    4.90e-2 -5.57e-2    6.43e-2 \\
 0.25   0.30    5.37e-2 -3.18e-2    3.85e-2 \\
 0.30   0.35    3.09e-2 -1.69e-2    2.11e-2 \\
 0.35   0.40    1.76e-2 -1.02e-2    1.24e-2 \\
 0.40   0.45    1.82e-2 -9.42e-3    1.12e-2 \\
 0.45   0.50    6.77e-3 -4.49e-3    5.24e-3 \\
 0.50   0.60    2.78e-3 -1.93e-3    2.20e-3 \\
}
\pgfplotstablesort[row sep=\\]\lhcfpPbV{
 ptmin  ptmax   rate    loerr       hierr   \\
 0.10   0.15    9.06e-2 -6.05e-2    6.61e-2 \\
 0.15   0.20    7.71e-2 -4.92e-2    5.54e-2 \\
 0.20   0.25    6.28e-2 -3.23e-2    3.84e-2 \\
 0.25   0.30    3.83e-2 -1.58e-2    2.31e-2 \\
 0.30   0.35    2.81e-2 -1.13e-2    1.36e-2 \\
 0.35   0.40    1.82e-2 -7.76e-3    8.27e-3 \\
 0.40   0.45    7.98e-3 -4.49e-3    4.33e-3 \\
 0.45   0.50    6.37e-3 -3.43e-3    3.05e-3 \\
 0.50   0.60    1.41e-3 -8.01e-3    6.56e-4 \\
}
\pgfplotstablesort[row sep=\\]\lhcfpPbVI{
 ptmin  ptmax   rate    loerr       hierr   \\
 0.00   0.05    3.56e-2 -5.14e-2    5.46e-2 \\
 0.05   0.10    1.02e-1 -4.63e-2    4.99e-2 \\
 0.10   0.15    8.53e-2 -4.15e-2    4.49e-2 \\
 0.15   0.20    8.25e-2 -3.45e-2    3.67e-2 \\
 0.20   0.25    3.96e-2 -2.11e-2    2.83e-2 \\
 0.25   0.30    2.19e-2 -9.62e-3    1.26e-2 \\
 0.30   0.35    1.14e-2 -6.25e-3    6.81e-3 \\
 0.35   0.40    5.53e-3 -3.93e-3    3.75e-3 \\
 0.40   0.50    2.67e-3 -1.57e-3    1.33e-3 \\
 0.50   0.60    1.31e-4 -2.45e-4    2.36e-4 \\
}
\pgfplotstablesort[row sep=\\]\lhcfpPbVII{
 ptmin  ptmax   rate    loerr       hierr   \\
 0.00   0.05    2.67e-2 -4.42e-2    4.64e-2 \\
 0.05   0.10    7.42e-2 -4.11e-2    4.31e-2 \\
 0.10   0.15    5.87e-2 -4.08e-2    4.19e-2 \\
 0.15   0.20    4.93e-2 -2.26e-2    2.13e-2 \\
 0.20   0.25    2.32e-2 -1.17e-2    1.16e-2 \\
 0.25   0.30    1.47e-2 -8.10e-3    7.54e-3 \\
 0.30   0.35    8.37e-3 -5.73e-3    4.31e-3 \\
 0.35   0.40    3.47e-3 -2.68e-3    1.77e-3 \\
 0.40   0.50    5.32e-4 -4.86e-4    3.39e-4 \\
}
\pgfplotstablesort[row sep=\\]\lhcfpPbVIII{
 ptmin  ptmax   rate    loerr       hierr   \\
 0.00   0.05    1.72e-2 -2.93e-2    3.07e-2 \\
 0.05   0.10    3.93e-2 -2.69e-2    2.83e-2 \\
 0.10   0.15    2.67e-2 -1.72e-2    1.81e-2 \\
 0.15   0.20    2.00e-2 -8.16e-3    8.44e-3 \\
 0.20   0.25    1.13e-2 -5.12e-3    5.06e-3 \\
 0.25   0.30    3.83e-3 -2.61e-3    1.85e-3 \\
 0.30   0.40    5.56e-4 -5.35e-4    3.86e-4 \\
}
\pgfplotstablesort[row sep=\\]\lhcfpPbIX{
 ptmin  ptmax   rate    loerr       hierr   \\
 0.00   0.05   -6.79e-3 -7.16e-3    7.43e-3 \\
 0.05   0.10    6.12e-3 -4.74e-3    3.51e-3 \\
 0.10   0.15    3.51e-3 -2.66e-3    1.54e-3 \\
 0.15   0.20    2.01e-3 -1.39e-3    8.46e-4 \\
 0.20   0.30    2.36e-4 -2.00e-4    1.08e-4 \\
}

%% file: abstract.tex
Understanding the behavior of large atomic nuclei (heavy ions) in high-energy collisions has been the focus of a concerted research effort over the past 10-15 years. Much of the latest progress in the field has centered around transverse momentum-dependent (or “unintegrated”) parton distributions: in particular the prediction of the high-energy behavior of these distributions, in the form of the Balitsky-JIMWLK equations, and the development of the hybrid factorization framework, which connects the unintegrated parton distributions to predictions for experimentally measured cross sections. With the advent of high-energy proton-nucleus collisions at RHIC and the LHC, we are able to experimentally test these predictions for the first time. In this dissertation, I show two case studies of these predictions, to illustrate the use of the hybrid factorization at leading and next-to-leading order.

First, as a simple example, I analyze the azimuthal angular correlation for a Drell-Yan process, the production of a lepton pair with an associated hadron. The correlation for back-to-back emission turns out to be determined by the low-momentum region of the unintegrated gluon distribution, and the correlation for parallel emission is determined by the high-momentum region. Accordingly, a proper prediction of the correlation at all angles requires a gluon distribution with physically realistic behavior at both high and low momenta. Furthermore, the properties of the central double peak that emerges in Drell-Yan production can provide some insight into the form of the gluon distribution.

I’ll then describe a numerical calculation of the cross section for inclusive hadron production, which incorporates all corrections up to next-to-leading order in the strong coupling. This calculation illustrates several obstacles presented by subleading terms, including the removal of divergences by renormalizing the integrated and unintegrated parton distributions. The results of the calculation are negative at high transverse momentum, which is surprising but may be mathematically reasonable, since the perturbative approximation to the cross section may break down under those kinematic conditions. However, it may be possible to make meaningful predictions for the nuclear modification ratio $R_{\pA}$ despite the negative cross section.

Moving beyond next-to-leading order, it may be possible to cure the negativity of the inclusive hadron cross section by altering the formulas used. I’ll show two possible methods of doing so: first, a straightforward resummation of selected higher-order terms corresponding to gluon loop diagrams is able to mitigate the negativity, though it requires some alterations of unclear theoretical origin. A more promising alternative seems to be use of exact kinematic definitions, incorporating terms which disappear in the infinite-energy limit; this constrains the kinematics to eliminate the region of phase space which most strongly contributes to the negativity. In this way, the calculation can be adapted to produce reasonable results at high transverse momentum.

%% file: acknowledgements.tex
First and foremost, I would like to thank my adviser, Anna Sta\'sto, for her support and guidance throughout my Ph.D. studies and the preparation of this dissertation.
I'm also indebted to my collaborator Bo-Wen Xiao for much patience explaining the details of saturation physics to me, and for giving me the opportunity to continue my studies as a postdoc.

I would be remiss not to express my gratitude for the thought-provoking feedback I've received from, and enlightening discussions I've had with, many professional friends and colleagues. (Even when they may not have realized it!) Let me acknowledge Elke-Caroline Aschenauer, Emil Avsar, Jeff Berger, Christian Cruz-Santiago, John Collins, Kevin Dusling, Will Horowitz, Jamal Jalilian-Marian, Tuomas Lappi, Heikki M\"antysaari, Risto Paatelainen, Matt Sievert, Jorge Sofo, Mark Strikman, and Feng Yuan.

On many occasions I have received extremely helpful assistance from anonymous internet users, facilitated by the Stack Exchange network of question and answer websites, especially the \TeX-\LaTeX, Physics, and Mathematics sites.

Finally, I would like to offer a special thanks to my family --- Zabeda, Katherine, and Sarah --- and my friends, of whom there are far too many to list here, for the support they've offered during the time I've spent at Penn State. (However: an extra special thanks to Michael, Jenn, Kathryne, and Liz. You know what you did to get your names here!)

%% file: introduction.tex
\chapter{Introduction}

Ever since quantum chromodynamics, the theory of strong interactions, was developed in the 1960's, high-energy physicists have endeavored to understand the internal structure of strongly bound particles.
This effort was driven, in large part, by experimental results from increasingly powerful particle accelerators.
But as accelerator technology has become more and more complex, so has the depth of structure displayed by the particles being collided.
Instead of having a fixed set of constituents, as one might classically expect, protons and atomic nuclei act as though they contain ever-increasing numbers of smaller and smaller particles as we probe them at higher and higher energies.

Physicists traditionally characterize this apparently changing composition using a set of \term[parton distribution]{parton distributions}, functions which characterize how the momentum is distributed among the constituents (partons) of the strongly bound particle.
Although these functions emerge from fundamentally nonperturbative physics, it is possible to describe how they vary with the parameters of the collision using the methods of perturbative QCD.
At high energies, which typically correlate to small momentum fractions $x$, a renormalization group equation derived by Balitsky, Fadin, Kuraev, and Lipatov~\cite{Fadin:1975cb,Lipatov:1976zz,Kuraev:1976ge,Kuraev:1977fs,Balitsky:1978ic} governs the evolution of the parton distributions as $x$ changes.
Physically, this equation, called the BFKL evolution equation, represents the process of parton branching, in which partons effectively divide into other partons with smaller fractions of the original hadron's momentum.
The BFKL equation predicts a rapid rise in the parton distributions, especially the gluon distribution, as $x$ becomes small.

However, this growth in the gluon distribution is not sustainable.
At very low values of $x$ it leads to violations of the Froissart bound~\cite{Froissart:1961ux,Lukaszuk:1967zz,MartinUnitarity}, a fundamental limit on the growth of cross sections with energy.
Violation of the Froissart bound is equivalent to breaking unitarity --- that is, to nonconservation of probability.
In the 1990's, Balitsky and Kovchegov~\cite{Balitsky:1995ub,Balitsky:1998kc,Balitsky:1998ya,Balitsky:2001re,Kovchegov:1999yj,Kovchegov:1999ua} investigated the dynamics of the gluon distribution as it approaches these very low values of $x$, finding an important contribution from gluon recombination, the inverse of the branching process that underlies BFKL evolution.
They derived a correction term to the BFKL equation that represents this recombination, a term which restricts the growth of the gluon distribution at small $x$, in a phenomenon known as \iterm{gluon saturation}.
Accelerator technology has only recently advanced to the point where we may be able to experimentally detect the effects of gluon saturation.
Naturally, one of the exciting topics in the world of modern particle physics is the search for these effects.

Detecting experimental signatures of saturation generally proceeds along these lines:
\begin{enumerate}
 \item Develop a theoretical model for the gluon distribution which exhibits saturation
 \item Insert that model into a formula which relates the gluon distribution to some measurable quantity, and calculate results for various parameters
 \item Identify features in the results which correspond to characteristics of the gluon distribution that identify saturation
 \item Compare the results to corresponding experimental data points to see if those features are present
\end{enumerate}
For the process to be successful, the formula relating the gluon distribution to a measurable quantity needs to be sensitive to small changes in the gluon distribution in the region of interest, and it also needs to be precise in the sense that it does not introduce significant theoretical uncertainties.
There is an ongoing effort in the saturation physics community to find measurements which satisfy these criteria.

In this dissertation, I'll describe two such measurements.
First, I'll examine the derivation of, and predictions for, the azimuthal angular correlation in Drell-Yan lepton pair-hadron production.
This quantity is essentially the cross section for the process \HepProcess{\Pproton\Pnucleus \HepTo \Plepton\APlepton\Phadron\Panything}, expressed as a function of the angular separation between two outgoing momenta: the total momentum of the lepton pair \HepProcess{\Plepton\APlepton}, and the momentum of the hadron \HepProcess{\Phadron}.
We'll see that the correlation exhibits characteristic features which are sensitive to details of the gluon distribution in the saturation regime, including a double peak for back-to-back emission.
Furthermore, examining the angular correlation in the back-to-back emission region allows us to nearly isolate the effect of the low-momentum part of the gluon distribution, and the correlation in the near-parallel emission region gives us a window on the high-momentum gluon distribution.

Afterwards, I'll describe the calculation of the inclusive hadron production cross section, \HepProcess{\Pproton\Pnucleus \HepTo \Phadron\Panything}, in the saturation formalism, with the complete next-to-leading order (NLO) corrections (except next-to-leading corrections to BK evolution).
Unlike the angular correlation calculation, which only incorporates leading order terms corresponding to tree-level diagrams, the NLO inclusive hadron production calculation involves loop diagrams, which come with their own set of theoretical challenges.
In particular, the formulas contain divergences in rapidity and momentum which need to be canceled by subtractions, and then analytical singularities need to be removed to produce formulas which can be evaluated numerically.
Although the inclusive hadron production cross section does not separate the low- and high-momentum regions of the gluon distribution as well as the correlation does, it does provide a very precise test since the NLO corrections act to reduce the theoretical uncertainty in the formula.

However, the LO+NLO calculation for the inclusive hadron cross section is only applicable in a limited kinematic regime.
We'll see that it produces negative results at high transverse momenta, so evidently the formula is not useful as a probe of the gluon distribution in those conditions.
This implies that further corrections to the formula are necessary to use the high-momentum cross section to accurately probe the gluon distribution.
I'll finish by discussing some of the first work toward that goal, which examines two types of higher-order corrections: a simple resummation of the most prominent contributions of higher-order loop diagrams, and an exact kinematic formula that matches the collinear factorization result which is known to be positive.
The exact kinematic formula, in particular, is quite successful at describing the experimental data at high transverse momentum where the simple LO+NLO calculation fails.

At each step, I'll present the theoretical analysis along with comparisons to experimental data.
Currently, the state of the art in experimental data for small-$x$ physics comes from the \term{BRAHMS} and \term{STAR} detectors at the Relativistic Heavy Ion Collider (\term{RHIC}).
These detectors share several common characteristics that make them suitable for the studies I will describe in this dissertation: they are capable of measuring particles emitted at forward angles close to the beamline, equivalent to high pseudorapidity $\pseudorapidity \sim 3-4$, and they have good resolution when measuring the momentum of emitted pions and other light hadrons at such forward pseudorapidities.
There are also detectors at the Large Hadron Collider (LHC) with these capabilities, namely \term{ALICE}, \term{ATLAS}, \term{CMS}, and \term{LHCf}, although most of the relevant experimental results from these detectors have not yet been published, so when comparing to data I will concentrate on the \term{BRAHMS} and \term{STAR} results.

The body of the dissertation is structured as follows.
Chapter~\ref{ch:saturation} lays out the basic principles of small-$x$ and saturation physics, together with its underlying background in quantum chromodynamics, and then chapter~\ref{ch:cgc} addresses specific concepts of saturation physics in more detail, in particular the color glass condensate model which describes target nucleons and nuclei under saturation conditions.
The remaining chapters describe the calculations previously mentioned.
Chapter~\ref{ch:correlation} shows the Drell-Yan lepton pair-hadron angular correlation, which illustrates the basic principles of calculations in small-$x$ physics.
Then, chapter~\ref{ch:crosssection} presents the cross section for single inclusive pion production incorporating \term{next-to-leading order} corrections, and describes some of the difficulties inherent in treating the higher-order contributions.
Chapter~\ref{ch:results} shows the numerical results of this calculation, and then chapter~\ref{ch:beyondnlo} discusses how additional contributions might be used to remedy some of the shortcomings of the LO+NLO formula.
Finally, chapter~\ref{ch:conclusion} summarizes the results and presents an outlook toward further development of these calculations.

%% file: structure.tex
\chapter{Hadron Structure From QCD}\label{ch:saturation}

\term{Quantum chromodynamics} (QCD), the theory that describes the behavior of strongly interacting particles, was first developed in the 1970s by physicists exploring generalizations of \term{quantum electrodynamics} (the quantum theory of electromagnetism).
Within a few short years, it became clear that constituents of the proton with small fractions $x$ of its momentum would play an increasingly important role in the physics as collision energy increased.
This chapter lays the groundwork for the discussion of small-$x$ physics in chapter~\ref{ch:cgc} by motivating the idea of a target hadron or nucleus as a dressed bound state of quarks, and introducing the integrated and unintegrated parton distributions used to describe it.

\section{Quark and Gluon Fields} 

QCD is based on the $\SUIII$ transformation group, whose \term{fundamental representation} is the set of all matrices $U$ that satisfy three properties:
\begin{itemize}
 \item Special
 \begin{equation}
  \det U = 1
 \end{equation}
 \item Unitary
 \begin{equation}
  \herm{U}U = 1
 \end{equation}
 \item Three-dimensional
 \begin{equation}
  U = \begin{pmatrix}
       U_{11} & U_{12} & U_{13} \\
       U_{21} & U_{22} & U_{23} \\
       U_{31} & U_{32} & U_{33}
      \end{pmatrix}
 \end{equation}
 where $U_{ij}$ (with two subscripts) are complex numbers.
\end{itemize}

To construct a physical theory from this group, we first need a set of quantum fields which $\SUIII$ transformations can act upon. Since the simplest nontrivial representation of $\SUIII$ is the three-dimensional \term{fundamental representation}, the simplest nontrivial set of fields we can use will have three members: $q_r$, $q_g$, and $q_b$.
\begin{align}\label{eq:SUIIItransformquarks}
 \begin{pmatrix}q'_r \\ q'_g \\ q'_b\end{pmatrix}
 &=
 \begin{pmatrix}
  U_{rr} & U_{rg} & U_{rb} \\
  U_{gr} & U_{gg} & U_{gb} \\
  U_{br} & U_{bg} & U_{bb}
 \end{pmatrix}
 \begin{pmatrix}q_r \\ q_g \\ q_b\end{pmatrix} \\
 q'_k &= U_{kj}q_j
\end{align}
where I use the usual convention of summing over repeated indices. These are the \term{quark fields}.

We then put these fields into a Lagrangian in combinations which are $\SUIII$ invariant: when performing an $\SUIII$ transformation on the fields, the Lagrangian should remain unchanged.
For a Dirac field like the quarks, the Lagrangian will be
\begin{equation}
 \mathcal{L} = \bar{q} i(\gamma^\mu\mathcal{D}_{\mu} - m)q
\end{equation}
where I've omitted the $\SUIII$ index for notational simplicity, but we should keep in mind that $q$ still refers to the $\SUIII$ vector of three quark fields.

While the mass term $\bar{q} im q$ is trivially $\SUIII$ invariant (any $\SUIII$ transformation made to the quark fields $q$ cancels out with the inverse transformation made to $\bar{q}$), the kinetic term $\bar{q} i(\gamma^\mu\mathcal{D}_\mu)q$ is not, at least not if we use the simple partial derivative $\mathcal{D}_\mu = \partial_\mu$.
Applying an $\SUIII$ transformation~\eqref{eq:SUIIItransformquarks} to that term, and expanding $\partial_\mu U q$ using the product rule, produces
\begin{equation}
 i\bar{q} \herm{U} \gamma^\mu\partial_\mu U q = 
 i\bar{q} \gamma^\mu \partial_\mu q + i\bar{q} \herm{U}\gamma^\mu(\partial_\mu U) q
\end{equation}
The first term is the same one, involving the partial derivative, that we had before the transformation, but this residual contribution $i\bar{q} \herm{U}\gamma^\mu(\partial_\mu U) q$ spoils the gauge invariance.

\iftime{include physical motivation (comparator) from Peskin \& Schroeder ch 4}To restore gauge invariance, we'll need to add to $\mathcal{D}_\mu$ an extra term $\Delta_\mu$ which transforms in a way that will cancel out that residual contribution.
Specifically, the product $\bar{q} \Delta_\mu q$ needs to change by $-\bar{q}\herm{U}\partial_\mu U q$ under an $\SUIII$ transformation:
\begin{equation}
 \bar{q} \Delta_\mu q \to \bar{q} \herm{U} \Delta_\mu U q - \bar{q} \herm{U}(\partial_\mu U) q
\end{equation}
Now, an arbitrary $\SUIII$ transformation $U$ can be written as $e^{i\alpha_c \generator^c}$, where $\alpha_c$ is an eight-component vector (corresponding to the eight generators $\generator^c$ of $\SUIII$).
From this, we can explicitly write $\partial_\mu U = i(\partial_\mu \alpha_c)\generator^c U$, which means the transformation of $\Delta_\mu$ has to satisfy
\begin{equation}
 \bar{q} \Delta_\mu q \to \bar{q} \herm{U} \Delta_\mu U q - \bar{q} \herm{U} i(\partial_\mu \alpha_c)\generator^c U q
\end{equation}
or
\begin{equation}
 \Delta_\mu \to \herm{U} [\Delta_\mu - i(\partial_\mu \alpha_c)\generator^c] U
\end{equation}
Given this form, it makes sense to write $\Delta_\mu = i\gs\gluonfield_{c\mu} \generator^c$, and specify that the \iterm{gauge field}, or \iterm{gluon field} for QCD specifically, transforms as
\begin{equation}
 \gluonfield_{c\mu} \to \gluonfield_{c\mu} - \frac{1}{\gs}\partial_\mu\alpha_c
\end{equation}
The covariant derivative then becomes
\begin{equation}
 \mathcal{D}_\mu = \partial_\mu + i\gs\generator^c \gluonfield_{c\mu}
\end{equation}
where $\gs$ is the coupling, $\generator^c$ is an $\suiii$ generator, and $\gluonfield_{c\mu}$ is a new field we will call the \term{gluon field} (or the gauge field).
In this way, the requirement of local $\SUIII$ gauge invariance, where the transformation matrix $U$ can be a function of spacetime position while still preserving the symmetry, leads to the existence of eight \term{gluon fields}.

As explained in standard references~\cite{PeskinSchroeder,HalzenMartin,PinkBook}, we can derive a kinetic term for the gluon fields by defining $F^{i}_{\mu\nu}$ from
\begin{equation}
 \commut{\mathcal{D}_\mu}{\mathcal{D}_\nu} = -i\gs F^c_{\mu\nu} \frac{\sigma^c}{2}
\end{equation}
or, explicitly,
\begin{equation}
 F^c_{\mu\nu} = \partial_\mu A^c_\nu - \partial_\nu A^c_\mu + \gs\epsilon^{cjk} A^j_\mu A^k_\nu
\end{equation}
The kinetic term, which must be a gauge-invariant function of $F$, works out to be $-\frac{1}{4}F^2$.
Putting this all together, we have the Lagrangian of QCD
\begin{equation}
 \mathcal{L} = \bar{q} i(\gamma^\mu\mathcal{D}_{\mu} - m)q - \frac{1}{4} \bigl(F^c_{\mu\nu}\bigr)^2
\end{equation}
with three quark fields and eight gluon fields.
Expanding the terms shows the following types of interactions:\iftime{add diagrams}
\begin{subequations}
\begin{align}
 & -m\bar{q}q
 & & \text{quark mass} \\
 & -\gs\bar{q}\gamma^\mu \generator^c \gluonfield_{c\mu} q
 & & \text{quark-quark-gluon vertex} \label{eq:quarkquarkgluonvertexterm} \\
 & \gs(\partial_\mu \gluonfield^c_\nu - \partial_\nu \gluonfield^c_\mu)\epsilon^{cjk} \gluonfield^{j\mu} \gluonfield^{k\nu}
 & & \text{triple gluon vertex} \label{eq:triplegluonvertexterm} \\
 & \gs^2\epsilon^{cjk} \gluonfield^{j\mu} \gluonfield^{k\nu} \epsilon^{cj'k'} \gluonfield^{j'}_\mu \gluonfield^{k'}_\nu
 & & \text{quadruple gluon vertex} \label{eq:quadgluonvertexterm} 
\end{align}
\end{subequations}
The gluon self-interactions~\eqref{eq:triplegluonvertexterm} and~\eqref{eq:quadgluonvertexterm} are a unique feature of a nonabelian theory; they don't occur in quantum electrodynamics, for example.

\subsection{Running Coupling}

One feature of QCD (or any renormalizable quantum field theory) that will play an important role in calculations is the running of the strong coupling.
So far we have been treating $\gs$, or equivalently $\alphas = \gs^2/4\pi$, as a constant, but in fact it has to vary with energy.
Here is the reasoning: suppose we have some observable quantity $O$ which depends on a single energy scale $Q$.
$O$ may be proportional to some power of $Q$, determined by its units, but any more complex energy dependence will have to occur in the form $Q/\mu$, where $\mu$ is the renormalization scale of the theory (the scale which separates high-energy behavior we can ignore from low-energy behavior which we cannot).
However, $O$ should be independent of $\mu$. So we need its expression to satisfy
\begin{equation}
 \biggl[\mu^2\pd{}{\mu^2} + \mu^2\pd{\alphas}{\mu^2}\pd{}{\alphas}\biggr] O = 0
\end{equation}
This can be rewritten as a differential equation for the coupling, which takes the form
\begin{equation}\label{eq:diffeqcoupling}
 \pd{\alphas}{\ln(Q^2/\mu^2)} = \mu^2\pd{\alphas}{\mu^2} = -b\alphas^2(1 + b'\alphas + b''\alphas^2 + \cdots)
\end{equation}
With this in hand, as shown in e.g. reference~\cite{PinkBook}, we can take the observable $O$, expressed in terms of the coupling at an energy scale $Q$, and write it as a perturbation series in the coupling constant at the renormalization scale, $\alphas(\mu)$:
\begin{equation}
 O(\alphas(Q^2)) = O_1\bigl[1 - c_1\alphas(\mu^2) + c_2 (\alphas(\mu^2))^2 + \cdots\bigr]
\end{equation}
with $O = O_1 \alphas + \cdots$.
This becomes relevant when we discuss the BFKL and BK equations, where we will interpret the running of the coupling as providing next-to-leading order corrections to the operator that implements the BFKL/BK evolution.

For calculations up to next-to-leading order, we can safely take the first term on the right from equation~\eqref{eq:diffeqcoupling} and solve it to obtain
\begin{equation}
 \alphas = \frac{1}{b\ln\frac{Q^2}{\mu^2}}
\end{equation}
A conventional choice for $\mu$ is the parameter $\Lambda_\text{QCD}$ which has a value around $\SI{200}{MeV}$, although as a renormalization scale parameter, one can make different, equally valid choices for $\mu$, corresponding to a theoretical uncertainty in calculations that use the running coupling.

\subsection{Dressed Particles}

At this point we have to make a distinction between \iterm{bare particles}, which are the particles corresponding to the fields in the Lagrangian, and \iterm{dressed particles}, which are the physical particles that propagate through space and time.

Physically, we can understand the difference as follows.
The \HepProcess{\Pquark\Pquark\Pgluon} vertex in equation~\eqref{eq:quarkquarkgluonvertexterm} tells us that a bare quark, for example, will occasionally emit gluons as it propagates.
Those gluons in turn emit other gluons, and some of these emitted gluons will turn into quark-antiquark pairs.
In the end, the quark will be surrounded by a ``cloud'' of copropagating particles.
But the particles from this cloud can interact with each other and be reabsorbed by the original particle.
If the relative amounts of quarks, antiquarks, and gluons are balanced, all these emissions and reabsorptions reach an equilibrium, and we have a whole cloud of particles propagating together, not just a single disturbance in a quark field.
This propagating cloud constitutes a \term{dressed quark}.

Mathematically, this situation arises because excitations in the \term{bare fields} of the QCD Lagrangian are not eigenstates of the Hamiltonian, not when interactions are included.
The eigenstates are linear combinations of bare fields, and these linear combinations correspond to \term{dressed particles}.

When a \term{dressed quark} interacts with another particle, there is some probability that the other particle actually interacts with a bare gluon, or a pair of bare gluons, or a bare quark and antiquark, or even a ``subcloud'' that corresponds to a \term{dressed gluon} inside the top-level \term{dressed quark}.
Accordingly, we'll say that there is some probability of finding, say, a gluon ``inside'' a quark, which will be given by the splitting wavefunction~$\psi$ later in this chapter.

At a higher level, we can actually think of a proton or any other hadron as a dressed bound state of three bare quarks.
The overarching goal of hadron structure research, as described in the following sections, is understanding the nature of this dressing.

\section{Structure of Protons and Nuclei}

Modern hadron structure research has its origin in 1969, with the development of the \iterm{parton model} by Feynman and Bj\"orken~\cite{Feynman:1989dd,Bjorken:1969ja,Feynman:1973xc}.
This model treats a target proton or nucleus as a collection of pointlike constituents which incoherently interact with the projectile.

As motivation for the parton model, consider the severe \term{Lorentz contraction} that affects protons and nuclei in a high-energy collider.
The proton or nucleus is contracted into an ellipsoid with longitudinal extent $\frac{1}{\gamma}$ of its radius, and due to time dilation, its internal dynamics --- the branching and recombining of bare particles in a \term{dressed proton} --- are slowed by a factor of $\gamma$.
For example, a target proton or nucleus at RHIC, in the reference frame of the projectile, has $\gamma = \frac{E}{m_p} = \frac{\SI{200}{GeV}}{\SI{938}{MeV}} = 213$.
In \pA{} collisions at the LHC, the corresponding value is $\gamma = \frac{\SI{2.76}{TeV}}{\SI{938}{MeV}} = 2940$.
These values of $\gamma$ are large enough that we can take the $\gamma\to\infty$ limit, which corresponds to analyzing the collision in the \iterm{infinite momentum frame}.
In this frame, the target acts as a static configuration of quantum fields on a flat disk, and as a result, we can consider the scattering event, the \iterm{hard process}, to be effectively instantaneous, equivalent to scattering off a pointlike constituent.

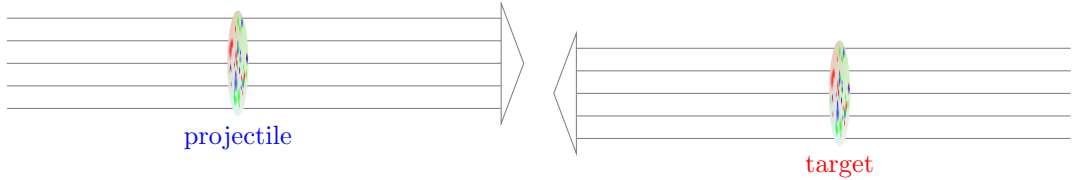
\begin{figure}
 \tikzsetnextfilename{squishedproton}
 \begin{tikzpicture}
  \draw[gray,thin,yshift=.2cm] \foreach \y in {-.6,-.3,...,.6} {(-0.48\textwidth,\y) -- (-.5,\y)};
  \draw[gray,thin,yshift=-.2cm] \foreach \y in {-.6,-.3,...,.6} {(0.48\textwidth,\y) -- (.5,\y)};
  \filldraw[draw=gray,thin,fill=white,yshift=.2cm] (-.2,0) -- (-.5,.8) -- (-.5,-.8) -- cycle;
  \filldraw[draw=gray,thin,fill=white,yshift=-.2cm] (.2,0) -- (.5,.8) -- (.5,-.8) -- cycle;
  \begin{scope}
   \begin{scope}[shift={(-4cm,.2cm)},xscale=0.2]
    \drawproton[splotch background]{20}{0}{2}
    \coordinate (left parton) at (parton-2-14);
    \coordinate (left bottom) at (0,-20pt);
   \end{scope}
   \begin{scope}[shift={(4cm,-.2cm)},xscale=0.2]
    \drawproton[splotch background]{20}{0}{2}
    \coordinate (right parton) at (parton-2-6);
    \coordinate (right bottom) at (0,-20pt);
   \end{scope}
   \node[below,blue] at (left bottom) {projectile};
   \node[below,red] at (right bottom) {target};
  \end{scope}
 \end{tikzpicture}
 \caption{A proton or nucleus behaves like a flat disk in a collision}
 \label{fig:squishedproton}
\end{figure}

\subsection{The Parton Model and Integrated Parton Distributions}\label{sec:partonmodel}

As the originators of hadron structure research did, let's consider \term{deep inelastic scattering} (DIS), a process which features an \emph{electron} projectile impacting a target proton or nucleus and breaking it up: \HepProcess{\Pelectron\Pnucleus \HepTo \Pelectron\Panything}.
Under the parton model, the behavior of the target hadron in a DIS event is determined by two pieces: the hard scattering event involving the target parton, and the distribution of those partons within the target hadron.\footnote{%
This division between hard scattering and the parton distribution is formalized in quantum field theory by factorization theorems~\cite{Collins:1982wa,Collins:1983ju,Collins:1988ig} which have nontrivial proofs.
However, we simply need to know that factorization works; most of the details are beyond the scope of this dissertation, and those that are not (e.g. the role of the factorization scale $\mu$) will be introduced later as needed.
See e.g. reference~\cite{Collins:1989gx} for a more involved discussion of factorization.
}

To understand the roles of these pieces, consider the simplified situation of scattering off a fixed electric charge distribution $\rho(\vec{r})$.
The hard scattering is represented by the cross section for scattering off a point elementary charge, called the \iterm{Mott cross section},\iftime{convert to $\uddc\sigma/\udc\xg\udc Q^2$}
\begin{equation}
 \biggl(\udd{\sigma}{\Omega}\biggr)_\text{Mott} = \frac{\alphaem^2 E^2}{4 k^4 \sin^4(\theta/2)}\biggl(1 - \frac{k^2}{E^2}\sin^2\frac{\theta}{2}\biggr) \label{eq:mottcrosssection}
\end{equation}
where $E$ is the incoming particle's energy and $k$ is the magnitude of its momentum, and $\theta$ is the angle by which it is scattered.
Now, we \emph{could} in principle represent the distribution of partons within the target with the charge distribution $\rho(\vec{r})$, and then the scattering amplitude would be a convolution of the Mott amplitude with the charge distribution:
\begin{equation}
 \scamp(\vec{r}) = \iiint \scamp_\text{Mott}(\vec{r} - \vec{r}') \rho(\vec{r'}) \udddc\vec{r}'
\end{equation}
with $\udd{\sigma}{\Omega} \sim \scamp$.
However, we already have the Mott cross section in terms of momentum, and we would like the end result in momentum space as well, so it's easier to just rewrite that whole relation in momentum space, by which the convolution becomes a simple product.
\begin{equation}
 \scamp(\vec{q}) = \scamp_\text{Mott}(\vec{q}) \eformfac(\vec{q})
\end{equation}
The function $\eformfac(\vec{q})$, called the \iterm{electric form factor}, is the Fourier transform of the charge distribution $\iiint \rho(\vec{r}) \exp(-i\vec{q}\cdot\vec{r})\udddc\vec{r}$. in this limit where the target's recoil can be neglected.
This simplified model gives the overall cross section as
\begin{equation}
 \udd{\sigma}{\Omega} = \biggl(\udd{\sigma}{\Omega}\biggr)_\text{Mott}\abs{\eformfac(\vec{q})}^2
\end{equation}
The first factor represents the hard scattering event, and the second factor represents the distribution of partons.
It's conventional to denote that second factor as $\structureF(\vec{q}) \defn \abs{\eformfac(\vec{q})}^2$, called the \iterm{structure function}.

For realistic collisions, this model fails for several reasons: the partons have magnetic moments in addition to their electric charges, spin interactions between the projectile and target partons mean we can no longer use the Mott cross section as the ``reference'' cross section for a single parton, and the target will recoil.
But the general framework, of expressing a cross section in terms of structure functions, still works.
These complications can be accomodated by generalizing the formula for the differential cross section of DIS to include two structure functions, $\structureF_1$ and $\structureF_2$.
\begin{equation}
 \frac{\uddc\sigma}{\udc\xg \udc Q^2} = \frac{4\pi\alphas^2}{Q^4}\biggl[[1 + (1 - y)^2]\structureF_1(\xg, Q^2) + \frac{1 - y}{\xg}\bigl[\structureF_2(\xg, Q^2) - 2\xg\structureF_1(\xg, Q^2)\bigr]\biggr]
\end{equation}
where $y = \frac{q\cdot p}{k\cdot p}$ in the language of figure~\ref{fig:kinematics}.
However, the specific form of the coefficients is not as important as the fact that the cross section can be expressed as multiples of two structure functions in this way.

\subsubsection{Collision Kinematics}

At this stage I need to take a brief interlude to explain the meaning of the kinematic variables $\xg$ and $Q^2$ which I've just introduced.

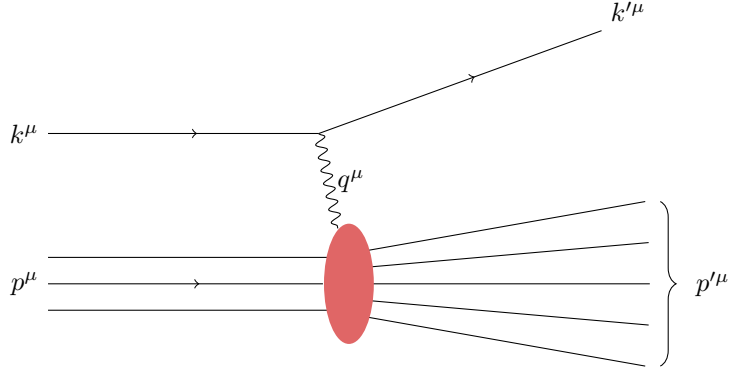
\begin{figure}
 \tikzsetnextfilename{kinematics}
 \begin{tikzpicture}[scale=2]
   \coordinate (interaction) at (-0.2,0);
   \coordinate (left electron) at (-2,0);
   \node[blob,transform shape] (target) at (0,-1) {};
   \draw[lepton] (left electron) node[left] {$k^\mu$} to (interaction);
   \draw[lepton] (interaction) to +(20:2) node[above right] {$k'^\mu$};
   \draw[photon] (interaction) -- (target) node[pos=0.5,right] {$q^\mu$};
   \draw[baryon] (left electron |- target) node[left] {$p^\mu$} to (target);
   \foreach \n in {-2,-1,0,1,2} {
    \draw ($(target)+\n*(0,0.1)$) -- +(5*\n:2) coordinate (endpoint \n);
   }
   \draw[decorate,decoration={brace,amplitude=6pt}] ($(endpoint 2)+(0.1,0)$) -- ($(endpoint -2)+(0.1,0)$) node[pos=0.5,right=10pt] {$p'^\mu$};
   \node[fill=gluondist!60!white,blob,transform shape] at (target) {};
 \end{tikzpicture}
 \caption[Illustration of a deep inelastic scattering event]{Illustration of a deep inelastic scattering event with the momentum variables used to describe it. $k^\mu$ and $k'^\mu$ are the momenta of the incoming and outgoing electron, and similarly with $p^\mu$ and $p'^\mu$}
 \label{fig:kinematics}
\end{figure}
When analyzing a \term{DIS} collision, we have four momenta to work with, as shown in figure~\ref{fig:kinematics}: the incoming and outgoing momenta of the electron, $k^\mu$ and $k'^\mu$ respectively, the incoming momentum of the proton $p^\mu$, and the outgoing momentum $p'^\mu$ of the assorted hadrons produced by the breakup of the proton.
Between the conservation law ($k^\mu + p^\mu = k'^\mu + p'^\mu$), the on-shell relations for incoming and outgoing momenta ($k^2 = k'^2 = m_e^2$, $p^2 = m_p^2$), and the freedom to choose a suitable reference frame, we can eliminate all but three degrees of freedom from these momenta.%
\footnote{Unfortunately, there is almost no consistency in the pedagogical literature when it comes to parametrizing those two degrees of freedom.~\cite[ch. 8]{GriffithsParticlePhysics}
Following a derivation in any of the standard textbooks can be rather difficult.}
(For elastic scattering, we can use $p'^2 = m_\text{proton}^2$ to get down to two.)

\begin{figure}
 \tikzsetnextfilename{xq2diagram}
 \begin{tikzpicture}
  \path[xq2shading] (-1.0,7.5) rectangle (10.5,-1.0);
  \fill[path fading=east,fading transform={xshift=-1.2cm,rotate=-20},rounded corners=5pt,yellow!60!black] (-1.0,7.5) rectangle (10.5,-1.0);
  \draw[->,every node/.append style={above,red!70!black,rotate=90,font={\small}}] (0,-0.5) -- (0,7) node[at={(0,0)},above right] {$x=1$} node[pos=0.5] {$\ln\frac{1}{x}$} node[pos=0.9] {small $x$};
  \draw[->,every node/.append style={below,cyan!70!black,font={\small}}] (-0.5,0) -- (10,0) node[at={(0,0)},below right] {small $Q$} node[pos=0.5] {$\ln\frac{Q^2}{Q_0^2}$} node[pos=0.9] {large $Q$};
  \node[black,text centered,font=\footnotesize,right] at (0.5,7) {saturation};

  \begin{scope}[scale=0.8,xshift=50pt,yshift=30pt]
   \foreach \iprotonx in {0,...,3} {
    \foreach \iprotonq in {0,...,3} {
     \begin{scope}[xshift={90*\iprotonq *1pt},yshift={60*\iprotonx *1pt},scale=0.6]
      \pgfmathsetmacro{\partonlevel}{\iprotonx}
      \pgfmathsetmacro{\protonradius}{25 * (1 + 0.3 * sqrt(\iprotonx + 4*\iprotonq/3))}
      \drawproton[background=white,parton size decay rate={0},initial parton size={6-\iprotonq}]{\protonradius}{\partonlevel}{\partonlevel}
     \end{scope}
    }
   }
  \end{scope}
 \end{tikzpicture}
 \caption[Effective structure of a target proton or nucleus with $\xg$ and $Q^2$]{Snapshots of the structure of a target proton or nucleus as $\xg$ and $Q^2$ vary. The upper left corner is the saturation regime, which we will encounter in the next chapter.}
 \label{fig:xq2diagram}
\end{figure}
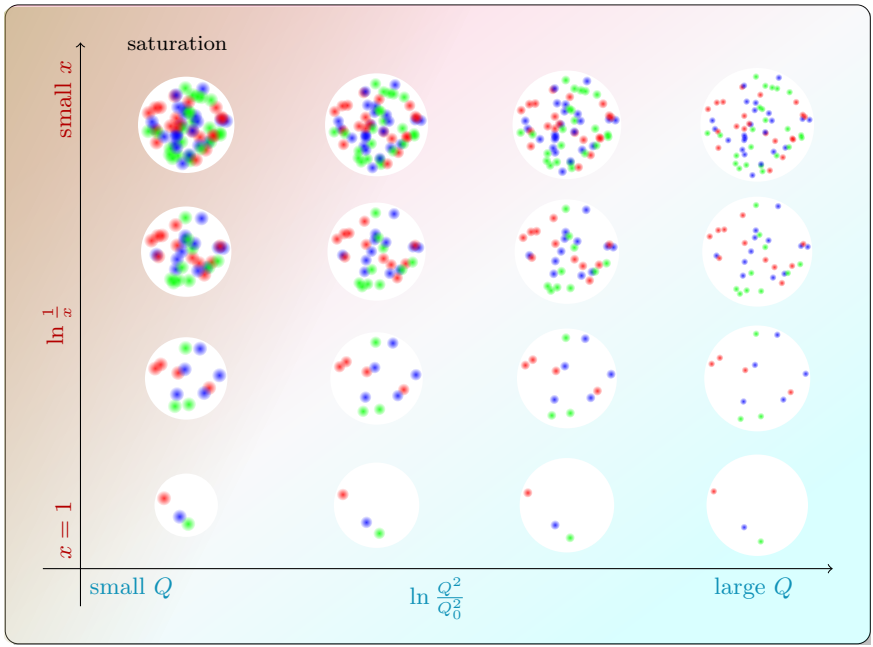

In the analysis of deep inelastic scattering, the parametrization of choice is the photon virtuality $Q^2$ and the Bj\"orken variable $x_\text{Bj}$, defined in terms of the momentum transfer $q^\mu \defn k^\mu - k'^\mu$:
\begin{align}
 Q^2 &\defn -q^2 &
 x_\text{Bj} &\defn \frac{Q^2}{2p\cdot q}
\end{align}
These variables are particularly useful in combination with the parton model because both of them have straightforward physical interpretations, at least in leading order calculations.
Suppose the photon interacts with a parton carrying a fraction $\xg$ of the proton's momentum, $p_q^\mu = \xg p^\mu$, and let us neglect the parton's mass.
We can write the squared four-momentum of the parton \emph{after} the interaction as
\begin{align}
 0 \approx p_q'^2 = (p_q + q)^2 \approx q^2 + 2p_q\cdot q = -2p\cdot q x_\text{Bj} + 2\xg p\cdot q = 2p\cdot q(\xg - x_\text{Bj})
\end{align}
which implies that $x_\text{Bj} = \xg$: the Bj\"orken variable represents the momentum fraction of the struck parton.

Knowing $\xg$ and $Q^2$ of a collision allows us to characterize the behavior of the target in that collision, as shown in figure~\ref{fig:xq2diagram}.

\subsubsection{Integrated Parton Distributions}

We now have the behavior of a target proton or nucleus in a high-energy collision parametrized in terms of two structure functions, $\structureF_1(\xg, Q^2)$ and $\structureF_2(\xg, Q^2)$.
If the target consisted only of quarks with a fixed fraction $\xg$ of the target's momentum, we would find~\cite[ch. 4]{PinkBook} that
\begin{equation}
 \structureF_2(\xg, Q^2) = 2\xg\structureF_1(\xg, Q^2) = \delta(1 - \xg)
\end{equation}
Experimentally, we know this is not the case; structure functions can be measured and we find that they form a broad distribution.
It will be convenient to separate the contributions from each of the different flavors of partons to the structure function, which we do by expressing it as a sum of parton distributions,
\begin{equation}\label{eq:integratedpdf}
 \structureF_2(\xg) = \sum_{\text{flavors }i} e_i^2 \xg \intpdf(\xg)
\end{equation}
These functions $\xg\intpdf(\xg)$ are the \iterm{integrated parton distributions} (PDFs).

PDFs cannot be completely calculated within perturbative QCD.
Although the $Q^2$ dependence of parton distributions is given by perturbation theory, in the form of the DGLAP evolution equations~\cite{Dokshitzer:1977sg,Gribov:1972ri,Lipatov:1974qm,Altarelli:1977zs} (see also~\cite{PeskinSchroeder,HalzenMartin,PinkBook,QCDHighEnergy,Salam:1999cn} for pedagogical descriptions), the initial conditions for these equations emerge from nonperturbative physics acting at low $Q^2$.
Thereform, the parton distributions used in calculations are developed by postulating a functional form, which satisfies the DGLAP equations, with several free parameters, and fitting that function to a wide variety of collider data.
Different research groups have different methods of performing the fit.\footnote{The Durham PDF server at \url{http://hepdata.cedar.ac.uk/pdfs/} is a good starting point for information about PDF fitting groups and their results.}
A representative set are shown in figure~\ref{fig:PDF}.
\begin{figure}
 \ifusepgf
  \includegraphics[scale=0.35]{d09-158f18a.eps}
  \includegraphics[scale=0.35]{d09-158f18b.eps}
 \else
  \includegraphics[scale=0.35]{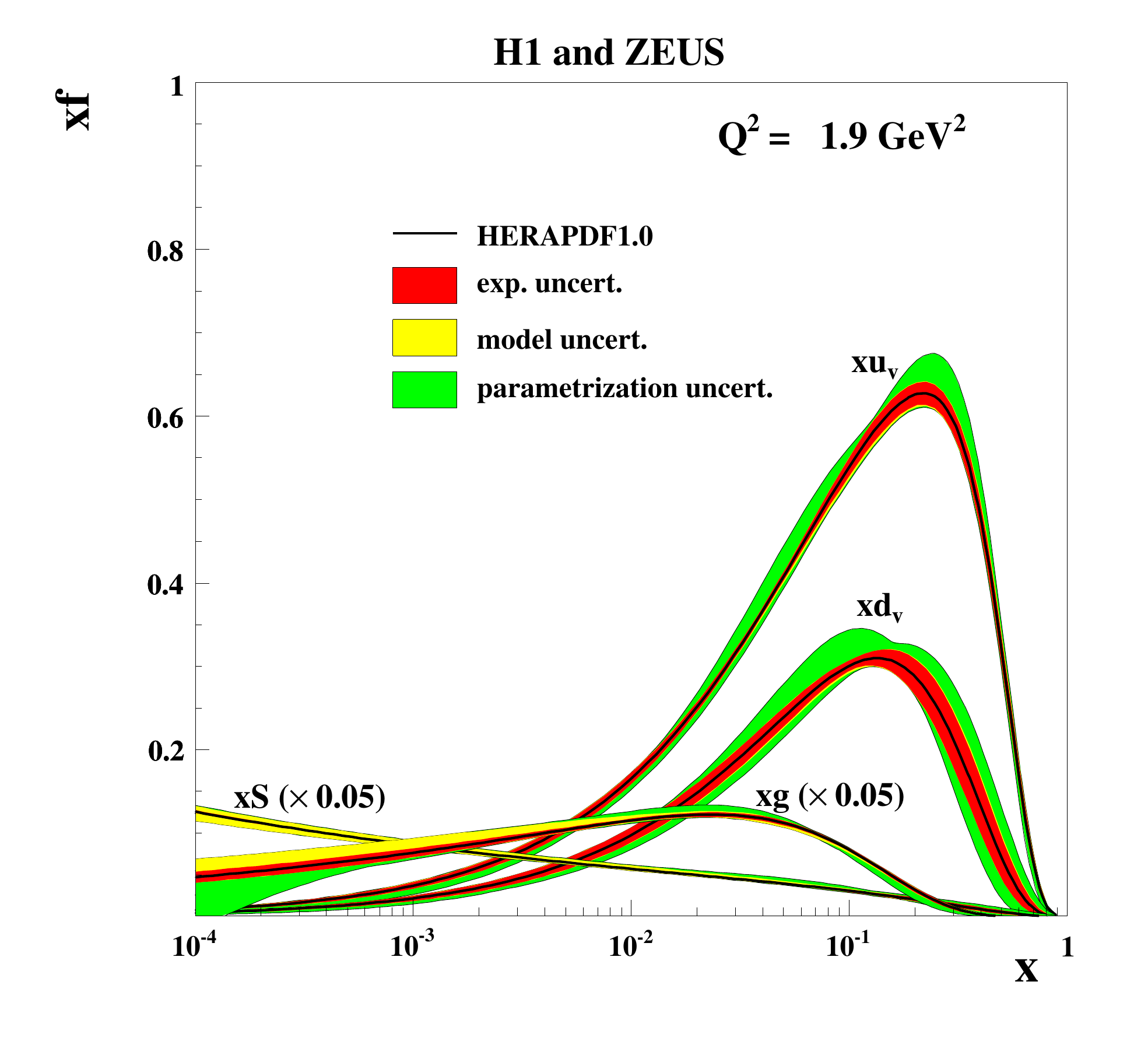}
  \includegraphics[scale=0.35]{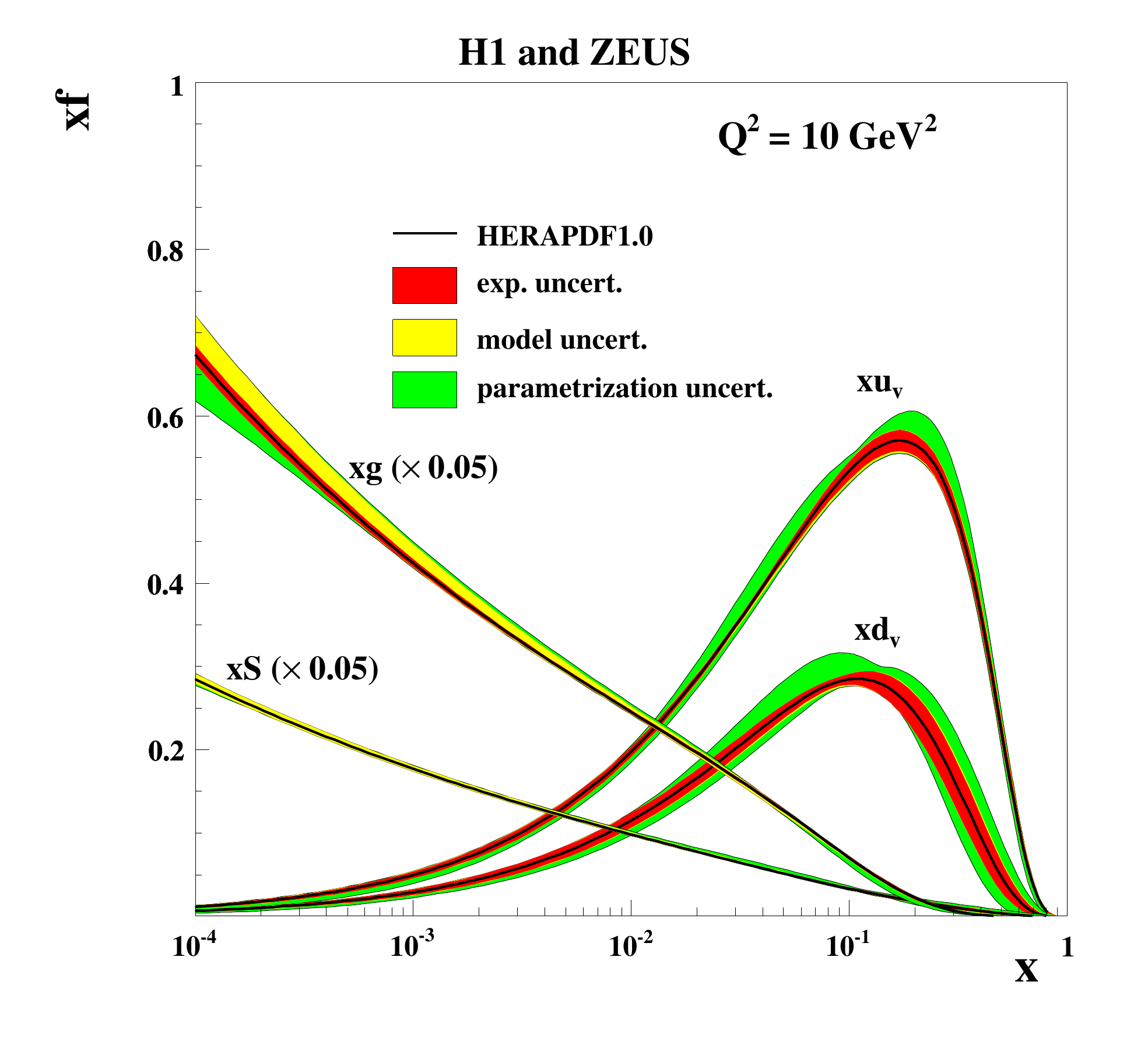}
 \fi
 \caption[Parton distributions fit to HERA data]{Parton distributions $\xg \intpdf(\xg, Q^2)$ derived from fits to measurements at the HERA collider. These are not the independent parton distributions for each parton flavor, but the valence quark distributions, $\xg u_v = \xg (u - \bar{u})$ and $\xg d_v = \xg (d - \bar{d})$ (differences between quark and antiquark distributions of the same flavor), and the sea quark distribution, $\sum_i (\intpdf + \bar f_{i/h})/2$ (average of all quark and antiquark distributions). The gluon PDF is kept separate, and it is clear that it is by far the largest one at small $x$.
 These plots were originally published in reference~\cite{Aaron:2009aa} and are reused under the terms of the Creative Commons Attribution 4.0 license.}
 \label{fig:PDF}
\end{figure}
At small values of $\xg \lesssim \num{1e-2}$, we can see from figure~\ref{fig:PDF} that the gluon distribution exceeds all others by several orders of magnitude.
Accordingly, small-$x$ physics focuses on determining the behavior of the gluon distribution.

\subsection{Unintegrated Parton Distributions}

Factorizing cross sections into hard factors and integrated parton distributions works well for inclusive measurements, where we measure the total cross section, but the procedure has proved unable to explain features of certain exclusive measurements, where the cross section is broken down by the kinematic properties of one final-state particle.
It was proposed in the 1990s~\cite{Sivers:1989cc,Sivers:1990fh} that parton distributions which depend on the transverse momentum transfer could provide the additional flexibility needed.
Since that time, a large body of work has been developed focusing on these transverse momentum-dependent (TMD) distributions and their associated factorization procedure.

These unintegrated gluon distributions can be built up from two fundamental building blocks~\cite{Dominguez:2010xd,Dominguez:2011wm}, the Weizs\"acker-Williams distribution~\cite{McLerran:1993ni,McLerran:1993ka}, which measures the number density of gluons, and the dipole distribution\note{any citation for this?}.
Both distributions have the same asymptotic behavior at large $\kperp$~\cite{Kharzeev:2003wz}, but they contribute to different physical processes. This dissertation focuses on the dipole gluon distribution specifically.

Probably the simplest expression of the unintegrated dipole gluon distribution is the derivative of the integrated gluon distribution from the previous section~\eqref{eq:integratedpdf},~\cite{Stasto:2014sea}\note{conflict between \cite{Stasto:2014sea,Dominguez:2011wm,Salam:1999cn}}
\begin{equation}\label{eq:dipoleFintGrelation}
 Q^2\dipoleF(\xg, Q^2) = \frac{2\pi^2\alphas}{\Nc}\pd{}{Q^2}[\xg \intg(\xg, Q^2)]
\end{equation}
There are a variety of other notations used in the literature, among them $\dipolephi$, which is related to $\dipoleF$ as
\begin{equation}\label{eq:dipoleFfromphi}
 \dipoleF(\vec\kperp) = \frac{1}{8\pi}\lapl\dipolephi(\xg, \vec\kperp) + \delta^{(2)}(\vec\kperp) 
\end{equation}
One will also occasionally see $\varphi = \frac{1}{2}\dipolephi$.

It's quite common to use the position space version of the distribution, $\dipoleS$, which is expressed as a Fourier integral
\begin{equation}\label{eq:dipoleFdefinition}
 \dipoleF(\vec{k}_\perp) = \int\frac{\uddc\vec{x}_\perp\uddc\vec{y}_\perp}{(2\pi)^2}e^{-i\vec{k}_\perp\cdot(\vec{x}_\perp - \vec{y}_\perp)}\dipoleS(\vec{x}_\perp,\vec{y}_\perp)
\end{equation}
as well as the dipole amplitude
\begin{equation}\label{eq:dipoleNdefinition}
 \dipoleN(\vec\xperp, \vec\yperp, \xg) = 1 - \dipoleS(\vec\xperp, \vec\yperp)
\end{equation}



\subsubsection{Analytic Models}

We've seen that parton distributions cannot be calculated using perturbative QCD, but we still need some concrete form for the PDF if we're going to use it in the calculation of a cross section.
Developing and refining these concrete forms is an active area of research.
The general procedure is to postulate some analytic expression with several free parameters that can then be determined from fits to collider data.
In this way, various researchers have developed models for the unintegrated gluon distribution, with varying levels of complexity.

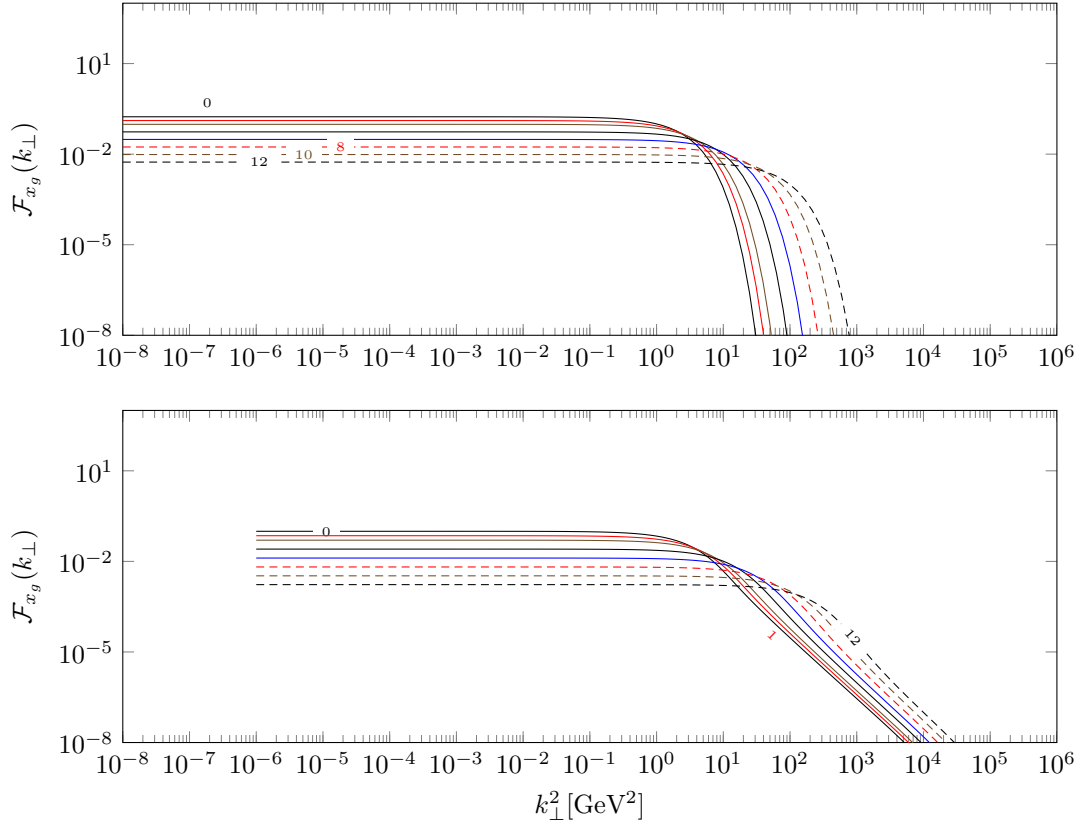
\begin{figure}
 \tikzsetnextfilename{dipoleFGBWMV}
 \begin{tikzpicture}
  \begin{groupplot}[
    group style={group size=1 by 2,xlabels at=edge bottom},
    xmode=log,ymode=log,domain=1e-8:1e2,samples=100,xmin=1e-8,xmax=1e6,ymin=1e-8,ymax=1e3,xlabel={$\kperp^2 [\si{GeV^2}]$},ylabel=$\dipoleF(\kperp)$,width=14cm,height=6cm
  ]
  \nextgroupplot
   \addplot+[black,no markers] {0.17285257*exp(-\x/1.84151091)} node[pos=0.04,above,font=\tiny] {$0$};
   \addplot+[no markers] {0.12959822*exp(-\x/2.45612863)};
   \addplot+[no markers] {0.09716776*exp(-\x/3.2758795)};
   \addplot+[no markers] {0.05462212*exp(-\x/5.82749003)};
   \addplot+[no markers,domain=1e-8:1e3] {0.03070541*exp(-\x/10.36657182)};
   \addplot+[no markers,domain=1e-8:1e3] {0.01726082*exp(-\x/18.44118321)} node[pos=0.1,font=\tiny,fill=white] {$8$};
   \addplot+[no markers,domain=1e-8:1e3] {0.00970304*exp(-\x/32.80517841)} node[pos=0.12,font=\tiny,fill=white] {$10$};
   \addplot+[no markers,domain=1e-8:1e3] {0.00545449*exp(-\x/58.35741224)} node[pos=0.12,font=\tiny,fill=white] {$12$};
  \nextgroupplot[restrict x to domain=-8:6,log basis x=10]
   \addplot+[black,no markers] file {datafiles/MV/MV-Pb85-000.dat} node[pos=0.04,font=\tiny,fill=white] {$0$};
   \addplot+[no markers] file {datafiles/MV/MV-Pb85-001.dat} node[pos=0.5,sloped,below,font=\tiny] {$1$};
   \addplot+[no markers] file {datafiles/MV/MV-Pb85-002.dat};
   \addplot+[no markers] file {datafiles/MV/MV-Pb85-004.dat};
   \addplot+[no markers] file {datafiles/MV/MV-Pb85-006.dat};
   \addplot+[no markers] file {datafiles/MV/MV-Pb85-008.dat};
   \addplot+[no markers] file {datafiles/MV/MV-Pb85-010.dat};
   \addplot+[no markers] file {datafiles/MV/MV-Pb85-012.dat} node[pos=0.45,sloped,font=\tiny,fill=white] {$12$};
  \end{groupplot}
 \end{tikzpicture}
 \caption[The GBW and MV gluon distributions in momentum space]{Plots of the GBW and MV gluon distributions in momentum space. The key feature of the MV distribution is the power tail at large $\kperp$, as opposed to the exponential falloff of the GBW model. These distributions use the same saturation scale of $\Qs^2 = \centrality \massnumber^{1/3} (x_0/\xg)^\lambda$ with $\centrality = 0.85$, $\massnumber = 208$, $x_0 = 0.000304$, and $\lambda = 0.288$.}
 \label{fig:dipoleFGBWMV}
\end{figure}

\paragraph{GBW Model}
One model which is particularly common due to its simplicity was developed by Golec-Biernat and W\"usthoff~\cite{GolecBiernat:1998js,GolecBiernat:1999qd}.
The model is defined by two components.
First, an exponential form for the gluon distribution
\begin{equation}\label{eq:GBWdipoleS}
 \dipoleS(\vec\rperp) \defn \exp\biggl(\frac{\rperp^2\Qs^2}{4}\biggr)
\end{equation}
where $\Qs$ is the \term{saturation scale}, a momentum scale that characterizes the transition between $\dipoleS \approx 1$ and $\dipoleS \ll 1$.
We'll see the significance of the saturation scale in the next chapter.
The other component is the dependence of the saturation momentum on $\xg$,
\begin{equation}\label{eq:GBWsatscale}
 \Qs^2 = Q_0^2\biggl(\frac{x_0}{\xg}\biggr)^\lambda
\end{equation}
$Q_0$ is the saturation momentum at some reference momentum fraction $x_0$, so one of $x_0$ and $Q_0$ can be chosen arbitrarily.
The choice made by Golec-Biernat and W\"usthoff, based on a fit to DIS data collected at HERA, is to set $Q_0 = \SI{1}{GeV}$ for a proton, for which the corresponding value of $x_0$ from the fit is $x_0 = 0.000304$.

For a nucleus, the formula for $\Qs^2$ is modified by adding a factor of $\centrality \massnumber^{1/3}$, where $\massnumber$ is the atomic mass number and $\centrality$ is the centrality coefficient, $\centrality = \sqrt{1 - \Rperp^2/R^2}$ with $\Rperp$ being the impact parameter and $R$ being the radius of the nucleus.
Effectively, $\massnumber^{1/3}$ is proportional to the number of nucleons that fit across a diameter of the nucleus, and thus $\centrality \massnumber^{1/3}$ can be thought of as a measure of the number of nucleons the projectile will encounter on its way through the nucleus.%
\footnote{A more precise argument for why the saturation scale of a nucleus scales as $\massnumber^{1/3}$ can be found in reference~\cite{QCDHighEnergy} and many papers on the subject.}

The momentum space distribution for the GBW model,
\begin{equation}\label{eq:GBWdipoleF}
 \dipoleF(\vec\kperp) = \frac{1}{\pi \Qs^2} \exp\biggl(-\frac{\kperp^2}{\Qs^2}\biggr)
\end{equation}
is plotted in figure~\ref{fig:dipoleFGBWMV}.
It exhibits a characteristic sharp exponential falloff at high $\kperp$.

\paragraph{MV Model}
The GBW model has one significant shortcoming: it falls off exponentially at high momenta $\kperp \gtrsim \Qs$, which conflicts with what we would expect from rather general arguments based on perturbative QCD.
To remedy this, we can use the MV model~\cite{McLerran:1993ni,McLerran:1993ka,McLerran:1994vd,McLerran:1997fk}, which modifies the exponent to produce a power law (not exponential) decay at high momentum:
\begin{equation}\label{eq:MVdipoleS}
 \dipoleS(\vec\rperp) \defn \exp\biggl[\frac{\rperp^2\Qs^2}{4}\ln\biggl(e + \frac{1}{\rperp\Lambda_\text{MV}}\biggr)\biggr]
\end{equation}
With this model as well, the $\xg$ dependence of the saturation scale $\Qs$ is unconstrained.
Conventionally we take it to have the same functional dependence as in the GBW model~\eqref{eq:GBWsatscale}.

There is no symbolic expression for the momentum space distribution $\dipoleF$ in the MV model, but it can be computed numerically.
The result is plotted in figure~\ref{fig:dipoleFGBWMV}.
Unlike the GBW model, this gluon distribution has a softer drop at high $\kperp$, not exponential but proportional to a power law $\kperp^{-a}$ with $a \approx 4.5$.

\paragraph{AAMQS Model}
Recent work by Albacete, Armesto, Milhano, Quiroga, and Salgado~\cite{Albacete:2010sy} has generalized the preceding two models by adding an anomalous dimension, $\gamma$.
\begin{equation}
 \dipoleS(\vec\rperp) \defn \exp\biggl[\frac{(\rperp^2\Qs^2)^\gamma}{4}\ln\biggl(e + \frac{1}{\rperp\Lambda_\text{MV}}\biggr)\biggr]
\end{equation}
The MV model can be considered a special case of this with $\gamma = 1$, and the GBW model a further specialization with $\Lambda_\text{MV} \to \infty$.

In practice, the AAMQS model is usually used as an initial condition to the BK evolution equation (to be introduced later), rather than as an independent model of the gluon distribution in its own right.
The best fit with this model yields $\gamma$ between $\approx 0.9$ and $\approx 1.4$ depending on the values of other parameters, suggesting that the additional degree of freedom in the model may play a useful role in describing data.
I mention it here because it appears frequently in the literature, although I will not use it for the fits in the remainder of this dissertation.

%% file: cgc.tex
\chapter{The Color Glass Condensate}\label{ch:cgc}

The previous chapter presented a very high-level overview of how the picture of hadron structure emerges from QCD.
In this chapter, I'm going to revisit those concepts, showing the details of how they emerge from the underlying quantum field theory.
These derivations will also serve as a demonstration of how the cross sections in later chapters can be calculated.

I'll begin with a discussion of \term{light cone perturbation theory}, a method of approaching quantum field theory that is commonly used in the field.
The focus of this chapter is on the \term{color glass condensate}, which is a model developed to describe the behavior of colliding nucleons and nuclei when their gluon distributions are large.
Having discussed the CGC prepares us to analyze the small-$x$ evolution of the unintegrated gluon distribution, given by the BK equation, which will play a key role in the results presented in the remainder of the dissertation.

\section{Light Cone Perturbation Theory}\label{sec:lcpt}

Calculations in saturation physics are often performed in \iterm{light cone perturbation theory},\footnote{For an accessible introduction to \term{light cone perturbation theory}, including a set of rules analogous to those used for Feynman diagrams, see reference~\cite{QCDHighEnergy}, which is based on the more detailed treatment in references~\cite{Lepage:1980fj,Brodsky:1989pv}.} which is similar to, but slightly different from, the ``traditional'' perturbation theory that uses Feynman diagrams.

The basis of light cone perturbation theory is the light cone coordinate system, in which any four-vector $a^\mu$ is expressed as the components $(a^+, a^-, \vec{a}_\perp)$:
\begin{align}
 a^+ &= a_t + a_z &
 a^- &= a_t - a_z &
 \vec{a}_\perp &= (a_x, a_y)
\end{align}
Light cone coordinates are described in more detail in appendix~\ref{ap:lightcone}.\iftime{maybe that should just be here}

This choice of coordinates is particularly convenient to describe two particles moving at the speed of light in opposite directions along the $z$ axis. For one particle, which we'll take to be the projectile, the minus component of position is constant, and the plus component can be used to parametrize the path, filling the role of time. And of course for the target, the reverse is true: $x^+$ is constant and $x^-$ parametrizes the path.\iftime{add a figure}

Light cone perturbation theory comes with its own rules for representing physical processes by Feynman-like diagrams.
For comparison, consider a Feynman diagram in plain old $\phi^3$ field theory.
\begin{center}
 \tikzsetnextfilename{phi3}
 \begin{tikzpicture}
  \coordinate[label=110:$x_1^\mu$] (v1) at (-1,0);
  \coordinate[label=70:$x_2^\mu$] (v2) at (1,0);
  \draw (-2,0) -- (v1);
  \draw (v1) to[out=40,in=140] (v2);
  \draw (v1) to[out=-40,in=-140] (v2);
  \draw (v2) -- (2,0);
 \end{tikzpicture}
\end{center}
This diagram represents the total contribution of \emph{all} possible processes in which the particle lines are connected this way, no matter where in spacetime the vertices occur.
In particular, even processes where vertex 2 is chronologically \emph{before} vertex 1 ($t_2 < t_1$) are included in the diagram's contribution.
On the other hand, there is a different procedure, \iterm{time ordered perturbation theory}, in which the vertices have to be ordered by time, so we use two diagrams to represent the same process: one for all the contributions where $t_2 > t_1$, and one for all the contributions where $t_2 < t_1$.
\begin{center}
 \tikzsetnextfilename{toptdiagrams}
 \begin{tikzpicture}
  \begin{scope}
   \coordinate[label=110:$x_1^\mu$] (v1) at (-1,0);
   \coordinate[label=70:$x_2^\mu$] (v2) at (1,0);
   \draw (-2,0) -- (v1);
   \draw (v1) to[out=40,in=140] (v2);
   \draw (v1) to[out=-40,in=-140] (v2);
   \draw (v2) -- (2,0);
   \coordinate (center left) at (0,0);
  \end{scope}
  \begin{scope}[xshift=6cm]
   \coordinate[label=-30:$x_1^\mu$] (v1) at (1,-1);
   \coordinate[label=150:$x_2^\mu$] (v2) at (-1,1);
   \draw (-2,-1) -- (v1);
   \draw (v2) to[out=-5,in=95] (v1);
   \draw (v2) to[out=-85,in=175] (v1);
   \draw (v2) -- (2,1);
   \coordinate (center right) at (0,0);
  \end{scope}
  \node at ($(center left)+(0,-1.4)$) {$t_2 > t_1$};
  \node at ($(center right)+(0,-1.4)$) {$t_2 < t_1$};
 \end{tikzpicture}
\end{center}
Light cone perturbation theory is quite similar, except that instead of being ordered by time, the vertices are ordered by the light cone time $x^+$.
The light cone version of the diagram on the left represents all processes where $x_2^+ > x_1^+$, and the one on the right, all processes where $x_2^+ < x_1^+$.

Coupled with this choice of coordinates and diagrams is the use of \iterm{light cone gauge} for the gluon field. In general, the light cone gauge is defined as
\begin{equation}
 \gaugevector^\mu A_\mu^a = 0
\end{equation}
for some four-vector $\eta$ which is lightlike, satisfying $\eta^\mu \eta_\mu = 0$. A common choice, which I'll be using in this work, is
\begin{equation}
 \gaugevector = (0, 2, \vec 0)
 \qquad\implies\qquad
 A^+ = 0
\end{equation}

\subsection{The Light Cone Wavefunction}
A quantum state in light cone perturbation theory is a superposition of Fock states, each of which corresponds to a definite number of particles of specific types with definite quantum numbers. For example, a state with one gluon with plus momentum $k^+$, transverse momentum $\vec\kperp$, polarization $\lambda$, and color indices $\vec a$ might be represented $\ket{k^+, \vec\kperp, \lambda, \vec a}$. More generally, let's denote a state with $n_g$ gluons and $n_q$ quarks, where the individual momenta are irrelevant, as $\ket{n_g, n_q}$.

All possible Fock states form a complete basis, so if we sum $\ket{n_g, n_q}\bra{n_g, n_q}$ over all possible particle numbers, and integrate over all possible momenta, polarizations, and color indices (represented by $\udc\phasespace$, for ``phase space'') at each particle number, the resulting operator is the identity operator
\begin{equation}
 1 = \sum_{n_g, n_q}\int\udc\phasespace_{n_g, n_q}\ket{n_g, n_q}\bra{n_g, n_q}
\end{equation}
So, if $\ket{p}$ is the state that represents, say, a proton, we can write
\begin{equation}\label{eq:identityprotonstate}
 \ket{p} = \sum_{n_g, n_q}\int\udc\phasespace_{n_g, n_q}\ket{n_g, n_q}\braket{n_g, n_q|p}
\end{equation}
This is reminiscent of the way we define a position-space wavefunction in basic quantum mechanics:
\begin{equation}
 \ket{\psi} = \int\udddc\vec{r} \ket{\vec r}\braket{\vec r|\psi}
\end{equation}
In this case we would define the complex-valued function $\psi(\vec r) \defn \braket{\vec r|\psi}$ as the wavefunction. Analogously, from equation~\eqref{eq:identityprotonstate}, we define the complex-valued function
\begin{equation}
 \Psi_p(n_g, n_q, \text{momenta}) \defn \braket{n_g, n_q|p}
\end{equation}
as the \iterm{light cone wavefunction}. Just as the normal wavefunction represents the amplitude of a state at a particular position, the light-cone wavefunction represents the amplitude of a state to have a particular particle content with specified momenta.

\subsection{Compound Diagrams}
\begin{figure}
 \tikzset{baseline={(0,0)}}
 \begin{align*}
  \tikzsetnextfilename{compounddiagram1}
  \begin{tikzpicture}
   \coordinate (interaction) at (0,0);
   \draw[quark] (-1,0) -- (interaction);
   \draw[quark] (interaction) -- +(40:1);
   \draw[gluon] (interaction) -- +(-40:1);
   \node[align=center] at (0,1.2) {diagram};
   \node[align=center] at (0,-1.2) {$\scamp_1$};
  \end{tikzpicture}
  \quad\times\quad
  \tikzsetnextfilename{compounddiagram2}
  \begin{tikzpicture}[xscale=-1]
   \coordinate (interaction) at (0,0);
   \draw[quark] (-1,0) -- (interaction);
   \draw[quark] (interaction) -- +(40:1);
   \draw[gluon] (interaction) -- +(-40:1);
   \node[align=center] at (0,1.2) {conjugate\\diagram};
   \node[align=center] at (0,-1.2) {$\conj{\scamp_1}$};
  \end{tikzpicture}
  \quad&=\quad
  \tikzsetnextfilename{compounddiagram3}
  \begin{tikzpicture}
   \coordinate (left interaction) at (-1,0);
   \coordinate (right interaction) at (1,0);
   \draw[dashed] (0,-0.7) -- (0,1);
   \draw[quark] (-2,0) -- (left interaction);
   \draw[quark] (left interaction) to[out=40,in=140] (right interaction);
   \draw[gluon] (left interaction) to[out=-40,in=-140] (right interaction);
   \draw[quark] (right interaction) -- (2,0);
   \node[align=center] at (0,-1.2) {$\conj{\scamp_1}\scamp_1$};
  \end{tikzpicture}
  \\
  \tikzsetnextfilename{compounddiagram4}
  \begin{tikzpicture}
   \coordinate (interaction) at (0,0);
   \draw[quark] (-1,0) -- (interaction);
   \draw[quark] (interaction) -- +(40:1);
   \draw[gluon] (interaction) -- +(-40:1);
   \node[align=center] at (0,-1.2) {$\scamp_1$};
  \end{tikzpicture}
  \quad\times\quad
  \tikzsetnextfilename{compounddiagram5}
  \begin{tikzpicture}[xscale=-1]
   \coordinate (interaction 1) at (180:0.5);
   \coordinate (interaction 2) at (60:0.5);
   \coordinate (interaction 3) at (-60:0.5);
   \draw[quark] (-1,0) -- (interaction 1);
   \draw[quark] (interaction 1) -- (interaction 3);
   \draw[quark] (interaction 3) -- (interaction 2);
   \draw[gluon] (interaction 1) -- (interaction 2);
   \draw[quark] (interaction 2) -- +(40:0.5);
   \draw[gluon] (interaction 3) -- +(-40:0.5);
   \node[align=center] at (0,-1.2) {$\conj{\scamp_2}$};
  \end{tikzpicture}
  \quad&=\quad
  \tikzsetnextfilename{compounddiagram6}
  \begin{tikzpicture}
   \coordinate (left interaction) at (-1,0);
   \coordinate (right interaction 1) at ($(1,0)+(0:0.5)$);
   \coordinate (right interaction 2) at ($(1,0)+(120:0.5)$);
   \coordinate (right interaction 3) at ($(1,0)+(-120:0.5)$);
   \draw[dashed] (0,-0.85) -- (0,1);
   \draw[quark] (-2,0) -- (left interaction);
   \draw[quark] (left interaction) to[out=60,in=150] (right interaction 2);
   \draw[gluon] (left interaction) to[out=-60,in=-150] (right interaction 3);
   \draw[quark] (right interaction 1) -- (right interaction 3);
   \draw[quark] (right interaction 3) -- (right interaction 2);
   \draw[gluon] (right interaction 1) -- (right interaction 2);
   \draw[quark] (right interaction 1) -- +(0.5,0);
   \node[align=center] at (0,-1.2) {$\conj{\scamp_2}\scamp_1$};
  \end{tikzpicture}
 \end{align*}
 \caption{Two examples of compound diagrams for the process \HepProcess{\Pquark \HepTo \Pquark\Pgluon}}
 \label{fig:compounddiagram}
\end{figure}
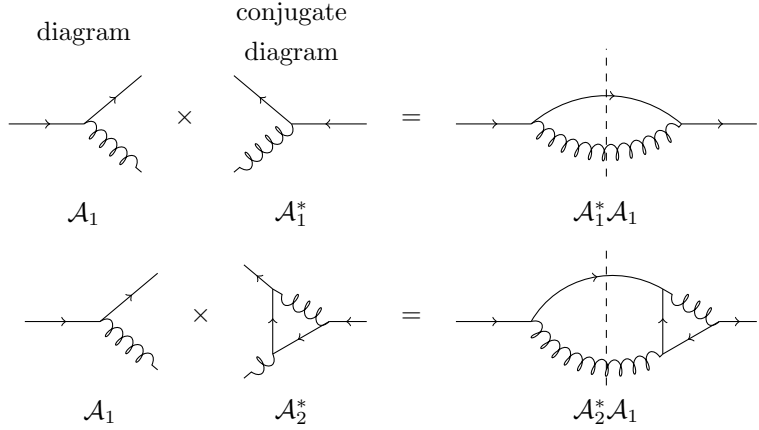

Each individual diagram in light cone perturbation theory (as in Feynman perturbation theory) represents a contribution $\scamp_i$ to the overall scattering amplitude $\scamp$.
\begin{equation}
 \scamp = \scamp_1 + \scamp_2 + \cdots
\end{equation}
When we calculate the cross section, we're going to multiply the amplitude by its complex conjugate, which produces a series of terms of the form $(\text{conjugate amplitude})\times(\text{amplitude})$.
\begin{equation}
 \abs{\scamp}^2 = (\conj{\scamp_1} + \conj{\scamp_2} + \cdots)(\scamp_1 + \scamp_2 + \cdots) = \conj{\scamp_1} \scamp_1 + \conj{\scamp_2} \scamp_2 + \cdots + \conj{\scamp_1} \scamp_2 + \conj{\scamp_2} \scamp_1 + \cdots
\end{equation}
If we adopt the convention that the mirror image of a diagram represents the complex conjugate of the corresponding term, then each of these $(\text{conjugate amplitude})\times(\text{amplitude})$ terms can be represented using a ``compound diagram'' that is constructed from one normal and one mirrored diagram.
Figure~\ref{fig:compounddiagram} shows a couple of examples.
In these compound diagrams, a dashed line, the \iterm{cut}, indicates where the normal diagram ends and the mirrored one begins, or equivalently, the intermediate state of the compound diagram which corresponds to the final state of the normal diagram.

We can apply the rules of perturbation theory to construct the mathematical expression corresponding to the compound diagram, and we'll get precisely the product of the normal and mirrored diagrams.
In this way, we can use the rules of our field theory to obtain the terms of $\abs{\scamp}^2$ directly, which is often easier than finding $\scamp$ and multiplying it by its complex conjugate.

There is one caveat: when constructing the expression that corresponds to a compound diagram, we would ordinarily allow the momenta of the intermediate particles to be off-shell (i.e. to not satisfy $p^\mu p_\mu = m^2$).
However, in reality, these ``intermediate'' particles are actually final-state particles, which will necessarily have on-shell momenta.
So we'll have to manually insert the constraint that the momentum is on-shell for each particle line that crosses the cut.

\section{Calculations in the Color Glass Condensate}

In roughly the past decade or so, the color glass condensate (CGC) model has emerged as a promising candidate for describing the behavior of protons and nuclei in high-energy collisions where the gluon distributions are large.
The model derives its name from its treatment of the target nucleon or nucleus as a densely packed gluon fluid, or ``color glass.''
The appeal of the CGC is its ability to efficiently handle situations where multiple gluon interactions play a large role in determining the physics.

In this section I'll introduce some of the underlying principles of CGC calculations and demonstrate their use in calculations.

\subsection{Wilson Lines}\label{sec:wilsonlines}

Much of the formalism of the CGC is based on Wilson lines.
A Wilson line is defined as the path-ordered exponential of a field along a path:
\begin{equation}\label{eq:wilsonline}
 U(\vec\xperp) = \pathorder\exp\biggl(i g_S\int\udc s \ud{x^\mu}{s} \generator^c \gluonfield_{c\mu}(x^\nu)\biggr)
\end{equation}
Physically, this represents the accumulated effect of the field $\gluonfield_{c\mu}$ on a particle passing through it along the path parametrized by $s$, where the fundamental interaction between the particle and the field contributes the factor of $\generator^c$.
For example, the Wilson line with $\generator^c$ in the fundamental representation of $\SUIII$ represents the superposition of all possible multiple gluon interactions between a propagating quark and the target nucleon or nucleus.
\begin{equation}
 \tikzsetnextfilename{wilsonlineresummation}
 \begin{tikzpicture}[baseline={(current bounding box.center)}]
  \matrix[nodes={align=center}]{
   \node{$U$};
   &
   \node{$=$};
   &
   \node{$U_0$};
   &
   \node{$+$};
   &
   \node{$U_1$};
   &
   \node{$+$};
   &
   \node{$U_2$};
   &
   \node{$+$};
   &
   \node{$U_3$};
   &
   \node{$+$};
   &
   \node{$\cdots$};
   \\
   \draw[quark] (-1,0) -- (1,0);
   \draw[gluon] (-0.5,0) -- +(0,-1);
   \draw[gluon] (0.5,0) -- +(0,-1);
   \node at (0,-.5) {$\cdots$};
   &
   \node{$=$};
   &
    \draw[quark] (-0.5,0) -- (0.5,0);
   &
   \node{$+$};
   &
    \draw[quark] (-0.5,0) -- (0.5,0);
    \draw[gluon] (0,0) -- +(0,-1);
   &
   \node{$+$};
   &
    \draw[quark] (-0.75,0) -- (0.75,0);
    \draw[gluon] (-0.25,0) -- +(0,-1);
    \draw[gluon] (0.25,0) -- +(0,-1);
   &
   \node{$+$};
   &
    \draw[quark] (-1,0) -- (1,0);
    \draw[gluon] (-0.5,0) -- +(0,-1);
    \draw[gluon] (0,0) -- +(0,-1);
    \draw[gluon] (0.5,0) -- +(0,-1);
   &
   \node{$+$};
   &
   \node{$\cdots$};
   \\
  };
 \end{tikzpicture}
\end{equation}
Alternatively, a Wilson line where the generator is in the adjoint representation, denoted $\tilde T^c$, represents the superposition of interactions between a propagating \emph{gluon} and the field of the nucleus.
In lieu of equation~\eqref{eq:wilsonline}, this gives us
\begin{equation}\label{eq:adjointwilsonline}
 W(\vec\xperp) = \pathorder\exp\biggl(i g_S\int\udc s\ud{x^\mu}{s} \tilde T^c A_{c\mu}(x^\nu)\biggr)
\end{equation}
Adjoint Wilson lines can be expressed in terms of fundamental Wilson lines using the identity~\eqref{eq:ident:wilsonlinereduce}.

All the degrees of freedom in the gluon field can also be represented as Wilson lines --- in other words, any function of $\gluonfield_{c\mu}$ can also be written in terms of $U$ (or its equivalent in a different representation).
But Wilson lines become particularly appealing when we write them as \iterm{Wilson loops}, traces of products of Wilson lines which form a closed path.
The fact that the path is closed, together with the trace operation, means that \term{Wilson loops} are automatically \term{gauge invariant}, and in fact \emph{any} gauge-invariant quantity in the theory can be expressed in terms of Wilson loops.~\cite[sec. 15.3]{PeskinSchroeder}

\subsection{Gluon Distributions as Wilson Line Correlators}

\begin{figure}
 \tikzsetnextfilename{dipolecoords}
 \begin{tikzpicture}[pole/.style={shape=circle,minimum width=2mm,minimum height=2mm}]
  \node[pole,fill=red,label={-60:$\vec\xperp$}] (x) at (1,-1) {};
  \node[pole,fill=green,label={150:$\vec\yperp$}] (y) at (-1,1) {};
  \node[pole,fill=blue,label={40:$\vec\bperp$}] (b) at (1.3,1.1) {};
  \draw[-latex] (x) -- (y) node[below left,pos=0.5] {$\vec\rperp$};
  \draw[-latex] (x) -- (b) node[right,pos=0.5] {$\vec\sperp$};
  \draw[-latex] (y) -- (b) node[above,pos=0.5] {$\vec\tperp$};
 \end{tikzpicture}
 \caption[Coordinates for color dipole splitting]{Coordinates for color dipoles. This represents the splitting of a single color dipole $\vec\rperp$ into two, $\vec\sperp$ and $\vec\tperp$.}
 \label{fig:dipolecoords}
\end{figure}

Unintegrated parton distributions are, of course, gauge invariant quantities, so we can express them in terms of Wilson loops.
In this framework, the dipole gluon distribution is a correlator of two Wilson lines,
\begin{equation}\label{eq:dipoleSdefinition}
 \dipoleS(\vec\xperp, \vec\yperp) = \frac{1}{\Nc}\cgcavg{\trace[U(\vec\xperp)\herm U(\vec\yperp)]}
\end{equation}
where $\cgcavg{\cdots}$ denotes a ``target averaging operation''~\cite[sec. 2.5]{Gelis:2012ri} which works like this: from the gluon field in the nucleus, extract the components corresponding to gluons having longitudinal momentum fraction $\xg$.
Then average over all possible configurations of these components of the field.
Here $\vec\xperp$ and $\vec\yperp$ can be thought of as representing the coordinates of the endpoints of a color dipole, such as a quark-antiquark pair.

We're also going to encounter the quadrupole gluon distribution, which is a correlator of four Wilson lines,
\begin{equation}\label{eq:quadrupoleSdefinition}
 \quadrupoleS(\vec\xperp, \vec\bperp, \vec\yperp) = \frac{1}{\Nc^2}\cgcavg{\trace[U(\vec\xperp)\herm U(\vec\bperp)]\trace[U(\vec\bperp)\herm U(\vec\yperp)]}
\end{equation}
As before, $\vec\xperp$ and $\vec\yperp$ represent the coordinates of the endpoints of a parent color dipole, which has emitted a new dipole at transverse coordinate $\vec\bperp$, as shown in figure~\ref{fig:dipolecoords}.
This dipole splitting process will be discussed in more detail in section~\ref{sec:gluondynamics}.

As with the dipole distribution, this can be written in momentum space,
\begin{equation}\label{eq:quadrupoleGdefinition}
 \quadrupoleG(\vec\kperp, \vec\kpperp) = \int\frac{\uddc\vec\xperp\uddc\vec\yperp\uddc\vec\bperp}{(2\pi)^4}e^{-i\vec\kperp\cdot\vec(\xperp - \vec\bperp)}e^{-i\vec\kpperp\cdot(\vec\bperp - \vec\yperp)}\quadrupoleS(\vec\xperp, \vec\bperp, \vec\yperp)
\end{equation}

\subsection{Constructing Diagrams in the CGC}
Calculations using the CGC start with the corresponding diagrams in light cone perturbation theory.
In most cases, they involve at least some combination of the following three components:
\begin{itemize}
 \item the splitting wavefunction $\psi(p^+, k^+, \vec\rperp)$, which is a light cone wavefunction giving the amplitude for a \term{dressed quark} with plus-component momentum $p_+$ to contain a gluon with plus-component momentum $k^+ \defn (1 - \xisym)p^+$ and a dressed quark with plus-component momentum $\xisym p^+$ with transverse separation $\vec\rperp$.
 Effectively, this represents the emission of the gluon from the quark.
 \begin{center}
  \tikzsetnextfilename{splittingwavefunction}
  \begin{tikzpicture}
   \node[fill=gray!30,interaction] (interaction) at (0,0) {$\psi$};
   \draw[quark] (-1,0) -- (interaction) node[pos=0,left] {$p^+$};
   \draw[quark] (interaction) -- (40:1);
   \draw[gluon] (interaction) -- (-40:1) node[pos=1,below right] {$k^+$};
  \end{tikzpicture}
 \end{center}
 \item the fundamental Wilson line $U(\vec\xperp)$, from equation~\eqref{eq:wilsonline}, which represents the interaction between a quark at transverse position $\vec\xperp$ and the chromodynamic field of the nucleus. Again, this can be modeled as the superposition of all possible numbers of gluons interacting with the quark.
 \begin{center}
  \tikzsetnextfilename{fundamentalwilsonline}
  \begin{tikzpicture}
   \draw[quark] (0,0) -- (2,0) node[pos=0,left] {$\vec\xperp$};
   \draw[gluon] (0.5,0) -- +(0,-1);
   \draw[gluon] (1.5,0) -- +(0,-1);
   \node at (1,-.5) {$\cdots$};
  \end{tikzpicture}
 \end{center}
 \item the adjoint Wilson line $W(\vec\yperp)$, from equation~\eqref{eq:adjointwilsonline}, which is the same thing with a gluon as the propagating particle.
 \begin{center}
  \tikzsetnextfilename{adointwilsonline}
  \begin{tikzpicture}
   \draw[gluon] (0,0) node[left] {$\vec\yperp$} -- (1,0) node[fill=gray!30,interaction] (inta) {} -- (2,0) node[fill=gray!30,interaction] (intb) {} -- (3,0);
   \draw[gluon] (inta) -- +(0,-1);
   \draw[gluon] (intb) -- +(0,-1);
   \node at (1.5,-.5) {$\cdots$};
  \end{tikzpicture}
 \end{center}
\end{itemize}

As an example of how to put these pieces together, consider the process \HepProcess{\Pquark\Pnucleus \HepTo \Pquark\Pgluon\Pnucleus}, in which a quark from the projectile interacts with the target nucleus and emits a gluon.
These two events could happen in either order.
If the interaction with the gluon field of the nucleus comes first, the process is represented by the diagram
\begin{center}
 \tikzsetnextfilename{interactwithgluonfield}
 \begin{tikzpicture}
   \node[fill=gray!30,interaction] (interaction) at (0,0) {$\psi$};
   \draw[quark] (-2,0) -- (-1,0) node[pos=0,left] {$p^+$} node[pos=1,interaction,fill=gray!30] (cgc interaction) {};
   \draw[quark] (cgc interaction) -- (interaction);
   \draw[gluon] (cgc interaction) -- +(0,-1);
   \draw[quark] (interaction) -- (40:1);
   \draw[gluon] (interaction) -- (-40:1) node[pos=1,below right] {$k^+$};
 \end{tikzpicture}
\end{center}
This will be represented by the factor
\begin{equation}
 \psi(p^+, k^+, \vec\rperp)U(\vec\xperp)
\end{equation}

Alternatively, the interaction could take place after the emission of the gluon, in which case the chromodynamic field interacts with \emph{both} the quark and the emitted gluon.
\begin{center}
 \tikzsetnextfilename{interactwithbothfields}
 \begin{tikzpicture}
   \node[fill=gray!30,interaction] (interaction) at (0,0) {$\psi$};
   \draw[quark] (-1,0) -- (interaction) node[pos=0,left] {$p^+$};
   \draw[quark] (interaction) to[out=40,in=180] +(2,0.5) coordinate (quark endpoint);
   \draw[gluon] (interaction) to[out=-40,in=180] +(2,-0.5) coordinate (gluon endpoint);
   \coordinate (cgc interaction) at (1.5,0);
   \coordinate (quark cgc interaction) at (cgc interaction |- quark endpoint);
   \coordinate (gluon cgc interaction) at (cgc interaction |- gluon endpoint);
   \node[fill=gray!30,blob,minimum height=1.5cm,fit=(quark cgc interaction) (gluon cgc interaction)] (cgc interaction blob) {};
   \draw[gluon] (cgc interaction blob) -- ($(cgc interaction blob.south)+(0,-0.7)$);
 \end{tikzpicture}
\end{center}
We can think of this as the construction shown in figure~\ref{fig:quarkgluonincgc}: a superposition of all different numbers of nuclear gluons that could interact with the propagating partons. Mathematically, the superposition in figure~\ref{fig:quarkgluonincgc} corresponds to the expression of the form
\begin{equation}
 U_0 W_0 + U_1 W_0 + U_0 W_1 + U_2 W_0 + U_1 W_1 + U_0 W_2 + \cdots = UW
\end{equation}
So the interaction of the nuclear chromodynamic field with a quark and gluon propagating together is the product of the fundamental and adjoint Wilson lines. Accordingly, the expression corresponding to this diagram is
\begin{equation}
 \psi(p^+, k^+, \vec\xperp - \vec\bperp)U(\vec\xperp)W(\vec\bperp)
\end{equation}
where $\vec\xperp - \vec\bperp$ appears in the splitting wavefunction because it is the transverse separation between the fundamental and adjoint Wilson lines.

\begin{figure}
 \tikzsetnextfilename{quarkgluonincgc}
 \begin{tikzpicture}[baseline={(current bounding box.center)},wilson label/.style={}]
  \matrix[nodes={align=center},row sep=8mm,column sep=8mm]{
    \node[wilson label] at (0,0.5) {$U_0 W_0$};
    \draw[quark] (-1,0) -- (1,0);
    \draw[gluon] (-1,-0.5) -- (1,-0.5);
   &
    \node[wilson label] at (0,0.5) {$U_1 W_0$};
    \draw[quark] (-1,0) -- (1,0);
    \draw[gluon] (-1,-0.5) -- (1,-0.5);
    \draw[gluon] (0,0) -- +(0,-1.5);
   &
    \node[wilson label] at (0,0.5) {$U_2 W_0$};
    \draw[quark] (-1,0) -- (1,0);
    \draw[gluon] (-1,-0.5) -- (1,-0.5);
    \draw[gluon] (-0.25,0) -- +(0,-1.5);
    \draw[gluon] (0.25,0) -- +(0,-1.5);
   &
    \node[wilson label] at (0,0.5) {$U_3 W_0$};
    \draw[quark] (-1,0) -- (1,0);
    \draw[gluon] (-1.0,-0.5) -- (1.0,-0.5);
    \draw[gluon] (-0.5,0) -- +(0,-1.5);
    \draw[gluon] (0,0) -- +(0,-1.5);
    \draw[gluon] (0.5,0) -- +(0,-1.5);
   &
    \node at (0,-0.5) {$\cdots$};
   \\
    \node[wilson label] at (0,0.5) {$U_0 W_1$};
    \draw[quark] (-1,0) -- (1,0);
    \draw[gluon] (-1,-0.5) -- (1,-0.5);
    \draw[gluon] (0,-0.5) -- +(0,-1);
   &
    \node[wilson label] at (0,0.5) {$U_1 W_1$};
    \draw[quark] (-1,0) -- (1,0);
    \draw[gluon] (-1,-0.5) -- (1,-0.5);
    \draw[gluon] (-0.25,0) -- +(0,-1.5);
    \draw[gluon] (0.25,-0.5) -- +(0,-1);
   &
    \node[wilson label] at (0,0.5) {$U_2 W_1$};
    \draw[quark] (-1,0) -- (1,0);
    \draw[gluon] (-1.0,-0.5) -- (1.0,-0.5);
    \draw[gluon] (-0.5,0) -- +(0,-1.5);
    \draw[gluon] (0,-0.5) -- +(0,-1);
    \draw[gluon] (0.5,0) -- +(0,-1.5);
   &
    \node[wilson label] at (0,0.5) {$U_3 W_1$};
    \draw[quark] (-1,0) -- (1,0);
    \draw[gluon] (-1.0,-0.5) -- (1.0,-0.5);
    \draw[gluon] (-0.75,0) -- +(0,-1.5);
    \draw[gluon] (-0.25,-0.5) -- +(0,-1);
    \draw[gluon] (0.25,0) -- +(0,-1.5);
    \draw[gluon] (0.75,0) -- +(0,-1.5);
   &
    \node at (0,-0.5) {$\cdots$};
   \\
    \node[wilson label] at (0,0.5) {$U_0 W_2$};
    \draw[quark] (-1,0) -- (1,0);
    \draw[gluon] (-1,-0.5) -- (1,-0.5);
    \draw[gluon] (-0.25,-0.5) -- +(0,-1);
    \draw[gluon] (0.25,-0.5) -- +(0,-1);
   &
    \node[wilson label] at (0,0.5) {$U_1 W_2$};
    \draw[quark] (-1,0) -- (1,0);
    \draw[gluon] (-1.0,-0.5) -- (1.0,-0.5);
    \draw[gluon] (-0.5,-0.5) -- +(0,-1);
    \draw[gluon] (0,0) -- +(0,-1.5);
    \draw[gluon] (0.5,-0.5) -- +(0,-1);
   &
    \node[wilson label] at (0,0.5) {$U_2 W_2$};
    \draw[quark] (-1,0) -- (1,0);
    \draw[gluon] (-1.0,-0.5) -- (1.0,-0.5);
    \draw[gluon] (-0.75,0) -- +(0,-1.5);
    \draw[gluon] (-0.25,-0.5) -- +(0,-1);
    \draw[gluon] (0.25,0) -- +(0,-1.5);
    \draw[gluon] (0.75,-0.5) -- +(0,-1);
   &
    \node at (0,-0.5) {$\ddots$};
   \\
    \node[wilson label] at (0,0.5) {$U_0 W_3$};
    \draw[quark] (-1,0) -- (1,0);
    \draw[gluon] (-1.0,-0.5) -- (1.0,-0.5);
    \draw[gluon] (-0.5,-0.5) -- +(0,-1);
    \draw[gluon] (0,-0.5) -- +(0,-1);
    \draw[gluon] (0.5,-0.5) -- +(0,-1);
   &
    \node[wilson label] at (0,0.5) {$U_1 W_3$};
    \draw[quark] (-1,0) -- (1,0);
    \draw[gluon] (-1.0,-0.5) -- (1.0,-0.5);
    \draw[gluon] (-0.75,-0.5) -- +(0,-1);
    \draw[gluon] (-0.25,0) -- +(0,-1.5);
    \draw[gluon] (0.25,-0.5) -- +(0,-1);
    \draw[gluon] (0.75,-0.5) -- +(0,-1);
   &
    \node at (0,-0.5) {$\ddots$};
   \\
    \node{$\vdots$};
   &
    \node{$\vdots$};
   \\
  };
 \end{tikzpicture}
 \caption{Decomposition of the interaction of a quark and gluon with the chromodynamic field of a target}
 \label{fig:quarkgluonincgc}
\end{figure}
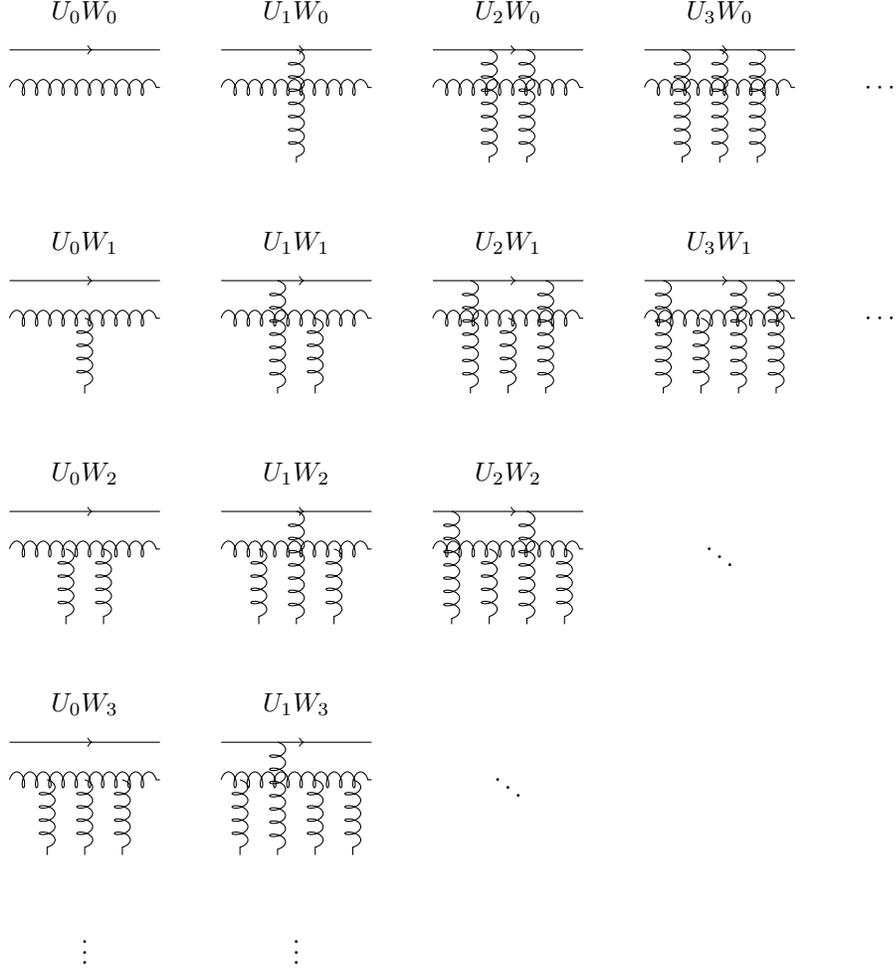

\subsection{Impact Parameter Independence}

A significant simplifying assumption, made in many studies of gluon distributions, is that the gluon field of the nucleus has an infinite extent in the transverse plane, and is homogeneous except for local fluctuations.
For these gluon distributions, this corresponds to the statement that the Wilson line correlators depend only on $\vec\rperp \defn \vec\xperp - \vec\yperp$ but are independent of the impact parameter $\vec\Rperp \defn \frac{1}{2}(\vec\xperp + \vec\yperp)$.%
\footnote{References~\cite{Berger:2010sh,Berger:2011ew,Berger:2011wx,Berger:2012wx,BergerThesis:2012} numerically investigate the consequences of dropping this assumption.}
If we make this assumption, the dipole distribution can be expressed as a function of $\vec\rperp$ only, and its Fourier transform~\eqref{eq:dipoleFdefinition} can be simplified to
\begin{equation}
 \dipoleF(\vec\kperp)
 = \int\frac{\uddc\vec\Rperp}{4}\int\frac{\uddc\vec\rperp}{(2\pi)^2} e^{-i\vec\kperp\cdot\vec\rperp}\dipoleS(\vec\rperp)
 = \Sperp\int\frac{\uddc\vec\rperp}{(2\pi)^2} e^{-i\vec\kperp\cdot\vec\rperp}\dipoleS(\vec\rperp)
\end{equation}
where $\Sperp$ is the transverse area of the target over which the interaction takes place.
We can express the inverse relation as
\begin{equation}\label{eq:dipoleSinvFT}
 \dipoleS(\vec\rperp) = \frac{1}{\Sperp} \int\uddc\vec\kperp e^{i\vec\kperp\cdot\vec\rperp}\dipoleF(\vec\kperp)
\end{equation}

For the quadrupole distribution, we can do the same thing with $\vec\sperp \defn \vec\xperp - \vec\bperp$ and $\vec\tperp \defn \vec\yperp - \vec\bperp$, where $\vec\bperp$ is the transverse coordinate of the emitted color pole).
The resulting inverse transform is
\begin{equation}\label{eq:quadrupoleSinvFT}
 \quadrupoleS(\vec\sperp, \vec\tperp) = \frac{1}{\Sperp}\int\uddc\vec\kperp\uddc\vec\kpperp e^{i\vec\kperp\cdot\vec\sperp}e^{i\vec\kpperp\cdot\vec\tperp}\quadrupoleG(\vec\kperp, \vec\kpperp)
\end{equation}
Assuming impact parameter independence also allows us to use the mean field approximation, which we'll encounter later~\eqref{eq:meanfieldappx}, to factor the momentum space quadrupole distribution as follows:
\begin{equation}
 \quadrupoleG(\vec\kperp, \vec\kpperp) = \dipoleF(\vec\kperp)\dipoleF(\vec\kpperp)
\end{equation}

\section{Dynamics of Small-\texorpdfstring{$x$}{x} Gluons}\label{sec:gluondynamics}

\begin{figure}
 \tikzsetnextfilename{ladderdiagram}
 \begin{tikzpicture}
  \draw[dashed] (0,-2.5) -- (0,2.5);
  \draw[quark] (-2,2) -- (2,2);
  \draw[quark] (2,-2) -- (-2,-2);
  \draw[gluon] (1,2) -- (1,-2);
  \draw[gluon] (-1,-2) -- (-1,2);
  \draw[gluon] (-1,1) -- (1,1);
  \draw[gluon] (-1,0) -- (1,0);
  \draw[gluon] (-1,-1) -- (1,-1);
 \end{tikzpicture}
 \caption[A ladder diagram]{A ``ladder diagram'' --- a compound diagram showing the interaction of gluons between two quarks. This can also be interpreted as in figure~\ref{fig:dipolesplitting}, as emission of a gluon from one of the quarks, where the gluon branches into other gluons with progressively smaller momentum fractions on its way to interacting with the other quark.}
 \label{fig:ladderdiagram}
\end{figure}
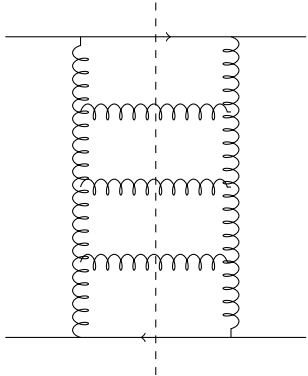
Even though parton distributions are fundamentally nonperturbative, it is possible to determine their behavior in high-energy collisions using perturbative QCD.
The effort to do so begins with the work of Balitsky, Fadin, Kuraev, and Lipatov in the late 1970s.
Their original derivations~\cite{Fadin:1975cb,Lipatov:1976zz,Kuraev:1976ge,Kuraev:1977fs,Balitsky:1978ic} focused on the behavior of cross sections, not specifically the unintegrated parton distributions, but the underlying physics is the same in both cases.
What they discovered is that the high-energy behavior of hadron scattering cross sections is primarily a result of \iterm{ladder diagrams}, of the type shown in figure~\ref{fig:ladderdiagram}.
These diagrams give contributions proportional to $(\alphas \ln\xg)^n$.
Even though $\alphas$ is small in these high-energy collisions, for sufficiently small values of $\xg$ the combination $\alphas\ln\xg$ can be large, so that a leading order ($n = 0$) or next-to-leading order ($n = 0,1$) approximation to the series won't be sufficient.
We need a \iterm{resummation} of these logarithmic contributions --- that is, a single expression giving the sum of the entire infinite series of diagrams --- to produce a good approximation to the cross section.

\begin{figure}
  \tikzsetnextfilename{dipolesplitting}
  \begin{tikzpicture}
   \path (0,0) -- (7,0) coordinate[pos=0.2] (branch point 1) node[pos=0,left] {target parton} coordinate[pos=1] (quark endpoint);
   \draw[quark,red] (0,0) -- (branch point 1);
   \path (branch point 1) +(5,-5) coordinate (interaction point);
   \draw[quark,green] (branch point 1) -- (quark endpoint);
   \draw[gluon,red!50!green] (branch point 1) .. controls +(0,-0.5) and +(-0.5,0) .. +(1,-1) coordinate (branch point 2);
   \draw[gluon,green!50!blue] (branch point 2) .. controls +(0.3,0.3) and +(-0.5,0) .. ++(1,0.5) -- +(3,0);
   \draw[gluon,red!50!blue] (branch point 2) .. controls +(0,-0.5) and +(-0.5,0) .. +(1,-1) coordinate (branch point 3);
   \draw[gluon,red!50!green] (branch point 3) .. controls +(0.3,0.3) and +(-0.5,0) .. ++(1,0.5) -- +(2,0);
   \draw[gluon,blue!50!green] (branch point 3) .. controls +(0,-0.5) and +(-0.5,0) .. +(1,-1) coordinate (branch point 4);
   \draw[gluon,red!50!blue] (branch point 4) .. controls +(0.3,0.3) and +(-0.5,0) .. ++(1,0.5) -- +(1,0);
   \draw[gluon,red!50!green] (branch point 4) .. controls +(0,-0.5) and +(-0.5,0) .. +(1,-1) coordinate (branch point 5);
   \draw[gluon,green!50!blue] (branch point 5) .. controls +(0.3,0.3) and +(-0.5,0) .. +(1,0.5);
   \draw[gluon,red!50!blue] (branch point 5) -- +(1,-1) coordinate (branch point 6);
   \node[below left=5mm] (large x label) at (branch point 1) {large $\xg$};
   \node[left=7mm] (small x label) at (branch point 5) {small $\xg$};
   \draw[-latex] (large x label) -- (small x label) node[pos=0.5,sloped,below] {loss of momentum};
   \path (interaction point) ++(-2,0) coordinate (start baryon);
   \draw[quark] (start baryon) to +(3,0);
   \node[left] at (start baryon) {projectile};
   
   \coordinate (upper edge) at ($(branch point 1)+(0,0.5)$);
   \coordinate (lower edge) at ($(start baryon)+(0,-0.5)$);
   
   \tikzset{
    dipole/.style={line width=1mm,opacity=0.7,line cap=round},
    cut connector/.style={help lines},
    cut/.style={densely dashed,help lines}
   }
   \begin{pgfonlayer}{background}
    \foreach \i in {1,...,5} {
     \pgfmathsetmacro{\j}{\i + 1}
     \coordinate (midpoint \i) at ($(branch point \i)!.5!(branch point \j)$);
     \draw[cut] (midpoint \i |- upper edge) -- (midpoint \i |- lower edge) coordinate (bottom dash endpoint \i);
    }
   \end{pgfonlayer}

   \begin{scope}[shift={(bottom dash endpoint 1)},shift={(-3cm,-1cm)},scale=0.5]
    \begin{pgfonlayer}{background}
     \node[draw=black,fill=protonbg,fill opacity=0.6,circle,minimum width=50pt] (dipole view) {};
     \draw[cut connector] (dipole view) -- (bottom dash endpoint 1);
    \end{pgfonlayer}
    \coordinate (parton-1) at (-15pt,4pt);
    \coordinate (parton-2) at (10pt,26pt);
    \coordinate (parton-3) at (17pt,-12pt);
    \coordinate (parton-4) at (30pt,-8pt);
    \coordinate (parton-5) at (-27pt,-10pt);
    \coordinate (parton-6) at (-18pt,16pt);
    \draw[dipole,red!50!green] (parton-1) -- (parton-2);
    \fill[green] (parton-1) circle(3pt);
    \fill[red] (parton-2) circle(1pt);
   \end{scope}
   \begin{scope}[shift={(bottom dash endpoint 2)},shift={(-2cm,-1.5cm)},scale=0.5]
    \begin{pgfonlayer}{background}
     \node[draw=black,fill=protonbg,fill opacity=0.6,circle,minimum width=50pt] (dipole view) {};
     \draw[cut connector] (dipole view) -- (bottom dash endpoint 2);
    \end{pgfonlayer}
    \coordinate (parton-1) at (-15pt,4pt);
    \coordinate (parton-2) at (10pt,26pt);
    \coordinate (parton-3) at (17pt,-12pt);
    \coordinate (parton-4) at (30pt,-8pt);
    \coordinate (parton-5) at (-27pt,-10pt);
    \coordinate (parton-6) at (-18pt,16pt);
    \draw[dipole,red!50!green] (parton-1) -- (parton-3);
    \draw[dipole,red!50!blue] (parton-2) -- (parton-3);
    \draw[dipole,green!50!blue] (parton-1) -- (parton-3);
    \fill[green] (parton-1) circle(3pt);
    \fill[red] (parton-2) circle(1pt);
    \fill[blue] (parton-3) circle(1pt);
   \end{scope}
   \begin{scope}[shift={(bottom dash endpoint 3)},shift={(-1cm,-2cm)},scale=0.5]
    \begin{pgfonlayer}{background}
     \node[draw=black,fill=protonbg,fill opacity=0.6,circle,minimum width=50pt] (dipole view) {};
     \draw[cut connector] (dipole view) -- (bottom dash endpoint 3);
    \end{pgfonlayer}
    \coordinate (parton-1) at (-15pt,4pt);
    \coordinate (parton-2) at (10pt,26pt);
    \coordinate (parton-3) at (17pt,-12pt);
    \coordinate (parton-4) at (30pt,-8pt);
    \coordinate (parton-5) at (-27pt,-10pt);
    \coordinate (parton-6) at (-18pt,16pt);
    \draw[dipole,green!50!blue] (parton-1) -- (parton-3);
    \draw[dipole,green!50!blue] (parton-3) -- (parton-4);
    \draw[dipole,red!50!green] (parton-2) -- (parton-4);
    \fill[green] (parton-1) circle(3pt);
    \fill[red] (parton-2) circle(1pt);
    \fill[blue] (parton-3) circle(1pt);
    \fill[green] (parton-4) circle(1pt);
   \end{scope}
   \begin{scope}[shift={(bottom dash endpoint 4)},shift={(0cm,-2.5cm)},scale=0.5]
    \begin{pgfonlayer}{background}
     \node[draw=black,fill=protonbg,fill opacity=0.6,circle,minimum width=50pt] (dipole view) {};
     \draw[cut connector] (dipole view) -- (bottom dash endpoint 4);
    \end{pgfonlayer}
    \coordinate (parton-1) at (-15pt,4pt);
    \coordinate (parton-2) at (10pt,26pt);
    \coordinate (parton-3) at (17pt,-12pt);
    \coordinate (parton-4) at (30pt,-8pt);
    \coordinate (parton-5) at (-27pt,-10pt);
    \coordinate (parton-6) at (-18pt,16pt);
    \draw[dipole,green!50!blue] (parton-1) -- (parton-3);
    \draw[dipole,red!50!green] (parton-2) -- (parton-4);
    \draw[dipole,red!50!green] (parton-4) -- (parton-5);
    \draw[dipole,red!50!blue] (parton-3) -- (parton-5);
    \fill[green] (parton-1) circle(3pt);
    \fill[red] (parton-2) circle(1pt);
    \fill[blue] (parton-3) circle(1pt);
    \fill[green] (parton-4) circle(1pt);
    \fill[red] (parton-5) circle(1pt);
   \end{scope}
   \begin{scope}[shift={(bottom dash endpoint 5)},shift={(1cm,-3cm)},scale=0.5]
    \tikzset{dipole/.style={line width=1mm,opacity=0.7,line cap=round}}
    \begin{pgfonlayer}{background}
     \node[draw=black,fill=protonbg,fill opacity=0.6,circle,minimum width=50pt] (dipole view) {};
     \draw[cut connector] (dipole view) -- (bottom dash endpoint 5);
    \end{pgfonlayer}
    \coordinate (parton-1) at (-15pt,4pt);
    \coordinate (parton-2) at (10pt,26pt);
    \coordinate (parton-3) at (17pt,-12pt);
    \coordinate (parton-4) at (30pt,-8pt);
    \coordinate (parton-5) at (-27pt,-10pt);
    \coordinate (parton-6) at (-18pt,16pt);
    \draw[dipole,green!50!blue] (parton-1) -- (parton-3);
    \draw[dipole,red!50!green] (parton-2) -- (parton-4);
    \draw[dipole,red!50!blue] (parton-3) -- (parton-5);
    \draw[dipole,red!50!blue] (parton-5) -- (parton-6);
    \draw[dipole,green!50!blue] (parton-4) -- (parton-6);
    \fill[green] (parton-1) circle(3pt);
    \fill[red] (parton-2) circle(1pt);
    \fill[blue] (parton-3) circle(1pt);
    \fill[green] (parton-4) circle(1pt);
    \fill[red] (parton-5) circle(1pt);
    \fill[blue] (parton-6) circle(1pt);
   \end{scope}
  \end{tikzpicture}
 \caption[A color dipole cascade, or gluon branching, process]{A color dipole cascade, which is another way of looking at one side of a ladder diagram (figure~\ref{fig:ladderdiagram}). As the gluon (the color dipole) branches, moving from left to right, the lowest gluon has progressively smaller fractions $\xg$ of the target's longitudinal momentum. Each circle at the bottom shows a view, in the transverse plane, of the color dipole cascade at the corresponding value of $\xg$ indicated by the dashed line through the top part of the diagram.}
 \label{fig:dipolesplitting}
\end{figure}
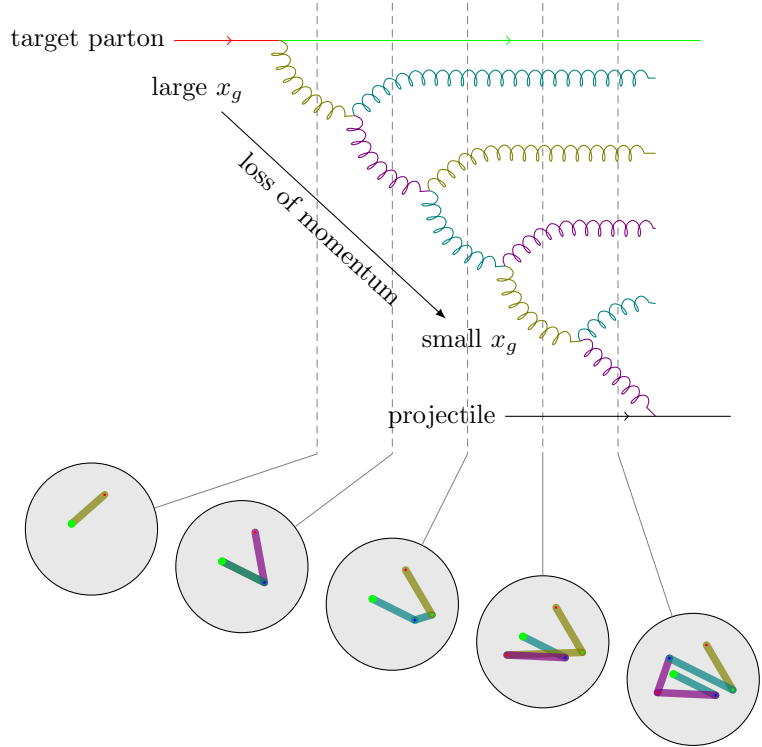

This resummation is provided, in a sense, by the BFKL equation, which is often written in terms of the dipole scattering amplitude~\eqref{eq:dipoleNdefinition} as
\begin{equation}\label{eq:BFKLN}
 \pd{\dipoleN(\vec\rperp, \xg)}{\ln(1/\xg)} =
 \frac{\alphas\Nc}{2\pi^2}\int\uddc\vec\bperp \frac{\rperp^2}{\sperp^2\tperp^2}\bigl[\dipoleN(\vec\sperp, \xg) + \dipoleN(\vec\tperp, \xg) - \dipoleN(\vec\rperp, \xg)\bigr]
\end{equation}
(this is, properly speaking, only the leading logarithmic, angle-independent approximation to the full BFKL equation).
Intuitively, this can be interpreted as the color dipole splitting process pictured in figure~\ref{fig:dipolesplitting}.
As we step toward smaller values of the gluon momentum fraction $\xg$, an existing color dipole of size $\vec\rperp$ (corresponding to a gluon, or a quark-antiquark pair) disappears and two new ones of sizes $\vec\sperp$ and $\vec\tperp$ take its place.

\subsection{Unitarity Violation}

The BFKL equation describes the behavior of parton distributions at small $\xg$, corresponding to high collision energy $\mandelstams$, under a range of kinematic conditions, but it's not difficult to see that it cannot be a \emph{complete} description.
It's possible to solve the equation for $\dipoleN$ in terms of the eigenfunctions of the integral kernel, which are simply powers of the dipole size, i.e.
\begin{equation}
 \int\uddc\vec\bperp \frac{\rperp^2}{\sperp^2\tperp^2} \bigl[\sperp^\gamma + \tperp^\gamma - \rperp^\gamma\bigr] = 2\pi\bigl[\underbrace{2\digamma(1) - \digamma(\gamma) - \digamma(1 - \gamma)}_{\chi(\gamma)}\bigr]\rperp^\gamma
\end{equation}
where $\digamma(x) = \ud{\ln\Gamma(x)}{x}$.
A general solution of~\eqref{eq:BFKLN} is a linear combination of these functions,
\begin{equation}
 \dipoleN(\vec\rperp, \xg) = \int_{-\infty}^{\infty} \udc\nu C(\nu, \xg) \rperp^\gamma
\end{equation}
and plugging this into~\eqref{eq:BFKLN} gives us the $\xg$ evolution of the coefficients:
\begin{equation}
 \pd{C(\gamma, \xg)}{\ln(1/\xg)} = \frac{\alphas\Nc}{\pi} C(\gamma, \xg)\chi(\gamma)
 \qquad\implies\qquad
 C(\gamma, \xg) = \biggl(\frac{1}{\xg}\biggr)^{\alphas\Nc\chi(\gamma)/\pi}
\end{equation}
So as $\xg$ gets smaller, $C(\gamma, \xg)$ grows exponentially.
The largest contribution to this exponential growth comes from $\gamma = \frac{1}{2}$, where $\chi(\gamma) = 4\ln 2$.
When we run through the mathematical machinery to convert $\dipoleN$ into a cross section, we can expect that, in the limit of large collision energy $\mandelstams$, the cross section will grow as $\sigma \sim \mandelstams^{4\ln 2 \alphas\Nc/\pi - 1}$.~\cite{QCDHighEnergy}

This growth conflicts with an elementary result of quantum field theory called the Froissart-Martin bound~\cite{Froissart:1961ux,Lukaszuk:1967zz,MartinUnitarity}, which says that the unitarity of the S-matrix (equivalent to conservation of probability) implies that cross sections cannot grow any faster than $(\ln\mandelstams)^2$ in the high-energy limit.
So there must be at least another term in the equation that acts to limit the growth of the solution as $\dipoleN$ becomes large.

One can solve this problem by incorporating a negative nonlinear term which can be interpreted as representing multiple scatterings of the probe off target partons, or as interactions between the partons in the target prior to the scattering.
This solution was first proposed by Gribov, Levin, and Ryskin~\cite{Gribov:1981ac} (reviewed in~\cite{Gribov:1984tu}), and eventually developed into the Balitsky-Kovchegov (BK) equation~\cite{Balitsky:1995ub,Balitsky:1998kc,Balitsky:1998ya,Balitsky:2001re,Kovchegov:1999yj,Kovchegov:1999ua}:
\begin{equation}\label{eq:BKN}
 \pd{\dipoleN(\vec\rperp, \xg)}{\ln(1/\xg)} =
 \frac{\alphas\Nc}{2\pi^2}\int\uddc\vec\bperp \frac{\rperp^2}{\sperp^2\tperp^2}\bigl[\dipoleN(\vec\sperp, \xg) + \dipoleN(\vec\tperp, \xg) - \dipoleN(\vec\rperp, \xg) - \dipoleN(\vec\sperp, \xg)\dipoleN(\vec\tperp, \xg)\bigr]
\end{equation}
The nonlinear term in this equation limits the growth of $\dipoleN$ as $\xg$ becomes small.
Because of that term, $\dipoleN(\vec\rperp, \xg) = 1$ is a fixed point of this equation, so the nonlinear evolution keeps $\dipoleN < 1$ which is sufficient to preserve the Froissart bound.

\subsection{Saturation}

The BK equation exhibits an interesting feature called \iterm{geometric scaling}, which we can see most easily by rewriting the equation in momentum space.
In terms of the momentum space distribution $\dipolephi$, the leading logarithmic BK equation takes the form~\cite{Kovchegov:1999ua}
\begin{equation}\label{eq:BKphichi}
 \pd{\dipolephi(\kperp, \xg)}{\ln(1/\xg)} = \frac{\alphas\Nc}{\pi}\biggl[\chi\biggl(-\pd{}{\ln\kperp^2}\biggr)\dipolephi(\kperp) - \dipolephi(\kperp)^2\biggr]
\end{equation}
We have seen that the high-energy behavior of the BK solution is dominated by $\chi(\gamma)$ near $\gamma = \frac{1}{2}$, so we can substitute in the expansion
\begin{equation}
 \chi(\gamma) = 4\ln 2 + \frac{\chi_2}{2}\biggl(\gamma - \frac{1}{2}\biggr)^2 + \orderof{\gamma^4}
\end{equation}
where $\chi_2 \defn \chi\bigl(\frac{1}{2}\bigr)$, to write~\eqref{eq:BKphichi} as~\cite{Munier:2003vc,Munier:2003sj,Munier:2004xu}
\begin{equation}
 \pd{\dipolephi(\kperp, \xg)}{\ln(1/\xg)} = \frac{\alphas\Nc}{\pi}\biggl[(4\ln 2)\dipolephi(\kperp, \xg) + \frac{\chi_2}{2}\biggl(\pd{}{\ln\kperp^2} + \frac{1}{2}\biggr)^2\dipolephi(\kperp, \xg) - \dipolephi(\kperp, \xg)^2\biggr]
\end{equation}
This general form is known as the Fisher-Komolgorov-Petrov-Piscounov (FKPP) equation~\cite{Fisher1937,Kolmogorov1937}.
Its asymptotic solutions, in the limit of large $\ln\frac{1}{\xg}$, can be written as a one-variable function $\kperp^2 / \Qs^2$, where the \iterm{saturation scale} $\Qs$ marks the transition between the high-momentum region where $\phi$ is small, and the low-momentum region where it is large.
For a proton, the dependence of this variable on $\xg$ works out to~\cite{Munier:2003vc}
\begin{equation}
 \Qs^2\biggl(\ln\frac{1}{\xg}\biggr) = Q_0^2 \biggl(\ln\frac{1}{\xg}\biggr)^{\frac{3}{2(1 - \gamma_0)}}\exp\biggl(-\frac{\alphas\Nc}{\pi}\frac{\chi(\gamma_0)}{1 - \gamma_0}\ln\frac{1}{\xg}\biggr)
\end{equation}
with $\gamma_0 = 1 - \frac{1}{2}\sqrt{1 + 32\ln 2/\chi_2}$ and $Q_0$ some constant.
Geometric scaling is the observation that $\phi$ depends on $\kperp$ and $\xg$ only through the combination $\kperp^2 / \Qs^2$.
This scaling behavior is also exhibited by deep inelastic scattering data from HERA~\cite{Stasto:2000er}, supporting the idea that the BK equation is physically relevant to these processes.\iftime{also show running coupling result}


For a nucleus, recall that $\Qs^2$ is enhanced by a factor of $\centrality \massnumber^{1/3}$, where $\centrality$ is the centrality parameter of the collision and $\massnumber$ is the mass number of the nucleus.

\subsection{Dipole Derivation}

Having explored the properties of BFKL and BK evolution, let's examine its derivation in more detail.
A commonly referenced derivation of the BFKL equation%
\footnote{In most introductory sources, such as~\cite{Salam:1999cn,PinkBook,QCDHighEnergy}, the BFKL equation is presented as a generalization of the DGLAP equation. However, this dissertation only concerns BFKL dynamics, and the derivation I present stands on its own, so I've omitted any discussion of DGLAP evolution.}
comes from the color dipole model, proposed by Mueller in reference~\cite{Mueller:1993rr}.
In this section I'll present a somewhat simplified argument from reference~\cite{Gelis:2012ri}, which is nevertheless equivalent to the dipole model derivation and illustrates its essential features.

\begin{figure}
 \tikzsetnextfilename{lophotonxsec}
 \begin{tikzpicture}[scale=2]
   \coordinate (upper quark) at (0,0.5);
   \coordinate (left photon) at (-2.5,0);
   \node[blob,transform shape] (nucleus) at (0,-2) {};
   \coordinate[label={right:$\psi$}] (left splitting) at ($(left photon)+(1,0)$);
   \coordinate[label={left:$\conj\psi$}] (right splitting) at ($(left splitting)!2!(left splitting -| upper quark)$);
   \coordinate (right photon) at ($(left photon)!2!(left photon -| upper quark)$);
   \coordinate (lower quark) at ($(upper quark)!2!(upper quark |- left photon)$);
   \coordinate[label={below:$U(\vec\xperp)$}] (upper interaction) at (nucleus |- upper quark);
   \coordinate[label={above:$\herm U(\vec\yperp)$}] (lower interaction) at (nucleus |- lower quark);

   
   \draw[photon] (left photon) node[left] {$p^+$} to (left splitting);
   \draw[photon] (right splitting) to (right photon) node[right] {$p^+$};
   \draw[quark] (left splitting) .. controls +(80:0.5) and +(180:1) .. (upper interaction) node[above left,pos=0.5] {$i$} coordinate[pos=0.3] (gluon emission) coordinate[pos=0.7] (gluon absorption);
   \draw[quark] (upper interaction) .. controls +(0:1) and +(100:0.5) .. (right splitting) node[above right,pos=0.5] {$j$};
   \draw[quark] (right splitting) .. controls +(-100:0.5) and +(0:1) .. (lower interaction) node[below right,pos=0.5] {$k$};
   \draw[quark] (lower interaction) .. controls +(180:1) and +(-80:0.5) .. (left splitting) node[below left,pos=0.5] {$l$};
   \draw[nucleus] (left photon |- nucleus) to (nucleus);
   \draw[nucleus] (nucleus) to (nucleus -| right photon);
%
   \node[blob,fill=gray,fill opacity=0.2,fit=(upper interaction) (lower interaction)] (gluon blob) {};
   \draw[small gluon] (gluon blob) to (nucleus);
%
   \node[fill=gluondist!60!white,blob,transform shape] at (nucleus) {};
   
 \end{tikzpicture}
 \caption{Leading order diagram for interaction of a photon with a nucleus}
 \label{fig:lophotonxsec}
\end{figure}
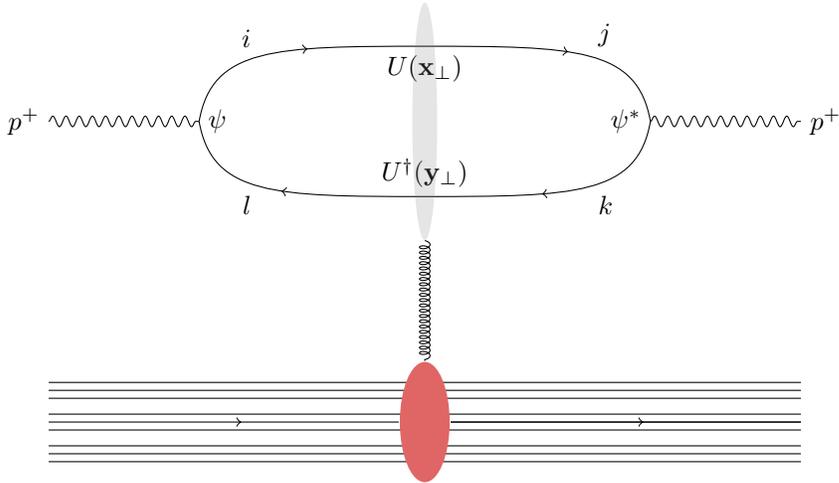

Consider the process in which a photon fluctuates into a quark-antiquark pair which interacts with the gluon field of the nucleus and then turns back into a photon.%
\note{Would this also apply without fluctating back into a photon?}
The leading order diagram for this process is shown in figure~\ref{fig:lophotonxsec}.
Following the rules of light cone perturbation theory, we construct the corresponding amplitude by starting at an arbitrary point on the quark loop and following it around, picking up a factor for each interaction we encounter as we go:
\begin{itemize}
 \item The splitting of the photon into the quark-antiquark pair contributes a factor of the photon splitting wavefunction $\psi_{li}(p^+, k^+, \vec\xperp - \vec\yperp)$, where $k^+$ is the longitudinal momentum of the quark (or we could choose it to be the antiquark)
 \item The interaction of the upper quark line with the gluon field of the nucleus contributes the Wilson line $U_{ij}(\vec\xperp)$
 \item The reforming of the photon from the quark-antiquark pair contributes the conjugated photon splitting wavefunction $\conj{\psi_{jk}}(p^+, k^+, \vec\xperp - \vec\yperp)$
 \item The interaction of the lower quark line with the gluon field contributes the conjugate Wilson line $\herm U_{kl}(\vec\yperp)$
\end{itemize}
The color indices on each factor come from the two lines that connect to it.
We don't include propagators for the photons because they are external lines.
Putting everything together, we get
\begin{equation}
 \psi_{li}(p^+, k^+, \vec\rperp)U_{ij}(\vec\xperp)\conj{\psi_{jk}}(p^+, k^+, \vec\rperp)\herm U_{kl}(\vec\yperp)
\end{equation}
Since photons are color-neutral, their splitting wavefunctions are diagonal in the color indices, $\psi_{ij} \propto \delta_{ij}$, so we can simplify this to
\begin{equation}\label{eq:lodipolexsec}
 \abs{\psi(p^+, k^+, \vec\rperp)}^2 \trace[U(\vec\xperp)\herm U(\vec\yperp)]
\end{equation}

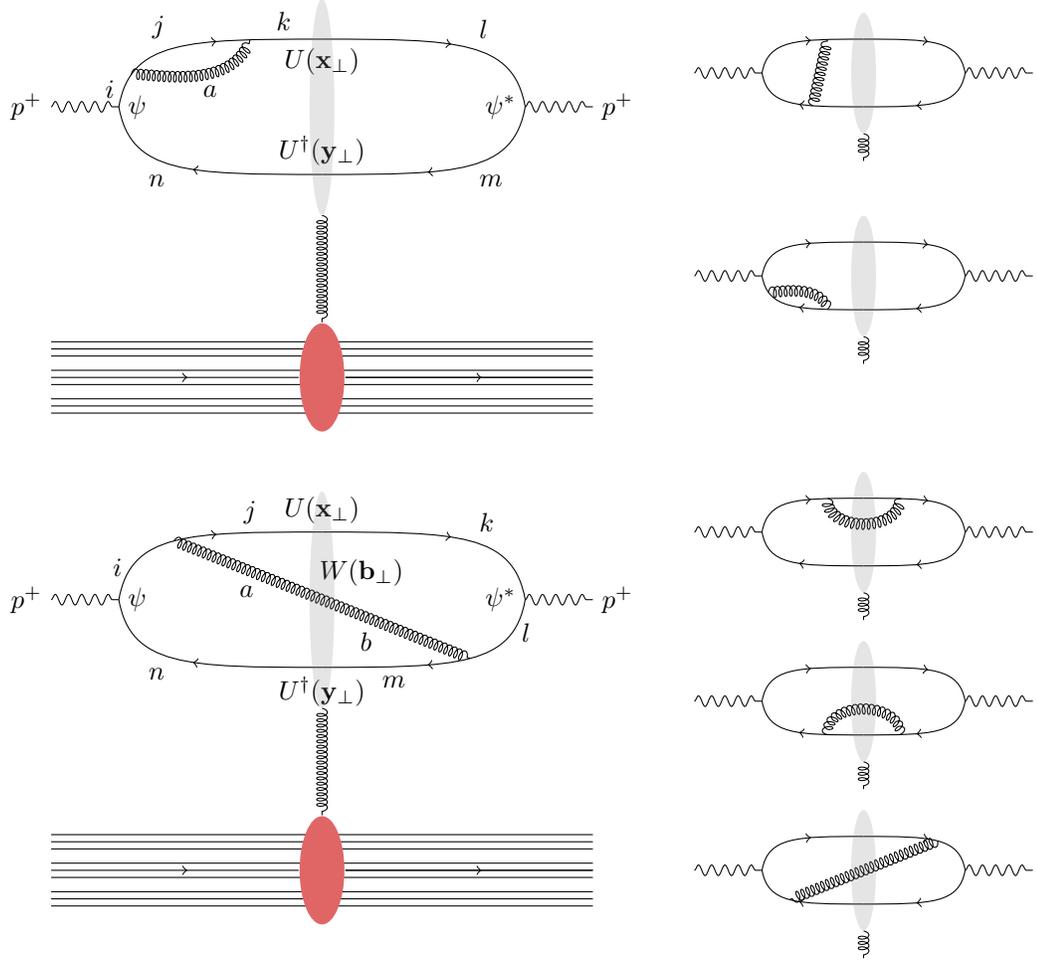
\begin{figure}
 \tikzsetnextfilename{nlophotonxsec-virtual}
 \begin{tikzpicture}[scale=0.9]
  \begin{scope}[scale=2]
   \coordinate (upper quark) at (0,0.5);
   \coordinate (left photon) at (-2,0);
   \node[blob,transform shape] (nucleus) at (0,-2) {};
   \coordinate[label={right:$\psi$}] (left splitting) at ($(left photon)+(0.5,0)$);
   \coordinate[label={left:$\conj\psi$}] (right splitting) at ($(left splitting)!2!(left splitting -| upper quark)$);
   \coordinate (right photon) at ($(left photon)!2!(left photon -| upper quark)$);
   \coordinate (lower quark) at ($(upper quark)!2!(upper quark |- left photon)$);
   \coordinate[label={below:$U(\vec\xperp)$}] (upper interaction) at (nucleus |- upper quark);
   \coordinate[label={above:$\herm U(\vec\yperp)$}] (lower interaction) at (nucleus |- lower quark);
   \draw[photon] (left photon) node[left] {$p^+$} to (left splitting);
   \draw[photon] (right splitting) to (right photon) node[right] {$p^+$};
   \draw[quark] (left splitting) .. controls +(80:0.5) and +(180:1) .. (upper interaction) node[left,pos=0.1] {$i$} coordinate[pos=0.2] (gluon emission) node[above left,pos=0.5] {$j$} coordinate[pos=0.8] (gluon absorption) node[above,pos=0.9] {$k$};
   \draw[quark] (upper interaction) .. controls +(0:1) and +(100:0.5) .. (right splitting) node[above right,pos=0.5] {$l$};
   \draw[quark] (right splitting) .. controls +(-100:0.5) and +(0:1) .. (lower interaction) node[below right,pos=0.5] {$m$};
   \draw[quark] (lower interaction) .. controls +(180:1) and +(-80:0.5) .. (left splitting) node[below left,pos=0.5] {$n$};
   \draw[nucleus] (left photon |- nucleus) to (nucleus);
   \draw[nucleus] (nucleus) to (nucleus -| right photon);
   \draw[small gluon] (gluon emission) .. controls +(0:0.2) and +(-110:0.4) .. (gluon absorption) node[below right,pos=0.5] {$a$};
   \node[blob,fill=gray,fill opacity=0.2,fit=(upper interaction) (lower interaction)] (gluon blob) {};
   \draw[small gluon] (gluon blob) to (nucleus);
   \node[fill=gluondist!60!white,blob,transform shape] at (nucleus) {};
  \end{scope}
  \begin{scope}[shift={(8cm,0.5cm)}]
   \coordinate (upper quark) at (0,0.5);
   \coordinate (left photon) at (-2.5,0);
   \coordinate (left splitting) at ($(left photon)+(1,0)$);
   \coordinate (right splitting) at ($(left splitting)!2!(left splitting -| upper quark)$);
   \coordinate (right photon) at ($(left photon)!2!(left photon -| upper quark)$);
   \coordinate (lower quark) at ($(upper quark)!2!(upper quark |- left photon)$);
   \coordinate (upper interaction) at (upper quark);
   \coordinate (lower interaction) at (lower quark);
   \draw[photon] (left photon) to (left splitting);
   \draw[photon] (right splitting) to (right photon);
   \draw[quark] (left splitting) .. controls +(80:0.5) and +(180:1) .. (upper interaction) coordinate[pos=0.8] (gluon absorption);
   \draw[quark] (upper interaction) .. controls +(0:1) and +(100:0.5) .. (right splitting);
   \draw[quark] (right splitting) .. controls +(-100:0.5) and +(0:1) .. (lower interaction);
   \draw[quark] (lower interaction) .. controls +(180:1) and +(-80:0.5) .. (left splitting) coordinate[pos=0.3] (gluon emission);
   \draw[small gluon] (gluon emission) -- (gluon absorption);
   \node[blob,fill=gray,fill opacity=0.2,fit=(upper interaction) (lower interaction)] (gluon blob) {};
   \draw[small gluon] (gluon blob) to +(0,-1.3);
  \end{scope}
  \begin{scope}[shift={(8cm,-2.5cm)}]
   \coordinate (upper quark) at (0,0.5);
   \coordinate (left photon) at (-2.5,0);
   \coordinate (left splitting) at ($(left photon)+(1,0)$);
   \coordinate (right splitting) at ($(left splitting)!2!(left splitting -| upper quark)$);
   \coordinate (right photon) at ($(left photon)!2!(left photon -| upper quark)$);
   \coordinate (lower quark) at ($(upper quark)!2!(upper quark |- left photon)$);
   \coordinate (upper interaction) at (upper quark);
   \coordinate (lower interaction) at (lower quark);
   \draw[photon] (left photon) to (left splitting);
   \draw[photon] (right splitting) to (right photon);
   \draw[quark] (left splitting) .. controls +(80:0.5) and +(180:1) .. (upper interaction);
   \draw[quark] (upper interaction) .. controls +(0:1) and +(100:0.5) .. (right splitting);
   \draw[quark] (right splitting) .. controls +(-100:0.5) and +(0:1) .. (lower interaction);
   \draw[quark] (lower interaction) .. controls +(180:1) and +(-80:0.5) .. (left splitting) coordinate[pos=0.8] (gluon emission) coordinate[pos=0.2] (gluon absorption);
   \draw[small gluon] (gluon emission) .. controls +(0:0.2) and +(110:0.4) .. (gluon absorption);
   \node[blob,fill=gray,fill opacity=0.2,fit=(upper interaction) (lower interaction)] (gluon blob) {};
   \draw[small gluon] (gluon blob) to +(0,-1.3);
  \end{scope}
 \end{tikzpicture}
 
 \vspace{5mm}
 
 \tikzsetnextfilename{nlophotonxsec-real}
 \begin{tikzpicture}[scale=0.9]
  \begin{scope}[scale=2]
   \coordinate (upper quark) at (0,0.5);
   \coordinate (left photon) at (-2,0);
   \node[blob,transform shape] (nucleus) at (0,-2) {};
   \coordinate[label={right:$\psi$}] (left splitting) at ($(left photon)+(0.5,0)$);
   \coordinate[label={left:$\conj\psi$}] (right splitting) at ($(left splitting)!2!(left splitting -| upper quark)$);
   \coordinate (right photon) at ($(left photon)!2!(left photon -| upper quark)$);
   \coordinate (lower quark) at ($(upper quark)!2!(upper quark |- left photon)$);
   \coordinate[label={above:$U(\vec\xperp)$}] (upper interaction) at (nucleus |- upper quark);
   \coordinate[label={below:$\herm U(\vec\yperp)$}] (lower interaction) at (nucleus |- lower quark);
   \draw[photon] (left photon) node[left] {$p^+$} to (left splitting);
   \draw[photon] (right splitting) to (right photon) node[right] {$p^+$};
   \draw[quark] (left splitting) .. controls +(80:0.5) and +(180:1) .. (upper interaction) node[left,pos=0.2] {$i$} coordinate[pos=0.5] (gluon emission) node[above,pos=0.8] {$j$};
   \draw[quark] (upper interaction) .. controls +(0:1) and +(100:0.5) .. (right splitting) node[above right,pos=0.5] {$k$};
   \draw[quark] (right splitting) .. controls +(-100:0.5) and +(0:1) .. (lower interaction) node[right,pos=0.2] {$l$} coordinate[pos=0.5] (gluon absorption) node[below,pos=0.8] {$m$};
   \draw[quark] (lower interaction) .. controls +(180:1) and +(-80:0.5) .. (left splitting) node[below left,pos=0.5] {$n$};
   \draw[nucleus] (left photon |- nucleus) to (nucleus);
   \draw[nucleus] (nucleus) to (nucleus -| right photon);
   \draw[small gluon] (gluon emission) -- (gluon absorption) node[below left,pos=0.3] {$a$} node[above right,pos=0.46] {$W(\vec\bperp)$} node[below left,pos=0.7] {$b$};
   \node[blob,fill=gray,fill opacity=0.2,fit=(upper interaction) (lower interaction)] (gluon blob) {};
   \draw[small gluon] (gluon blob) to (nucleus);
   \node[fill=gluondist!60!white,blob,transform shape] at (nucleus) {};
  \end{scope}
  \begin{scope}[shift={(8cm,1cm)}]
   \coordinate (upper quark) at (0,0.5);
   \coordinate (left photon) at (-2.5,0);
   \coordinate (left splitting) at ($(left photon)+(1,0)$);
   \coordinate (right splitting) at ($(left splitting)!2!(left splitting -| upper quark)$);
   \coordinate (right photon) at ($(left photon)!2!(left photon -| upper quark)$);
   \coordinate (lower quark) at ($(upper quark)!2!(upper quark |- left photon)$);
   \coordinate (upper interaction) at (upper quark);
   \coordinate (lower interaction) at (lower quark);
   \draw[photon] (left photon) to (left splitting);
   \draw[photon] (right splitting) to (right photon);
   \draw[quark] (left splitting) .. controls +(80:0.5) and +(180:1) .. (upper interaction) coordinate[pos=0.8] (gluon emission);
   \draw[quark] (upper interaction) .. controls +(0:1) and +(100:0.5) .. (right splitting) coordinate[pos=0.2] (gluon absorption);
   \draw[quark] (right splitting) .. controls +(-100:0.5) and +(0:1) .. (lower interaction);
   \draw[quark] (lower interaction) .. controls +(180:1) and +(-80:0.5) .. (left splitting);
   \draw[small gluon] (gluon emission) .. controls +(-90:0.5) and +(-90:0.5) .. (gluon absorption);
   \node[blob,fill=gray,fill opacity=0.2,fit=(upper interaction) (lower interaction)] (gluon blob) {};
   \draw[small gluon] (gluon blob) to +(0,-1.3);
  \end{scope}
  \begin{scope}[shift={(8cm,-1.5cm)}]
   \coordinate (upper quark) at (0,0.5);
   \coordinate (left photon) at (-2.5,0);
   \coordinate (left splitting) at ($(left photon)+(1,0)$);
   \coordinate (right splitting) at ($(left splitting)!2!(left splitting -| upper quark)$);
   \coordinate (right photon) at ($(left photon)!2!(left photon -| upper quark)$);
   \coordinate (lower quark) at ($(upper quark)!2!(upper quark |- left photon)$);
   \coordinate (upper interaction) at (upper quark);
   \coordinate (lower interaction) at (lower quark);
   \draw[photon] (left photon) to (left splitting);
   \draw[photon] (right splitting) to (right photon);
   \draw[quark] (left splitting) .. controls +(80:0.5) and +(180:1) .. (upper interaction);
   \draw[quark] (upper interaction) .. controls +(0:1) and +(100:0.5) .. (right splitting);
   \draw[quark] (right splitting) .. controls +(-100:0.5) and +(0:1) .. (lower interaction) coordinate[pos=0.8] (gluon absorption);
   \draw[quark] (lower interaction) .. controls +(180:1) and +(-80:0.5) .. (left splitting) coordinate[pos=0.2] (gluon emission);
   \draw[small gluon] (gluon emission) .. controls +(90:0.5) and +(90:0.5) .. (gluon absorption);
   \node[blob,fill=gray,fill opacity=0.2,fit=(upper interaction) (lower interaction)] (gluon blob) {};
   \draw[small gluon] (gluon blob) to +(0,-1.3);
  \end{scope}
  \begin{scope}[shift={(8cm,-4cm)}]
   \coordinate (upper quark) at (0,0.5);
   \coordinate (left photon) at (-2.5,0);
   \coordinate (left splitting) at ($(left photon)+(1,0)$);
   \coordinate (right splitting) at ($(left splitting)!2!(left splitting -| upper quark)$);
   \coordinate (right photon) at ($(left photon)!2!(left photon -| upper quark)$);
   \coordinate (lower quark) at ($(upper quark)!2!(upper quark |- left photon)$);
   \coordinate (upper interaction) at (upper quark);
   \coordinate (lower interaction) at (lower quark);
   \draw[photon] (left photon) to (left splitting);
   \draw[photon] (right splitting) to (right photon);
   \draw[quark] (left splitting) .. controls +(80:0.5) and +(180:1) .. (upper interaction);
   \draw[quark] (upper interaction) .. controls +(0:1) and +(100:0.5) .. (right splitting) coordinate[pos=0.5] (gluon emission);
   \draw[quark] (right splitting) .. controls +(-100:0.5) and +(0:1) .. (lower interaction);
   \draw[quark] (lower interaction) .. controls +(180:1) and +(-80:0.5) .. (left splitting) coordinate[pos=0.5] (gluon absorption);
   \draw[small gluon] (gluon emission) -- (gluon absorption);
   \node[blob,fill=gray,fill opacity=0.2,fit=(upper interaction) (lower interaction)] (gluon blob) {};
   \draw[small gluon] (gluon blob) to +(0,-1.3);
  \end{scope}
 \end{tikzpicture}
 \caption[Next-to-leading order diagrams for a dipole interacting with a gluon field]{Next-to-leading order diagrams contributing to the interaction of a color dipole with the gluon field of a nucleus. In the top set of diagrams, the emitted gluon does not interact with the nucleus, whereas in the bottom set of diagrams it does, adding an adjoint Wilson line to the amplitude. In the small diagrams on the right, the nucleus itself is not shown to save space.}
 \label{fig:nlophotonxsec}
\end{figure}

The next-to-leading order corrections to this are represented by diagrams with a gluon line, shown in figure~\ref{fig:nlophotonxsec}.
Each emission or absorption of a gluon brings in a vertex factor of $2\gs\generator^a\frac{\vec\epsilon_{\lambda}\cdot\vec\qperp}{\qperp^2}$, where $\vec\qperp$ is the transverse momentum difference between the quark and gluon; however, to be consistent with our other expressions we should exchange transverse momentum for transverse position.
Performing the Fourier transform gets us to the splitting factor~\cite{Gelis:2012ri}\note{Why is this not the same as the splitting wavefunction in e.g. Chirilli, Xiao, Yuan~\cite{Chirilli:2012jd}?}
\begin{equation}
 \int\frac{\uddc\vec\qperp}{(2\pi)^2}e^{i\vec\qperp\cdot\vec\rperp} 2\gs\generator^a\frac{\vec\epsilon_{\lambda}\cdot\vec\qperp}{\qperp^2}
 =
 \frac{2i\gs}{2\pi}\generator^a\frac{\vec\epsilon_{\lambda}\cdot\vec\rperp}{\rperp^2}
\end{equation}
It's also necessary to sum over gluon polarizations using equation~\eqref{eq:ident:polarization}.

Putting all this together, the top left diagram in figure~\ref{fig:nlophotonxsec} gives the expression
\begin{multline}
 \int^{k^+}\frac{\udc q^+}{q^+} \int\uddc\vec\bperp \psi_{ni}(p^+, k^+, \vec\rperp)\biggl(\frac{2i\gs}{2\pi}\generator_{ij}^a\frac{\vec\epsilon_{\lambda}\cdot\vec\sperp}{\sperp^2}\biggr) \\
 \times \biggl(-\frac{2i\gs}{2\pi}\generator_{jk}^a\frac{\vec\epsilon_{\lambda}\cdot\vec\sperp}{\sperp^2}\biggr)U_{kl}(\vec\xperp)\conj{\psi_{lm}}(p^+, k^+, \vec\rperp)\herm{U_{mn}}(\vec\yperp)
\end{multline}
where all repeated indices are summed over.
The upper limit on the $q^+$ integral comes from the constraint that the gluon momentum can't be greater than that of the quark it was emitted from.

\note{this argument is highly questionable}
Performing the integral over $q^+$ gives $\ln k^+ + \text{constant}$.
Now, the quark momentum $k^+$ is related to the momentum of the target gluons it interacts with: high-momentum quarks will interact with high-momentum gluons, and same for low momentum.
We can therefore conclude that $k^+ \propto \xg$, and rewrite $\ln k^+ + \text{constant}$ as $-\ln(1/\xg) + \text{constant}$.
A suitable shift in reference frame then allows us to eliminate the constant.

Taking into account that $\psi_{ij} \propto \delta_{ij}$, and that the polarizations are \note{how can we say this when gluons are virtual?}orthogonal to the gluon momentum so that the last term in~\eqref{eq:ident:polarization} vanishes, this becomes
\begin{equation}
 -\frac{4\alphas}{\pi} \ln\frac{1}{\xg} \abs{\psi(\vec\rperp)}^2\trace[\generator^a \generator^a U(\vec\xperp)\herm{U}(\vec\yperp)] \int\frac{\uddc\vec\bperp}{(2\pi)^2}\frac{1}{\sperp^2}
\end{equation}
where $\alphas = \frac{\gs^2}{4\pi}$ and $\vec\sperp = \vec\xperp - \vec\bperp$.
By a similar procedure, the bottom left diagram gives
\begin{multline}
 \int^{k^+}\frac{\udc q^+}{q^+} \int\uddc\vec\bperp \psi_{ni}(p^+, k^+, \vec\rperp) \biggl(\frac{2i\gs}{2\pi}\generator_{ij}^a\frac{\vec\epsilon_{\lambda}\cdot\vec\sperp}{\sperp^2}\biggr) \\
 \times U_{jk}(\vec\xperp) \conj{\psi_{kl}}(p^+, k^+, \vec\rperp) \biggl(-\frac{2i\gs}{2\pi}\generator_{lm}^b\frac{\vec\epsilon_{\lambda}\cdot\vec\tperp}{\tperp^2}\biggr) \herm{U_{mn}}(\vec\yperp) W^{ab}(\vec\bperp)
\end{multline}
which reduces to
\begin{equation}
 -\frac{4\alphas}{\pi} \ln\frac{1}{\xg} \abs{\psi(\vec\rperp)}^2\trace[\generator^a U(\vec\xperp) \generator^b \herm{U}(\vec\yperp)] \int\frac{\uddc\vec\bperp}{(2\pi)^2}\frac{\vec\sperp\cdot\vec\tperp}{\sperp^2\tperp^2}W^{ab}(\vec\bperp)
\end{equation}

When all seven diagrams in figure~\ref{fig:nlophotonxsec} are taken into account, the total NLO contribution, from explicit gluon emission, works out to
\begin{multline}\label{eq:nlodipolexsec}
 -\frac{\alphas}{2\pi^2}\ln\frac{1}{\xg}\abs{\psi(\vec\rperp)}^2\int\uddc\vec\bperp \frac{\rperp^2}{\sperp^2\tperp^2}\Bigl\{\Nc\trace[U(\vec\xperp)\herm U(\vec\yperp)] \\
 - \trace[U(\vec\xperp)\herm U(\vec\bperp)]\trace[U(\vec\bperp)\herm U(\vec\yperp)]\Bigr\}
\end{multline}

Recall that the leading order contribution~\eqref{eq:lodipolexsec} was
\begin{equation*}
 \abs{\psi(p^+, k^+, \vec\rperp)}^2 \trace[U(\vec\xperp)\herm U(\vec\yperp)]
\end{equation*}
This expression includes the contribution from emission of gluons with momentum fractions \emph{less} than $\xg$, which are resummed by the Wilson lines.
Therefore, the incremental contribution from emission of gluons with momentum fraction \emph{equal} to $\xg$, within a range $\Delta\bigl[\ln\frac{1}{\xg}\bigr]$, will be
\begin{equation}\label{eq:bkleftside}
 \abs{\psi(p^+, k^+, \vec\rperp)}^2\pd{\trace[U(\vec\xperp)\herm U(\vec\yperp)]}{\ln(1/\xg)}\Delta\biggl[\ln\frac{1}{\xg}\biggr]
\end{equation}
However, this is precisely the same quantity represented by the single gluon emission contributions~\eqref{eq:nlodipolexsec}, if we take the difference between two closely spaced values of $\xg$:
\begin{multline}\label{eq:bkrightside}
 -\frac{\alphas}{2\pi^2}\Delta\biggl[\ln\frac{1}{\xg}\biggr] \abs{\psi(\vec\rperp)}^2\int\uddc\vec\bperp \frac{\rperp^2}{\sperp^2\tperp^2}\Bigl\{\Nc\trace[U(\vec\xperp)\herm U(\vec\yperp)] \\
 - \trace[U(\vec\xperp)\herm U(\vec\bperp)]\trace[U(\vec\bperp)\herm U(\vec\yperp)]\Bigr\}
\end{multline}
Equating expressions~\eqref{eq:bkleftside} and~\eqref{eq:bkrightside}, we find the following evolution equation for the Wilson loop:
\begin{multline}
 \pd{\trace[U(\vec\xperp)\herm U(\vec\yperp)]}{\ln(1/\xg)}
 =
 -\frac{\alphas}{2\pi^2} \int\uddc\vec\bperp \frac{\rperp^2}{\sperp^2\tperp^2}\Bigl\{\Nc\trace[U(\vec\xperp)\herm U(\vec\yperp)] \\
 - \trace[U(\vec\xperp)\herm U(\vec\bperp)]\trace[U(\vec\bperp)\herm U(\vec\yperp)]\Bigr\}
\end{multline}

\subsubsection{CGC Target Averaging}

Technically, this equation only applies to one specific quantum state of the gluon field of the nucleus.
But we can't prepare a nucleus in a specific state; in a collider we get an average over all possible states.
So to get an equation that describes physically measurable results, we need to use the target averaging procedure $\cgcavg{\cdots}$ to get
\begin{multline}
 \pd{\cgcavg{\trace[U(\vec\xperp)\herm U(\vec\yperp)]}}{\ln(1/\xg)} = -\frac{\alphas}{2\pi^2}\int\uddc\vec\bperp \frac{\rperp^2}{\sperp^2\tperp^2}\Bigl\{\Nc\cgcavg{\trace[U(\vec\xperp)\herm U(\vec\yperp)]} \\
 - \cgcavg{\trace[U(\vec\xperp)\herm U(\vec\bperp)]\trace[U(\vec\bperp)\herm U(\vec\yperp)]}\Bigr\}
\end{multline}
or, in terms of the dipole~\eqref{eq:dipoleSdefinition} and quadrupole~\eqref{eq:quadrupoleSdefinition} gluon distributions,
\begin{equation}
 \pd{\dipoleS(\vec\rperp)}{\ln(1/\xg)} = -\frac{\alphas\Nc}{2\pi^2}\int\uddc\vec\bperp \frac{\rperp^2}{\sperp^2\tperp^2}\Bigl[\dipoleS(\vec\rperp) - \quadrupoleS(\vec\sperp, \vec\tperp)\Bigr]
\end{equation}
This differential equation will allow us to compute the evolution of $\dipoleS$ in $\xg$, provided that we can compute $\quadrupoleS$.
So we might be inspired to look for a corresponding equation for $\quadrupoleS$.
This equation does exist, but it expresses the evolution of $\quadrupoleS$ in terms of $\sextupoleS$, a correlator of two fundamental and \emph{two} adjoint Wilson lines, $\trace[U\herm{U} W \herm{W}]$.
And the evolution equation for $\sextupoleS$ would be expressed in terms of $\trace[U\herm{U}WW\herm{W}]$, and so on.
This series of equations involving progressively larger Wilson loops is called the Balitsky-JIMWLK hierarchy~\cite{Balitsky:1995ub,JalilianMarian:1997jx,JalilianMarian:1997gr,Iancu:2000hn,Iancu:2001ad,Weigert:2000gi}.

In practice, to avoid the full complexity of the Balitsky-JIMWLK hierarchy, we can use the \iterm{mean field approximation}, which is simply the assumption that the two Wilson loops in the quadrupole gluon distribution are uncorrelated, and thus the target average of the product, $\cgcavg{\trace[U\herm{U}]\trace[U\herm{U}]}$ factors into the product of two averages, $\cgcavg{\trace[U\herm{U}]}\cgcavg{\trace[U\herm{U}]}$.
The validity of this assumption in the limit of large $\Nc$ is supported by numerical studies~\cite{Rummukainen:2003ns,Kovchegov:2008mk,Alvioli:2012ba}.
This allows us to reduce the quadrupole gluon distribution as follows:
\begin{equation}\label{eq:meanfieldappx}
 \quadrupoleS(\vec\xperp, \vec\bperp, \vec\yperp) = \dipoleS(\vec\xperp, \vec\bperp)\dipoleS(\vec\bperp, \vec\yperp)
\end{equation}
so that the evolution equation is a self-contained differential equation for the dipole distribution
\begin{equation}\label{eq:BKS}
 \pd{\dipoleS(\vec\rperp)}{\ln(1/\xg)} = -\frac{\alphas\Nc}{\pi}\int\uddc\vec\bperp \frac{\rperp^2}{\sperp^2\tperp^2}\Bigl[\dipoleS(\vec\rperp) - \dipoleS(\vec\sperp)\dipoleS(\vec\tperp)\Bigr]
\end{equation}
This is called the Balitsky-Kovchegov (BK) equation~\cite{Balitsky:1995ub,Balitsky:1998kc,Balitsky:1998ya,Balitsky:2001re,Kovchegov:1999yj,Kovchegov:1999ua} in the leading logarithmic (LL) approximation.

The designation as LL reflects the fact that there is a single factor of $\alphas$ for each factor of $\ln\frac{1}{\xg}$.
One can go beyond this and incorporate subleading terms which are proportional to $\alphas\biggl(\alphas\ln\frac{1}{\xg}\biggr)^n$, producing the next-to-leading logarithmic (NLL) BK equation~\cite{Fadin:1998py,Ciafaloni:1998gs,Balitsky:2008zza}, though the numerical implementation of the full NLL BK evolution is quite complex.
Some progress has been made toward such an implementation but it is not yet ready for use in phenomenology.~\cite{HeikkiPrivate2014}

It turns out that incorporating selected NLL corrections, specifically those which result from the running of the QCD coupling, is expected to give a reasonable approximation to the full NLL evolution.
Many numerical studies, including those to be discussed in the remainder of this dissertation, use this rcBK (BK with running coupling) gluon distribution.
A widely used sample implementation is available in e.g. reference~\cite{Albacete:2009fh}.

\subsection{Numerical Analysis of BK Evolution}\label{sec:bknumerics}

\begin{figure}
 \tikzsetnextfilename{dipoleFplot}
 \begin{tikzpicture}
  \begin{groupplot}[
    group style={group size=1 by 3,xlabels at=edge bottom},
    xmode=log,ymode=log,domain=1e-8:1e2,samples=100,xmin=1e-8,xmax=1e6,ymin=1e-8,ymax=1e3,xlabel={$\kperp^2 [\si{GeV^2}]$},ylabel=$\dipoleF(\kperp)$,width=14cm,height=6cm
  ]
  \nextgroupplot
   \addplot+[black,no markers] {0.172853*exp(-\x/1.84151)} node[pos=0.04,above,font=\tiny] {$0$};
   \addplot+[no markers] {0.170575*exp(-\x/1.86610)}; 
   \addplot+[no markers] {0.168328*exp(-\x/1.89101)}; 
   \addplot+[no markers] {0.161762*exp(-\x/1.96777)}; 
   \addplot+[no markers] {0.151383*exp(-\x/2.10269)}; 
   \addplot+[no markers,domain=1e-8:1e2] {0.0458855*exp(-\x/6.93704)} node[pos=0.1,font=\tiny,fill=white] {$100$};
   \addplot+[no markers,domain=1e-8:1e3] {0.0121808*exp(-\x/26.1321)} node[pos=0.12,font=\tiny,fill=white] {$200$};
   \addplot+[no markers,domain=1e-8:1e4] {0.00323352*exp(-\x/98.4406)} node[pos=0.14,font=\tiny,fill=white] {$300$};
  \nextgroupplot
   \addplot+[black,no markers] file {datafiles/Fg/BK-Pb85-000.dat} node[pos=0.04,font=\tiny,fill=white] {$0$};
   \addplot+[no markers] file {datafiles/Fg/BK-Pb85-001.dat} node[pos=0,below,font=\tiny] {$1$} node[pos=0.5,sloped,below,font=\tiny] {$1$};
   \addplot+[no markers] file {datafiles/Fg/BK-Pb85-002.dat} node[pos=0,below,font=\tiny] {$2$};
   \addplot+[no markers] file {datafiles/Fg/BK-Pb85-005.dat} node[pos=0,left,font=\tiny] {$5$};
   \addplot+[no markers] file {datafiles/Fg/BK-Pb85-010.dat} node[pos=0,below,font=\tiny] {$10$};
   \addplot+[no markers] file {datafiles/Fg/BK-Pb85-100.dat} node[pos=0,below,font=\tiny] {$100$} node[pos=0.5,sloped,font=\tiny,fill=white] {$100$};
   \addplot+[no markers] file {datafiles/Fg/BK-Pb85-200.dat} node[pos=0,left,font=\tiny] {$200$} node[pos=0.55,sloped,font=\tiny,fill=white] {$200$};
   \addplot+[no markers] file {datafiles/Fg/BK-Pb85-300.dat} node[pos=0,right,font=\tiny] {$300$} node[pos=0.6,sloped,font=\tiny,fill=white] {$300$};
  \nextgroupplot
   \addplot+[black,no markers] file {datafiles/Fg/BK-Pb85-run-000.dat} node[pos=0.04,font=\tiny,fill=white] {$0$};
   \addplot+[no markers] file {datafiles/Fg/BK-Pb85-run-001.dat} node[pos=0,below,font=\tiny] {$1$} node[pos=0.5,sloped,below,font=\tiny] {$1$};
   \addplot+[no markers] file {datafiles/Fg/BK-Pb85-run-002.dat} node[pos=0,below,font=\tiny] {$2$};
   \addplot+[no markers] file {datafiles/Fg/BK-Pb85-run-005.dat} node[pos=0,below,font=\tiny] {$5$};
   \addplot+[no markers] file {datafiles/Fg/BK-Pb85-run-010.dat} node[pos=0,below,font=\tiny] {$10$};
   \addplot+[no markers] file {datafiles/Fg/BK-Pb85-run-100.dat} node[pos=0,below,font=\tiny] {$100$} node[pos=0.5,sloped,font=\tiny,fill=white] {$100$};
   \addplot+[no markers] file {datafiles/Fg/BK-Pb85-run-200.dat} node[pos=0,below right,font=\tiny] {$200$} node[pos=0.5,sloped,font=\tiny,fill=white] {$200$};
   \addplot+[no markers] file {datafiles/Fg/BK-Pb85-run-300.dat} node[pos=0,below,font=\tiny] {$300$} node[pos=0.5,sloped,font=\tiny,fill=white] {$300$};
  \end{groupplot}
 \end{tikzpicture}
 \caption[GBW, BK, and rcBK gluon distributions, solved in momentum space]{The gluon distribution from the GBW model (top), and the output of the numerical BK evolution with fixed coupling (middle) and running coupling (bottom), for central ($c = 0.85$) \pPb{} ($A = 208$) collisions. Each curve shows $\dipoleF(\kperp)$ for $\xg = x_\text{init}\exp(-n\delta Y)$, where $n$ is the number that labels the curve and $\delta Y = \frac{1}{400}\ln(10^8) \approx 0.04605$ is the step size in rapidity used by the numerical integrator. The curve labeled $0$ is the initial condition. Notable features include the transient ``spike'' at the low $\kperp^2$ region of the solution to the BK evolution (which is a numerical artifact), and the significant enhancement at large $\kperp^2$ relative to the GBW model. This figure is adapted from reference~\cite{Stasto:2012ru}.}
 \label{fig:dipoleFplot}
\end{figure}
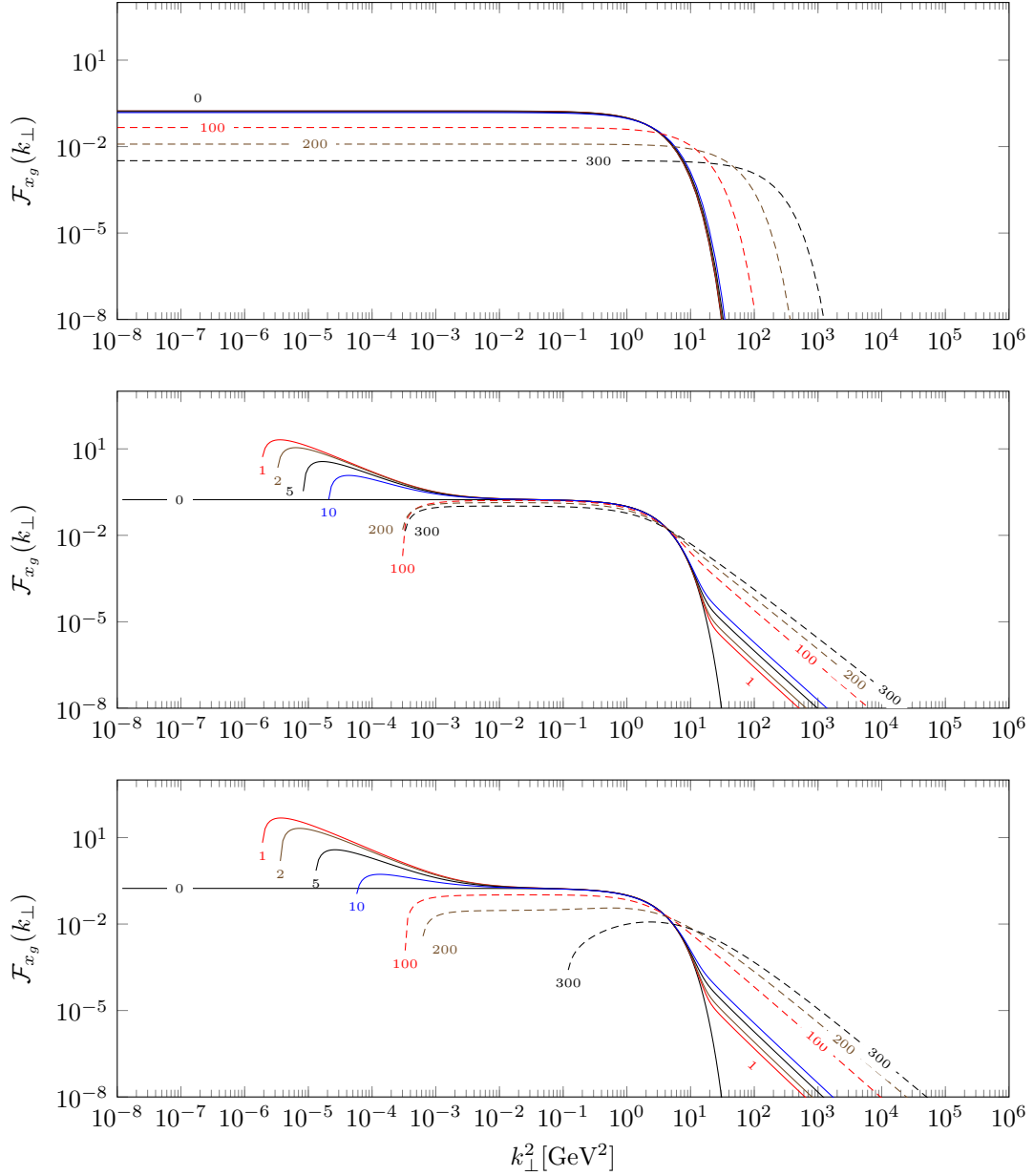

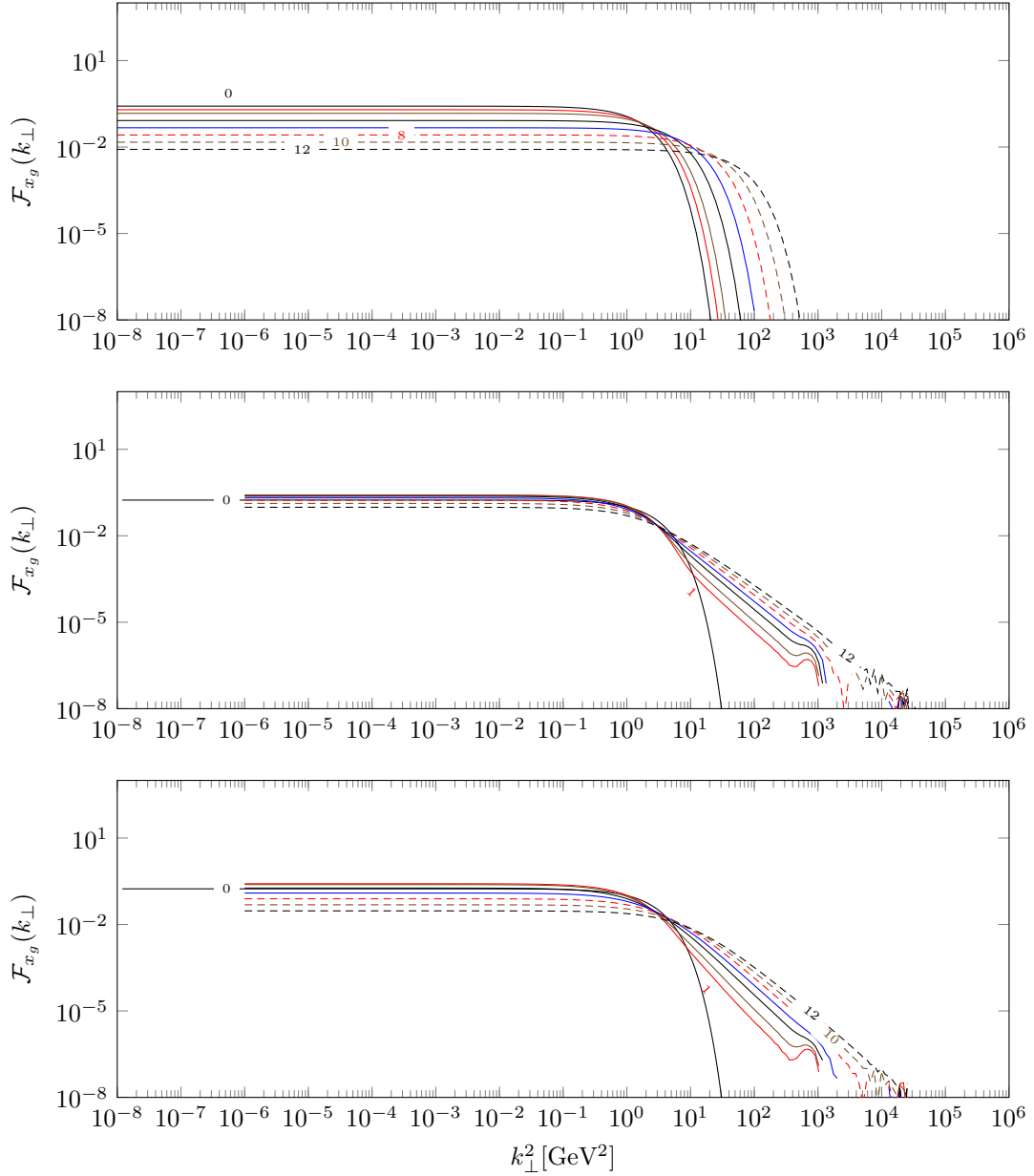
\begin{figure}
 \tikzsetnextfilename{dipoleNmomentumspace}
 \begin{tikzpicture}
  \begin{groupplot}[
    group style={group size=1 by 3,xlabels at=edge bottom},
    xmode=log,ymode=log,domain=1e-8:1e2,samples=100,xmin=1e-8,xmax=1e6,ymin=1e-8,ymax=1e3,xlabel={$\kperp^2 [\si{GeV^2}]$},ylabel=$\dipoleF(\kperp)$,width=14cm,height=6cm
  ]
  \nextgroupplot
   \addplot+[black,no markers] {0.262366*exp(-\x/1.21323)} node[pos=0.04,above,font=\tiny] {$0$};
   \addplot+[no markers] {0.196712*exp(-\x/1.61816)};
   \addplot+[no markers] {0.147487*exp(-\x/2.15823)};
   \addplot+[no markers] {0.0829086*exp(-\x/3.83929)};
   \addplot+[no markers] {0.0466064*exp(-\x/6.82974)};
   \addplot+[no markers,domain=1e-8:1e3] {0.0261995*exp(-\x/12.1495)} node[pos=0.1,font=\tiny,fill=white] {$8$};
   \addplot+[no markers,domain=1e-8:1e3] {0.0147278*exp(-\x/21.6128)} node[pos=0.12,font=\tiny,fill=white] {$10$};
   \addplot+[no markers,domain=1e-8:1e3] {0.00827914*exp(-\x/38.4472)} node[pos=0.14,font=\tiny,fill=white] {$12$};
  \nextgroupplot[restrict x to domain=-8:6,log basis x=10]
   \addplot+[black,no markers] file {datafiles/Fg/BK-Pb85-000.dat} node[pos=0.04,font=\tiny,fill=white] {$0$};
   \addplot+[no markers] file {datafiles/kovp/BK-Pb85-001.dat} node[pos=0.5,sloped,below,font=\tiny] {$1$};
   \addplot+[no markers] file {datafiles/kovp/BK-Pb85-002.dat};
   \addplot+[no markers] file {datafiles/kovp/BK-Pb85-004.dat};
   \addplot+[no markers] file {datafiles/kovp/BK-Pb85-006.dat};
   \addplot+[no markers] file {datafiles/kovp/BK-Pb85-008.dat};
   \addplot+[no markers] file {datafiles/kovp/BK-Pb85-010.dat};
   \addplot+[no markers] file {datafiles/kovp/BK-Pb85-012.dat} node[pos=0.45,sloped,font=\tiny,fill=white] {$12$};
  \nextgroupplot[restrict x to domain=-8:6,log basis x=10]
   \addplot+[black,no markers] file {datafiles/Fg/BK-Pb85-run-000.dat} node[pos=0.04,font=\tiny,fill=white] {$0$};
   \addplot+[no markers] file {datafiles/kovp/BK-Pb85-run-001.dat} node[pos=0.45,sloped,below,font=\tiny] {$1$};
   \addplot+[no markers] file {datafiles/kovp/BK-Pb85-run-002.dat};
   \addplot+[no markers] file {datafiles/kovp/BK-Pb85-run-004.dat};
   \addplot+[no markers] file {datafiles/kovp/BK-Pb85-run-006.dat};
   \addplot+[no markers] file {datafiles/kovp/BK-Pb85-run-008.dat};
   \addplot+[no markers] file {datafiles/kovp/BK-Pb85-run-010.dat} node[pos=0.5,sloped,font=\tiny,fill=white] {$10$};
   \addplot+[no markers] file {datafiles/kovp/BK-Pb85-run-012.dat} node[pos=0.51,sloped,font=\tiny,fill=white] {$12$};
  \end{groupplot}
 \end{tikzpicture}
 \caption[GBW, BK, and rcBK gluon distributions, solved in position space and Fourier transformed]{Analogous to figure~\ref{fig:dipoleFplot}, these plots show the dipole gluon distribution in momentum space, $\dipoleF$, computed by solving the equation in position space~\eqref{eq:BKN} and numerically Fourier transforming it into momentum space. The top plot shows the analytic result from the GBW model, the middle one shows the result from the BK equation with fixed coupling $\alphas = 0.1$~\eqref{eq:BKN}, and the bottom one shows the result with running coupling corrections incorporated. Each curve in this plot shows $\dipoleF(\kperp)$ for $\xg = x_\text{init}\exp(-n)$, where $n$ is the number that labels the curve, so the numbers labeling the curves are not directly comparable to those in figure~\ref{fig:dipoleFplot}. In addition, these curves are for minimum bias ($\centrality = 0.56$) \pPb{} ($\massnumber = 208$) collisions, so the normalization is not the same as in figure~\ref{fig:dipoleFplot}, though the trend of the small-$x$ evolution is clear. The most notable feature of these plots is the numerical instability at high $\kperp$ which results from the Fourier transform.}
 \label{fig:dipoleNmomentumspace}
\end{figure}

Given the importance of small-$x$ evolution in calculating cross sections, it should come as no surprise that there are many, many numerical studies of the solutions to both the BFKL and BK equation (as well as the Balitsky-JIMWLK hierarchy, but that goes beyond the scope of this dissertation).
One can either solve the equation in position space, using the form~\eqref{eq:BKN} directly, or convert it to momentum space, in terms of $\dipolephi$, and solve that.
The latter approach is taken in reference~\cite{GolecBiernat:2001if}, where the momentum space BK equation is written as
\begin{equation}\label{eq:BKphi}
 \pd{\dipolephi(\kperp, \xg)}{\ln(1/\xg)} = \frac{\alphas\Nc}{\pi}\int_0^\infty\frac{\udc\kpperp^2}{\kpperp^2}\biggl[\frac{\kpperp^2\dipolephi(\kpperp, \xg) - \kperp^2\dipolephi(\kperp, \xg)}{\abs{\kperp^2 - \kpperp^2}} + \frac{\kperp^2\dipolephi(\kperp, \xg)}{\sqrt{4\kpperp^4 + \kperp^4}}\biggr] - \frac{\alphas\Nc}{\pi}\dipolephi(\kperp)^2
\end{equation}
Then a numerical Fourier transform allows us to convert either solution to the other.

Figures~\ref{fig:dipoleFplot} and~\ref{fig:dipoleNmomentumspace} show two computations of $\dipoleF$: in figure~\ref{fig:dipoleFplot}, the solution is computed from the momentum space BK equation~\eqref{eq:BKphi} and then $\dipoleF$ is calculated using equation~\eqref{eq:dipoleFfromphi}, whereas the plots in figure~\ref{fig:dipoleNmomentumspace} come from the solution to the position space BK equation~\eqref{eq:BKN} which is then Fourier transformed.
These two specific sets of plots have different normalization, so they are not directly comparable, but looking at the mid-$\kperp$ region we can see some slight differences in how they depart from the initial condition as $\xg$ gets smaller.

One important difference between the two solutions is the present of numerical instabilities at large $\kperp$ in the Fourier transform from the position space solution (figure~\ref{fig:dipoleNmomentumspace}).
The instability results from floating-point roundoff error in the Fourier transform.
This limits the validity of this solution to moderate values of $\kperp$, though this will not be a problem for the calculations we will use this solution for in chapter~\ref{ch:crosssection}.
Conversely, in the direct momentum-space solution (figure~\ref{fig:dipoleFplot}), we see a spike at low $\kperp$, which is a numerical artifact that owes its existence to the lower boundary condition on $\kperp$.
In this case, the solution can only be used above a certain momentum, though again, this is not a problem for the calculations in chapter~\ref{ch:correlation} because it only samples the gluon distribution in the moderate $\kperp$ range.

We can characterize the speed of the BK evolution by finding the \term{saturation scale} $\Qs$ as a function of $\ln\frac{1}{\xg}$.
In the \term{GBW model}, of course, this is purely an exponential dependence,
\begin{equation}
 \Qs^2\biggl(\ln\frac{1}{\xg}\biggr) = Q_0^2 x_0^\lambda \exp\biggl(\lambda\ln\frac{1}{\xg}\biggr)
\end{equation}
The analytic solution to the BK equation predicts~\cite{Munier:2003vc,Munier:2003sj,Mueller:2002zm} a more complicated form,
\begin{equation}
 \Qs^2\biggl(\ln\frac{1}{\xg}\biggr) = Q_0^2 \biggl(\ln\frac{1}{\xg}\biggr)^{\frac{3}{2(1 - \gamma_0)}}\exp\biggl(-\frac{\alphas\Nc}{\pi}\frac{\chi(\gamma_0)}{1 - \gamma_0}\ln\frac{1}{\xg}\biggr)
\end{equation}

\begin{figure}
 \tikzsetnextfilename{satscale}
 \begin{tikzpicture}
  \begin{axis}[ymode=log,xlabel={$\ln\frac{1}{\xg}$},ylabel=$\Qs$,legend pos=north west]
   \addplot+[domain=0:20,no markers] {0.56*pow(208,1./3.)*pow(0.000304,0.288)*exp(0.288*\x)};
   \addplot+[mark size=1pt] table[x=Yg,y=Qs] {datafiles/Qsdependence/Qs_kovp_Lambda2_01.dat};
   \legend{GBW,rcBK}
  \end{axis}
 \end{tikzpicture}
 \caption[GBW and rcBK saturation scales]{Saturation scales for the GBW model and the solution to the BK equation with running coupling, solved in position space and Fourier transformed into momentum space, extracted by finding $\kperp$ at which $\dipoleF(\kperp) = 0.5\dipoleF(0)$. The flat portion at the left end of the plot is where the initial condition of the BK evolution is still strongly influencing the result. We can see that the evolution is slower in the BK solution.}
 \label{fig:satscale}
\end{figure}
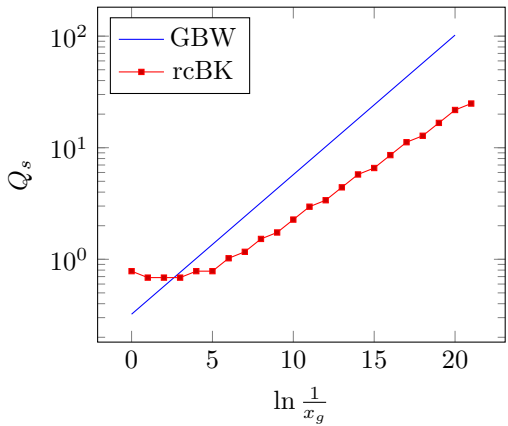

\begin{figure}
 \tikzsetnextfilename{kmax}
 \begin{tikzpicture}
  \begin{axis}[width=11cm,height=6cm,legend style={at={(1.1,0.5)},anchor=west},legend cell align=left,legend columns=1,ymode=log,xlabel=$Y$,ylabel=$k_\text{max}$,y unit=GeV,
  cycle list={{blue,mark=square*},{brown,mark=diamond*},{black,thick,mark=*},{green!90!black,mark=triangle*},{red,mark=pentagon*}}]
   \foreach \as in {as200,lam0588,GBW,lam0010,as062} {%
    \addplot+[mark repeat=100] file {datafiles/Qsdependence/\as.peak.dat};
   }
   \legend{$\alpha_s = 0.2$,$\Lambda_\text{QCD}^2 = 0.0588$,GBW,$\Lambda_\text{QCD}^2 = 0.0010$,$\alpha_s = 0.062$}
  \end{axis}
 \end{tikzpicture}
 \caption[Saturation scale as a function of $\xg$ for GBW and several BK solutions]{This plot, reproduced from reference~\cite{Stasto:2012ru}, shows how adjustment of the parameters of the BK evolution alters the $\xg$ dependence of the saturation scale. Each curve shows the peak of the momentum distribution, $k_\text{max} = \max k\phi(k^2, Y)$, computed from the analytic formula for the GBW model, from the fixed coupling BK evolution for selected values of $\alpha_s$, and from the running coupling BK evolution with selected values of $\Lambda_\text{QCD}^2$. For the BK evolution curves, the slope in the upper range of rapidities decreases as $\alpha_s$ or $\Lambda_\text{QCD}^2$ decreases, and the closest match to the slope of the GBW model curve at $Y > 15$ is achieved with $\alpha_s = 0.062$ for fixed coupling or $\Lambda_\text{QCD}^2 = 0.001$ for running coupling. The jagged ``steps'' in the curves reflect the finite spacing of the momentum grid used in the evolution.}
 \label{fig:kmax}
\end{figure}
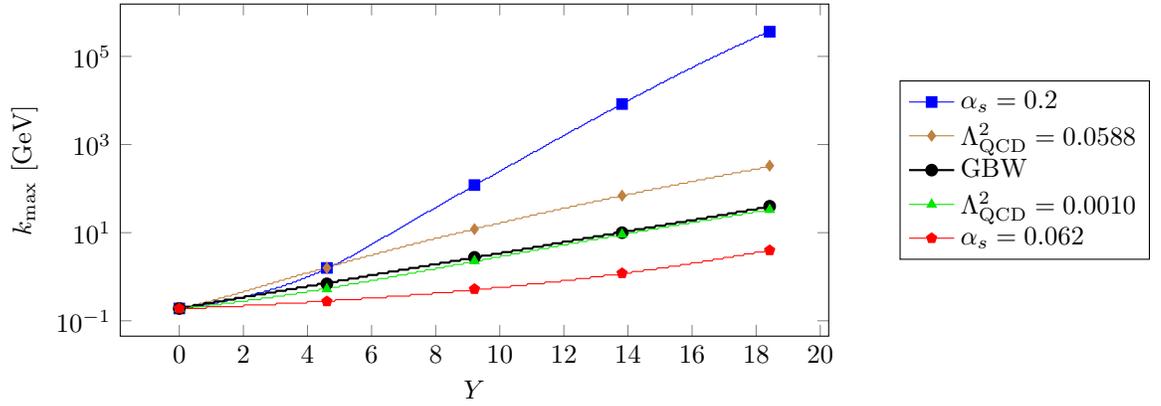

To do this calculation we need a way to extract the saturation scale from the solution to the BK equation.
There are a couple of complementary prescriptions we can use, depending on which form of the equation we're working from.
A simple method is to define the saturation scale as the momentum at which the solution crosses some particular threshold; specifically, in position space,
\begin{equation}
 \dipoleS(1/\Qs) = S_0
\end{equation}
or in momentum space,
\begin{equation}
 \dipoleF(\Qs) = F_0 \dipoleF(0)
\end{equation}
Alternatively, one can define the saturation scale as the momentum which maximizes the product $\kperp\dipolephi(\kperp)$, as in reference~\cite{GolecBiernat:2001if}.
The results from the various prescriptions differ slightly in their overall normalization, but they all predict essentially the same trend.

%% file: correlation.tex
\chapter{Drell-Yan Correlations}\label{ch:correlation}

In the previous chapters, we saw how the color glass condensate provides a theoretical model of the effective structure of protons and nuclei.
Of course, the model is useless without the ability to connect it to measurable quantities, namely cross sections.
With that goal in mind, this chapter shows the basic process of computing the cross section for the Drell-Yan process \HepProcess{\Pproton\Pnucleus \HepTo \Plepton\APlepton\Phadron\Panything}, from the unintegrated gluon distribution.

Drell-Yan processes, first predicted in 1970~\cite{Drell:1970wh}, are those in which the decay of a virtual photon or Z boson produces a lepton and antilepton. Lepton pair production, as it's called, is particularly appealing as a probe of hadron structure for a few reasons:
\begin{itemize}
 \item Electrons or muons and their antiparticles make very distinctive signatures in the detector, so they're easy to identify
 \item \iftime{citation for this?}Not much else produces lepton-antilepton pairs, so background contamination is low
 \item Photons and Z bosons are uncharged, so they are not subject to strong interactions on their way out of the collision and therefore final state interactions with the hadron remnants are absent
 \item The decay of the photon or Z into the lepton-antilepton pair is represented by a factor which can be exactly calculated from electroweak theory
\end{itemize}
Accordingly, lepton pair production gives us a fairly direct window on the unintegrated dipole gluon distribution.
The correlation, as opposed to a simple cross section, is particularly sensitive to the transverse momentum dependence of the distribution, making it especially suitable as a probe of saturation.

\section{Kinematics of \texorpdfstring{$\pA$}{pA} collisions}\label{sec:pakinematics}

\begin{figure}
 \tikzsetnextfilename{correlationdiagram}
 \begin{tikzpicture}
  \node[circle,fill=red!40] (interaction) at (0,0) {};
  \draw[quark] (interaction) +(-4,0) -- (interaction) node[pos=0.5,below] {$\xp\DYprotonp^+$};
  \path (interaction) +(0,-2) node[ellipse,minimum width=3mm,minimum height=1cm,fill=gray!50] (gluon creation) {};
  \draw[gluon] (gluon creation) -- (interaction) node[pos=0.5,right,align=left] {$\xg\DYnucleusp^-$\\$\vec\DYcgcp$};
  \path (interaction) ++(2,0) coordinate (photon emission) +(-15:1.5) node[circle,minimum width=5mm,fill=gray!50] (fragmentation) {$\DYfragmentationfrac$};
  \draw[quark] (interaction) -- (photon emission) node[pos=0.5,above] {$\DYtotalp^\mu$};
  \draw[photon] (photon emission) -- +(45:1) coordinate (photon splitting) node[pos=0.5,above left] {$\DYphotonp$};
  \draw[lepton] (photon splitting) +(70:1.5) -- (photon splitting);
  \draw[lepton] (photon splitting) -- +(20:2);
  \draw[quark] (photon emission) -- (fragmentation) node[pos=0.5,below] {$\DYquarkp^\mu$};
  \draw[quark] (fragmentation) -- +(-15:3) node[right] {$\DYpionp^\mu$};
  \foreach \th in {-20,-10} \draw (fragmentation) -- +(\th:2);
  \draw[nucleus] (gluon creation) +(-4,0) node[left] {$\DYnucleusp^\mu$} to (gluon creation);
  \draw[nucleus] (gluon creation) to +(4,0) node[right] {$\DYanythingp^\mu$};
 \end{tikzpicture}
 \caption{Feynman diagram for \HepProcess{\Pproton\Pnucleus \HepTo \Plepton\APlepton\Ppi\Panything} at leading order}
 \label{fig:correlationdiagram}
\end{figure}

In a proton-nucleus collision, we assume the proton comes in from the $-z$ direction and the nucleon (which is part of a nucleus) comes in from the $+z$ direction, with purely longitudinal momenta
\begin{align}
 \DYprotonp^\mu &= (E_p, 0, 0, \DYprotonp) &
 \DYnucleusp^\mu &= (E_A, 0, 0, -\DYnucleusp)
\end{align}
respectively. Here $\DYprotonp$ and $\DYnucleusp$ represent the magnitudes of the proton's 3-momentum $\vec\DYprotonp$ and the nucleon's 3-momentum $\vec\DYnucleusp$, respectively --- note that $\DYnucleusp$ refers to the momentum of \emph{one} nucleon, not the entire nucleus.

It's easiest to analyze the collision in the longitudinal center of mass frame, in which the center of mass of the proton and nucleon (not nucleus) is at rest: $\DYprotonp[z] + \DYnucleusp[z] = 0$. Since the incoming proton and nucleon have zero transverse momentum and are traveling at near light speed, we can neglect the masses of the proton and nucleon and \emph{approximately} say
\begin{subequations}
\begin{align}
 E_p &= \DYprotonp = \DYprotonp[z] &
 E_A &= \DYnucleusp = -\DYnucleusp[z] \\
 \DYprotonp^+ &= 2E_p = 2\DYprotonp[z] &
 \DYnucleusp^+ &= 0 \label{eq:pluscomponent} \\
 \DYprotonp^- &= 0 &
 \DYnucleusp^- &= 2E_A = -2\DYnucleusp[z] \label{eq:minuscomponent}
\end{align}%
\label{eq:pamomenta}%
\end{subequations}%
according to the light-cone coordinate definitions in appendix~\ref{ap:lightcone}. Given our assumption that $\vec{p}_{p\perp} = \vec{p}_{A\perp} = 0$, this implies that $E_p = E_A = \DYprotonp = \DYnucleusp$. This enables us to write the Mandelstam variable $s$ for the proton and nucleon (also denoted $s_{NN}$ in many sources) as
\begin{equation}
 s = (\DYprotonp^\mu + \DYnucleusp^\mu)^2 = \DYprotonp^+ \DYnucleusp^-
\end{equation}
which in combination with equations~\eqref{eq:pamomenta} means that
\begin{equation}\label{eq:equaltosqrts}
 \DYprotonp^+ = \DYnucleusp^- = \sqrt{s}
\end{equation}

Our eventual goal is to determine the momentum fractions $\xp$, $\xg$, and $\DYfragmentationfrac$, which we need for the parton distribution and gluon distribution, in terms of observable quantities.
For this reaction, those are $\ppionperp$, $\sqs$, $\pseudorapidity_{\Ppi}$, $\pseudorapidity_{\Pphoton}$, and $\vphotonmass$, where $\pseudorapidity$ is the pseudorapidity, defined as
\begin{equation}
 \pseudorapidity = \frac{1}{2}\ln\tan\frac{\theta}{2}
\end{equation}
Pseudorapidity and rapidity are related by the formula
\begin{equation}
 \rapidity = \ln\frac{\sqrt{m^2 + \pperp^2\cosh^2\pseudorapidity} + \pperp\sinh\pseudorapidity}{\sqrt{m^2 + \pperp^2}}
\end{equation}
In situations where $m\ll\pperp\cosh\pseudorapidity$, such as those considered in this dissertation, this reduces to $\rapidity = \pseudorapidity$, so we can consider them essentially interchangeable for the processes considered here.

Calculating the momentum fractions proceeds from the law of conservation of momentum, with the assumption that the plus component of the total momentum is contributed by the initial quark and the minus and transverse components of the total momentum are contributed by the gluon:
\begin{subequations}\label{eq:momentumconservation}
\begin{align}
 \xp \DYprotonp^+ &= \DYtotalp^+ = \DYphotonp^+ + \DYquarkp^+ \label{eq:plusconservation} \\
 \xg \DYnucleusp^- &= \DYtotalp^- = \DYphotonp^- + \DYquarkp^- \label{eq:minusconservation} \\
 \vec\DYcgcpperp &= \vec\DYtotalpperp = \vec\DYphotonpperp + \vec\DYquarkpperp \label{eq:perpconservation}
\end{align}
\end{subequations}
We'll also need the final state momentum fractions, defined as $\DYphotonpfrac \defn \DYphotonp^+/\DYprotonp^+$ and $\DYpionpfrac \defn \DYpionp^+/\DYprotonp^+$.
Using equations~\eqref{eq:momentumpm} and~\eqref{eq:equaltosqrts}, these can be written as
\begin{align}\label{eq:defnzh}
 \DYphotonpfrac &= \sqrt{\frac{\DYphotonpperp^2 + \vphotonmass^2}{\mandelstams}}e^{\rapidity_{\Pphoton}} &
 \DYpionpfrac &= \frac{\DYpionpperp}{\sqrt{s}}e^{\rapidity_{\Ppi}}
\end{align}
assuming that the pion mass is negligible.
Now, we define $\DYfragmentationfrac \defn \frac{\DYpionp^+}{\DYquarkp^+}$, the momentum fraction of the pion relative to the quark it fragments from, and take the pion to be collinear to the quark, so that all components of their momenta obey this proportionality: $\DYpionp^\mu = \DYfragmentationfrac\DYquarkp^\mu$.
Then in terms of these variables, the plus component of the conservation equation~\eqref{eq:plusconservation} leads to
\begin{equation}
 \xp = \DYphotonpfrac + \frac{\DYpionpfrac}{\DYfragmentationfrac} = \sqrt{\frac{\DYphotonpperp^2 + \vphotonmass^2}{\mandelstams}}e^{\rapidity_{\Pphoton}} + \frac{\ppionperp}{\DYfragmentationfrac\sqs}e^{\rapidity_{\Ppi}}
\end{equation}
Similarly, we know that $p^- = p^+ e^{-2\rapidity}$~\eqref{eq:lightconerapidityrelation}, and using that on the right side of equation~\eqref{eq:minusconservation}, we get
\begin{equation}\label{eq:xgexclusive}
 \xg = \DYphotonpfrac e^{-2\rapidity_{\Pphoton}} + \frac{\DYpionpfrac}{\DYfragmentationfrac}e^{-2\rapidity_{\Ppi}} = \sqrt{\frac{\DYphotonpperp^2 + \vphotonmass^2}{\mandelstams}}e^{-\rapidity_{\Pphoton}} + \frac{\ppionperp}{\DYfragmentationfrac\sqs}e^{-\rapidity_{\Ppi}}
\end{equation}

For the inclusive cross section, we'll need to integrate over the entire allowed phase space of the final state particles emerging from fragmentation.
Our code parametrizes that space with the ``reduced'' photon momentum fraction $\DYphotontotalpfrac \defn \frac{\DYphotonp^+}{\DYtotalp^+} = \frac{\DYphotonpfrac}{\xp}$, which represents the fraction of the total plus momentum carried by the photon. and the total transverse momentum $\vec\DYtotalpperp$, instead of the transverse pion momentum $\DYpionpperp$ and rapidity $\rapidity_{\Ppi}$ and the momentum fraction $\DYfragmentationfrac$, so we'll need to remove the latter variables from the formulas.

To achieve this parametrization, we can use the fact that
\begin{align}\label{eq:zmomentumdivision}
 \DYphotontotalpfrac\DYtotalp^+ &= \DYphotonp^+ = \DYphotonpfrac\DYprotonp^+ &
 (1 - \DYphotontotalpfrac)\DYtotalp^+ &= \DYquarkp^+ = \frac{\DYpionp^+}{\DYfragmentationfrac} = \frac{\DYpionpfrac\DYprotonp^+}{\DYfragmentationfrac}
\end{align}
Combining the equation on the left with~\eqref{eq:plusconservation} and the definition of $\DYphotonpfrac$~\eqref{eq:defnzh}, we find that
\begin{equation}
 \xp = \frac{\DYtotalp^+}{\DYprotonp^+} = \frac{\DYphotonpfrac}{\DYphotontotalpfrac} = \frac{1}{\DYphotontotalpfrac}\sqrt{\frac{\DYphotonpperp^2 + \vphotonmass^2}{\mandelstams}}e^{\rapidity_{\Pphoton}}
\end{equation}

For $\xg$, we can use the relation $e^{-\rapidity_{\Ppi}} = \frac{\DYpionpperp}{\DYpionp^+} = \frac{\DYquarkpperp}{\DYquarkp^+}$, which follows from equation~\eqref{eq:momentumpm} and the proportionality of $\DYpionp^\mu$ and $\DYquarkp^\mu$, to express equation~\eqref{eq:xgexclusive} as
\begin{equation}
 \xg = \sqrt{\frac{\DYphotonpperp^2 + \vphotonmass^2}{\mandelstams}}e^{-\rapidity_{\Pphoton}} + \frac{\DYquarkpperp^2}{\DYquarkp^+\sqs}
\end{equation}
We know from equation~\eqref{eq:zmomentumdivision} on the right that $\DYquarkp^+ = (1 - \DYphotontotalpfrac)\DYtotalp^+$, and from~\eqref{eq:plusconservation} that $\DYtotalp^+ = \xp\DYprotonp^+$.
Putting these together with the assumption that $\DYprotonp^+ = \sqs$, we find that
\begin{equation}
 \xg = \sqrt{\frac{\DYphotonpperp^2 + \vphotonmass^2}{\mandelstams}}e^{-\rapidity_{\Pphoton}} + \frac{\DYquarkpperp^2}{(1-\DYphotontotalpfrac)\xp\mandelstams}
\end{equation}
Adding in momentum conservation~\eqref{eq:perpconservation} gets us to the final form of the expression,
\begin{equation}
 \xg = \sqrt{\frac{\DYphotonpperp^2 + \vphotonmass^2}{\mandelstams}}e^{-\rapidity_{\Pphoton}} + \frac{(\vec\DYtotalpperp - \vec\DYphotonpperp)^2}{(1-\DYphotontotalpfrac)\xp\mandelstams}
\end{equation}

\section{Hybrid Factorization}

The final ingredient we will need for the calculation of the Drell-Yan process is the hybrid factorization formalism.
Earlier, in section~\ref{sec:partonmodel}, I briefly mentioned that our ability to split a calculation into the hard scattering process (the low-level interaction), which is calculable from perturbative QCD, and the parton distribution, which is not, relies on the existence of factorization theorems.
In fact, factorization can be implemented in several different ways.
The traditional approach~\cite{Collins:1982wa,Collins:1983ju,Collins:1988ig}, the \iterm{collinear factorization} framework, treats both projectile and target on an equal footing, representing them both with integrated parton distributions.
However, the collinear factorization approach with integrated PDFs is not well suited to describe dense systems which are within the saturation regime.

An alternative approach, called hybrid factorization~\cite{Dumitru:2005gt}, was developed specifically for asymmetric proton-nucleus collisions.
In hybrid factorization, the target is treated as a color glass condensate and represented using an unintegrated parton distribution.
This framework is designed for the kinematic region where $\rapidity \gtrsim 2.5$, the larger the better --- this is called the ``forward rapidity'' region because large values of $\rapidity$, defined as
\begin{equation}
 \rapidity = \frac{1}{2}\ln\frac{E + p_z}{E - p_z}
\end{equation}
correspond to emission of particles near the direction of the projectile.
At these forward rapidities, the interaction is heavily weighted toward small-$x$ gluons from the target, $\xg\sim\num{e-4}$ or even lower, and relatively large-$x$ partons from the projectile, with $\xp\sim\num{e-1}$ or $\num{e-2}$.

In either approach, factorization necessitates ignoring the dynamics of the interacting fields above a certain momentum scale, in the same sense as renormalization group calculations.
Physical quantities are, of course, independent of this \iterm{factorization scale}, designated $\mu$, but when we calculate a perturbative approximation to fixed order (leading order or next-to-leading order etc.), we will find that the resulting formula does depend on the factorization scale.
At leading order, the dependence arises entirely through nonperturbative quantities, the parton distributions and fragmentation functions, but beyond leading order, the perturbative part of the calculation can have its own dependence on $\mu$.
We'll see this borne out in the calculations of chapter~\ref{ch:crosssection}.

There is some loose physical motivation for choosing a particular value for the factorization scale in any given calculation~\cite{Collins:1990bu,Maltoni:2007tc}, but in general, the dependence on the arbitrary scale $\mu$ should be viewed as a theoretical source of uncertainty in the final result.
Incorporating more terms in the perturbative series tends to reduce the dependence on $\mu$, as we'll see in chapter~\ref{ch:results}.
This provides a powerful motivation to extend perturbative quantities beyond leading order.

\section{Calculation of Drell-Yan Correlation}
Let's now turn to the procedure for CGC calculations described in chapter~\ref{ch:cgc}, which we can use to calculate the cross section for the Drell-Yan process.
The calculation begins with the process \HepProcess{\Pquark\Pnucleus \HepTo \Pquark\Pphoton\Pnucleus}, whose cross section is calculated in momentum space in reference~\cite{Gelis:2002ki} and in transverse coordinate space in~\cite{Dominguez:2011wm}.
However, these calculations only treat the case of a real photon, $\vphotonmass = 0$, whereas we need the result for photon virtuality $\vphotonmass^2 > 0$.
The modifications to the calculation of~\cite{Dominguez:2011wm} required to account for nonzero virtuality are described in reference~\cite{Dominguez:2011br}.

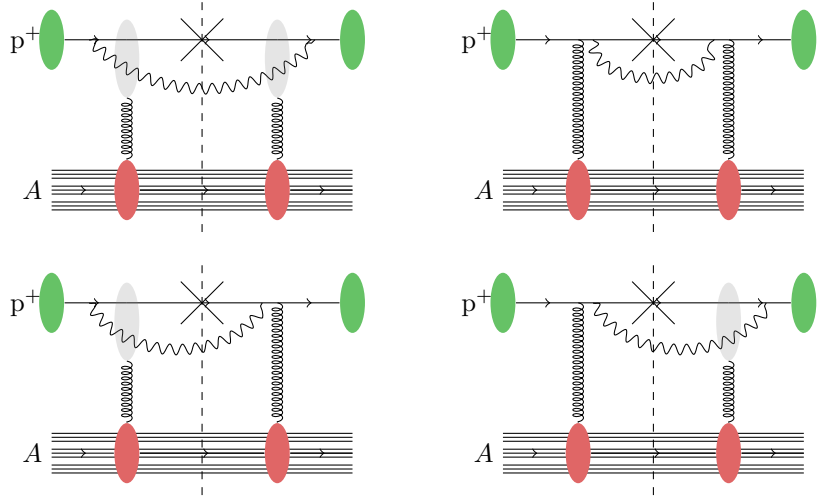
\begin{figure}
 \tikzsetnextfilename{photonemissiondiagram}
 \begin{tikzpicture}
  \begin{scope}
   \coordinate (origin) at (0,0);
   \node[blob] (left proton) at (-2,0) {};
   \node[blob] (left nucleus) at (-1,-2) {};
   \coordinate (left interaction) at (left nucleus |- origin);
   \coordinate (right interaction) at ($(left interaction)!2!(left interaction -| origin)$);
   \node[blob] (right nucleus) at ($(left nucleus)!2!(left nucleus -| origin)$) {};
   \node[blob] (right proton) at ($(left proton)!2!(left proton -| origin)$) {};

   \coordinate (left photon attach) at ($(left proton)!.5!(left interaction)$);
   \coordinate (right photon attach) at ($(right proton)!.5!(right interaction)$);
   
   \node[left] at (left proton) {$\prbr$};
   \draw[quark] (left proton) to (left interaction);
   \draw[quark] (left interaction) to (right interaction);
   \draw[quark] (right interaction) to (right proton);
   \draw[nucleus] (left proton |- left nucleus) node[left] {$A$} to (left nucleus);
   \draw[nucleus] (left nucleus) to (right nucleus);
   \draw[nucleus] (right nucleus) to (right nucleus -| right proton);
   \draw[photon] (left photon attach) to[out=-45,in=225] (right photon attach);
   
   \coordinate (left gluon intermediate) at ($(left interaction)!.5!90:(left proton)$);
   \coordinate (right gluon intermediate) at ($(right interaction)!.5!270:(right proton)$){};
   \node[blob,fill=gray,fill opacity=0.2,fit=(left interaction) (left gluon intermediate)] (left gluon blob) {};
   \node[blob,fill=gray,fill opacity=0.2,fit=(right interaction) (right gluon intermediate)] (right gluon blob) {};
   \draw[small gluon] (left gluon blob) to (left nucleus);
   \draw[small gluon] (right gluon blob) to (right nucleus);
   
   \node[fill=partondist!60!white,blob] at (left proton) {};
   \node[fill=gluondist!60!white,blob] at (left nucleus) {};
   \node[fill=partondist!60!white,blob] at (right proton) {};
   \node[fill=gluondist!60!white,blob] at (right nucleus) {};
   
   \draw[dashed] (origin) +(0,0.5) -- (origin |- left nucleus) -- +(0,-0.6);
   \draw (origin) +(45:.4) -- +(45:-.4) +(-45:.4) -- +(-45:-.4);
  \end{scope}

  \begin{scope}[xshift=6cm]
   \coordinate (origin) at (0,0);
   \node[blob] (left proton) at (-2,0) {};
   \node[blob] (left nucleus) at (-1,-2) {};
   \coordinate (left interaction) at (left nucleus |- origin);
   \coordinate (right interaction) at ($(left interaction)!2!(left interaction -| origin)$);
   \node[blob] (right nucleus) at ($(left nucleus)!2!(left nucleus -| origin)$) {};
   \node[blob] (right proton) at ($(left proton)!2!(left proton -| origin)$) {};

   \coordinate (left photon attach) at ($(left proton)!1.2!(left interaction)$);
   \coordinate (right photon attach) at ($(right proton)!1.2!(right interaction)$);
   
   \node[left] at (left proton) {$\prbr$};
   \draw[quark] (left proton) to (left interaction);
   \draw[quark] (left interaction) to (right interaction);
   \draw[quark] (right interaction) to (right proton);
   \draw[nucleus] (left proton |- left nucleus) node[left] {$A$} to (left nucleus);
   \draw[nucleus] (left nucleus) to (right nucleus);
   \draw[nucleus] (right nucleus) to (right nucleus -| right proton);
   \draw[photon] (left photon attach) to[out=-90,in=270] (right photon attach);
   
   \draw[small gluon] (left interaction) to (left nucleus);
   \draw[small gluon] (right interaction) to (right nucleus);
   
   \node[fill=partondist!60!white,blob] at (left proton) {};
   \node[fill=gluondist!60!white,blob] at (left nucleus) {};
   \node[fill=partondist!60!white,blob] at (right proton) {};
   \node[fill=gluondist!60!white,blob] at (right nucleus) {};

   \draw[dashed] (origin) +(0,0.5) -- (origin |- left nucleus) -- +(0,-0.6);
   \draw (origin) +(45:.4) -- +(45:-.4) +(-45:.4) -- +(-45:-.4);
  \end{scope}

  \begin{scope}[yshift=-3.5cm]
   \coordinate (origin) at (0,0);
   \node[blob] (left proton) at (-2,0) {};
   \node[blob] (left nucleus) at (-1,-2) {};
   \coordinate (left interaction) at (left nucleus |- origin);
   \coordinate (right interaction) at ($(left interaction)!2!(left interaction -| origin)$);
   \node[blob] (right nucleus) at ($(left nucleus)!2!(left nucleus -| origin)$) {};
   \node[blob] (right proton) at ($(left proton)!2!(left proton -| origin)$) {};

   \coordinate (left photon attach) at ($(left proton)!.5!(left interaction)$);
   \coordinate (right photon attach) at ($(right proton)!1.2!(right interaction)$);
   
   \node[left] at (left proton) {$\prbr$};
   \draw[quark] (left proton) to (left interaction);
   \draw[quark] (left interaction) to (right interaction);
   \draw[quark] (right interaction) to (right proton);
   \draw[nucleus] (left proton |- left nucleus) node[left] {$A$} to (left nucleus);
   \draw[nucleus] (left nucleus) to (right nucleus);
   \draw[nucleus] (right nucleus) to (right nucleus -| right proton);
   \draw[photon] (left photon attach) to[out=-60,in=240] (right photon attach);
   
   \coordinate (left gluon intermediate) at ($(left interaction)!.5!90:(left proton)$);
   \node[blob,fill=gray,fill opacity=0.2,fit=(left interaction) (left gluon intermediate)] (left gluon blob) {};
   \draw[small gluon] (left gluon blob) to (left nucleus);
   \draw[small gluon] (right interaction) to (right nucleus);
   
   \node[fill=partondist!60!white,blob] at (left proton) {};
   \node[fill=gluondist!60!white,blob] at (left nucleus) {};
   \node[fill=partondist!60!white,blob] at (right proton) {};
   \node[fill=gluondist!60!white,blob] at (right nucleus) {};

   \draw[dashed] (origin) +(0,0.5) -- (origin |- left nucleus) -- +(0,-0.6);
   \draw (origin) +(45:.4) -- +(45:-.4) +(-45:.4) -- +(-45:-.4);
  \end{scope}

  \begin{scope}[xshift=6cm,yshift=-3.5cm]
   \coordinate (origin) at (0,0);
   \node[blob] (left proton) at (-2,0) {};
   \node[blob] (left nucleus) at (-1,-2) {};
   \coordinate (left interaction) at (left nucleus |- origin);
   \coordinate (right interaction) at ($(left interaction)!2!(left interaction -| origin)$);
   \node[blob] (right nucleus) at ($(left nucleus)!2!(left nucleus -| origin)$) {};
   \node[blob] (right proton) at ($(left proton)!2!(left proton -| origin)$) {};

   \coordinate (left photon attach) at ($(left proton)!1.2!(left interaction)$);
   \coordinate (right photon attach) at ($(right proton)!.5!(right interaction)$);
   
   \node[left] at (left proton) {$\prbr$};
   \draw[quark] (left proton) to (left interaction);
   \draw[quark] (left interaction) to (right interaction);
   \draw[quark] (right interaction) to (right proton);
   \draw[nucleus] (left proton |- left nucleus) node[left] {$A$} to (left nucleus);
   \draw[nucleus] (left nucleus) to (right nucleus);
   \draw[nucleus] (right nucleus) to (right nucleus -| right proton);
   \draw[photon] (left photon attach) to[out=-60,in=240] (right photon attach);
   
   \draw[small gluon] (left interaction) to (left nucleus);
   \coordinate (right gluon intermediate) at ($(right interaction)!.5!270:(right proton)$);
   \node[blob,fill=gray,fill opacity=0.2,fit=(right interaction) (right gluon intermediate)] (right gluon blob) {};
   \draw[small gluon] (right gluon blob) to (right nucleus);
   
   \node[fill=partondist!60!white,blob] at (left proton) {};
   \node[fill=gluondist!60!white,blob] at (left nucleus) {};
   \node[fill=partondist!60!white,blob] at (right proton) {};
   \node[fill=gluondist!60!white,blob] at (right nucleus) {};

   \draw[dashed] (origin) +(0,0.5) -- (origin |- left nucleus) -- +(0,-0.6);
   \draw (origin) +(45:.4) -- +(45:-.4) +(-45:.4) -- +(-45:-.4);
  \end{scope}
 \end{tikzpicture}
 \caption{Compound diagrams for a quark emitting a photon and interacting with a nucleus}
 \label{fig:photonemissiondiagram}
\end{figure}
Following the general procedure described in chapter~\ref{ch:saturation}, we start with the four diagrams contributing to this process at leading order, as shown in figure~\ref{fig:photonemissiondiagram}.
They differ only in whether the interaction with the nucleus occurs before or after the emission of the photon.
For the first diagram, putting together the splitting wavefunction and the Wilson line on each side of the cut, we get
\begin{multline}
 \frac{\udc^6\sigma}{\udddc\vec{p}_{\Pphoton}\udddc\vec{k}_{\Pquark}} =
 \iint\frac{\uddc\vec\xperp}{(2\pi)^2} \iint\frac{\uddc\vec\xpperp}{(2\pi)^2} \iint\frac{\uddc\vec\bperp}{(2\pi)^2} \iint\frac{\uddc\vec\bpperp}{(2\pi)^2}
 e^{-i\vec\DYphotonpperp\cdot(\vec\xperp - \vec\xpperp)} e^{-i\vec\DYquarkpperp\cdot(\vec\bperp - \vec\bpperp)} \\
 \times \sum_{\lambda\alpha\beta} \conj{(\psi_{\alpha\beta}^{\lambda})}(\vec\xpperp - \vec\bpperp)\psi_{\alpha\beta}^{\lambda}(\vec\xperp - \vec\bperp)
 \braket{\trace U\bigl(\DYphotontotalpfrac\vec\xperp + (1 - \DYphotontotalpfrac)\vec\bperp\bigr) \herm{U}\bigl(\DYphotontotalpfrac\vec\xpperp + (1 - \DYphotontotalpfrac)\vec\bpperp\bigr)}_{\xg}
\end{multline}
where $\lambda$ represents the polarization of the photon, and $\alpha$ and $\beta$ are the helicities of the initial and final quarks, respectively.
The averaging operation $\braket{\cdots}_{\xg}$ averages over the unknown color charge distribution of the nucleus.
The exponential factors have their origin in the LSZ reduction formula.\footnote{More detail on this aspect of the derivation is in references~\cite{Gelis:2002ki,Gelis:2002fw}, and see also~\cite{QFTItzyksonZuber}.}\iftime{be more explicit about the derivation}
At the end of this expression, we see the dipole gluon distribution in the form
\begin{equation}
 \braket{\trace U(\vec\xperp) \herm{U}(\vec\yperp))}_{\xg} = \dipoleS(\vec\xperp, \vec\yperp)
\end{equation}

Adding in the other diagrams from figure~\ref{fig:photonemissiondiagram} gives the expression\note{why minus signs?}
\begin{multline}
 \frac{\udc^6\sigma}{\udddc\vec{p}_{\Pphoton}\udddc\vec{k}_{\Pquark}} =
 \iint\frac{\uddc\vec\xperp}{(2\pi)^2} \iint\frac{\uddc\vec\xpperp}{(2\pi)^2} \iint\frac{\uddc\vec\bperp}{(2\pi)^2} \iint\frac{\uddc\vec\bpperp}{(2\pi)^2}
 e^{-i\vec\DYphotonpperp\cdot(\vec\xperp - \vec\xpperp)} e^{-i\DYquarkpperp\cdot(\vec\bperp - \vec\bpperp)} \\
 \times \sum_{\lambda\alpha\beta} \conj{(\psi_{\alpha\beta}^{\lambda})}(\vec\xpperp - \vec\bpperp)\psi_{\alpha\beta}^{\lambda}(\vec\xperp - \vec\bperp)
 \bigl[\dipoleS\bigl(\DYphotontotalpfrac\vec\xperp + (1 - \DYphotontotalpfrac)\vec\bperp, \DYphotontotalpfrac\vec\xpperp + (1 - \DYphotontotalpfrac)\vec\bpperp\bigr) + \dipoleS(\vec\bperp, \vec\bpperp) \\
 - \dipoleS\bigl(\vec\bperp, \DYphotontotalpfrac\vec\xpperp + (1 - \DYphotontotalpfrac)\vec\bpperp\bigr) - \dipoleS\bigl(\DYphotontotalpfrac\vec\xperp + (1 - \DYphotontotalpfrac)\vec\bperp, \vec\bpperp\bigr)\bigr]
\end{multline}
Following reference~\cite{Dominguez:2011wm}, we make the substitutions $\vec{u} = \vec\xperp - \vec\bperp$, $\vec{u}' = \vec\xpperp - \vec\bpperp$, and
\begin{align*}
 \text{first term:}  & & \vec{v} &= \DYphotontotalpfrac\vec\xperp + (1 - \DYphotontotalpfrac)\vec\bperp & \vec{v}' &= \DYphotontotalpfrac\vec\xpperp + (1 - \DYphotontotalpfrac)\vec\bpperp \\
 \text{second term:} & & \vec{v} &= \vec\bperp & \vec{v}' &= \vec\bpperp \\
 \text{third term:}  & & \vec{v} &= \vec\bperp & \vec{v}' &= \DYphotontotalpfrac\vec\xpperp + (1 - \DYphotontotalpfrac)\vec\bpperp \\
 \text{fourth term:} & & \vec{v} &= \DYphotontotalpfrac\vec\xperp + (1 - \DYphotontotalpfrac)\vec\bperp & \vec{v}' &= \vec\bpperp
\end{align*}
which results in
\begin{multline}\label{eq:DYxsecbeforeintegrals}
 \frac{\udc^6\sigma}{\udddc\vec{p}_{\Pphoton}\udddc\vec{k}_{\Pquark}} =
 \iint\frac{\uddc\vec{u}}{(2\pi)^2} \iint\frac{\uddc\vec{u}'}{(2\pi)^2} \iint\frac{\uddc\vec{v}}{(2\pi)^2} \iint\frac{\uddc\vec{v}'}{(2\pi)^2}
 e^{-i\vec\DYtotalpperp\cdot(\vec{v} - \vec{v}')} \dipoleS(\vec{v}, \vec{v}') \\
 \times \sum_{\lambda\alpha\beta} \conj{(\psi_{\alpha\beta}^{\lambda})}(\vec{u}')\psi_{\alpha\beta}^{\lambda}(\vec{u})
 \bigl[
    e^{-i(\vec\DYphotonpperp - \DYphotontotalpfrac\vec\DYtotalpperp)\cdot\vec u}e^{i(\vec\DYphotonpperp - \DYphotontotalpfrac\vec\DYtotalpperp)\cdot\vec u'}
  + e^{-i\vec\DYphotonpperp\cdot\vec u}e^{i\vec\DYphotonpperp\cdot\vec u'} \\
  - e^{-i\vec\DYphotonpperp\cdot\vec u}e^{i(\vec\DYphotonpperp - \DYphotontotalpfrac\vec\DYtotalpperp)\cdot\vec u'}
  - e^{-i(\vec\DYphotonpperp - \DYphotontotalpfrac\vec\DYtotalpperp)\cdot\vec u}e^{i\vec\DYphotonpperp\cdot\vec u'}
 \bigr]
\end{multline}
In this expression, the integrals over $\vec{v}$ and $\vec{v}'$ give us the momentum space dipole gluon distribution, according to equation~\eqref{eq:dipoleFdefinition}, and the integrals over $\vec{u}$ and $\vec{u}'$ implement the Fourier transform of the splitting wavefunction,
\begin{equation}\label{eq:splittingft}
 \iint\frac{\uddc\vec{u}}{(2\pi)^2} \iint\frac{\uddc\vec{u}'}{(2\pi)^2} \sum_{\lambda\alpha\beta} \conj{(\psi_{\alpha\beta}^{\lambda})}(\vec{u}') \psi_{\alpha\beta}^{\lambda}(\vec{u}) e^{-i\vec\qiperp\cdot\vec{u}} e^{-i\vec\qiiperp\cdot\vec{u}'}
\end{equation}
for different values of $\vec\qiperp$ and $\vec\qiiperp$ in each term.

This is the first point at which the nonzero virtuality becomes relevant.
The explicit form of the splitting wavefunction for a virtual photon is given in~\cite{Dominguez:2011br} as
\begin{equation}
 \psi_{\alpha\beta}^{\lambda}(\DYtotalp^+, \DYphotonp^+, \vec u) = 
 2\pi \sqrt{\frac{2}{\DYphotonp^+}}
 \begin{cases}
  (1 - \DYphotontotalpfrac) \vphotonmass \besselKzero(\epsM u), & \lambda = 0\quad\text{(longitudinal)} \\
  i\epsM\besselKone(\epsM u) \frac{\vec{u}\cdot\vec{\epsilon}_\perp^{(1)}}{u^2}\bigl(\delta_{\alpha-}\delta_{\beta-} + (1 - \DYphotontotalpfrac)\delta_{\alpha+}\delta_{\beta+}\bigr), & \lambda = 1\quad\text{(transverse)} \\
  i\epsM\besselKone(\epsM u) \frac{\vec{u}\cdot\vec{\epsilon}_\perp^{(2)}}{u^2}\bigl(\delta_{\alpha+}\delta_{\beta+} + (1 - \DYphotontotalpfrac)\delta_{\alpha-}\delta_{\beta-}\bigr), & \lambda = 2\quad\text{(transverse)}
 \end{cases}
\end{equation}
where $\epsM^2 = (1 - \DYphotontotalpfrac)\vphotonmass^2$.
We can plug that into~\eqref{eq:splittingft} and perform the sums over $\lambda$, $\alpha$, and $\beta$.
Using the identity~\eqref{eq:ident:polarization}, equation~\eqref{eq:splittingft} reduces to
\begin{multline}\label{eq:explicitsplittingft}
 \iint\frac{\uddc\vec{u}}{(2\pi)^2} \iint\frac{\uddc\vec{u}'}{(2\pi)^2} 4\pi^2 \frac{2}{\DYphotonp^+} \besselKzero(\epsM u)\besselKzero(\epsM u') \vphotonmass^2 (1 - \DYphotontotalpfrac)^2 e^{-i\vec\qiperp\cdot\vec{u}} e^{i\vec\qiiperp\cdot\vec{u}'} \\
 + \iint\frac{\uddc\vec{u}}{(2\pi)^2} \iint\frac{\uddc\vec{u}'}{(2\pi)^2} 4\pi^2 \frac{2}{\DYphotonp^+} \besselKone(\epsM u)\besselKone(\epsM u') \epsM^2 \\
 \times \frac{\vec{u}'\cdot\vec{u} - (\vec{u}'\cdot\vec\DYphotonp)(\vec{u}\cdot\vec\DYphotonp)/\DYphotonp^2}{u'^2 u^2}\bigl[(1 - \DYphotontotalpfrac)^2 + 1\bigr]e^{-i\vec\qiperp\cdot\vec{u}} e^{i\vec\qiiperp\cdot\vec{u}'}
\end{multline}
where the first line comes from the longitudinal component and the second and third lines from the transverse components.

\iftime{Figure out how to do integrals in second/third line of~\eqref{eq:explicitsplittingft}}

Performing the Fourier integrals, the cross section for \HepProcess{\Pquark\Pnucleus \HepTo \Pquark\Pphoton\Panything} becomes~\cite{Dominguez:2011br}
\begin{multline}\label{eq:partonlevelquarkphotonxsec}
 \frac{\udc^8\sigma}{\udc\rapidity_{\Pphoton}\udc\rapidity_{\Pquark} \uddc\vec\DYphotonpperp\uddc\vec\DYquarkpperp\uddc\vec b} =
 \frac{\alphaem e_q^2}{2\pi^2} (1 - \DYphotontotalpfrac)\dipoleF(\vec\DYcgcpperp) \\
  \times\Biggl\{\bigl[1 + (1 - \DYphotontotalpfrac)^2\bigr]\frac{\DYphotontotalpfrac^2 \DYcgcpperp^2}{\bigl[(\vec\DYphotonpperp - \DYphotontotalpfrac\vec\DYcgcpperp)^2 + \epsM^2\bigr]\bigl[\DYphotonpperp^2 + \epsM^2\bigr]} \\
  - \DYphotontotalpfrac^2\epsM^2\biggl[\frac{1}{\DYphotonpperp^2 + \epsM^2} - \frac{1}{(\vec\DYphotonpperp - z\vec\DYcgcpperp)^2 + \epsM^2}\biggr]^2
 \Biggr\}
\end{multline}
where $e_q$ is the charge of the initial-state quark.

To obtain the cross section for $\HepProcess{\Pproton\Pnucleus \HepTo \Plepton\APlepton\Ppizero\Panything}$, we need to multiply~\eqref{eq:partonlevelquarkphotonxsec} by the integrated parton distribution, the fragmentation function, and the factor representing the splitting of the virtual photon into a lepton pair, which is \note{why is this the factor?}$\frac{\alphaem}{3\pi}\frac{\udc\vphotonmass^2}{\vphotonmass^2}$.
Then we integrate over all possible values of the unknown momentum fraction $\DYfragmentationfrac$.
\begin{multline}\label{eq:doubleinclusiveleptonpairxsec}
 \frac{\udc^9\sigma}{\udc\rapidity_{\Pphoton}\udc\rapidity_{\Ppi} \uddc\vec\DYphotonpperp\uddc\vec\ppionperp\uddc\vec b\,\udc\vphotonmass^2} =
 \frac{\alphaem^2 e_q^2}{6\pi^3 \vphotonmass^2} \int_{\frac{\DYpionpfrac}{1 - \DYphotonpfrac}}^1\frac{\udc\DYfragmentationfrac}{\DYfragmentationfrac^2} \sum_q \xp q(\xp, \mu) D_{\Ppi/q}(\DYfragmentationfrac, \mu)(1 - \DYphotontotalpfrac)\dipoleF(\vec\gluonkperp) \\
  \times\Biggl\{\bigl[1 + (1 - \DYphotontotalpfrac)^2\bigr]\frac{\DYphotontotalpfrac^2 \gluonkperp^2}{\bigl[(\vec\DYphotonpperp - \DYphotontotalpfrac\vec\gluonkperp)^2 + \epsM^2\bigr]\bigl[\DYphotonpperp^2 + \epsM^2\bigr]} \\
  - \DYphotontotalpfrac^2\epsM^2\biggl[\frac{1}{\DYphotonpperp^2 + \epsM^2} - \frac{1}{(\vec\DYphotonpperp - \DYphotontotalpfrac\vec\gluonkperp)^2 + \epsM^2}\biggr]^2
 \Biggr\}
\end{multline}

The other cross section we need, for the denominator of the correlation, is the inclusive cross section for \HepProcess{\Pproton\Pnucleus \HepTo \Ppizero\Panything}, which we can get by integrating over all possible combinations of the pion momentum and everything else (\Panything) that could come out of the reaction --- or, equivalently, we can just ignore the fact that the final-state quark fragments at all, and just integrate over all possible momenta of that quark.
It's most straightforward to get this by starting again from equation~\eqref{eq:partonlevelquarkphotonxsec} and inserting the integrated parton distribution (but not the fragmentation function) and the lepton splitting factor $\frac{\alphaem}{3\pi}\frac{\udc\vphotonmass^2}{\vphotonmass^2}$.

Then we integrate over the unknown momentum of the quark, represented by $\rapidity_{\Pquark}$ and $\vec\DYquarkp$.
We're assuming the quark to be \note{why?}on-shell, $\DYquarkp^\mu\DYquarkp[\mu] = m_{\Pquark}^2$, which means there are only three degrees of freedom in this momentum.
Now, since the photon momentum $\vec\DYphotonp$ is a constant, we can change variables from $\DYquarkp$ to $\vec{q} = \vec\DYquarkp + \vec\DYphotonp$, which is a simple shift.
We can then take advantage of the transformations\footnote{To derive the Jacobian determinant for $\rapidity_{\Pquark}$, substitute $p^+ = \sqrt{p_z^2 + p_\perp^2 + m^2} + p_z$ into equation~\eqref{eq:rapiditymomentumpm} and differentiate. Only $\pd{\rapidity}{p_z} = \frac{1}{E}$ survives into the final determinant; the other components wind up being multiplied by zero.}
\begin{align}
 \udc\DYphotontotalpfrac \uddc\vec\gluonkperp &= \frac{\DYphotontotalpfrac}{E_q}\udddc\vec{q} &
 \udc\rapidity_{\Pquark} \uddc\vec\gluonkperp &= \frac{1}{E_{\Pquark}}\udddc\vec{q}
\end{align}
to work out the inclusive cross section to
\begin{multline}\label{eq:singleinclusiveleptonpairxsec}
 \frac{\udc^6\sigma}{\udc\rapidity_{\Pphoton} \uddc\vec\DYphotonpperp \uddc\vec b\,\udc\vphotonmass^2} =
 \frac{\alphaem^2 e_q^2}{6\pi^3 \vphotonmass^2} \sum_q \int_{\DYphotonpfrac}^1 \frac{\udc\DYphotontotalpfrac}{\DYphotontotalpfrac} \xp q(\xp, \mu) \iint\uddc\vec\gluonkperp \dipoleF(\vec\gluonkperp) \\
  \times\Biggl\{\bigl[1 + (1 - z)^2\bigr]\frac{z^2 \gluonkperp^2}{\bigl[(\vec\DYphotonpperp - \DYphotontotalpfrac\vec\gluonkperp)^2 + \epsM^2\bigr]\bigl[\DYphotonpperp^2 + \epsM^2\bigr]} \\
  - \DYphotontotalpfrac^2\epsM^2\biggl[\frac{1}{\DYphotonpperp^2 + \epsM^2} - \frac{1}{(\vec\DYphotonpperp - \DYphotontotalpfrac\vec\gluonkperp)^2 + \epsM^2}\biggr]^2
 \Biggr\}
\end{multline}
which is consistent with other inclusive results derived in references~\cite{Kopeliovich:2000fb,Baier:2004tj,Gelis:2002nn,Gelis:2002fw}.

The correlation is defined as the ratio of the double inclusive to single inclusive cross sections. For this work, we integrate over transverse momentum and assume there is no impact parameter dependence, thus considering the correlation only as a function of rapidity, virtual photon mass, and angle.
\begin{equation}\label{eq:leptonpaircorrelation}
 C^\text{DY}(\rapidity_{\Pphoton}, \rapidity_{\Ppizero}, \vphotonmass, \Delta\phi) =
 \frac
 {\displaystyle \int_{\DYphotonpperp > \pTcut} \uddc\vec\DYphotonpperp \int_{\ppionperp > \pTcut} \uddc\vec\ppionperp \frac{\udc^9\sigma}{\udc\rapidity_{\Pphoton}\udc\rapidity_{\Ppizero} \uddc\vec\DYphotonpperp\uddc\vec\ppionperp\uddc\vec b\,\udc\vphotonmass^2}}
 {\displaystyle \int_{\DYphotonpperp > \pTcut} \uddc\vec\DYphotonpperp \frac{\udc^6\sigma}{\udc\rapidity_{\Pphoton} \uddc\vec\DYphotonpperp \uddc\vec b\,\udc\vphotonmass^2}}
\end{equation}

\section{Results}

Figures~\ref{fig:corr_rhic_2} and~\ref{fig:corr_lhc_2} show the predictions of this model for kinematic conditions characteristic of RHIC and the LHC, respectively.
The RHIC results in figure~\ref{fig:corr_rhic_2} use $\sqs = \SI{200}{GeV}$, $\massnumber = 197$, and $\pTcut = \SI{1.5}{GeV}$, and the LHC results use $\sqs = \SI{8800}{GeV}$, $\massnumber = 208$, and $\pTcut = \SI{1.5}{GeV}$.

There are two characteristics that are immediately evident: the double-peak structure surrounding back-to-back emission ($\Delta\phi \approx \pi$), and the variation in the parallel-emission behavior around $\Delta\phi \approx 0,2\pi$.
While the double peak is roughly similar for all three models of the gluon distribution, for most kinematic conditions the BK equation with either fixed or running coupling gives a large excess for parallel emission relative to the GBW model.
Since the BK solution differs from the GBW model primarily in the asymptotic behavior of $\dipoleF$ at large $\kgperp$, power versus exponential respectively, we can see that the parallel emission comes primarily from large-transverse momentum gluons, whereas the back-to-back emission comes mostly from low-transverse momentum gluons.
This makes sense intuitively, since if the lepton pair and the hadron are to emerge in nearly the same direction, they'll have to have at least $2\pTcut$ of total transverse momentum to be detected, but back-to-back emission can come from a system with little or no transverse momentum.

Figure~\ref{fig:corr_momentum_breakdown} shows an explicit breakdown of the momenta $\DYcgcpperp$ that contribute to the correlation at each angle, which confirms this interpretation: the correlation for parallel emission comes from higher momentum than the central peak.
Overall, the correlation at the LHC receives contributions from higher momenta than at RHIC, due to there being more total energy $\sqs$ available in the collision.

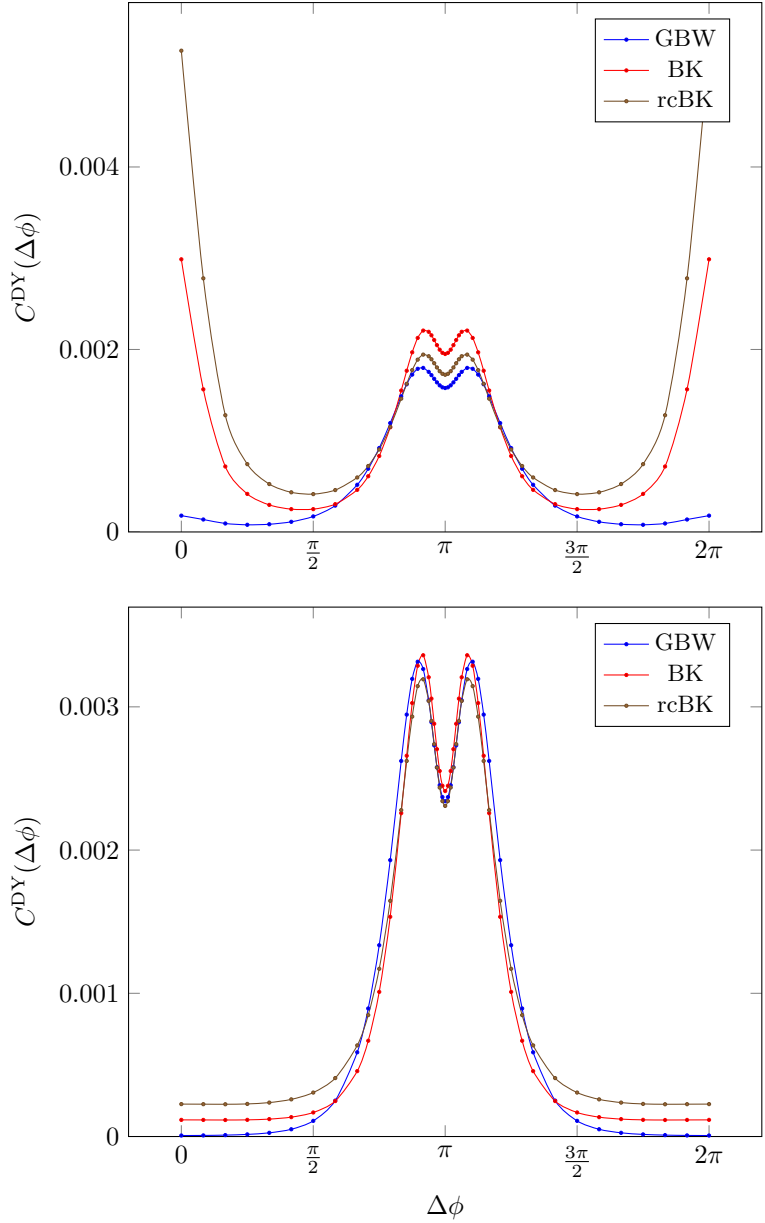
\begin{figure}
 \tikzsetnextfilename{corr_rhic_2}
 \begin{tikzpicture}
  \begin{groupplot}[group style={rows=2,xlabels at=edge bottom,ylabels at=edge left},width=10cm,correlation graph,legend style={at={(0.95,0.97)},anchor=north east}]
  \nextgroupplot
   \addplot+[smooth] file {datafiles/rhic.252585.p12666.dat}; 
   \addplot+[smooth] file {datafiles/rhic.252585.p27336.dat}; 
   \addplot+[smooth] file {datafiles/rhic.252585.p20033.dat}; 
   \legend{GBW,BK,rcBK};
  \nextgroupplot
   \addplot+[smooth] file {datafiles/rhic.252585.p17509.dat}; 
   \addplot+[smooth] file {datafiles/rhic.252585.p27337.dat}; 
   \addplot+[smooth] file {datafiles/rhic.252585.p30742.dat}; 
   \legend{GBW,BK,rcBK};
  \end{groupplot}
 \end{tikzpicture}
 \caption[Azimuthal angular correlation results for RHIC]{The angular correlations between the virtual photons and pions at RHIC, at medium rapidity, $\rapidity_{\Pphoton} = \rapidity_{\Ppizero} = 2.5$. The upper graph shows the correlation for a virtual photon mass of $M = \SI{0.5}{GeV}$, and the lower one, for $M = \SI{4}{GeV}$. In each case, the three curves for GBW, fixed coupling BK, and running coupling BK, exhibit basically the same double-peak structure around $\Delta\phi = \pi$, but they show differing behavior near $\Delta\phi = 0,2\pi$, the near side correlation. This relates to the large-$k^2$ behavior of the corresponding gluon distributions. Reproduced from figure 5 of~\cite{Stasto:2012ru}.}
 \label{fig:corr_rhic_2}
\end{figure}

\begin{figure}
 \tikzsetnextfilename{corr_lhc_2}
 \begin{tikzpicture}
  \begin{groupplot}[group style={rows=2,xlabels at=edge bottom,ylabels at=edge left},width=10cm,correlation graph,legend style={at={(0.95,0.97)},anchor=north east}]
  \nextgroupplot
   \addplot+[smooth] file {datafiles/lhc.404085.p7033.dat};  
   \addplot+[smooth] file {datafiles/lhc.404085.p2584.dat};  
   \addplot+[smooth] file {datafiles/lhc.404085.p28301.dat}; 
   \legend{GBW,BK,rcBK};
  \nextgroupplot
   \addplot+[smooth] file {datafiles/lhc.404085.p24755.dat}; 
   \addplot+[smooth] file {datafiles/lhc.404085.p13089.dat}; 
   \addplot+[smooth] file {datafiles/lhc.404085.p6263.dat};  
   \legend{GBW,BK,rcBK};
  \end{groupplot}
 \end{tikzpicture}
 \caption[Azimuthal angular correlation results for the LHC]{The virtual photon-pion angular correlations at LHC at rapidity $Y_{\gamma} = Y_{\pi} = 4$. The upper and lower graphs show $M = \SI{4}{GeV}$ and $M = \SI{8}{GeV}$, respectively. As in figure~\ref{fig:corr_rhic_2}, the key difference is in the near side correlation, arising due to differences in the high-momentum region of the gluon distribution. Reproduced from figure 6 of~\cite{Stasto:2012ru}.}
 \label{fig:corr_lhc_2}
\end{figure}
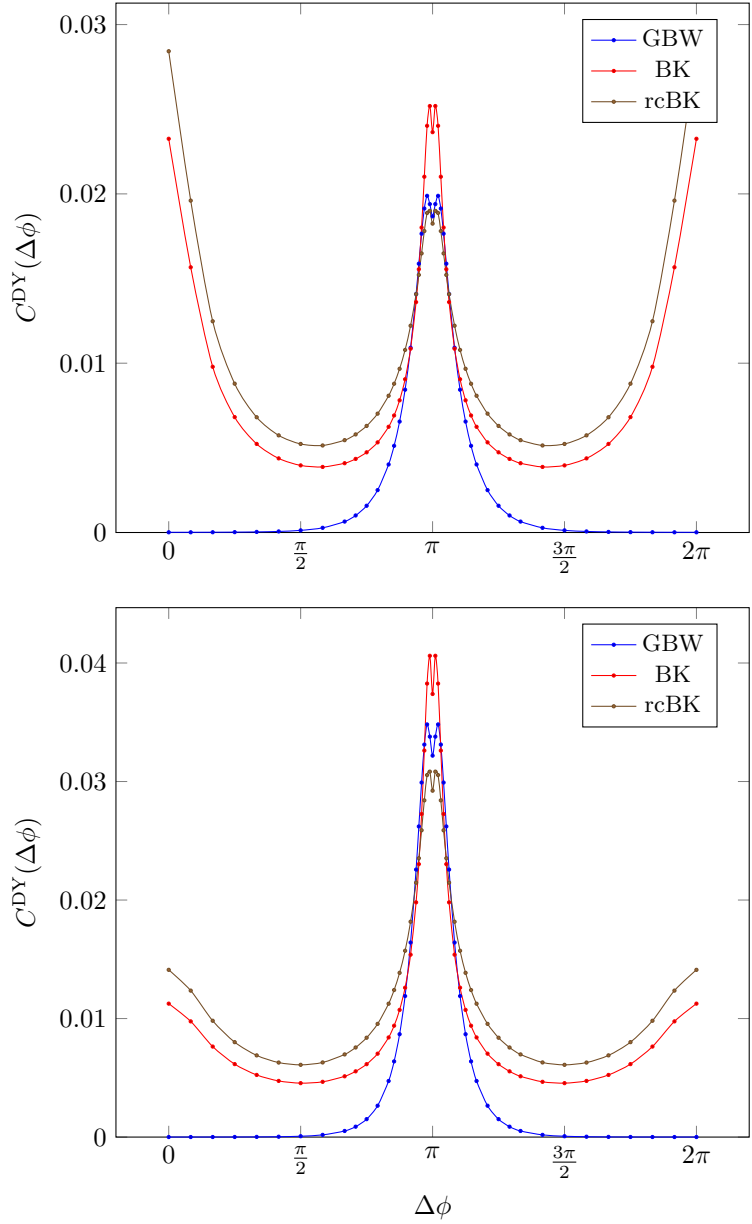

\begin{figure}
 \tikzsetnextfilename{corr_momentum_breakdown}
 \begin{tikzpicture}
  \begin{groupplot}[yticklabel={},group style={columns=2,xlabels at=edge bottom,ylabels at=edge left,group name=ktdist plots,horizontal sep=1.3cm},width=7cm,ktdist correlation graph]
  \nextgroupplot[ymax=0.004,title=RHIC]
   \foreach \i in {7,...,14}
    \addplot+[smooth] file {datafiles/ktsplit/rhic.mixed.\i.dat} \closedcycle;
  \nextgroupplot[ymax=0.06,title=LHC]
   \foreach \i in {7,...,14}
    \addplot+[smooth] file {datafiles/ktsplit/lhc.mixed.\i.dat} \closedcycle;
  \end{groupplot}
  \pgfplotsktdistlegend{(ktdist plots c2r1.east)}
 \end{tikzpicture}
 \caption[Contributions to the correlation split by momentum]{Breakdown of the contributions to the correlation, split by the transverse momentum $\DYcgcpperp$ acquired by the quark as it interacts with the gluon field of the target. Momentum ranges are in $\si{GeV}$. The plot on the left corresponds to the top plot of figure~\ref{fig:corr_rhic_2}, with $\vphotonmass = \SI{0.5}{GeV}$ at RHIC, and the one on the right corresponds to the top plot of figure~\ref{fig:corr_lhc_2} with $\vphotonmass = \SI{4}{GeV}$ using LHC parameters.
 In both cases, there is a fairly sharp transition between the momenta that contribute to the central back-to-back emission peak and those that contribute to the parallel emission peak. For the RHIC plot on the left, that transition is around $\SI{1}{GeV}$, and on the right, between $\SI{10}{GeV}$ and $\SI{100}{GeV}$.}
 \label{fig:corr_momentum_breakdown}
\end{figure}
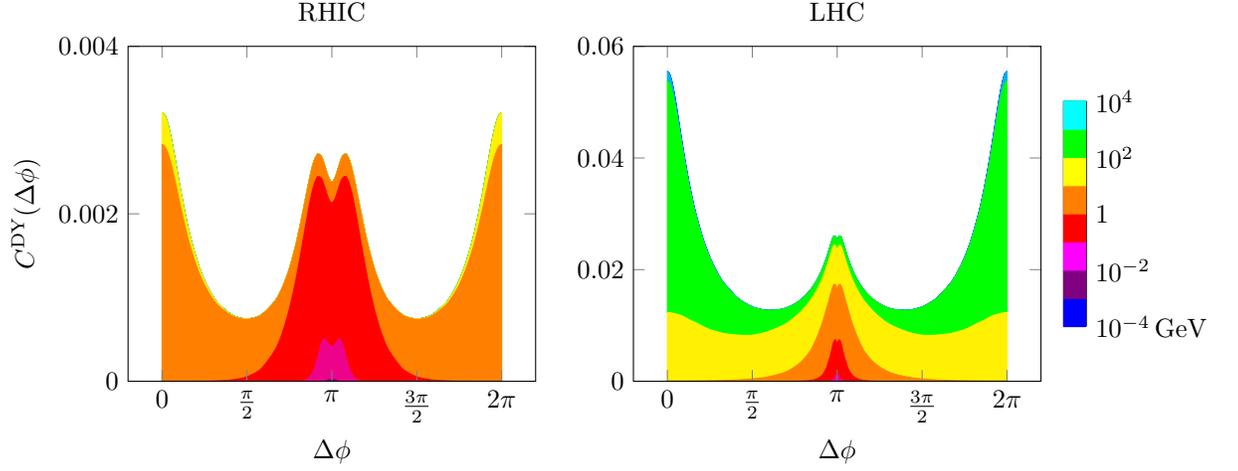

\subsection{Peak Characteristics}
The double peak of figures~\ref{fig:corr_rhic_2} and~\ref{fig:corr_lhc_2} is particularly interesting because it is unique to the Drell-Yan process; it doesn't appear in similar calculations for the correlation between produced hadrons or jets~\cite{Marquet:2007vb}\iftime{find other refs}.
We can trace the origin of the central ``dip'' to the kinematic factor which appears in equations~\eqref{eq:doubleinclusiveleptonpairxsec} and~\eqref{eq:singleinclusiveleptonpairxsec}:
\begin{multline}
 \bigl[1 + (1 - z)^2\bigr]\frac{z^2 \gluonkperp^2}{\bigl[(\vec\DYphotonpperp - z\vec\gluonkperp)^2 + \epsM^2\bigr]\bigl[\DYphotonpperp^2 + \epsM^2\bigr]} \\
  - z^2\epsM^2\biggl[\frac{1}{\DYphotonpperp^2 + \epsM^2} - \frac{1}{(\vec\DYphotonpperp - z\vec\gluonkperp)^2 + \epsM^2}\biggr]^2
\end{multline}
This approaches zero as $\gluonkperp \to 0$.
If we were analyzing the parton-level cross section~\eqref{eq:partonlevelquarkphotonxsec} as a function of $\DYphotonpperp$ and $\kqperp$, we would expect to see the correlation drop to zero when $\vec\DYphotonpperp + \vec\kqperp = \vec0$.
But the double inclusive result~\eqref{eq:doubleinclusiveleptonpairxsec} involves an integration over $\DYfragmentationfrac$, so even when evaluated at specific values of $\vec\DYphotonpperp$ and $\vec\ppionperp$, it incorporates a range of values of $\vec\gluonkperp = \frac{\vec\ppionperp}{\DYfragmentationfrac} + \vec\DYphotonpperp$.
Thus there won't be any point $(\vec\DYphotonpperp, \vec\ppionperp)$ where the correlation is exactly zero; effectively, the fragmentation blurs the peak structure.
We also integrate over transverse momentum when calculating the correlation~\eqref{eq:leptonpaircorrelation}, which smooths out the peak even more.

Depending on the kinematic conditions, the central suppression may be small enough that a true double peak doesn't appear.
This appears in reference~\cite{JalilianMarian:2012bd}, for example, which shows the same analysis for real photons, which is the $\vphotonmass \to 0$ limit of this calculation, and in figure~\ref{fig:doublepeakvmass} we can see that the peak depth decreases as we move toward lower photon masses.\iftime{add more points to plot down to $\vphotonmass\to 0$}

\begin{figure}
 \tikzsetnextfilename{doublepeakwidthvmass}
 \begin{tikzpicture}
  \begin{axis}[legend pos=north west,xlabel={$\vphotonmass$},ylabel={$\Delta\phi_\text{peak} - \pi$}]
   \addplot+ table { 
    M peakoffset
    3.5 0.505242244580697
    4.  0.5231020754291658
    4.5 0.5412438109809554
    5.  0.5696079385302859
    5.5 0.6139499351830389
    6.  0.7074384876008883
   };
   \addplot+ table { 
    M peakoffset
    3.5 0.39617642631907657
    4.  0.41358641517298667
    4.5 0.4345054838900131
    5.  0.46003774207045156
    5.5 0.5007418643445982
    6.  0.5723321804822441
   };
   \legend{GBW,BK};
  \end{axis}
 \end{tikzpicture}
 \tikzsetnextfilename{doublepeakdepthvmass}
 \begin{tikzpicture}
  \begin{axis}[legend pos=south west,xlabel={$\vphotonmass$},ylabel={$\bigl(\Delta\phi_\text{peak} - \Delta\phi(\pi)\bigr)/\Delta\phi(\pi)$}]
   \addplot+ table { 
    M peakdepth
    3.5 0.7418593086834159
    4.  0.9029418942018631
    4.5 1.0565303364025218
    5.  1.1392863248087637
    5.5 1.021735708663326
    6.  0.5477350382149863
   };
   \addplot+ table { 
    M peakdepth
    3.5 0.618835998312987
    4.  0.7549967325055231
    4.5 0.8855514195389754
    5.  0.9578430820251274
    5.5 0.8504956788909221
    6.  0.4139689642988043
   };
   \legend{GBW,BK};
  \end{axis}
 \end{tikzpicture}
 \caption[Variation of peak half-width and depth with $\vphotonmass$]{Plots of the half-width of the double peak, on the left, and the depth of the dip, on the right, as a function of virtual photon mass.}
 \label{fig:doublepeakvmass}
\end{figure}
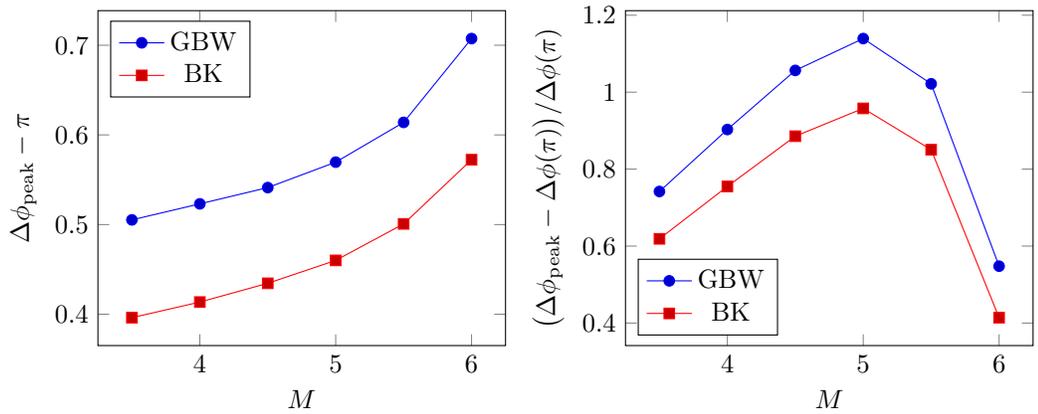

Because the correlation at the back-to-back emission peak comes from low-momentum regions of the gluon distribution, we can reasonably expect the characteristics of the peak to reflect the presence and nature of saturation.
In particular, the results in figures~\ref{fig:corr_rhic_2} and~\ref{fig:corr_lhc_2} show that the results for the running coupling BK equation, for which the evolution of the saturation scale with $\ln\frac{1}{\xg}$ is slower, leads to a suppression of the peak height relative to the fixed coupling.
A larger saturation scale $\Qs$ seems to produce a taller peak.

Figure~\ref{fig:doublepeakvmass} supports the same conclusion by showing that the fixed coupling BK equation produces a peak which is both shallower and thinner than that of the GBW model.
Looking back at figure~\ref{fig:kmax}, we see that the fixed coupling BK solution has a smaller saturation scale than the GBW model, so again, we see that a large saturation scale makes the double peak larger.

\section{Conclusions}

There are two main ideas to take away from this chapter.
First, the calculation of the lepton pair-hadron angular correlation serves as an example of how the CGC and hybrid formalism are used in practice.
This lays the groundwork for the next-to-leading order calculation to be shown in the next chapter.

Secondly, the correlation is an interesting numerical result in its own right.
It's clear that it is a sensitive probe of the gluon distribution in both high- and low-momentum regions.
We see an enhancement of the correlation around parallel emission, which gives us information about the high-momentum tail of the gluon distribution, and also the unique double peak structure for near-parallel emission, whose properties can reveal the characteristics of the gluon distribution in the low-momentum saturation regime.

%% file: crosssection.tex
\chapter{NLO Inclusive Pion Production}\label{ch:crosssection}

Leading order calculations, like that in the last chapter, typically give a decent approximation to measured cross sections.
But there are several reasons to look to the next order of perturbation theory in hopes of improving the results.
Of course, one generally expects that adding more terms to a perturbation series will give a more accurate approximation of the exact (all-order) result.
Beyond that, though, the cross section depends on the renormalization and factorization scales $\mu$ through terms which are subleading in $\alphas$.
If we are able to calculate some of those terms explicitly, the remaining dependence on $\mu$, which is unknown and thus must be considered a theoretical uncertainty, will be reduced.

Calculating the subleading corrections to a cross section is not a trivial task.
Beyond leading order, the number of diagrams that contribute grows drastically, often roughly exponentially with the order.
There are also additional kinematic degrees of freedom, corresponding to the momenta of emitted particles, which make the calculation of each diagram more complicated, and can introduce divergences which need to be regulated.

In this chapter I'll review the calculations for the inclusive process \HepProcess{\Pproton\Pnucleus \HepTo \Ppi\Panything} including all next-to-leading order corrections.
The leading order calculation was performed in reference~\cite{Dumitru:2002qt}, and~\cite{Dumitru:2005gt,Fujii:2011fh,Albacete:2012xq,Rezaeian:2012ye} added various subleading corrections.
However, the complete LO+NLO result was only derived in~\cite{Chirilli:2012jd} (summarized in~\cite{Chirilli:2011km}), and numerically implemented in~\cite{Stasto:2013cha} with the program SOLO, ``Saturation at One-Loop Order''~\cite{SOLO}.

\section{Kinematics}\label{sec:xseckinematics}
As in section~\ref{sec:pakinematics}, we take the proton and nucleon to have momenta in the $\pm z$ directions,
\begin{align}
 \xsecprotonp^\mu &= (E_p, 0, 0, \xsecprotonp) &
 \xsecnucleusp^\mu &= (E_A, 0, 0, -\xsecnucleusp)
\end{align}
leading to $\xsecprotonp^+ = \xsecnucleusp^- = \sqrt{s}$ (equation~\eqref{eq:equaltosqrts}). And, again, the fundamental collision involves one parton from the proton with plus component momentum $\xp\xsecprotonp^+$, and one parton from the nucleus, with a minus component of momentum $\xg\xsecnucleusp^-$ and transverse momentum $\vec{k}_{g\perp}$. $p^\mu$ represents the momentum of the produced hadron.

\subsection{Leading Order}
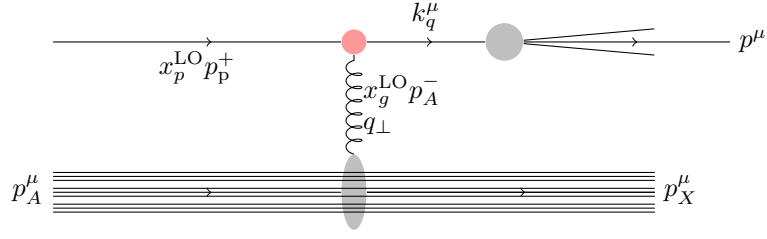
\begin{figure}
 \tikzsetnextfilename{qqlodiagram}
 \begin{tikzpicture}
  \node[circle,fill=red!40] (interaction) at (0,0) {};
  \draw[quark] (interaction) +(-4,0) -- (interaction) node[pos=0.5,below] {$\xplo\xsecprotonp^+$};
  \path (interaction) +(0,-2) node[ellipse,minimum width=3mm,minimum height=1cm,fill=gray!50] (gluon creation) {};
  \draw[gluon] (gluon creation) -- (interaction) node[pos=0.5,right,align=left] {$\xglo\xsecnucleusp^-$\\$\xseccgcpperp$};
  \path (interaction) +(2,0) node[circle,minimum width=5mm,fill=gray!50] (fragmentation) {};
  \draw[quark] (interaction) -- (fragmentation) node[pos=0.5,above] {$\xsecquarkp^\mu$};
  \draw[quark] (fragmentation) -- +(3,0) node[right] {$\xsechadronp^\mu$};
  \foreach \th in {-5,5} \draw (fragmentation) -- +(\th:2);
  \draw[nucleus] (gluon creation) +(-4,0) node[left] {$\xsecnucleusp^\mu$} to (gluon creation);
  \draw[nucleus] (gluon creation) to +(4,0) node[right] {$\xsecanythingp^\mu$};
 \end{tikzpicture}
 \caption{Feynman diagram for \HepProcess{\Pproton\Pnucleus \HepTo \Phadron\Panything} at leading order in the quark-quark channel}
 \label{fig:qqlodiagram}
\end{figure}

Applying conservation of momentum at the interaction vertex of figure~\ref{fig:qqlodiagram}, assuming that $p_p^- = \xsecnucleusp^+ = 0$ (equations~\eqref{eq:pluscomponent} and~\eqref{eq:minuscomponent}), gives
\begin{align}\label{eq:loconservation}
 \xplo\xsecprotonp^+ &= k^+ &
 \xglo \xsecnucleusp^- &= k^-
\end{align}
From equation~\eqref{eq:momentumpm}, assuming the quark mass is negligible ($k^\mu k_\mu = 0$), these become
\begin{align}
 \xplo\xsecprotonp^+ &= \kperp e^Y &
 \xglo\xsecnucleusp^- &= \kperp e^{-Y}
\end{align}
and then using equation~\eqref{eq:equaltosqrts} and $\kperp = \frac{\pperp}{z}$ we get
\begin{align}
 \xplo &= \frac{\pperp}{z\sqrt{s}} e^Y &
 \xglo &= \frac{\pperp}{z\sqrt{s}} e^{-Y}
\end{align}
Evidently, the measured momentum of the detected hadron determines everything except one unknown momentum fraction. We can choose any of $\z$, $\xplo$, or $\xglo$ to parametrize this unknown degree of freedom, and consequently, the formula for the cross section will involve a single integral over the chosen variable.

\subsection{Next-to-Leading Order}

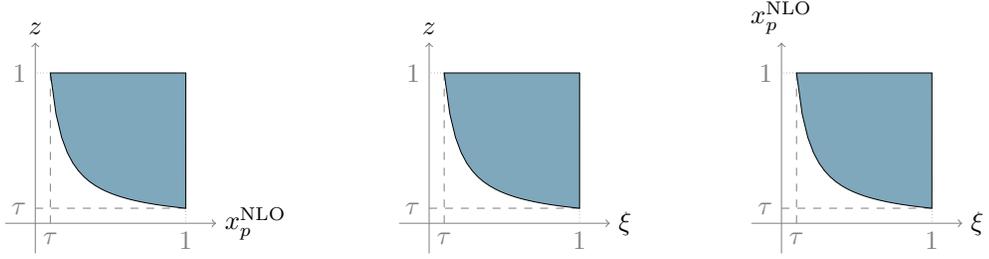
\begin{figure}
 \tikzsetnextfilename{integrationdomain1}
 \begin{tikzpicture}[scale=2]
  \draw[->,help lines] (-0.2,0) -- (1.2,0) node[right,black] {$\xpnlo$};
  \draw[->,help lines] (0,-0.2) -- (0,1.2) node[above,black] {$\z$};
  \draw[dashed,help lines] (0,0.1) node[left] {$\tau$} -- (1,0.1);
  \draw[dashed,help lines] (0.1,0) node[below] {$\tau$} -- (0.1,1);
  \draw[densely dotted,help lines] (1,0.1) -- (1,0) node[below] {$1$};
  \draw[densely dotted,help lines] (0.1,1) -- (0,1) node[left] {$1$};
  \filldraw[fill=cyan!50!black!60!white,draw=black,line width=0.2pt] plot[domain=.1:1] (\x,.1/\x) -- (1,1) -- cycle;
 \end{tikzpicture}
 \hspace{1cm}
 \tikzsetnextfilename{integrationdomain2}
 \begin{tikzpicture}[scale=2]
  \draw[->,help lines] (-0.2,0) -- (1.2,0) node[right,black] {$\xsecquarkpfrac$};
  \draw[->,help lines] (0,-0.2) -- (0,1.2) node[above,black] {$\z$};
  \draw[dashed,help lines] (0,0.1) node[left] {$\tau$} -- (1,0.1);
  \draw[dashed,help lines] (0.1,0) node[below] {$\tau$} -- (0.1,1);
  \draw[densely dotted,help lines] (1,0.1) -- (1,0) node[below] {$1$};
  \draw[densely dotted,help lines] (0.1,1) -- (0,1) node[left] {$1$};
  \filldraw[fill=cyan!50!black!60!white,draw=black,line width=0.2pt] plot[domain=.1:1] (\x,.1/\x) -- (1,1) -- cycle;
 \end{tikzpicture}
 \hspace{1cm}
 \tikzsetnextfilename{integrationdomain3}
 \begin{tikzpicture}[scale=2]
  \draw[->,help lines] (-0.2,0) -- (1.2,0) node[right,black] {$\xsecquarkpfrac$};
  \draw[->,help lines] (0,-0.2) -- (0,1.2) node[above,black] {$\xpnlo$};
  \draw[dashed,help lines] (0,0.1) node[left] {$\tau$} -- (1,0.1);
  \draw[dashed,help lines] (0.1,0) node[below] {$\tau$} -- (0.1,1);
  \draw[densely dotted,help lines] (1,0.1) -- (1,0) node[below] {$1$};
  \draw[densely dotted,help lines] (0.1,1) -- (0,1) node[left] {$1$};
  \filldraw[fill=cyan!50!black!60!white,draw=black,line width=0.2pt] plot[domain=.1:1] (\x,.1/\x) -- (1,1) -- cycle;
 \end{tikzpicture}
 \caption[Domains of integration in terms of various kinematic variables]{Domains of integration over the kinematic variables $\z$, $\xpnlo$, and $\xi$. Any two of the variables are sufficient to parametrize the integration domain, and the others can be taken as functions of whichever two are chosen as primary. The different images show the shape of the integration region for various choices of the two primary variables.}
 \label{fig:integrationdomain}
\end{figure}

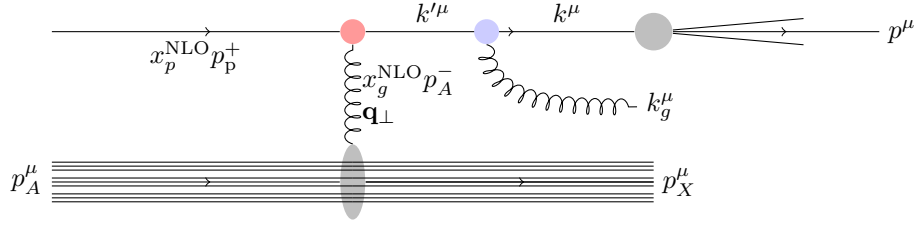
\begin{figure}
 \tikzsetnextfilename{qqnlodiagram}
 \begin{tikzpicture}
  \node[circle,fill=red!40] (interaction) at (0,0) {};
  \draw[quark] (interaction) +(-4,0) -- (interaction) node[pos=0.5,below] {$\xpnlo\xsecprotonp^+$};
  \path (interaction) +(0,-2) node[ellipse,minimum width=3mm,minimum height=1cm,fill=gray!50] (gluon creation) {};
  \draw[gluon] (gluon creation) -- (interaction) node[pos=0.5,right,align=left] {$\xgnlo\xsecnucleusp^-$\\$\vec{q}_\perp$};
  \path (interaction) +(4,0) node[circle,minimum width=5mm,fill=gray!50] (fragmentation) {};
  \draw[quark] (interaction) -- (fragmentation) node[pos=0.25,above] {$k'^\mu$} node[pos=0.75,above] {$k^\mu$} node[pos=0.45,circle,minimum width=2mm,fill=blue!20] (gluon emission) {};
  \draw[quark] (fragmentation) -- +(3,0) node[right] {$p^\mu$};
  \foreach \th in {-5,5} \draw (fragmentation) -- +(\th:2);
  \draw[nucleus] (gluon creation) +(-4,0) node[left] {$\xsecnucleusp^\mu$} to (gluon creation);
  \draw[nucleus] (gluon creation) to +(4,0) node[right] {$p_X^\mu$};
  \draw[gluon] (gluon emission) to[out=270,in=180] +(2,-1) node[right] {$k_g^\mu$};
 \end{tikzpicture}
 \caption{A representative Feynman diagram for \HepProcess{\Pproton\Pnucleus \HepTo \Phadron\Panything} at next-to-leading order in the quark-quark channel}
 \label{fig:qqnlodiagram}
\end{figure}

At next-to-leading order, the momentum of the emitted gluon is also unknown, which adds an additional degree of freedom, so the formula will involve an additional integral.
It's most convenient to express the NLO formulas in terms of a new variable, $\xsecquarkpfrac$, representing the fraction of $+$-component momentum retained by the original parton after it emits a gluon. For example, in figure~\ref{fig:qqnlodiagram},
\begin{align}\label{eq:xidef}
 \xsecquarkp^+ &= \xsecquarkpfrac \xsectotalp^+ &
 \xsecgluonp^+ &= (1 - \xsecquarkpfrac) \xsectotalp^+
\end{align}
We can then choose any two of $\z$, $\xpnlo$, $\xgnlo$, and $\xsecquarkpfrac$ to parametrize the region of integration for this double integral.
The integration region for various choices is shown in figure~\ref{fig:integrationdomain}.

As a momentum fraction, $\xsecquarkpfrac$ clearly cannot exceed $1$. To determine the lower bound on this new integral, though, we will need to express $\xsecquarkpfrac$ in terms of the other kinematic variables. Starting with momentum conservation,
\begin{subequations}
\begin{align}
 \xpnlo\xsecprotonp^+ &= \xsectotalp^+ = \xsecquarkp^+ + \xsecgluonp^+ \label{eq:nloconservationplus}\\
 \xgnlo\xsecnucleusp^- &= \xsectotalp^- = \xsecquarkp^- + \xsecgluonp^- \label{eq:nloconservationminus}\\
 \vec\xseccgcpperp &= \vec\xsectotalpperp = \vec\xsecquarkpperp + \vec\xsecgluonpperp \label{eq:nloconservationperp}
\end{align}
\end{subequations}
and plugging the definition~\eqref{eq:xidef} into equation~\eqref{eq:nloconservationplus} leads to
\begin{equation}
 \xpnlo\xsecprotonp^+ = k^+ + (1 - \xi)\xpnlo\xsecprotonp^+
\end{equation}
Solving this for $\xsecquarkpfrac$ and simplifying gives
\begin{align}
 \xsecquarkpfrac
 &= \frac{k^+}{\xpnlo\xsecprotonp^+} \label{eq:firstsimplifiedxi} \\
 &= \frac{\sqrt{\kperp^2 + m_k^2}}{\xpnlo\xsecprotonp^+}e^Y & &\text{equation~\eqref{eq:momentumpm}} \\
 &= \frac{\kperp}{\xpnlo\xsecprotonp^+} e^Y & & m_k \approx 0 \\
 &= \frac{\pperp}{\z \xpnlo\xsecprotonp^+} e^Y & & \kperp = \frac{\pperp}{\z} \\
 &= \frac{\pperp}{\z \xpnlo \sqrt{s}}e^Y & &\text{equation~\eqref{eq:equaltosqrts}}
\end{align}
Or, inverting the relation, we have
\begin{equation}\label{eq:xpnlo}
 \xpnlo = \frac{\pperp}{\z\xsecquarkpfrac\sqs}e^{\rapidity}
\end{equation}

Now looking at~\eqref{eq:nloconservationminus}, since the radiated gluon must be on shell ($k_g^2 = 0$) we have, for a gluon with nonzero momentum,
\begin{equation}
 k_g^- = \frac{\kgperp^2}{k_g^+} = \frac{\kgperp^2}{(1 - \xi)\xpnlo\xsecprotonp^+} \label{eq:simplifykgminus}
\end{equation}
and assuming that the quark is massless, $m_k\approx 0$, in combination with equation~\eqref{eq:firstsimplifiedxi} we have
\begin{equation}
 k^- \approx \frac{\kperp^2}{k_+} = \frac{\kperp^2}{\xi \xpnlo\xsecprotonp^+} \label{eq:simplifykminus}
\end{equation}
Both of these contain the combination $\xpnlo\xsecprotonp^+$ in the denominator, which can be rewritten as
\begin{equation}
 \frac{1}{\xpnlo\xsecprotonp^+} = \frac{z\xi\sqrt{s}}{\pperp}e^{-Y}\frac{1}{\sqrt{s}} = \frac{z\xi}{\pperp}e^{-Y} \label{eq:reciprocalxppplus}
\end{equation}
Then plugging equation~\eqref{eq:reciprocalxppplus} into~\eqref{eq:simplifykgminus} and~\eqref{eq:simplifykminus} and those into equation~\eqref{eq:nloconservationminus} gives
\begin{equation}
 \xgnlo\xsecnucleusp^- = \frac{z \kperp^2}{\pperp}e^{-Y} + \frac{z\xi \kgperp^2}{(1 - \xi)\pperp}e^{-Y}
\end{equation}
Since $z = \frac{\pperp}{\kperp}$, this simplifies to
\begin{equation}
 \xgnlo\xsecnucleusp^- = \kperp e^{-Y} + \frac{\xi \kgperp^2}{(1 - \xi)\kperp}e^{-Y}
\end{equation}
From conservation of transverse momentum, equation~\eqref{eq:nloconservationperp}, we have $\kgperp^2 = (\vec{q}_\perp - \vec\kperp)^2$, so
\begin{equation}
 \xgnlo\xsecnucleusp^- = \kperp e^{-Y} + \frac{(\vec{q}_\perp - \vec\kperp)^2}{\kperp}\frac{\xi}{1 - \xi}e^{-Y}
\end{equation}
and finally, setting $\xsecnucleusp^- = \sqrt{s}$, we can solve for $\xgnlo$ as
\begin{equation}\label{eq:xgnlo}
 \xgnlo = \frac{\kperp}{\sqrt{s}} e^{-Y} + \frac{(\vec\qperp - \vec\kperp)^2}{\kperp\sqrt{s}}\frac{\xi}{1 - \xi}e^{-Y} = \xglo\biggl(1 + \frac{(\vec\qperp - \vec\kperp)^2}{\kperp^2}\frac{\xsecquarkpfrac}{1 - \xsecquarkpfrac}\biggr)
\end{equation}

In practice, it is sometimes beneficial to assume the gluon transverse momentum $\vec\qperp$ is negligible or zero. This is approximately valid because the gluon distribution $\dipoleF$ falls off exponentially outside $\qperp \lesssim \Qs$, so the contribution of the region $\qperp \gtrsim \Qs$ to the cross section can be neglected. When $\Qs \ll \kperp$ that means we drop terms subleading in $\qperp/\kperp$, resulting in
\begin{equation}
 \xgnlo
 = \frac{\kperp}{\sqrt{s}} e^{-Y} + \frac{\kperp}{\sqrt{s}}\frac{\xi}{1 - \xi}e^{-Y}
 = \frac{\kperp}{\sqrt{s}}\frac{e^{-Y}}{1 - \xi}
\end{equation}

As a momentum fraction, $\xgnlo$ has to satisfy $\xgnlo < 1$.
Using~\eqref{eq:xgnlo}, that means the gluon momentum fraction $\xsecquarkpfrac$ is kinematically limited to
\begin{equation}\label{eq:xilimit}
 \xsecquarkpfrac < \frac{1 - \xglo}{1 - \xglo + \xglo(\vec\qperp - \vec\kperp)^2/\kperp^2}
\end{equation}
This excludes $\xsecquarkpfrac = 1$ unless $\vec\kperp = \vec\qperp$; physically, it corresponds to the fact that the gluon can only carry all the longitudinal momentum if the final state quark (prior to fragmentation) carries all the transverse momentum.
For simplicity, we ignore this limit for the basic computation described in chapter~\ref{ch:results} and reintroduce it in chapter~\ref{ch:beyondnlo}.

\section{NLO Cross Section Expressions}

As previously mentioned, several papers~\cite{Dumitru:2002qt,Dumitru:2005gt,Fujii:2011fh,Albacete:2012xq,Rezaeian:2012ye} spanning the decade from 2002 to 2012 calculated progressively more precise formulas for the $\pA\to hX$ cross section.
However, reference~\cite{Chirilli:2012jd} was the first to incorporate the complete NLO corrections, the formulas implemented by SOLO.
In this section I'll review the derivation of selected terms to demonstrate the process.

\subsection{Leading order}

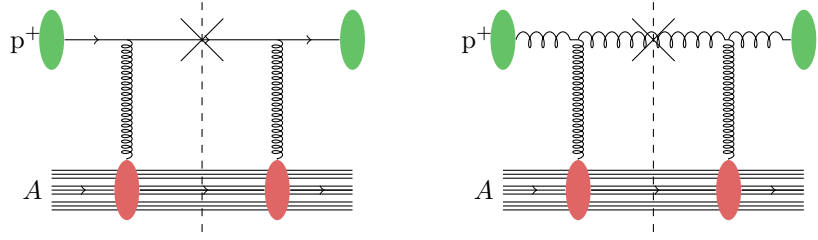
\begin{figure}
 \tikzsetnextfilename{locompounddiagrams}
 \begin{tikzpicture}
  \begin{scope}
   \coordinate (origin) at (0,0);
   \node[blob] (left proton) at (-2,0) {};
   \node[blob] (left nucleus) at (-1,-2) {};
   \coordinate (left interaction) at (left nucleus |- origin);
   \coordinate (right interaction) at ($(left interaction)!2!(left interaction -| origin)$);
   \node[blob] (right nucleus) at ($(left nucleus)!2!(left nucleus -| origin)$) {};
   \node[blob] (right proton) at ($(left proton)!2!(left proton -| origin)$) {};

   \node[left] at (left proton) {$\prbr$};
   \draw[quark] (left proton) to (left interaction);
   \draw[quark] (left interaction) to (right interaction);
   \draw[quark] (right interaction) to (right proton);
   \draw[nucleus] (left proton |- left nucleus) node[left] {$A$} to (left nucleus);
   \draw[nucleus] (left nucleus) to (right nucleus);
   \draw[nucleus] (right nucleus) to (right nucleus -| right proton);
   
   \draw[small gluon] (left interaction) to (left nucleus);
   \draw[small gluon] (right interaction) to (right nucleus);
   
   \node[fill=partondist!60!white,blob] at (left proton) {};
   \node[fill=gluondist!60!white,blob] at (left nucleus) {};
   \node[fill=partondist!60!white,blob] at (right proton) {};
   \node[fill=gluondist!60!white,blob] at (right nucleus) {};
   
   \draw[dashed] (origin) +(0,0.5) -- (origin |- left nucleus) -- +(0,-0.6);
   \draw (origin) +(45:.4) -- +(45:-.4) +(-45:.4) -- +(-45:-.4);
  \end{scope}
  \begin{scope}[xshift=6cm]
   \coordinate (origin) at (0,0);
   \node[blob] (left proton) at (-2,0) {};
   \node[blob] (left nucleus) at (-1,-2) {};
   \coordinate (left interaction) at (left nucleus |- origin);
   \coordinate (right interaction) at ($(left interaction)!2!(left interaction -| origin)$);
   \node[blob] (right nucleus) at ($(left nucleus)!2!(left nucleus -| origin)$) {};
   \node[blob] (right proton) at ($(left proton)!2!(left proton -| origin)$) {};

   \node[left] at (left proton) {$\prbr$};
   \draw[gluon] (left proton) to (left interaction);
   \draw[gluon] (left interaction) to (right interaction);
   \draw[gluon] (right interaction) to (right proton);
   \draw[nucleus] (left proton |- left nucleus) node[left] {$A$} to (left nucleus);
   \draw[nucleus] (left nucleus) to (right nucleus);
   \draw[nucleus] (right nucleus) to (right nucleus -| right proton);
   
   \draw[small gluon] (left interaction) to (left nucleus);
   \draw[small gluon] (right interaction) to (right nucleus);
   
   \node[fill=partondist!60!white,blob] at (left proton) {};
   \node[fill=gluondist!60!white,blob] at (left nucleus) {};
   \node[fill=partondist!60!white,blob] at (right proton) {};
   \node[fill=gluondist!60!white,blob] at (right nucleus) {};
   
   \draw[dashed] (origin) +(0,0.5) -- (origin |- left nucleus) -- +(0,-0.6);
   \draw (origin) +(45:.4) -- +(45:-.4) +(-45:.4) -- +(-45:-.4);
  \end{scope}
 \end{tikzpicture}
 \caption{Diagrams contributing to \HepProcess{\Pproton\Pnucleus \HepTo \Phadron\Panything} at leading order}
 \label{fig:locompounddiagrams}
\end{figure}

The leading order contributions, shown in figure~\ref{fig:locompounddiagrams}, can be calculated in much the same way as in chapter~\ref{ch:correlation}.
We can compute the amplitude for interaction between a quark and the gluon field of the nucleus, \HepProcess{\Pquark\Pnucleus \HepTo \Pquark\Panything}, as the Wilson line~\eqref{eq:wilsonline} in the fundamental representation (or its Fourier transform, to get it as a function of momentum):
\begin{equation}
 \scamp = \int\frac{\uddc\vec\xperp}{2\pi} e^{-i\vec\kperp\cdot\vec\xperp} U(\vec\xperp)
\end{equation}
To get the cross section, we multiply this by its complex conjugate to get
\begin{equation}
 \abs{\scamp}^2 = \int\frac{\uddc\vec\xperp\uddc\vec\yperp}{(2\pi)^2} e^{-i\vec\kperp\cdot\vec\xperp} e^{i\vec\kperp\cdot\vec\yperp} \trace U(\vec\xperp) \herm{U}(\vec\yperp)
\end{equation}
We then average over the unknown color charge distribution of the nucleus (just as we would average over initial spins in a QED calculation). The averaging operation is represented \note{where does $\frac{1}{\Nc}$ really come from?} by $\frac{1}{\Nc}\cgcavg{\cdots}$, and the resulting parton-level cross section is the momentum space dipole gluon distribution $\dipoleF$ from equation~\eqref{eq:dipoleFdefinition}.
\begin{align}
 \frac{\udddc\sigma_\text{LO}^{qA\to qA}}{\udc\rapidity\uddc\vec\kperp}
 &= \frac{1}{\Nc}\int\frac{\uddc\vec\xperp\uddc\vec\yperp}{(2\pi)^2} e^{-i\vec\kperp\cdot(\vec\xperp - \vec\yperp)} \Braket{\trace U(\vec\xperp) \herm{U}(\vec\yperp)}_{\xg} \\
 &= \frac{1}{\Nc}\int\frac{\uddc\vec\xperp\uddc\vec\yperp}{(2\pi)^2} e^{-i\vec\kperp\cdot(\vec\xperp - \vec\yperp)} \dipoleS(\vec\xperp, \vec\yperp) \\
 &= \dipoleF(\vec\kperp) 
\end{align}
The subscript $\xg$ indicates the longitudinal momentum fraction of the gluons involved in the interaction. This gives the cross section for one flavor of quark interacting with the gluon field of the nucleus.

We also have to include the contribution for a gluon interacting with the nucleus, \HepProcess{\Pgluon\Pnucleus \HepTo \Pgluon\Panything}.
This is the same but using the adjoint Wilson line~\eqref{eq:adjointwilsonline}.
Accordingly, the square of the amplitude is\footnote{Here all repeated indices, in particular $a$ and $b$, are summed over. This is a variant of the Einstein summation convention.}
\begin{equation}
 \abs{\scamp}^2 = \int\frac{\uddc\vec\xperp\uddc\vec\yperp}{(2\pi)^2} e^{-i\vec\kperp\cdot\vec\xperp} e^{i\vec\kperp\cdot\vec\yperp} \trace W^{ab}(\vec\xperp) \herm{W}^{ab}(\vec\yperp)
\end{equation}
Using identity~\eqref{eq:ident:wilsonlinereduce} the product becomes
\begin{equation}
 4\trace\Bigl\{\trace\bigl[T^a U(\vec\xperp) T^b \herm U(\vec\xperp)\bigr]\trace\bigl[\herm U(\vec\yperp) T^b U(\vec\yperp) T^a\bigr]\Bigr\}
\end{equation}
With the matrix indices explicit, this is
\begin{equation}
 4\trace\Bigl\{T_{ij}^a U_{jk}(\vec\xperp) T_{kl}^b \herm U_{li}(\vec\xperp) \herm U_{pq}(\vec\yperp) T_{qr}^b U_{rs}(\vec\yperp) T_{sp}^a\Bigr\}
\end{equation}
Using the identity $T_{ij}^a T_{kl}^b = \frac{1}{2}\delta_{il}\delta_{jk} - \frac{1}{2\Nc}\delta_{ij}\delta_{kl}$, after a slightly tedious calculation and much relabeling of indices this can be reduced to
\begin{multline}
 \trace\Bigl[
 U_{jk}(\vec\xperp) \herm U_{kj}(\vec\yperp) U_{il}(\vec\yperp) \herm U_{li}(\vec\xperp)
 - \frac{1}{\Nc} U_{jk}(\vec\xperp) \herm U_{ki}(\vec\xperp) U_{il}(\vec\yperp) \herm U_{lj}(\vec\yperp) \\
 - \frac{1}{\Nc} U_{il}(\vec\xperp) \herm U_{lj}(\vec\yperp) U_{jk}(\vec\yperp) \herm U_{ki}(\vec\xperp)
 + \frac{1}{\Nc^2} \underbrace{U_{ik}(\vec\xperp) \herm U_{ki}(\vec\xperp)}_{\Nc} \underbrace{U_{jl}(\vec\yperp) \herm U_{lj}(\vec\yperp)}_{\Nc}
 \Bigr]
\end{multline}
In the large-$\Nc$ limit, the middle two terms vanish, and we're left with
\begin{equation}
 \trace\bigl[U(\vec\xperp) \herm U(\vec\yperp)\bigr] \trace\bigl[U(\vec\yperp) \herm U(\vec\xperp)\bigr] - 1
\end{equation}
which, after color-averaging with $\frac{1}{\Nc^2 - 1}\cgcavg{\cdots}$, leaves us with
\begin{align}
 \adjointF(\kperp)
 &\defn \frac{\udddc\sigma_\text{LO}^{gA\to gA}}{\udc\rapidity\uddc\vec\kperp} \\
 &= \frac{1}{\Nc^2 - 1}\int\frac{\uddc\vec\xperp\uddc\vec\yperp}{(2\pi)^2} e^{-i\vec\kperp\cdot(\vec\xperp - \vec\yperp)} \Braket{\trace\bigl[U(\vec\xperp) \herm U(\vec\yperp)\bigr] \trace\bigl[U(\vec\yperp) \herm U(\vec\xperp)\bigr] - 1}_{\xg} \notag\\
 &= \frac{1}{\Nc^2 - 1}\int\frac{\uddc\vec\xperp\uddc\vec\yperp}{(2\pi)^2} e^{-i\vec\kperp\cdot(\vec\xperp - \vec\yperp)} \bigl(\dipoleS(\vec\xperp, \vec\yperp) \dipoleS(\vec\yperp, \vec\xperp) - 1\bigr)
\end{align}
In the last line I've used the \term{mean-field limit} to assume we can consider the two contracted Wilson line pairs to be uncorrelated.
If we separate the two terms inside parentheses, the $-1$ leads to a delta function $\delta^{(2)}(\vec\kperp)$ after integration, but in practice we always evaluate the gluon distribution at finite momentum, so we can safely drop that term.

To construct the final cross section for \HepProcess{\Pproton\Pnucleus \HepTo \Phadron\Panything}, we multiply the parton-level quark cross section by the quark distribution function $\xp q_i(\xp)$ and the fragmentation function $D_{\Phadron/i}(\z)/\z^2$, and similarly for the gluon cross section, and integrate over the longitudinal momentum space parametrized by $\z$, $\xp$, or $\xg$ as described in section~\ref{sec:xseckinematics}:
\begin{equation}
 \frac{\udddc\sigma_\text{LO}^{\pA\to hX}}{\udc\rapidity\uddc\vec{p}_\perp} = \int_\tau^1\frac{\udc\z}{\z^2}\Biggl[\sum_{\text{flavors }q}\xp q(\xp) D_{\Phadron/q}(\z) \dipoleF(\vec\kperp) + \xp g(\xp) D_{\Phadron/g}(\z) \adjointF(\xp, \kperp)\Biggr]
\end{equation}
where $\tau = \frac{\pperp}{\sqs}e^{\rapidity}$.

\subsection{Next-to-Leading Order Example}

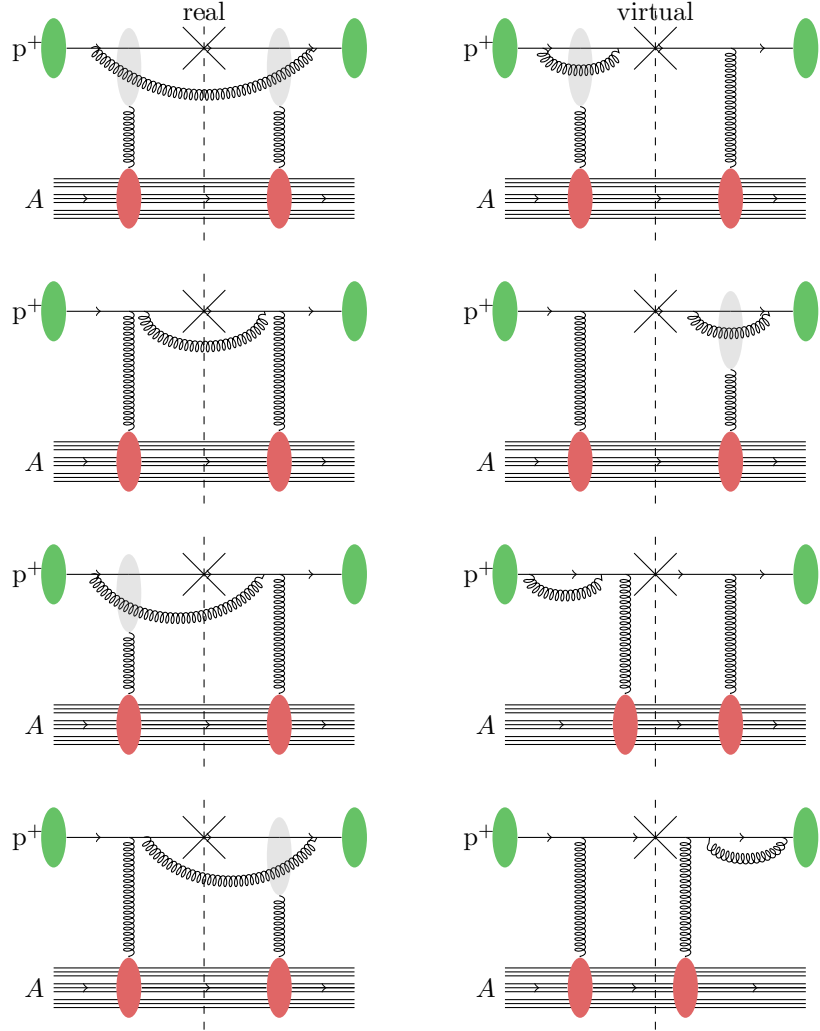
\begin{figure}
 \tikzsetnextfilename{nloqqcompounddiagrams}
 \begin{tikzpicture}[every node/.append style=transform shape]
  \node at (0cm,.5cm) {real};
  \node at (6cm,.5cm) {virtual};
  \begin{scope}
   \coordinate (origin) at (0,0);
   \node[blob] (left proton) at (-2,0) {};
   \node[blob] (left nucleus) at (-1,-2) {};
   \coordinate (left interaction) at (left nucleus |- origin);
   \coordinate (right interaction) at ($(left interaction)!2!(left interaction -| origin)$);
   \node[blob] (right nucleus) at ($(left nucleus)!2!(left nucleus -| origin)$) {};
   \node[blob] (right proton) at ($(left proton)!2!(left proton -| origin)$) {};

   \coordinate (left gluon attach) at ($(left proton)!.5!(left interaction)$);
   \coordinate (right gluon attach) at ($(right proton)!.5!(right interaction)$);
   
   \node[left] at (left proton) {$\prbr$};
   \draw[quark] (left proton) to (left interaction);
   \draw[quark] (left interaction) to (right interaction);
   \draw[quark] (right interaction) to (right proton);
   \draw[nucleus] (left proton |- left nucleus) node[left] {$A$} to (left nucleus);
   \draw[nucleus] (left nucleus) to (right nucleus);
   \draw[nucleus] (right nucleus) to (right nucleus -| right proton);
   \draw[small gluon] (left gluon attach) to[out=-45,in=225] (right gluon attach);
   
   \coordinate (left gluon intermediate) at ($(left interaction)!.5!90:(left proton)$);
   \coordinate (right gluon intermediate) at ($(right interaction)!.5!270:(right proton)$){};
   \node[blob,fill=gray,fill opacity=0.2,fit=(left interaction) (left gluon intermediate)] (left gluon blob) {};
   \node[blob,fill=gray,fill opacity=0.2,fit=(right interaction) (right gluon intermediate)] (right gluon blob) {};
   \draw[small gluon] (left gluon blob) to (left nucleus);
   \draw[small gluon] (right gluon blob) to (right nucleus);
   
   \node[fill=partondist!60!white,blob] at (left proton) {};
   \node[fill=gluondist!60!white,blob] at (left nucleus) {};
   \node[fill=partondist!60!white,blob] at (right proton) {};
   \node[fill=gluondist!60!white,blob] at (right nucleus) {};
   
   \draw[dashed] (origin) +(0,0.5) -- (origin |- left nucleus) -- +(0,-0.6);
   \draw (origin) +(45:.4) -- +(45:-.4) +(-45:.4) -- +(-45:-.4);
  \end{scope}

  \begin{scope}[yshift=-3.5cm]
   \coordinate (origin) at (0,0);
   \node[blob] (left proton) at (-2,0) {};
   \node[blob] (left nucleus) at (-1,-2) {};
   \coordinate (left interaction) at (left nucleus |- origin);
   \coordinate (right interaction) at ($(left interaction)!2!(left interaction -| origin)$);
   \node[blob] (right nucleus) at ($(left nucleus)!2!(left nucleus -| origin)$) {};
   \node[blob] (right proton) at ($(left proton)!2!(left proton -| origin)$) {};

   \coordinate (left gluon attach) at ($(left proton)!1.2!(left interaction)$);
   \coordinate (right gluon attach) at ($(right proton)!1.2!(right interaction)$);
   
   \node[left] at (left proton) {$\prbr$};
   \draw[quark] (left proton) to (left interaction);
   \draw[quark] (left interaction) to (right interaction);
   \draw[quark] (right interaction) to (right proton);
   \draw[nucleus] (left proton |- left nucleus) node[left] {$A$} to (left nucleus);
   \draw[nucleus] (left nucleus) to (right nucleus);
   \draw[nucleus] (right nucleus) to (right nucleus -| right proton);
   \draw[small gluon] (left gluon attach) to[out=-90,in=270] (right gluon attach);
   
   \draw[small gluon] (left interaction) to (left nucleus);
   \draw[small gluon] (right interaction) to (right nucleus);
   
   \node[fill=partondist!60!white,blob] at (left proton) {};
   \node[fill=gluondist!60!white,blob] at (left nucleus) {};
   \node[fill=partondist!60!white,blob] at (right proton) {};
   \node[fill=gluondist!60!white,blob] at (right nucleus) {};

   \draw[dashed] (origin) +(0,0.5) -- (origin |- left nucleus) -- +(0,-0.6);
   \draw (origin) +(45:.4) -- +(45:-.4) +(-45:.4) -- +(-45:-.4);
  \end{scope}

  \begin{scope}[yshift=-7cm]
   \coordinate (origin) at (0,0);
   \node[blob] (left proton) at (-2,0) {};
   \node[blob] (left nucleus) at (-1,-2) {};
   \coordinate (left interaction) at (left nucleus |- origin);
   \coordinate (right interaction) at ($(left interaction)!2!(left interaction -| origin)$);
   \node[blob] (right nucleus) at ($(left nucleus)!2!(left nucleus -| origin)$) {};
   \node[blob] (right proton) at ($(left proton)!2!(left proton -| origin)$) {};

   \coordinate (left gluon attach) at ($(left proton)!.5!(left interaction)$);
   \coordinate (right gluon attach) at ($(right proton)!1.2!(right interaction)$);
   
   \node[left] at (left proton) {$\prbr$};
   \draw[quark] (left proton) to (left interaction);
   \draw[quark] (left interaction) to (right interaction);
   \draw[quark] (right interaction) to (right proton);
   \draw[nucleus] (left proton |- left nucleus) node[left] {$A$} to (left nucleus);
   \draw[nucleus] (left nucleus) to (right nucleus);
   \draw[nucleus] (right nucleus) to (right nucleus -| right proton);
   \draw[small gluon] (left gluon attach) to[out=-60,in=240] (right gluon attach);
   
   \coordinate (left gluon intermediate) at ($(left interaction)!.5!90:(left proton)$);
   \node[blob,fill=gray,fill opacity=0.2,fit=(left interaction) (left gluon intermediate)] (left gluon blob) {};
   \draw[small gluon] (left gluon blob) to (left nucleus);
   \draw[small gluon] (right interaction) to (right nucleus);
   
   \node[fill=partondist!60!white,blob] at (left proton) {};
   \node[fill=gluondist!60!white,blob] at (left nucleus) {};
   \node[fill=partondist!60!white,blob] at (right proton) {};
   \node[fill=gluondist!60!white,blob] at (right nucleus) {};

   \draw[dashed] (origin) +(0,0.5) -- (origin |- left nucleus) -- +(0,-0.6);
   \draw (origin) +(45:.4) -- +(45:-.4) +(-45:.4) -- +(-45:-.4);
  \end{scope}

  \begin{scope}[yshift=-10.5cm]
   \coordinate (origin) at (0,0);
   \node[blob] (left proton) at (-2,0) {};
   \node[blob] (left nucleus) at (-1,-2) {};
   \coordinate (left interaction) at (left nucleus |- origin);
   \coordinate (right interaction) at ($(left interaction)!2!(left interaction -| origin)$);
   \node[blob] (right nucleus) at ($(left nucleus)!2!(left nucleus -| origin)$) {};
   \node[blob] (right proton) at ($(left proton)!2!(left proton -| origin)$) {};

   \coordinate (left gluon attach) at ($(left proton)!1.2!(left interaction)$);
   \coordinate (right gluon attach) at ($(right proton)!.5!(right interaction)$);
   
   \node[left] at (left proton) {$\prbr$};
   \draw[quark] (left proton) to (left interaction);
   \draw[quark] (left interaction) to (right interaction);
   \draw[quark] (right interaction) to (right proton);
   \draw[nucleus] (left proton |- left nucleus) node[left] {$A$} to (left nucleus);
   \draw[nucleus] (left nucleus) to (right nucleus);
   \draw[nucleus] (right nucleus) to (right nucleus -| right proton);
   \draw[small gluon] (left gluon attach) to[out=-60,in=240] (right gluon attach);
   
   \draw[small gluon] (left interaction) to (left nucleus);
   \coordinate (right gluon intermediate) at ($(right interaction)!.5!270:(right proton)$);
   \node[blob,fill=gray,fill opacity=0.2,fit=(right interaction) (right gluon intermediate)] (right gluon blob) {};
   \draw[small gluon] (right gluon blob) to (right nucleus);
   
   \node[fill=partondist!60!white,blob] at (left proton) {};
   \node[fill=gluondist!60!white,blob] at (left nucleus) {};
   \node[fill=partondist!60!white,blob] at (right proton) {};
   \node[fill=gluondist!60!white,blob] at (right nucleus) {};

   \draw[dashed] (origin) +(0,0.5) -- (origin |- left nucleus) -- +(0,-0.6);
   \draw (origin) +(45:.4) -- +(45:-.4) +(-45:.4) -- +(-45:-.4);
  \end{scope}

  \begin{scope}[xshift=6cm]
   \coordinate (origin) at (0,0);
   \node[blob] (left proton) at (-2,0) {};
   \node[blob] (left nucleus) at (-1,-2) {};
   \coordinate (left interaction) at (left nucleus |- origin);
   \coordinate (right interaction) at ($(left interaction)!2!(left interaction -| origin)$);
   \node[blob] (right nucleus) at ($(left nucleus)!2!(left nucleus -| origin)$) {};
   \node[blob] (right proton) at ($(left proton)!2!(left proton -| origin)$) {};

   \coordinate (left gluon attach) at ($(left proton)!.5!(left interaction)$);
   \coordinate (right gluon attach) at ($(left gluon attach)!2!(left interaction)$);
   
   \node[left] at (left proton) {$\prbr$};
   \draw[quark] (left proton) to (left interaction);
   \draw[quark] (left interaction) to (right interaction);
   \draw[quark] (right interaction) to (right proton);
   \draw[nucleus] (left proton |- left nucleus) node[left] {$A$} to (left nucleus);
   \draw[nucleus] (left nucleus) to (right nucleus);
   \draw[nucleus] (right nucleus) to (right nucleus -| right proton);
   \draw[small gluon] (left gluon attach) to[out=-90,in=270] (right gluon attach);
   
   \coordinate (left gluon intermediate) at ($(left interaction)!.5!90:(left proton)$);
   \node[blob,fill=gray,fill opacity=0.2,fit=(left interaction) (left gluon intermediate)] (left gluon blob) {};
   \draw[small gluon] (left gluon blob) to (left nucleus);
   \draw[small gluon] (right interaction) to (right nucleus);
   
   \node[fill=partondist!60!white,blob] at (left proton) {};
   \node[fill=gluondist!60!white,blob] at (left nucleus) {};
   \node[fill=partondist!60!white,blob] at (right proton) {};
   \node[fill=gluondist!60!white,blob] at (right nucleus) {};
   
   \draw[dashed] (origin) +(0,0.5) -- (origin |- left nucleus) -- +(0,-0.6);
   \draw (origin) +(45:.4) -- +(45:-.4) +(-45:.4) -- +(-45:-.4);
  \end{scope}

  \begin{scope}[xshift=6cm,yshift=-3.5cm]
   \coordinate (origin) at (0,0);
   \node[blob] (left proton) at (-2,0) {};
   \node[blob] (left nucleus) at (-1,-2) {};
   \coordinate (left interaction) at (left nucleus |- origin);
   \coordinate (right interaction) at ($(left interaction)!2!(left interaction -| origin)$);
   \node[blob] (right nucleus) at ($(left nucleus)!2!(left nucleus -| origin)$) {};
   \node[blob] (right proton) at ($(left proton)!2!(left proton -| origin)$) {};

   \coordinate (right gluon attach) at ($(right proton)!.5!(right interaction)$);
   \coordinate (left gluon attach) at ($(right gluon attach)!2!(right interaction)$);
   
   \node[left] at (left proton) {$\prbr$};
   \draw[quark] (left proton) to (left interaction);
   \draw[quark] (left interaction) to (right interaction);
   \draw[quark] (right interaction) to (right proton);
   \draw[nucleus] (left proton |- left nucleus) node[left] {$A$} to (left nucleus);
   \draw[nucleus] (left nucleus) to (right nucleus);
   \draw[nucleus] (right nucleus) to (right nucleus -| right proton);
   \draw[small gluon] (left gluon attach) to[out=-90,in=270] (right gluon attach);
   
   \coordinate (right gluon intermediate) at ($(right interaction)!.5!270:(right proton)$);
   \node[blob,fill=gray,fill opacity=0.2,fit=(right interaction) (right gluon intermediate)] (right gluon blob) {};
   \draw[small gluon] (right gluon blob) to (right nucleus);
   \draw[small gluon] (left interaction) to (left nucleus);
   
   \node[fill=partondist!60!white,blob] at (left proton) {};
   \node[fill=gluondist!60!white,blob] at (left nucleus) {};
   \node[fill=partondist!60!white,blob] at (right proton) {};
   \node[fill=gluondist!60!white,blob] at (right nucleus) {};
   
   \draw[dashed] (origin) +(0,0.5) -- (origin |- left nucleus) -- +(0,-0.6);
   \draw (origin) +(45:.4) -- +(45:-.4) +(-45:.4) -- +(-45:-.4);
  \end{scope}

  \begin{scope}[xshift=6cm,yshift=-7cm]
   \coordinate (origin) at (0,0);
   \node[blob] (left proton) at (-2,0) {};
   \node[blob] (left nucleus) at (-0.4,-2) {};
   \coordinate (left interaction) at (left nucleus |- origin);
   \node[blob] (right nucleus) at (1,-2) {};
   \coordinate (right interaction) at (right nucleus |- origin);
   \node[blob] (right proton) at ($(left proton)!2!(left proton -| origin)$) {};

   \coordinate (left gluon attach) at ($(left proton)!.2!(left interaction)$);
   \coordinate (right gluon attach) at ($(left proton)!.8!(left interaction)$);
   
   \node[left] at (left proton) {$\prbr$};
   \draw[quark] (left proton) to (left interaction);
   \draw[quark] (left interaction) to (right interaction);
   \draw[quark] (right interaction) to (right proton);
   \draw[nucleus] (left proton |- left nucleus) node[left] {$A$} to (left nucleus);
   \draw[nucleus] (left nucleus) to (right nucleus);
   \draw[nucleus] (right nucleus) to (right nucleus -| right proton);
   \draw[small gluon] (left gluon attach) to[out=-90,in=270] (right gluon attach);
   
   \draw[small gluon] (left interaction) to (left nucleus);
   \draw[small gluon] (right interaction) to (right nucleus);
   
   \node[fill=partondist!60!white,blob] at (left proton) {};
   \node[fill=gluondist!60!white,blob] at (left nucleus) {};
   \node[fill=partondist!60!white,blob] at (right proton) {};
   \node[fill=gluondist!60!white,blob] at (right nucleus) {};
   
   \draw[dashed] (origin) +(0,0.5) -- (origin |- left nucleus) -- +(0,-0.6);
   \draw (origin) +(45:.4) -- +(45:-.4) +(-45:.4) -- +(-45:-.4);
  \end{scope}

  \begin{scope}[xshift=6cm,yshift=-10.5cm]
   \coordinate (origin) at (0,0);
   \node[blob] (left proton) at (-2,0) {};
   \node[blob] (left nucleus) at (-1,-2) {};
   \coordinate (left interaction) at (left nucleus |- origin);
   \node[blob] (right nucleus) at (0.4,-2) {};
   \coordinate (right interaction) at (right nucleus |- origin);
   \node[blob] (right proton) at ($(left proton)!2!(left proton -| origin)$) {};

   \coordinate (left gluon attach) at ($(right proton)!.2!(right interaction)$);
   \coordinate (right gluon attach) at ($(right proton)!.8!(right interaction)$);
   
   \node[left] at (left proton) {$\prbr$};
   \draw[quark] (left proton) to (left interaction);
   \draw[quark] (left interaction) to (right interaction);
   \draw[quark] (right interaction) to (right proton);
   \draw[nucleus] (left proton |- left nucleus) node[left] {$A$} to (left nucleus);
   \draw[nucleus] (left nucleus) to (right nucleus);
   \draw[nucleus] (right nucleus) to (right nucleus -| right proton);
   \draw[small gluon] (left gluon attach) to[out=-90,in=270] (right gluon attach);
   
   \draw[small gluon] (left interaction) to (left nucleus);
   \draw[small gluon] (right interaction) to (right nucleus);
   
   \node[fill=partondist!60!white,blob] at (left proton) {};
   \node[fill=gluondist!60!white,blob] at (left nucleus) {};
   \node[fill=partondist!60!white,blob] at (right proton) {};
   \node[fill=gluondist!60!white,blob] at (right nucleus) {};
   
   \draw[dashed] (origin) +(0,0.5) -- (origin |- left nucleus) -- +(0,-0.6);
   \draw (origin) +(45:.4) -- +(45:-.4) +(-45:.4) -- +(-45:-.4);
  \end{scope}
 \end{tikzpicture}
 \caption[Diagrams contribution to \HepProcess{\Pproton\Pnucleus \HepTo \Phadron\Panything} in the quark-quark channel at next-to-leading order]{Diagrams contributing to the quark-quark channel at next-to-leading order. The left column shows real diagrams, in which the emitted gluon is an unmeasured final-state particle, and the right column shows virtual diagrams, in which the gluon is emitted and reabsorbed.}
 \label{fig:nloqqcompounddiagrams}
\end{figure}

The next-to-leading order contributions to the cross section arise from Feynman diagrams in which a gluon is emitted and/or absorbed by the parton coming from the proton. Naturally, this leads to a greatly increased number of Feynman diagrams; for example, figure~\ref{fig:nloqqcompounddiagrams} shows only those diagrams in which the initial- and final-state partons are both a quark, the $qq$ (or $q\to qg$ where the hadron comes from the final-state quark) channel. In addition to those, there are diagrams in which the initial- and final-state partons are both a gluon, the $gg$ channel (which comprises both $g\to gg$, and $g\to gq$ where the produced hadron comes from the gluon rather than the quark). There are also off-diagonal channels in which the initial- and final-state partons are different: the $qg$ channel ($q\to qg$ where the hadron comes from the final-state gluon) and $gq$ channel ($g\to gq$ where the hadron comes from the final-state quark).

The formulas for all these contributions are derived in references~\cite{Chirilli:2012jd}, which uses earlier results from~\cite{Marquet:2007vb,Dominguez:2011wm}.
Let's look at the quark-quark channel, for which the diagrams are shown in figure~\ref{fig:nloqqcompounddiagrams}, as an example.

The derivation proceeds in much the same way as that of the Drell-Yan process of chapter~\ref{ch:correlation}.
Using the general framework of light-cone perturbation theory and the CGC rules described in chapter~\ref{ch:saturation}, it's straightforward to see that the real diagrams, on the left side of figure~\ref{fig:nloqqcompounddiagrams}, correspond to the expression
\begin{multline}\label{eq:unintegratedrealqqresult}
 \frac{\udc^6 \sigma}{\udddc\vec\xsecquarkp\udddc\vec\xsecgluonp} =
 \alphas \CF \delta(\xp\xsecprotonp^+ - \xsecquarkp^+ - \xsecgluonp^+)
 \int \frac{\uddc\vec\xperp}{(2\pi)^2} \frac{\uddc\vec\xpperp}{(2\pi)^2} \frac{\uddc\vec\bperp}{(2\pi)^2} \frac{\uddc\vec\bpperp}{(2\pi)^2} \\
 \times
 e^{-i\vec\xsecgluonpperp\cdot(\vec\xperp - \vec\xpperp)} e^{-i\vec\xsecquarkpperp\cdot(\vec\bperp - \vec\bpperp)}
 \sum_{\lambda\alpha\beta}\conj{(\psi_{\alpha\beta}^\lambda)}(\vec\upperp)\psi_{\alpha\beta}^{\lambda}(\vec\uperp) \\
 \times
 \bigl[\sextupoleS(\vec\bperp, \vec\xperp, \vec\bpperp, \vec\xpperp) + \dipoleS(\vec\vperp, \vec\vpperp) - \tripoleS(\vec\bperp, \vec\xperp, \vec\vpperp) - \tripoleS(\vec\vperp, \vec\xpperp, \vec\bpperp)\bigr]
\end{multline}
where
\begin{align}
 \vec\uperp &= \vec\xperp - \vec\bperp &
 \vec\vperp &= (1 - \xsecquarkpfrac)\vec\xperp + \xsecquarkpfrac\vec\bperp
\end{align}
and similarly for the primed quantities, and
\begin{align}
 \tripoleS(\vec\xperp, \vec\bperp, \vec\yperp) &= \frac{1}{\CF\Nc}\Braket{\trace\bigl[U(\vec\xperp)\generator^d \herm{U}(\vec\yperp)\generator^c\bigr] W^{cd}(\vec\bperp)}_{\xg} \\
 \sextupoleS(\vec\xperp, \vec\yperp, \vec\xpperp, \vec\ypperp) &= \frac{1}{\CF\Nc}\Braket{\trace\bigl[U(\vec\xperp) \herm{U}(\vec\xpperp) \generator^d \generator^c\bigr] \bigl[W(\vec\yperp)\herm{W}(\vec\ypperp)\bigr]^{cd}}_{\xg}
\end{align}
As explained in chapter~\ref{ch:saturation}, each product of the form $UW$ arises from the interaction of the gluon field of the nucleus with the quark and the emitted gluon, whereas ``isolated'' fundamental Wilson lines (a $U$ without an associated factor of $W$) comes from interaction of the quark with the nucleus prior to emission of the gluon.
Accordingly, we can tell that the $\sextupoleS$ term arises from the top diagram on the left in figure~\ref{fig:nloqqcompounddiagrams}, the $\dipoleS$ term comes from the second diagram down, and each of the two $\tripoleS$ terms comes from one of the bottom diagrams on the left side of the figure.

Since we are computing an inclusive cross section, we'll integrate over the phase space of the unmeasured particle, here the emitted gluon.
After integrating over $\vec\xsecgluonp$, we get a factor of $\delta^{(2)}(\vec\xperp - \vec\xpperp)$, which simplifies $\sextupoleS(\vec\bperp, \vec\xperp, \vec\bpperp, \vec\xpperp) \to \dipoleS(\vec\bperp, \vec\bpperp)$. The expression~\eqref{eq:unintegratedrealqqresult} becomes
\begin{multline}\label{eq:realqqresult}
 \frac{\udc^3 \sigma}{\udddc\vec\xsecquarkp} =
 \alphas \CF 
 \int \frac{\uddc\vec\xperp}{(2\pi)^2} \frac{\uddc\vec\bperp}{(2\pi)^2} \frac{\uddc\vec\bpperp}{(2\pi)^2}
 e^{-i\vec\xsecquarkpperp\cdot(\vec\bperp - \vec\bpperp)}
 \sum_{\lambda\alpha\beta}\conj{(\psi_{\alpha\beta}^\lambda)}(\vec\upperp)\psi_{\alpha\beta}^{\lambda}(\vec\uperp) \\
 \times
 \bigl[\dipoleS(\vec\bperp, \vec\bpperp) + \dipoleS(\vec\vperp, \vec\vpperp) - \tripoleS(\vec\bperp, \vec\xperp, \vec\vpperp) - \tripoleS(\vec\vperp, \vec\xperp, \vec\bpperp)\bigr]
\end{multline}
where now
\begin{align}
 \vec\upperp &= \vec\xperp - \vec\bpperp &
 \vec\vpperp &= (1 - \xsecquarkpfrac)\vec\xperp + \xsecquarkpfrac\vec\bpperp
\end{align}

For the virtual diagrams, in the right column of figure~\ref{fig:nloqqcompounddiagrams}, the same procedure results in the expression
\begin{multline}\label{eq:virtualqqresult}
 \frac{\udddc\sigma}{\udddc\vec\xsecquarkp} =
 -2 \alphas \CF 
 \int \frac{\uddc\vec\vperp}{(2\pi)^2} \frac{\uddc\vec\vpperp}{(2\pi)^2} \frac{\uddc\vec\uperp}{(2\pi)^2}
 e^{-i\vec\xsecquarkpperp\cdot(\vec\vperp - \vec\vpperp)} \\
 \times
 \sum_{\lambda\alpha\beta}\conj{(\psi_{\alpha\beta}^\lambda)}(\vec\uperp)\psi_{\alpha\beta}^{\lambda}(\vec\uperp)
 \bigl[\dipoleS(\vec\vperp, \vec\vpperp) - \tripoleS(\vec\bperp, \vec\xperp, \vec\vpperp)\bigr]
\end{multline}
In this expression we only have one momentum, the quark momentum, because only the quark line passes through the cut, and this removes the need for the delta function because the initial quark momentum doesn't occur anywhere in the expression.
The integral over the loop momentum is already incorporated into the definition of the Wilson line.
And the factor of $2$ arises because the top two diagrams in the right column of figure~\ref{fig:nloqqcompounddiagrams} give the same contribution (specifically, the $\tripoleS$ term), and likewise for the bottom two diagrams (the $\dipoleS$ term).

Putting together both the real~\eqref{eq:realqqresult} and virtual~\eqref{eq:virtualqqresult} contributions gives the complete expression for the quark-quark channel.
\begin{multline}
 \frac{\udc^3 \sigma}{\udddc\vec\xsecquarkp} =
 \alphas \CF 
 \int \frac{\uddc\vec\uperp}{(2\pi)^2} \frac{\uddc\vec\vperp}{(2\pi)^2} \frac{\uddc\vec\vpperp}{(2\pi)^2}
 e^{-i\vec\xsecquarkpperp\cdot(\vec\bperp - \vec\bpperp)}
 \sum_{\lambda\alpha\beta}\conj{(\psi_{\alpha\beta}^\lambda)}(\vec\upperp)\psi_{\alpha\beta}^{\lambda}(\vec\uperp) \\
 \times
 \bigl[\dipoleS(\vec\bperp, \vec\bpperp) - \dipoleS(\vec\vperp, \vec\vpperp) + \tripoleS(\vec\bperp, \vec\xperp, \vec\vpperp) - \tripoleS(\vec\vperp, \vec\xperp, \vec\bpperp)\bigr]
\end{multline}
We can substitute in the definition of the splitting wavefunction,
\begin{equation}
 \psi_{\alpha\beta}^{\lambda}(\xp\xsecprotonp^+, \xsecquarkp^+, \vec\rperp) = 2\pi i\sqrt{\frac{2}{\xsecquarkp^+}}
 \begin{cases}
  \frac{\vec\rperp\cdot\epsilon_{\perp}^{(1)}}{\rperp^2}(\delta_{\alpha-}\delta_{\beta-} + \xsecquarkpfrac\delta_{\alpha+}\delta_{\beta+}), & \lambda = 1 \\
  \frac{\vec\rperp\cdot\epsilon_{\perp}^{(2)}}{\rperp^2}(\delta_{\alpha+}\delta_{\beta+} + \xsecquarkpfrac\delta_{\alpha-}\delta_{\beta-}), & \lambda = 2
 \end{cases}
\end{equation}
and simplify the gluon distributions using the identity
\begin{equation}
 \tripoleS(\vec\xperp, \vec\bperp, \vec\yperp) = \frac{\Nc}{2\CF}\biggl[\quadrupoleS(\vec\xperp, \vec\bperp, \vec\yperp) - \frac{1}{\Nc^2}\dipoleS(\vec\xperp, \vec\yperp)\biggr]
\end{equation}
to get
\begin{multline}
 \frac{\udc^3 \sigma}{\udddc\vec\xsecquarkp} =
 \alphas
 \int \frac{\uddc\vec\uperp}{(2\pi)^2} \frac{\uddc\vec\vperp}{(2\pi)^2} \frac{\uddc\vec\vpperp}{(2\pi)^2}
 e^{-i\vec\xsecquarkpperp\cdot(\vec\bperp - \vec\bpperp)}
 \sum_{\lambda\alpha\beta}\conj{(\psi_{\alpha\beta}^\lambda)}(\vec\upperp)\psi_{\alpha\beta}^{\lambda}(\vec\uperp) \\
 \times
 \Biggl(\CF\dipoleS(\vec\bperp, \vec\bpperp) - \CF\dipoleS(\vec\vperp, \vec\vpperp)
 - \frac{1}{2\Nc}\dipoleS(\vec\bperp, \vec\vpperp)
 + \frac{1}{2\Nc}\dipoleS(\vec\vperp, \vec\bpperp)\\
 + \frac{\Nc}{2}\quadrupoleS(\vec\bperp, \vec\xperp, \vec\vpperp)
 - \frac{\Nc}{2}\quadrupoleS(\vec\vperp, \vec\xperp, \vec\bpperp)
 \Biggr)
\end{multline}
which is in terms of only two gluon distributions, $\dipoleS$ and $\quadrupoleS$.
If we further use the mean field approximation~\eqref{eq:meanfieldappx}, that reduces to only $\dipoleS$; we do this in the SOLO code, but in this document I'll keep $\quadrupoleS$ explicit.

The last step in finding the cross section for \HepProcess{\Pproton\Pnucleus \HepTo \Phadron\Panything} in the quark-quark channel is to convolve with the parton distributions of the proton and the fragmentation functions.
\begin{equation}
 \frac{\udddc\sigma}{\udddc\vec\xsechadronp} = \int_\tau^1 \frac{\udc\z}{\z^2} \int_{\tau/\z}^1\udc\xsecquarkpfrac \xp q(\xp) D_{h/j}(\z) \frac{\udddc\sigma}{\udddc\vec\xsecquarkp}
\end{equation}

\subsection{Removal of Divergences}\label{sec:divergence}

In reference~\cite{Chirilli:2012jd}, the quark-quark channel cross section is given as the following two expressions:
\begin{multline}
 \frac{\alphas}{2\pi^2}\int_\tau^1\int_{\frac{\tau}{\z}}^1\frac{\udc\z\udc\xsecquarkpfrac}{\z^2}\xp q(\xp)D_{\Phadron/q}(\z)\frac{1 + \xsecquarkpfrac^2}{1 - \xsecquarkpfrac}
 \biggl[\CF\int\uddc\vec\xseccgcpperp \mathcal{I}(\vec\xsecquarkpperp, \vec\xseccgcpperp) \\
 + \frac{\Nc}{2}\int\uddc\vec\xseccgcpperp\uddc\vec\kgiperp\mathcal{J}(\vec\xsecquarkpperp, \vec\xseccgcpperp, \vec\kgiperp)\biggr]
\end{multline}
from the real diagrams and
\begin{multline}
 -\frac{\alphas}{2\pi^2}\int_\tau^1\int_0^1\frac{\udc\z\udc\xsecquarkpfrac}{\z^2}\xp q(\xp)D_{\Phadron/q}(\z)\frac{1 + \xsecquarkpfrac^2}{1 - \xsecquarkpfrac}
 \biggl[\CF\int\uddc\vec\xseccgcpperp \mathcal{I}(\vec\xseccgcpperp, \vec\xsecquarkpperp) \\
 + \frac{\Nc}{2}\int\uddc\vec\xseccgcpperp\uddc\vec\kgiperp\mathcal{J}(\vec\xseccgcpperp, \vec\xsecquarkpperp, \vec\kgiperp)\biggr]
\end{multline}
from the virtual diagrams, where
\begin{align}
 \mathcal{I}(\vec\xsecquarkpperp, \vec\xseccgcpperp)
 &= \dipoleF(\vec\xseccgcpperp) \biggl[
  \frac{\vec\xsecquarkpperp - \vec\xseccgcpperp}{(\vec\xsecquarkpperp - \vec\xseccgcpperp)^2}
  - \frac{\vec\xsecquarkpperp - \xsecquarkpfrac\vec\xseccgcpperp}{(\vec\xsecquarkpperp - \xsecquarkpfrac\vec\xseccgcpperp)^2}\biggr]^2
 \\
 \mathcal{J}(\vec\xsecquarkpperp, \vec\xseccgcpperp, \vec\kgiperp)
 &=
 \begin{multlined}[t][.6\textwidth]
  \biggl[
   \dipoleF(\vec\xseccgcpperp)\delta^{(2)}(\vec\xseccgcpperp - \vec\kgiperp)
   - \quadrupoleG(\vec\xseccgcpperp, \vec\kgiperp)
  \biggr]\\
  \times \frac{2(\vec\xsecquarkpperp - \xsecquarkpfrac\vec\xseccgcpperp)\cdot(\vec\xsecquarkpperp - \vec\kgiperp)}{(\vec\xsecquarkpperp - \xsecquarkpfrac\vec\xseccgcpperp)^2(\vec\xsecquarkpperp - \vec\kgiperp)^2}
 \end{multlined}
\end{align}
It's clear that these formulas are divergent in two ways: the \iterm{rapidity divergence}, where the integrand becomes undefined at $\xsecquarkpfrac\to 1$, and the \iterm{collinear divergence} when $\vec\xsecquarkpperp = \vec\xseccgcpperp$.
Regulating these divergences constitutes the largest obstacle in computing the full NLO corrections to the cross section.

Broadly speaking, here is how the process works. A simplified version of the quark-quark channel result looks like
\begin{multline}\label{eq:simplifiedqq}
 \biggl(\xsec*\biggr)_{qq} = \int_\tau^1 \frac{\udc\z}{\z^2} \xp q^{(0)}(\xp) D_{\Phadron/q}^{(0)}(\z) \dipoleF^{(0)}(\vec\xseccgcpperp) \\
 + \frac{\alphas}{2\pi}\int_\tau^1\frac{\udc\z}{\z^2} \int_{0}^1\udc\xsecquarkpfrac\,\xp q^{(0)}(\xp) D_{\Phadron/q}^{(0)}(\z)\frac{A\dipoleF^{(0)}(\vec\xseccgcpperp) + B\quadrupoleG^{(0)}(\vec\xseccgcpperp, \vec\xsecgluonpperp)}{1 - \xsecquarkpfrac}
\end{multline}
where $A$ and $B$ are nonsingular coefficients. The leading order term is finite on its own, but the next-to-leading order term is divergent at $\xsecquarkpfrac\to 1$. (In this simplified example, I'll omit the collinear divergence until later.)

We can rewrite this as
\begin{multline}
 \biggl(\xsec*\biggr)_{qq} = \int_\tau^1 \frac{\udc\z}{\z^2} \xp q^{(0)}(\xp) D_{\Phadron/q}^{(0)}(\z) \\
 \times \Biggl[\dipoleF^{(0)}(\vec\kperp) + \frac{\alphas}{2\pi} \int_{0}^1\udc\xsecquarkpfrac\frac{A\dipoleF^{(0)}(\vec\xseccgcpperp) + B\quadrupoleG^{(0)}(\vec\xseccgcpperp, \vec\xsecgluonpperp)}{1 - \xsecquarkpfrac}\Biggr]
\end{multline}
$\dipoleF^{(0)}(\vec\xseccgcpperp)$ is not actually a physical quantity, so there's no reason for it to be finite.
We can then assume that $\dipoleF^{(0)}(\vec\xseccgcpperp)$ is divergent in a way that cancels out the rapidity divergence, so the total of the terms in square brackets is finite.

At this point we could define the renormalized gluon distribution $\dipoleF(\vec\xseccgcpperp)$ to be exactly the term in brackets, so that the quark-quark cross section would just be
\begin{equation}
 \biggl(\xsec*\biggr)_{qq} = \int_\tau^1 \frac{\udc\z}{\z^2} \xp q^{(0)}(\xp) D_{\Phadron/q}^{(0)}(\z)\dipoleF(\vec\xseccgcpperp)
\end{equation}
but that would make $\dipoleF$ dependent on the exact coefficients $A$ and $B$.
If we do that, we can't generalize $\dipoleF$ to any other calculation.
Instead, we should define the renormalized distribution as the sum of $\dipoleF^{(0)}$ and some other divergent term that cancels out the singularity in the remaining part of the expression. That will leave us with
\begin{equation}
 \biggl(\xsec*\biggr)_{qq} = \int_\tau^1 \frac{\udc\z}{\z^2} \xp q^{(0)}(\xp) D_{\Phadron/q}^{(0)}(\z)\bigl[\dipoleF(\vec\xseccgcpperp) + C(\z, \xsecquarkpfrac, \vec\xseccgcpperp)\bigr]
\end{equation}
$C$ is called the \iterm{coefficient function}.
The procedure for how we choose to split the term in brackets between $\dipoleF$ and $C$ is a \iterm{renormalization scheme}.\footnote{For a detailed accessible description of subtractions, see sections 4.3 and 7.1 of reference~\cite{PinkBook}.}

When there are multiple divergences, namely the rapidity divergence and collinear divergence, we can generalize the procedure to incorporate these divergences into the parton distribution and fragmentation function as well.
In doing so, it's worth considering the physical nature of the divergences~\cite{Chirilli:2012jd}:
\begin{itemize}
 \item
 The rapidity divergence occurs when the quark momentum fraction $\xsecquarkpfrac$ goes to $1$, so that the emitted gluon's momentum becomes negligible in the rest frame of the target.
 These extremely low-momentum gluon emissions are physically indistinguishable from the target's own gluon field.
 When we renormalize the rapidity divergence using the target gluon distribution, we are effectively absorbing the low-momentum gluon emissions into the structure of the target.
 \item
 The collinear divergence, on the other hand, occurs when the quark and the emitted gluon are traveling in the same direction with nearly equal momenta.
 However, we know that a propagating quark is dressed with a cloud of virtual gluons (and other particles).
 An emitted gluon with nearly the same momentum as the quark will effectively be part of that cloud, so it makes sense to absorb the collinear divergence into the function that represents the nonperturbative dynamics of the propagating physical quark.
 This is either the fragmentation function, if the gluon is emitted after the interaction, or the integrated parton distribution, if it's emitted before the interaction.
\end{itemize}

With this in mind, let's look at a slightly less simplified version of the quark-quark channel cross section, just like~\eqref{eq:simplifiedqq} except that this contains the collinear divergence as well,
\begin{multline}\label{eq:simplifiedqqtwodiv}
 \biggl(\xsec*\biggr)_{qq} = \int_\tau^1 \frac{\udc\z}{\z^2} \xp q^{(0)}(\xp) D_{\Phadron/q}^{(0)}(\z) \dipoleF^{(0)}(\vec\xseccgcpperp) \\
 + \frac{\alphas}{2\pi}\int_\tau^1\frac{\udc\z}{\z^2} \int_{\tau/\z}^1\udc\xsecquarkpfrac\,\xp q^{(0)}(\xp) D_{\Phadron/q}^{(0)}(\z)\frac{A\dipoleF^{(0)}(\vec\xseccgcpperp) + B\quadrupoleG^{(0)}(\vec\xseccgcpperp, \vec\xsecgluonpperp)}{(1 - \xsecquarkpfrac)(\vec\xseccgcpperp - \vec\xsecgluonpperp)^2}
\end{multline}
With all three of the functions $\xp q^{(0)}(\xp)$, $D_{\Phadron/q}^{(0)}(\z)$, and $\dipoleF^{(0)}(\vec\xseccgcpperp)$ playing a role in removing the divergences, we should wind up with an expression of the form
\begin{multline}
 \biggl(\xsec*\biggr)_{qq} = \int_\tau^1 \frac{\udc\z}{\z^2}
 \Biggl[\xp q^{(0)}(\xp) - \frac{\alphas}{2\pi} \int_{\xp}^1 \udc\xi' \frac{C'\xp q(\xp/\xi')}{(\vec\xseccgcpperp - \vec\xsecgluonpperp)^2}\Biggr] \\
 \times \Biggl[D_{\Phadron/q}^{(0)}(\z) - \frac{\alphas}{2\pi} \int_{\z}^1\udc\xi'' \frac{C'' D_{\Phadron/q}(\z/\xi'')}{(\vec\xseccgcpperp - \vec\xsecgluonpperp)^2}\Biggr] \\
 \times \Biggl[\dipoleF^{(0)}(\vec\kperp) + \frac{\alphas}{2\pi} \int_{0}^1\udc\xsecquarkpfrac\frac{A\dipoleF^{(0)}(\vec\kperp) + B\quadrupoleG^{(0)}(\vec\kperp, \vec\xseccgcpperp)}{1 - \xsecquarkpfrac}\Biggr]
\end{multline}
where $C'$ and $C''$ are nonsingular coefficients (like $A$ and $B$).
For simplicity, let me express this as
\begin{equation}
 \biggl(\xsec*\biggr)_{qq} = \int_\tau^1 \frac{\udc\z}{\z^2}(q_0 - q_1)(D_0 - D_1)(F_0 + F_1)
\end{equation}
where $q_0$ and $q_1$ stand for the bare parton distribution and the subtraction, respectively, and similarly for $D$ and $F$.
Distributing we find
\begin{multline}\label{eq:distributed}
 \biggl(\xsec*\biggr)_{qq} = \int_\tau^1 \frac{\udc\z}{\z^2}\bigl[
   q_0 D_0 F_0 + q_0 D_0 F_1
 - q_1 D_0 F_0 - q_1 D_0 F_1\\
 - q_0 D_1 F_0 - q_0 D_1 F_1
 + q_1 D_1 F_0 + q_1 D_1 F_1
 \bigr]
\end{multline}
Note that each term is at an order in $\alphas$ corresponding to the sum of its subscripts (e.g. the last term is $\orderof{\alphas^3}$).

Now, if we go back to~\eqref{eq:simplifiedqqtwodiv}, which is the bare result,
\begin{align}
 \biggl(\xsec*\biggr)_{qq} &= \int_\tau^1 \frac{\udc\z}{\z^2}\bigl[q_0 D_0 F_0 \\
\intertext{and perform the rapidity subtraction}
 & + q_0 D_0 F_1 \\
\intertext{and then incorporate the parton distribution and fragmentation function}
 &- q_1 D_0 F_0 - q_0 D_1 F_0 \\
\intertext{and the rapidity subtractions that have to be incorporated into those subtractions}
 &- q_1 D_0 F_1 - q_0 D_1 F_1\bigr]
\end{align}
then we reproduce the distributed result~\eqref{eq:distributed} up to terms which are finite and subleading in $\alphas$ (specifically $\orderof{\alphas^2}$).
This is the basic procedure by which subtracting divergences gives a finite result.

Performing the full subtractions for the real and virtual contributions involves a fairly tedious calculation which is the subject of reference~\cite{Chirilli:2012jd}.
That work uses the $\msbar$ \term{renormalization scheme}, which entails regularizing integrals by dimensional regularization and bringing a term proportional to $-\frac{1}{\epsilon} + \eulergamma$ into the gluon distribution; the details of $\msbar$ renormalization are described in many standard references.
The result works out to
\begin{multline}
 \xsec 1nqq = \frac{\alphas}{2\pi}\int_\tau^1\int_{\frac{\tau}{\z}}^1\frac{\udc\z\udc\xsecquarkpfrac}{\z^2}\xp q(\xp)D_{\Phadron/q}(\z)
 \biggl[\int\frac{\uddc\vec\xperp\uddc\vec\yperp}{(2\pi)^2}\dipoleS(\vec\rperp)\h 12qq \\
 + \int\frac{\uddc\vec\xperp\uddc\vec\bperp\uddc\vec\yperp}{(2\pi)^4}\quadrupoleS(\vec\sperp, \vec\tperp)\h 14qq\biggr]
\end{multline}
where the hard factors $\h12qq$ and $\h14qq$ are given in equations~\eqref{eq:xsec12qq} and~\eqref{eq:xsec14qq}.

\section{Adapting Formulas for Numerical Implementation}

As they are given in reference~\cite{Chirilli:2012jd}, the formulas are integrals over various sets of position coordinates.
A computational implementation needs code to evaluate the integrands and also to perform the integration.
The integrands involve various singular terms proportional to $\delta(1 - \xsecquarkpfrac)$ and the plus prescription $\frac{1}{(1 - \xsecquarkpfrac)_+}$, which need to be rewritten in terms of well-behaved (nonsingular) functions in order to be numerically evaluated.
There are also Fourier factors like $e^{-i\vec\kperp\cdot\vec\rperp}$ in the integrals, which are numerically difficult to evaluate, so in the interest of accuracy it will be useful to integrate out those factors symbolically where possible.
This section describes the details of that process.

\subsection{Eliminating Singularities in \texorpdfstring{$\xsecquarkpfrac$}{xi}}\label{sec:xisingularities}
Each term in the cross section can be written as a sum of three parts: one proportional to a plus prescription-regularized fraction, one proportional to a delta function, and one ``normal'' contribution which has no singularities that will prevent its numerical evaluation.
\begin{equation}\label{eq:xsectermbreakdown}
 \int_\tau^1\udc\z \int_{\frac{\tau}{\z}}^1\udc\xsecquarkpfrac\biggl[\frac{\Fs*(\z,\xsecquarkpfrac)}{(1 - \xsecquarkpfrac)_+} + \Fd*(\z,\xsecquarkpfrac)\delta(1 - \xsecquarkpfrac) + \Fn*(\z,\xsecquarkpfrac)\biggr]
\end{equation}

To make the $\Fs*$ term numerically tractable, we need to simplify it using the definition of the plus prescription,
\begin{equation}\label{eq:plusprescription}
 f(x)_+\udc x \defn f(x)\udc x - \delta(1 - x)\udc x \int_0^1 f(y)\udc y
\end{equation}
When this is integrated from some $a\in[0,1]$ to $1$, that integral can be simplified as follows:
\begin{align}
 \int_a^1\frac{g(x)}{(1 - x)_+}\udc x &= \int_a^1\frac{g(x)}{1 - x}\udc x - g(1)\int_0^1\frac{1}{1 - y}\udc y \\
                                      &= \int_a^1\frac{g(x)}{1 - x}\udc x - g(1)\int_a^1\frac{1}{1 - x}\udc x - g(1)\int_0^a\frac{1}{1 - x}\udc x \\
                                      &= \int_a^1\frac{g(x) - g(1)}{1 - x}\udc x + g(1)\ln(1 - a)
\end{align}
Applying that to the $\Fs*$ term from~\eqref{eq:xsectermbreakdown}, we get
\begin{equation}
 \int_\tau^1\udc z \int_{\frac{\tau}{z}}^1\udc\xsecquarkpfrac \frac{\Fs*(z,\xsecquarkpfrac)}{(1 - \xsecquarkpfrac)_+}
 = \int_\tau^1\udc z\int_{\frac{\tau}{z}}^1\udc\xsecquarkpfrac\,\frac{\Fs*(z,\xsecquarkpfrac) - \Fs*(z,1)}{1 - \xsecquarkpfrac} + \int_\tau^1\udc z\,\Fs*(z,1)\ln\biggl(1 - \frac{\tau}{z}\biggr)
\end{equation}

The $\Fd*$ term is easier, as it only requires us to do the integration over $\xsecquarkpfrac$:
\begin{equation}
 \int_\tau^1\udc\z \int_{\frac{\tau}{\z}}^1\udc\xsecquarkpfrac \Fd*(\z,\xsecquarkpfrac)\delta(1 - \xsecquarkpfrac)
 = \int_\tau^1\udc\z \Fd*(\z,1)
\end{equation}

And of course, the $\Fn*$ term of~\eqref{eq:xsectermbreakdown} requires no symbolic integration.

Putting the three terms together, we get the following expression, which is equivalent to~\eqref{eq:xsectermbreakdown} but is suitable for numerical evaluation:
\begin{equation}
 \int_\tau^1\udc\z\int_{\frac{\tau}{\z}}^1\udc\xsecquarkpfrac\biggl[\frac{\Fs*(\z,\xsecquarkpfrac) - \Fs*(\z,1)}{1 - \xsecquarkpfrac} + \Fn*(\z,\xsecquarkpfrac)\biggr] + \int_\tau^1\udc\z\biggl[\Fs*(\z,1)\ln\biggl(1 - \frac{\tau}{\z}\biggr) + \Fd*(\z,1)\biggr]\label{eq:contributions}
\end{equation}
It is easy to split each hard factor up into the parts $\Fs*$, $\Fn*$, and $\Fd*$ and then the integration can be done automatically with an implementation of this formula.

\subsection{Reducing Additional Degrees of Freedom}
In general, each term of the cross section depends on several other variables beyond $\z$ and $\xsecquarkpfrac$. Which variables \note{awkward phrasing}those are varies from one term to another, though, and even depends on how a given term is expressed.

Broadly speaking there are three different ways that a term might be expressed:
\begin{description}
 \item[Cartesian position space] the term will be expressed as an integral over $\z$, $\xsecquarkpfrac$, and transverse displacements
 \begin{align}
  \vec\rperp &= \vec\xperp - \vec\yperp &
  \vec\sperp &= \vec\xperp - \vec\bperp &
  \vec\tperp &= \vec\yperp - \vec\bperp
 \end{align}
 This is closest to the original form given in reference~\cite{Chirilli:2012jd}, but is not particularly useful for computation because it includes Fourier factors like $e^{-i\vec\kperp\cdot\vec\rperp}$, which are numerically unstable.
 \item[Radial position space] the term will be expressed as an integral over $\z$, $\xsecquarkpfrac$, and the radial part of transverse positions $\rperp$, $\sperp$, and $\tperp$. In order to express the term in this way, we'll assume angular independence of the gluon distribution $\dipoleS$.
 \item[Momentum space] the term will be expressed as an integral over $\z$, $\xsecquarkpfrac$, and some number of momentum variables $\kgperp$, $\kgiperp$, etc. This is the most accurate form for numerical computation, but it requires us to do some symbolic Fourier integrals, which are not possible for every term. Also, the details of how to convert a given term into momentum space vary significantly from one term to another, even when it is possible.
\end{description}

\subsubsection{Dipole-Type Terms}\label{sec:dipolehardfactor}
Dipole-type terms involve integrals over only two transverse position coordinates, $\vec\xperp$ and $\vec\yperp$. A generic term of this type can be written as
\begin{multline}
 \xsec n2ji = \biggl(\frac{\alpha_s}{2\pi}\biggr)^n\int\frac{\udc\z\udc\xsecquarkpfrac}{\z^2}\xp f_i(\xp,\mu)D_{\Phadron/j}(\z, \mu) \\
 \times \int\frac{\uddc\vec\xperp\uddc\vec\yperp}{(2\pi)^2} \genericS(\vec\xperp, \vec\yperp) e^{-i C_0 \vec\kperp\cdot(\vec\xperp - \vec\yperp)} \hampl n2ji(\vec\xperp, \vec\yperp) \label{eq:unintegrateddipolecontribution}
\end{multline}
where $n = 0$ for leading order or $1$ for next-to-leading order, and $i$ and $j$ represent parton flavors (gluon, up, down, strange, and heavy quarks). $C_0$ is some factor independent of the position and momentum coordinates, and $\genericS$ represents one or more factors of the dipole gluon distribution $\dipoleS$. The hard factor $\h n2ji$ from reference~\cite{Chirilli:2012jd} corresponds to the combination
\begin{equation}
 \h n2ji = e^{-i C_0 \vec\kperp\cdot(\vec\xperp - \vec\yperp)} \hampl n2ji
\end{equation}

To manipulate these terms, we first simplify the expression by changing coordinates to
\begin{align}
 \vec{R}_\perp &= \vec\xperp + \vec\yperp &
 \vec\rperp &= \vec\xperp - \vec\yperp
\end{align}
In this work, we assume that $\dipoleS$, and thus $\genericS$, is independent\footnote{$\vec{R}_\perp$ is the impact parameter, conventionally denoted $\vec{b}$. Incorporating impact parameter dependence can have a noticeable effect on cross sections, as described in refs.~\autocite{Berger:2010sh,Berger:2011ew,Berger:2012wx}, but that will not be explored here.} of $\vec{R}_\perp$ and the angle of $\vec\rperp$, so we can write $\genericS(\vec\xperp, \vec\yperp)$ as $\genericS(r_\perp)$. It's also easy to verify, by inspection of the expressions in ref.~\autocite{Chirilli:2012jd}, that all the hard factors $\h n2ji$ are independent of $\vec{R}_\perp$. Accordingly, we can change variables
\begin{equation}
 \uddc\vec\xperp\uddc\vec\yperp = \frac{1}{4}\uddc\vec{R}_\perp\uddc\vec\rperp
\end{equation}
and integrate out the $\vec{R}_\perp$ dependence, defining
\begin{equation}
 \Sperp \defn \int\frac{\uddc\vec{R}_\perp}{4}
\end{equation}
$\Sperp$ can be interpreted as the cross-sectional area of the target\note{which is also the elastic scattering cross section in the limit of very large $\sqrt{s}$}. After this integration, the term~\eqref{eq:unintegrateddipolecontribution} becomes
\begin{equation}
 \xsec n2ji = \Sperp\biggl(\frac{\alphas}{2\pi}\biggr)^n\int\frac{\udc\z\udc\xsecquarkpfrac}{\z^2}\xp f_i(\xp,\mu)D_{\Phadron/j}(\z, \mu)\int\frac{\uddc\vec\rperp}{(2\pi)^2}\genericS(\rperp) e^{-i C_0 \vec\kperp\cdot\vec\rperp} \hampl n2ji(\vec\rperp)\label{eq:integrateddipolecontribution}
\end{equation}

Recall from section~\ref{sec:xisingularities} that a term can be expressed as a combination of a plus-regularized contribution, a delta-function contribution, and a normal contribution, as in equation~\eqref{eq:xsectermbreakdown}. We can obtain the three functions $\Fs n2ji$, $\Fd n2ji$, and $\Fn n2ji$ for each term by breaking down the term into these three contributions and pulling out all the integrals --- not only those over $\z$ and $\xsecquarkpfrac$, but also over position and momentum variables. Different terms (and different representations of those terms) will be integrated over different variables, so the third and additional arguments of $\Fs n2ji$ etc. will be used to indicate what is being integrated over. When the third argument is a 2D position vector, that indicates the Cartesian representation; when it is the magnitude of a 2D position vector, that indicates the radial representation; when it is a 2D momentum vector, that indicates the momentum-space representation.

\paragraph{Cartesian position space}
It's straightforward to break down equation~\eqref{eq:integrateddipolecontribution} into the three functions $\Fs*$, $\Fn*$, and $\Fd*$ from section~\ref{sec:xisingularities} as follows:
\begin{subequations}\label{eq:masterformula2}
\begin{align}
 \frac{\Fs n2ji(\z,\xsecquarkpfrac, \vec\rperp)}{(1 - \xsecquarkpfrac)_+} &= \Sperp\frac{\xp f_i D_{\Phadron/j}}{z^2}\frac{\genericS(r_\perp)}{(2\pi)^2} \biggl(\frac{\alpha_s}{2\pi}\biggr)^n e^{-i C_0\vec\kperp\cdot\vec\rperp} \bigl(\text{plus-regulated part of }\hampl n2ji\bigr) \\
 \Fn n2ji(\z,\xsecquarkpfrac, \vec\rperp) &= \Sperp\frac{\xp f_i D_{\Phadron/j}}{z^2}\frac{\genericS(r_\perp)}{(2\pi)^2} \biggl(\frac{\alpha_s}{2\pi}\biggr)^n e^{-i C_0\vec\kperp\cdot\vec\rperp} \bigl(\text{normal part of }\hampl n2ji\bigr) \\
 \Fd n2ji(\z,\xsecquarkpfrac, \vec\rperp)\delta(1 - \xsecquarkpfrac) &= \Sperp\frac{\xp f_i D_{\Phadron/j}}{z^2}\frac{\genericS(r_\perp)}{(2\pi)^2} \biggl(\frac{\alpha_s}{2\pi}\biggr)^n e^{-i C_0\vec\kperp\cdot\vec\rperp} \bigl(\text{delta function part of }\hampl n2ji\bigr)\label{eq:masterformula2delta}
\end{align}
\end{subequations}
The entire term~\eqref{eq:integrateddipolecontribution} can then be written as
\begin{equation}
 \xsec n2ji = \int\udc\z\udc\xsecquarkpfrac \int\uddc\vec\rperp\biggl[\frac{\Fs n2ji(\z,\xsecquarkpfrac, \vec\rperp)}{(1 - \xsecquarkpfrac)_+} + \Fn n2ji(\z,\xsecquarkpfrac, \vec\rperp) + \Fd n2ji(\z,\xsecquarkpfrac, \vec\rperp)\delta(1 - \xsecquarkpfrac)\biggr]
\end{equation}
or, using equation~\eqref{eq:contributions}, as
\begin{multline}\label{eq:dipolemasterformula}
 \xsec n2ji
 = \int_\tau^1\udc\z\int_{\tau/\z}^1\udc\xsecquarkpfrac \int\uddc\vec\rperp \biggl[\frac{\Fs n2ji(\z, \xsecquarkpfrac, \vec\rperp) - \Fs n2ji(\z, 1, \vec\rperp)}{1 - \xsecquarkpfrac} + \Fn n2ji(\z, \xsecquarkpfrac, \vec\rperp)\biggr] \\
 + \int_\tau^1\udc\z\int\uddc\vec\rperp \biggl[\Fs n2ji(\z, 1, \vec\rperp)\ln\biggl(1 - \frac{\tau}{\z}\biggr) + \Fd n2ji(\z, 1, \vec\rperp)\biggr]
\end{multline}
This is the integral actually performed by SOLO when this term is evaluated in position space. The functions $\Fs n2ji$, $\Fn n2ji$, and $\Fd n2ji$ can be picked out from the corresponding expression for $\h n2ji$ in reference~\cite{Chirilli:2012jd} using~\eqref{eq:masterformula2}, transcribed directly into the program, and then equation~\eqref{eq:dipolemasterformula} is implemented by a generic integration routine.

\paragraph{Radial position space}
It turns out that the non-Fourier part of each dipole-type term, $\hampl n2ji$, is independent of the angle of $\vec\rperp$.
This means we can integrate over the angle in equation~\eqref{eq:integrateddipolecontribution},
\begin{equation}\label{eq:radialdipolehardfactorintegral}
 \iint\frac{\uddc\vec\rperp}{(2\pi)^2} \genericS(\rperp) e^{-i C_0 \vec\kperp\cdot\vec\rperp} \hampl n2ji(r_\perp)
 = \frac{1}{2\pi}\int_0^\infty\udc \rperp \rperp \genericS(\rperp) \besselJzero(C_0 \kperp\rperp) \hampl n2ji(r_\perp)
\end{equation}
and again, pick out the pieces of the term as
\begin{subequations}\label{eq:masterformula2radial}
\begin{align}
 \frac{\Fs n2ji(\z,\xsecquarkpfrac, \rperp)}{(1 - \xsecquarkpfrac)_+} &= \frac{S_\perp}{\z^2} \xp f_i D_{\Phadron/j}\frac{\dipoleS(\vec\rperp)}{2\pi}\biggl(\frac{\alpha_s}{2\pi}\biggr)^n\rperp\besselJzero(\kperp\rperp)\bigl(\text{plus-regulated part of }\hampl n2ji\bigr) \\
 \Fn n2ji(\z,\xsecquarkpfrac, \rperp) &= \frac{S_\perp}{\z^2} \xp f_iD_{\Phadron/j}\frac{\dipoleS(\vec\rperp)}{2\pi}\biggl(\frac{\alpha_s}{2\pi}\biggr)^n\rperp\besselJzero(\kperp\rperp)\bigl(\text{normal part of }\hampl n2ji\bigr) \\
 \Fd n2ji(\z,\xsecquarkpfrac, \rperp)\delta(1 - \xsecquarkpfrac) &= \frac{S_\perp}{\z^2} \xp f_iD_{\Phadron/j}\frac{\dipoleS(\vec\rperp)}{2\pi}\biggl(\frac{\alpha_s}{2\pi}\biggr)^n\rperp\besselJzero(\kperp\rperp)\bigl(\text{delta function part of }\hampl n2ji\bigr)
\end{align}
\end{subequations}
The entire term, analogous to equation~\eqref{eq:dipolemasterformula}, will be
\begin{multline}\label{eq:dipolemasterformularadial}
 \biggl(\frac{\udddc\sigma}{\udc Y\uddc \pperp}\biggr)_\text{dip}
 = \int_\tau^1\udc\z\int_{\tau/\z}^1\udc\xsecquarkpfrac \int_0^\infty\udc\rperp \biggl[\frac{\Fs n2ji(\z, \xsecquarkpfrac, \rperp) - \Fs n2ji(\z, 1, \rperp)}{1 - \xsecquarkpfrac} + \Fn n2ji(\z, \xsecquarkpfrac, \rperp)\biggr] \\
 + \int_\tau^1\udc\z\int_0^\infty\udc\rperp \biggl[\Fs n2ji(\z, 1, \rperp)\ln\biggl(1 - \frac{\tau}{\z}\biggr) + \Fd n2ji(\z, 1, \rperp)\biggr]
\end{multline}

Essentially, the change from Cartesian to radial coordinates, equation~\eqref{eq:radialdipolehardfactorintegral}, amounts to making the replacement
\begin{equation}\label{eq:maptoradial}
 e^{-i C_0 \vec\kperp\cdot\vec\rperp}\to 2\pi\rperp \besselJzero(C_0 \kperp\rperp)
\end{equation}
and adjusting the variable of integration accordingly. In practice, this is the easiest way to determine the radial formulas: start with the Cartesian position-space expression from~\cite{Chirilli:2012jd}, replace the Fourier factors with Bessel functions \`a la~\eqref{eq:maptoradial}, and then break the formula down into $\Fs n2ji$, $\Fn n2ji$, and $\Fd n2ji$.

This formulation improves numerical accuracy compared to the Cartesian implementation, by virtue of requiring one fewer oscillatory integral.

\paragraph{Momentum space}
\note{momentum-space terms have to be calculated individually}In some cases, we can express a contribution to the cross section in terms of the momentum space gluon distribution as defined in equation~\eqref{eq:dipoleFdefinition}. The mathematical procedure for doing this is fairly specific to each individual term, but it will work out to some convolution of $\dipoleF$ and the Fourier transform of the rest of the term, $\hampl n2ji$:
\begin{equation}\label{eq:hardfactorfouriertransform}
 \hfampl n2ji(\kperp) = \iint\frac{\uddc\vec\rperp}{(2\pi)^2} e^{i\vec\kperp\cdot\vec\rperp} \hampl n2ji(\vec\rperp)
\end{equation}

For example, suppose $\genericS = \dipoleS$ in equation~\eqref{eq:radialdipolehardfactorintegral}. We can express $\dipoleS$ as
\begin{equation}\label{eq:dipoledistributionfouriertransform}
 \dipoleS(\vec\rperp)
 = \frac{1}{S_\perp}\iint\uddc\vec\qperp e^{i\vec\qperp\cdot\vec\rperp}\dipoleF(\vec\qperp)
\end{equation}
and plug it in to the position integral of equation~\eqref{eq:integrateddipolecontribution} to get
\begin{equation}\label{eq:momentumdipolehardfactorintegral}
 \iint\frac{\uddc\vec\rperp}{(2\pi)^2}\dipoleS(\vec\rperp) e^{-i C_0 \vec\kperp\cdot\vec\rperp} \hampl n2ji(\rperp)
 = \frac{1}{S_\perp}\iint\uddc\vec\qperp \dipoleF(\vec\qperp)\hfampl n2ji(\vec\qperp - C_0\vec\kperp)
\end{equation}
Or if $\genericS = \bigl[\dipoleS\bigl]^2$, we get
\begin{multline}\label{eq:momentumdipolesquaredhardfactorintegral}
 \iint\frac{\uddc\vec\rperp}{(2\pi)^2}\bigl[\dipoleS(\vec\rperp)\bigr]^2 e^{-i C_0 \vec\kperp\cdot\vec\rperp} \hampl n2ji(\rperp) \\
 = \frac{1}{\Sperp^2}\iint\uddc\vec\qiperp \dipoleF(\vec\qiperp)\iint\uddc\vec\qiiperp \dipoleF(\vec\qiiperp) \hfampl n2ji(C_0 \vec\kperp - \vec\qiperp - \vec\qiiperp)
\end{multline}
or so on. In general, there will be as many factors of $\dipoleF$ as there were of $\dipoleS$, and an extra 2D integration for every $\dipoleF$ beyond the first.

In the end, we'll get an expression for the contribution to the cross section of the form:
\begin{multline}\label{eq:dipolemasterformulamomentum}
 \biggl(\frac{\udddc\sigma}{\udc Y\uddc\vec \pperp}\biggr)_\text{dip}
 = \int_\tau^1\udc\z\int_{\tau/\z}^1\udc\xsecquarkpfrac \iint\uddc\vec\qperp \biggl[\frac{\Fs n2ji(\z, \xsecquarkpfrac, \vec\qperp) - \Fs n2ji(\z, 1, \vec\qperp)}{1 - \xsecquarkpfrac} + \Fn n2ji(\z, \xsecquarkpfrac, \vec\qperp)\biggr] \\
 + \int_\tau^1\udc\z\iint\uddc\vec\qperp \biggl[\Fs n2ji(\z, 1, \vec\qperp)\ln\biggl(1 - \frac{\tau}{\z}\biggr) + \Fd n2ji(\z, 1, \vec\qperp)\biggr]
\end{multline}
where\footnote{While the functions $\Fs n2ji$, $\Fn n2ji$, and $\Fd n2ji$ do depend on $\vec\kperp$, recall that $\vec\kperp = \frac{\vec\pperp}{z}$, so the dependence on $\vec\kperp$ does not add any additional degrees of freedom beyond those encapsulated in $z$. This is why the dependence on $\vec\kperp$ is not explicit in equations~\eqref{eq:masterformula2momentum}.}
\begin{subequations}\label{eq:masterformula2momentum}
\begin{align}
 \frac{\Fs n2ji(\z,\xsecquarkpfrac, \vec\qperp)}{(1 - \xsecquarkpfrac)_+} &= \frac{\xp f_i D_{\Phadron/j}}{\z^2} \dipoleF(\vec\qperp)\biggl(\frac{\alpha_s}{2\pi}\biggr)^n\bigl(\text{plus-regulated part of }\hfampl n2ji\bigr) \\
 \Fn n2ji(\z,\xsecquarkpfrac, \vec\qperp) &= \frac{\xp f_i D_{\Phadron/j}}{\z^2} \dipoleF(\vec\qperp)\biggl(\frac{\alpha_s}{2\pi}\biggr)^n\bigl(\text{normal part of }\hfampl n2ji\bigr) \\
 \Fd n2ji(\z,\xsecquarkpfrac, \vec\qperp)\delta(1 - \xsecquarkpfrac) &= \frac{\xp f_i D_{\Phadron/j}}{\z^2} \dipoleF(\vec\qperp)\biggl(\frac{\alpha_s}{2\pi}\biggr)^n\bigl(\text{delta function part of }\hfampl n2ji\bigr)
\end{align}
\end{subequations}
As mentioned, the number of momentum variables involved depends on how many factors of $\dipoleS$ there are.

The advantage of this formulation is that, in momentum space, $\Fs n2ji$, $\Fn n2ji$, and $\Fd n2ji$ don't oscillate.
General-purpose integration algorithms can be wildly inaccurate when applied to oscillating functions.
Although specialized algorithms~\cite{Levin1996,Ooura1999,Inverarity2002} do exist for performing Fourier-type integrals, it's still easier and more accurate to use equation~\eqref{eq:dipolemasterformulamomentum} so we can work with a relatively smooth, non-oscillatory integrand.

However, it's not always possible to express a given term in momentum space. Unless the Fourier transform of $\hampl n2ji$~\eqref{eq:hardfactorfouriertransform} can be computed symbolically, we won't have a momentum-space expression to use. One could compute that Fourier transform numerically, but that's no different from what we do in the radial position-space representation.

Furthermore, in deriving equation~\eqref{eq:masterformula2momentum} we haven't really \emph{eliminated} the Fourier factor; we've just ``shifted'' it into the Fourier transform of the dipole gluon distribution, equation~\eqref{eq:dipoledistributionfouriertransform}. We still need to compute $\dipoleF$ from $\dipoleS$ before calculating the cross section with equation~\eqref{eq:dipolemasterformulamomentum}. Still, the Fourier transform of $\dipoleS$ tends to be ``better behaved'' than the hard factor $\h n2ji$. $\dipoleS$ is a relatively smooth function which generally takes a roughly Gaussian form, and in fact under the assumption of angular independence, we can perform one integration symbolically:
\begin{equation}
 \dipoleF(\vec\kperp)
 = \Sperp\iint\uddc\vec\rperp e^{-i\vec\kperp\cdot\vec\rperp}\dipoleS(\vec\rperp)
 = 2\pi\Sperp\int_0^\infty\udc\rperp \rperp\besselJzero(\kperp\rperp)\dipoleS(\vec\rperp)
\end{equation}
This is still an oscillatory integral, but a one-dimensional one, so standard methods are much better able to handle it than they would a more complicated multidimensional integral.

In fact, it's possible to compute the solution to the BK equation in momentum space directly. Theoretically we should be able to use that in equations~\eqref{eq:masterformula2momentum}, but as described in section~\ref{sec:bknumerics}, there are some difficulties ensuring that the solutions in momentum space and position space are consistent.

\subsubsection{Quadrupole-Type Terms}
The other terms in~\cite{Chirilli:2012jd} are quadrupole-type, which involve integrals over three transverse position coordinates, $\vec\xperp$, $\vec\bperp$, and $\vec\yperp$. These terms can be written
\begin{multline}
 \xsec n4ji = \biggl(\frac{\alpha_s}{2\pi}\biggr)^n\int\frac{\udc\z\udc\xsecquarkpfrac}{\z^2}\xp f_i(\xp,\mu)D_{\Phadron/j}(\z, \mu) \int\frac{\uddc\vec\xperp\uddc\vec\bperp\uddc\vec\yperp}{(2\pi)^4} \\
 \times \genericS(\vec\xperp, \vec\bperp, \vec\yperp) e^{-i C_1 \vec\kperp\cdot(\vec\xperp - \vec\bperp)} e^{-i C_2 \vec\kperp\cdot(\vec\yperp - \vec\bperp)} \hampl n4ji(\vec\xperp, \vec\bperp, \vec\yperp) \label{eq:unintegratedquadrupolecontribution}
\end{multline}
where $C_1$ and $C_2$ are, like $C_0$, coefficients independent of position and momentum.
For this type of term, the hard factor $\h n4ji$ as given in~\cite{Chirilli:2012jd} corresponds to the combination
\begin{equation}
 \h n4ji = e^{-i C_1 \vec\kperp\cdot(\vec\xperp - \vec\bperp)} e^{-i C_2 \vec\kperp\cdot(\vec\yperp - \vec\bperp)} \hampl n4ji
\end{equation}
Handling these terms proceeds along the same general lines as the dipole-type terms, with a few differences to account for the additional coordinates. First we change variables to
\begin{align}
 \vec{R}_\perp &= \vec\xperp + \vec\yperp &
 \vec\sperp &= \vec\xperp - \vec\bperp &
 \vec\tperp &= \vec\yperp - \vec\bperp
\end{align}
and $\vec\rperp = \vec\sperp - \vec\tperp$. As in the dipole case, $\hampl n4ji$ is independent of $\vec{R}_\perp$ so we can integrate over that degree of freedom.
\begin{multline}
 \xsec n4ji = \Sperp\biggl(\frac{\alpha_s}{2\pi}\biggr)^n\int\frac{\udc\z\udc\xsecquarkpfrac}{\z^2}\xp f_i(\xp,\mu)D_{\Phadron/j}(\z, \mu) \int\frac{\uddc\vec\sperp\uddc\vec\tperp}{(2\pi)^4} \\
 \times \genericS(\vec\sperp, \vec\tperp) e^{-i C_1 \vec\kperp\cdot\vec\sperp} e^{-i C_2 \vec\kperp\cdot\vec\tperp} \hampl n4ji(\vec\sperp, \vec\tperp) \label{eq:integratedquadrupolecontribution}
\end{multline}

Now, $\genericS \hampl n4ji$ is a scalar, so it depends on $\vec\sperp$ and $\vec\tperp$ through scalar functions of those variables: $\sperp^2$, $\tperp^2$, and $\vec\sperp\cdot\vec\tperp = \frac{1}{2}(\sperp^2 + \tperp^2 - \rperp^2)$. This means we can't integrate over the angular variables of both $\vec\sperp$ and $\vec\tperp$, for the following reason. In Cartesian position space, $\genericS \hampl n4ji$ is written as a function of four variables, $s_{x\perp}$, $s_{y\perp}$, $t_{x\perp}$, and $t_{y\perp}$. We can express it as a function of three variables, $\rperp$, $\sperp$, and $\tperp$, but not two. So we can only hope to integrate over one angle. In principle, we could do that integration, but we're not going to have a simple substitution rule to do so, as we did with the dipole-type terms (equation~\eqref{eq:maptoradial}).

In practice, all the quadrupole-type terms in~\cite{Chirilli:2012jd} can be expressed in momentum space, so it's not worth the effort to find the radial representation at all.\iftime{find radial representation of quadrupole terms}\footnote{If one were inclined to do so, it could serve as a potentially useful check on the momentum-space implementation.}

\paragraph{Cartesian position space}
As with the dipole terms, we can practically copy the Cartesian position space formula from reference~\cite{Chirilli:2012jd}.
\begin{subequations}\label{eq:masterformula4}
\begin{align}
 \frac{\Fs n4ji(\z,\xsecquarkpfrac, \vec\rperp)}{(1 - \xsecquarkpfrac)_+} &=
 \begin{multlined}[t][.6\textwidth]
  \Sperp\frac{\xp f_i D_{\Phadron/j}}{\z^2}\frac{\genericS(\rperp, \sperp, \tperp)}{(2\pi)^4} \biggl(\frac{\alpha_s}{2\pi}\biggr)^n \\
  \times e^{-i C_1 \vec\kperp\cdot\vec\sperp} e^{-i C_2 \vec\kperp\cdot\vec\tperp} \bigl(\text{plus-regulated part of }\hampl n4ji\bigr)
 \end{multlined} \\
 \Fn n4ji(\z,\xsecquarkpfrac, \vec\rperp) &=
 \begin{multlined}[t][.6\textwidth]
  \Sperp\frac{\xp f_i D_{\Phadron/j}}{\z^2}\frac{\genericS(\rperp, \sperp, \tperp)}{(2\pi)^4} \biggl(\frac{\alpha_s}{2\pi}\biggr)^n \\
  \times e^{-i C_1 \vec\kperp\cdot\vec\sperp} e^{-i C_2 \vec\kperp\cdot\vec\tperp} \bigl(\text{normal part of }\hampl n4ji\bigr)
 \end{multlined}\\
 \Fd n4ji(\z,\xsecquarkpfrac, \vec\rperp)\delta(1 - \xsecquarkpfrac) &=
 \begin{multlined}[t][.6\textwidth]
  \Sperp\frac{\xp f_i D_{\Phadron/j}}{\z^2}\frac{\genericS(\rperp, \sperp, \tperp)}{(2\pi)^4} \biggl(\frac{\alpha_s}{2\pi}\biggr)^n \\
  \times e^{-i C_1 \vec\kperp\cdot\vec\sperp} e^{-i C_2 \vec\kperp\cdot\vec\tperp} \bigl(\text{delta function part of }\hampl n4ji\bigr)
 \end{multlined}
 \label{eq:masterformula4delta}
\end{align}
\end{subequations}
and by analogy with equation~\eqref{eq:dipolemasterformula}, the term becomes
\begin{multline}\label{eq:quadrupolemasterformula}
 \xsec n4ji
 = \int_\tau^1\udc\z\int_{\tau/\z}^1\udc\xsecquarkpfrac \int\uddc\vec\sperp\uddc\vec\tperp \\
 \times \biggl[\frac{\Fs n4ji(\z, \xsecquarkpfrac, \vec\sperp, \vec\tperp) - \Fs n4ji(\z, 1, \vec\sperp, \vec\tperp)}{1 - \xsecquarkpfrac} + \Fn n4ji(\z, \xsecquarkpfrac, \vec\sperp, \vec\tperp)\biggr] \\
 + \int_\tau^1\udc\z\int\uddc\vec\sperp\uddc\vec\tperp \biggl[\Fs n4ji(\z, 1, \vec\sperp, \vec\tperp)\ln\biggl(1 - \frac{\tau}{\z}\biggr) + \Fd n4ji(\z, 1, \vec\sperp, \vec\tperp)\biggr]
\end{multline}

\paragraph{Momentum space}
Also by analogy with the dipole terms, we can Fourier transform quadrupole-type terms into momentum space.
The resulting expression includes the momentum space hard factor:
\begin{equation}\label{eq:quadrupolehardfactorfouriertransform}
 \hfampl n4ji(\kperp) = \iint\frac{\uddc\vec\sperp}{(2\pi)^2}\iint\frac{\uddc\vec\tperp}{(2\pi)^2} e^{-i\vec\kperp\cdot\vec\sperp} e^{-i\vec\kperp\cdot\vec\tperp} \hampl n4ji(\vec\sperp, \vec\tperp)
\end{equation}
and either $\quadrupoleG$ or factors of $\dipoleF$, depending on the gluon distribution in each individual term.
As a straightforward example, the structure
\begin{equation}
 \int\frac{\uddc\vec\sperp}{(2\pi)^2}\frac{\uddc\vec\tperp}{(2\pi)^2}\quadrupoleS(\vec\sperp, \vec\tperp) e^{-i C_1 \vec\kperp\cdot\vec\sperp} e^{-i C_2 \vec\kperp\cdot\vec\tperp} \hampl n4ji(\vec\sperp, \vec\tperp)
\end{equation}
becomes
\begin{equation}
 \int\frac{\uddc\vec\qiperp}{(2\pi)^2}\frac{\uddc\vec\qiiperp}{(2\pi)^2} \quadrupoleG(\vec\qiperp, \vec\qiiperp) \hfampl n4ji(\vec\qiperp - C_1\vec\kperp, \vec\qiiperp - C_2\vec\kperp)
\end{equation}
Of course the exact nature of the expression varies significantly from term to term.

In the end the contribution of a quadrupole-type term takes the form
\begin{multline}\label{eq:quadrupolemasterformulamomentum}
 \biggl(\frac{\udddc\sigma}{\udc Y\uddc\vec \pperp}\biggr)_\text{quad}
 = \int_\tau^1\udc\z\int_{\tau/\z}^1\udc\xsecquarkpfrac \int\uddc\vec\qiperp\uddc\vec\qiiperp \\
 \times\biggl[\frac{\Fs n4ji(\z, \xsecquarkpfrac, \vec\qiperp, \vec\qiiperp) - \Fs n4ji(\z, 1, \vec\qiperp, \vec\qiiperp)}{1 - \xsecquarkpfrac}
 + \Fn n4ji(\z, \xsecquarkpfrac, \vec\qiperp, \vec\qiiperp)\biggr] \\
 + \int_\tau^1\udc\z\int\uddc\vec\qiperp\uddc\vec\qiiperp \biggl[\Fs n4ji(\z, 1, \vec\qiperp, \vec\qiiperp)\ln\biggl(1 - \frac{\tau}{\z}\biggr) + \Fd n4ji(\z, 1, \vec\qiperp, \vec\qiiperp)\biggr]
\end{multline}
with
\begin{subequations}\label{eq:masterformula4momentum}
\begin{align}
 \frac{\Fs n4ji(\z,\xsecquarkpfrac, \vec\qperp)}{(1 - \xsecquarkpfrac)_+} &= \frac{\xp f_i D_{\Phadron/j}}{\z^2} \quadrupoleG(\vec\qperp)\biggl(\frac{\alpha_s}{2\pi}\biggr)^n\bigl(\text{plus-regulated part of }\hfampl n2ji\bigr) \\
 \Fn n4ji(\z,\xsecquarkpfrac, \vec\qperp) &= \frac{\xp f_i D_{\Phadron/j}}{\z^2} \quadrupoleG(\vec\qperp)\biggl(\frac{\alpha_s}{2\pi}\biggr)^n\bigl(\text{normal part of }\hfampl n2ji\bigr) \\
 \Fd n4ji(\z,\xsecquarkpfrac, \vec\qperp)\delta(1 - \xsecquarkpfrac) &= \frac{\xp f_i D_{\Phadron/j}}{\z^2} \quadrupoleG(\vec\qperp)\biggl(\frac{\alpha_s}{2\pi}\biggr)^n\bigl(\text{delta function part of }\hfampl n2ji\bigr)
\end{align}
\end{subequations}
As mentioned, the number of momentum variables involved depends on how many factors of $\dipoleS$ there are.

\section{Summary of Terms}

The formulas in reference~\cite{Chirilli:2012jd} generically take the form
\begin{equation}
 \frac{\udddc\sigma}{\udc\rapidity\uddc\vec\pperp} = \xp f(\xp, \mu)\otimes \mathcal{S}\otimes\mathcal{H}\otimes D(\z, \mu^2)
\end{equation}
where $\mathcal{S}$ is some combination of dipole or quadrupole gluon distributions, and $\mathcal{H}$ is a hard factor, representing the perturbative part of the calculation. It's straightforward to extract the Cartesian position space formulas from the expressions given in~\cite{Chirilli:2012jd}, but an accurate numerical implementation requires the radial or momentum space formulas, which need to be derived. The purpose of this section is to show those derivations.

Table~\ref{tbl:hardfactorformulas} summarizes which equations give the contribution corresponding to each hard factor in each implementation, for easy reference.

\begin{table}
 \rowcolors{2}{white}{gray!10}
 \renewcommand{\arraystretch}{1.5}
 \begin{tabularx}{\textwidth}{lXXX}
  \toprule
  Term &
  Cartesian &
  Radial &
  Momentum \\
  \midrule
  $\h02qq$ &
  \eqref{eq:h02qq:cartesian:xsec} with \eqref{eq:h02qq:cartesian:Fd}
  &
  &
  \eqref{eq:h02qq:momentum:xsec} with \eqref{eq:h02qq:momentum:Fd}
  \\
  $\h12qq$ &
  \eqref{eq:h12qq:cartesian:xsec} with \eqref{eq:h12qq:cartesian:Fs}, \eqref{eq:h12qq:cartesian:Fd}
  &
  \eqref{eq:h12qq:radial:xsec} with \eqref{eq:h12qq:radial:Fs}, \eqref{eq:h12qq:radial:Fd}
  &
  \\
  $\h14qq$ &
  \eqref{eq:h14qq:cartesian:xsec} \newline
  with \eqref{eq:h14qq:cartesian:Fs}, \eqref{eq:h14qq:cartesian:Fd:quadrupole}, \eqref{eq:h14qq:cartesian:Fd:dipole}
  &
  &
  \eqref{eq:h14qq:momentum:xsec} \newline
  with \eqref{eq:h14qq:momentum:Fs}, \eqref{eq:h14qq:momentum:Fd}
  \\
  $\h02gg$ &
  \eqref{eq:h02gg:cartesian:xsec} with \eqref{eq:h02gg:cartesian:Fd}
  &
  \eqref{eq:h02gg:radial:xsec} with \eqref{eq:h02gg:radial:Fd}
  &
  \eqref{eq:h02gg:momentum:xsec} with \eqref{eq:h02gg:momentum:Fd}
  \\
  $\h12gg$ &
  \eqref{eq:h12gg:cartesian:xsec} \newline
  with \eqref{eq:h12gg:cartesian:Fs}, \eqref{eq:h12gg:cartesian:Fn}, \eqref{eq:h12gg:cartesian:Fd}
  &
  \eqref{eq:h12gg:radial:xsec} \newline
  with \eqref{eq:h12gg:radial:Fs}, \eqref{eq:h12gg:radial:Fn}, \eqref{eq:h12gg:radial:Fd}
  &
  \\
  $\h12q{\bar q}$ &
  \eqref{eq:h12qqbar:cartesian:xsec} with \eqref{eq:h12qqbar:cartesian:Fd}
  &
  &
  \eqref{eq:h12qqbar:momentum:xsec} with \eqref{eq:h12qqbar:momentum:Fd}
  \\
  $\h16gg$ &
  \eqref{eq:h16gg:cartesian:xsec} \newline
  with \eqref{eq:h16gg:cartesian:Fs}, \eqref{eq:h16gg:cartesian:Fd:quadrupole}, \eqref{eq:h16gg:cartesian:Fd:dipole}
  &
  &
  \eqref{eq:h16gg:momentum:xsec} \newline
  with \eqref{eq:h16gg:momentum:Fs}, \eqref{eq:h16gg:momentum:Fd}
  \\
  $\h{1,1}2gq$ &
  \eqref{eq:h112gq:cartesian:xsec} with \eqref{eq:h112gq:cartesian:Fn}
  &
  \eqref{eq:h112gq:radial:xsec} with \eqref{eq:h112gq:radial:Fn}
  &
  \\
  $\h{1,2}2gq$ &
  \eqref{eq:h122gq:cartesian:xsec} with \eqref{eq:h122gq:cartesian:Fn}
  &
  \eqref{eq:h122gq:radial:xsec} with \eqref{eq:h122gq:cartesian:Fn}
  &
  \\
  $\h14gq$ &
  \eqref{eq:h14gq:cartesian:xsec} with \eqref{eq:h14gq:cartesian:Fn}
  &
  &
  \eqref{eq:h14gq:momentum:xsec} with \eqref{eq:h14gq:momentum:Fn}
  \\
  $\h{1,1}2qg$ &
  \eqref{eq:h112qg:cartesian:xsec} with \eqref{eq:h112qg:cartesian:Fn}
  &
  \eqref{eq:h112qg:radial:xsec} with \eqref{eq:h112qg:radial:Fn}
  &
  \\
  $\h{1,2}2qg$ &
  \eqref{eq:h122qg:cartesian:xsec} with \eqref{eq:h122qg:cartesian:Fn}
  &
  \eqref{eq:h122qg:radial:xsec} with \eqref{eq:h122qg:radial:Fn}
  &
  \\
  $\h14qg$ &
  \eqref{eq:h14qg:cartesian:xsec} with \eqref{eq:h14qg:cartesian:Fn}
  &
  &
  \eqref{eq:h14qg:momentum:xsec} with \eqref{eq:h14qg:momentum:Fn}
  \\
  \bottomrule
 \end{tabularx}
 \caption{Summary of formulas for each hard factor}
 \label{tbl:hardfactorformulas}
\end{table}

\subsection{\texorpdfstring{$\h02qq$}{H(0)2qq}}
In reference~\cite{Chirilli:2012jd}, the term from the leading order $qq$ channel is given in equations (41) and (42) as
\begin{equation}\label{eq:xsec02qq}
 \xsec02qq = \Sperp\int\frac{\udc\z\udc\xsecquarkpfrac}{\z^2}\int\frac{\uddc\vec\rperp}{(2\pi)^2} \sum_q \xp q(\xp, \mu)D_{\Phadron/q}(\z, \mu) \dipoleS(\vec\rperp) e^{-i\vec\kperp\cdot\vec\rperp}\delta(1 - \xsecquarkpfrac)
\end{equation}
This can be expressed as
\begin{equation}\label{eq:h02qq:cartesian:xsec}
 \xsec02qq = \int_\tau^1\udc\z \int\uddc\vec\rperp \Fd02qq(\z, 1, \vec\rperp)
\end{equation}
where
\begin{equation}\label{eq:h02qq:cartesian:Fd}
 \Fd02qq(\z, \xsecquarkpfrac, \vec\rperp) = \frac{\Sperp}{(2\pi)^2}\frac{1}{\z^2} \sum_q \xp q(\xp, \mu)D_{\Phadron/q}(\z, \mu) \dipoleS(\vec\rperp)e^{-i\vec\kperp\cdot\vec\rperp}
\end{equation}
We can translate this into the radial formulation easily using equation~\eqref{eq:maptoradial}, obtaining
\begin{equation}\label{eq:h02qq:radial:Fd}
 \Fd02qq(\z, \xsecquarkpfrac, \rperp) = \frac{S_\perp}{z^2} \sum_q \xp q(\xp, \mu)D_{\Phadron/q}(\z, \mu)\frac{\dipoleS(\vec\rperp)}{2\pi}\biggl(\frac{\alpha_s}{2\pi}\biggr)^n \rperp \besselJzero(\kperp\rperp)
\end{equation}
which gets integrated as
\begin{equation}\label{eq:h02qq:radial:xsec}
 \xsec02qq = \int_\tau^1\udc\z \int_0^\infty\udc\rperp \Fd02qq(\z, 1, \rperp)
\end{equation}
And for the momentum space implementation, equation~\eqref{eq:momentumdipolehardfactorintegral} gives us $\hfampl02qq(\vec\kperp) = \delta(1 - \xsecquarkpfrac)\delta^{(2)}(\vec\kperp)$, and then~\eqref{eq:masterformula2momentum} we get
\begin{equation}\label{eq:h02qq:momentum:Fd}
 \Fd02qq(\z, \xsecquarkpfrac, \vec\kperp) = \frac{1}{\z^2} \sum_q \xp q(\xp, \mu)D_{\Phadron/q}(\z, \mu)\dipoleF(\vec\kperp)
\end{equation}
This case is particularly simple because no momentum integration at all is needed:
\begin{equation}\label{eq:h02qq:momentum:xsec}
 \xsec02qq = \int_\tau^1\udc\z \Fd02qq(\z, 1, \vec\kperp)
\end{equation}

In each case $\Fs*$ and $\Fn*$ are, of course, zero.

\subsection{\texorpdfstring{$\h02gg$}{H(0)2gg}}
The expression for the leading-order gluon contribution comes from equations (74) and (75) of~\cite{Chirilli:2012jd}:
\begin{equation}\label{eq:xsec02gg}
 \xsec02gg = \Sperp\int\frac{\udc\z\udc\xsecquarkpfrac}{\z^2}\int\frac{\uddc\vec\rperp}{(2\pi)^2} \xp g(\xp, \mu)D_{\Phadron/g}(\z, \mu) \bigl[\dipoleS(\vec\rperp)\bigr]^2 e^{-i\vec\kperp\cdot\vec\rperp}\delta(1 - \xsecquarkpfrac)
\end{equation}
The Cartesian position space implementation is straightforward,
\begin{equation}\label{eq:h02gg:cartesian:xsec}
 \xsec02gg = \int_\tau^1\udc\z \int\uddc\vec\rperp \Fd02gg(\z, 1, \vec\rperp)
\end{equation}
where
\begin{equation}\label{eq:h02gg:cartesian:Fd}
 \Fd02gg(\z, \xsecquarkpfrac, \vec\rperp) = \frac{\Sperp}{(2\pi)^2}\frac{1}{\z^2} \xp g(\xp, \mu)D_{\Phadron/g}(\z, \mu)[\dipoleS(\vec\rperp)]^2 e^{-i\vec\kperp\cdot\vec\rperp}
\end{equation}
and again using~\eqref{eq:maptoradial} for the radial implementation, we find
\begin{equation}\label{eq:h02gg:radial:xsec}
 \xsec02gg = \int_\tau^1\udc\z \int\udc\rperp \rperp \Fd02gg(\z, 1, \rperp)
\end{equation}
where
\begin{equation}\label{eq:h02gg:radial:Fd}
 \Fd02gg(\z, \xsecquarkpfrac, \rperp) = \frac{S_\perp}{z^2} \xp g(\xp, \mu)D_{\Phadron/g}(\z, \mu)\frac{[\dipoleS(\vec\rperp)]^2}{2\pi}\rperp\besselJzero(\kperp\rperp)
\end{equation}

For the momentum space implementation, we get an extra integration, as in~\eqref{eq:momentumdipolesquaredhardfactorintegral}.
We wind up with
\begin{equation}\label{eq:h02gg:momentum:xsec}
 \xsec02gg = \int_\tau^1\udc\z\iint\uddc\vec\qperp \Fd02gg(\z, 1, \vec\qperp)
\end{equation}
where
\begin{equation}\label{eq:h02gg:momentum:Fd}
 \Fd02gg(\z, \xsecquarkpfrac, \vec\qperp) = \frac{1}{\Sperp\z^2} \xp g(\xp, \mu)D_{\Phadron/g}(\z, \mu)\dipoleF(\vec\qperp)\dipoleF(\vec\kperp - \vec\qperp)
\end{equation}
It will be typical for most terms that the momentum-space expressions have to be derived on a case-by-case basis. There are not as many patterns that can be exploited, as there are with the Cartesian expressions.

\subsection{\texorpdfstring{$\h12qq$}{H(1)2qq}}
From equations (41) and (43) of~\cite{Chirilli:2012jd}, the $\h12qq$ contribution to the cross section in the large-$N_c$ limit is given as
\begin{multline}\label{eq:xsec12qq}
 \xsec12qq = \int\frac{\udc\z\udc\xsecquarkpfrac}{\z^2} \sum_q \xp q(\xp, \mu)D_{\Phadron/q}(z, \mu) \int\frac{\uddc\vec\rperp}{(2\pi)^2} \frac{\alphas}{2\pi}
 C_F \biggl[\frac{1 + \xsecquarkpfrac^2}{(1 - \xsecquarkpfrac)_+} + \frac{3}{2}\delta(1 - \xsecquarkpfrac)\biggr] \\
 \times \ln\frac{c_0^2}{r_\perp^2 \mu^2}\biggl(e^{-i\vec\kperp\cdot\vec\rperp} + \frac{1}{\xsecquarkpfrac^2}e^{-i\frac{\vec\kperp}{\xsecquarkpfrac}\cdot\vec\rperp}\biggr) - 3C_F\delta(1 - \xsecquarkpfrac)e^{-i\vec\kperp\cdot\vec\rperp}\ln\frac{c_0^2}{r_\perp^2 \kperp^2}
\end{multline}
This expression consists of individual terms which follow the pattern of equation~\eqref{eq:unintegrateddipolecontribution}, so we can just substitute into the right side of~\eqref{eq:masterformula2} to get
\begin{multline}\label{eq:h12qq:cartesian:Fs}
 \Fs12qq(\z, \xsecquarkpfrac, \vec\rperp) = \frac{1}{4\pi^2}\frac{\alpha_s}{2\pi}\frac{C_F S_\perp}{z^2} \sum_q \xp q(\xp, \mu)D_{\Phadron/q}(z, \mu)\dipoleS(\vec\rperp)(1 + \xsecquarkpfrac^2) \\
 \times\ln\frac{c_0^2}{r_\perp^2 \mu^2}\biggl(e^{-i\vec\kperp\cdot\vec\rperp} + \frac{1}{\xsecquarkpfrac^2}e^{-i\frac{\vec\kperp}{\xsecquarkpfrac}\cdot\vec\rperp}\biggr)
\end{multline}
and
\begin{multline}\label{eq:h12qq:cartesian:Fd:unsimplified}
 \Fd12qq(\z,\xsecquarkpfrac, \vec\rperp) = \frac{3}{4\pi^2}\frac{\alpha_s}{2\pi}\frac{C_F S_\perp}{z^2} \sum_q \xp q(\xp, \mu)D_{\Phadron/q}(z, \mu)\dipoleS(\vec\rperp) \\
 \times\biggl[\frac{1}{2} \ln\frac{c_0^2}{r_\perp^2 \mu^2}\biggl(e^{-i\vec\kperp\cdot\vec\rperp} + \frac{1}{\xsecquarkpfrac^2}e^{-i\frac{\vec\kperp}{\xsecquarkpfrac}\cdot\vec\rperp}\biggr)
 - e^{-i\vec\kperp\cdot\vec\rperp}\ln\frac{c_0^2}{r_\perp^2 \kperp^2}\biggr]
\end{multline}
As a delta contribution, this last equation will only ever be evaluated at $\xsecquarkpfrac = 1$, so we can plug that in immediately and get a simpler expression:
\begin{equation}\label{eq:h12qq:cartesian:Fd}
 \Fd12qq(\z, 1, \vec\rperp) = \frac{3}{4\pi^2}\frac{\alpha_s}{2\pi}\frac{C_F S_\perp}{z^2} \sum_q \xp q(\xp, \mu)D_{\Phadron/q}(z, \mu)\dipoleS(\vec\rperp)e^{-i\vec\kperp\cdot\vec\rperp}\ln\frac{\kperp^2}{\mu^2}
\end{equation}
Implementing this expression instead of the general one~\eqref{eq:h12qq:cartesian:Fd:unsimplified} reduces the chances of mistakes in the implementation and reduces any numerical errors in the evaluation.

Expressions~\eqref{eq:h12qq:cartesian:Fs} and~\eqref{eq:h12qq:cartesian:Fd} go into~\eqref{eq:dipolemasterformula}
\begin{multline}\label{eq:h12qq:cartesian:xsec}
 \xsec12qq
 = \int_\tau^1\udc\z\int_{\tau/\z}^1\udc\xsecquarkpfrac \int\uddc\vec\rperp \frac{\Fs12qq(\z, \xsecquarkpfrac, \vec\rperp) - \Fs12qq(\z, 1, \vec\rperp)}{1 - \xsecquarkpfrac} \\
 + \int_\tau^1\udc\z\int\uddc\vec\rperp \biggl[\Fs12qq(\z, 1, \vec\rperp)\ln\biggl(1 - \frac{\tau}{\z}\biggr) + \Fd12qq(\z, 1, \vec\rperp)\biggr]
\end{multline}
to produce the final contribution to the cross section.

To transform into the radial representation, we use equation~\eqref{eq:maptoradial} to get
\begin{align}
 \Fs12qq(\z, \xsecquarkpfrac, \rperp) &= \frac{1}{2\pi}\frac{\alpha_s}{2\pi}\frac{C_F S_\perp}{z^2} \sum_q \xp q(\xp, \mu)D_{\Phadron/q}(z, \mu)\dipoleS(\vec\rperp) \notag\\
 &\quad \times(1 + \xsecquarkpfrac^2) \rperp \ln\frac{c_0^2}{r_\perp^2 \mu^2} \biggl[\besselJzero(\kperp\rperp) + \frac{1}{\xsecquarkpfrac^2}\besselJzero\biggl(\frac{\kperp\rperp}{\xsecquarkpfrac}\biggr)\biggr] \label{eq:h12qq:radial:Fs}\\
 \Fd12qq(z, 1, \rperp) &= \frac{3}{2\pi}\frac{\alpha_s}{2\pi}\frac{C_F S_\perp}{z^2} \sum_q \xp q(\xp, \mu)D_{\Phadron/q}(z, \mu)\dipoleS(\vec\rperp) \rperp\besselJzero(\kperp\rperp) \ln\frac{\kperp^2}{\mu^2} \label{eq:h12qq:radial:Fd}
\end{align}
and the contribution to the cross section will be
\begin{multline}\label{eq:h12qq:radial:xsec}
 \xsec12qq
 = \int_\tau^1\udc\z\int_{\tau/\z}^1\udc\xsecquarkpfrac \int\uddc\vec\rperp \frac{\Fs12qq(\z, \xsecquarkpfrac, \vec\rperp) - \Fs12qq(\z, 1, \vec\rperp)}{1 - \xsecquarkpfrac} \\
 + \int_\tau^1\udc\z\int\uddc\vec\rperp \biggl[\Fs12qq(\z, 1, \vec\rperp)\ln\biggl(1 - \frac{\tau}{\z}\biggr) + \Fd12qq(\z, 1, \vec\rperp)\biggr]
\end{multline}

The logarithmic factors $\ln\frac{\czero^2}{\rperp^2\mu^2}$ can't be Fourier transformed symbolically, so there is no momentum-space expression for this term.

\subsection{\texorpdfstring{$\h14qq$}{H(1)4qq}}\label{sec:h14qq}
From equations (41) and (45) of~\cite{Chirilli:2012jd}, this contribution to the cross section is
\begin{multline}\label{eq:xsec14qq}
 \xsec14qq = -4\pi N_c \Sperp\frac{\alphas}{2\pi} \int\frac{\udc\z\udc\xsecquarkpfrac}{\z^2} \sum_q \xp q(\xp, \mu)D_{\Phadron/q}(\z, \mu) \\ 
 \times \int\frac{\uddc\vec\sperp\uddc\vec\tperp}{(2\pi)^4} \quadrupoleS(\vec\sperp, \vec\tperp) e^{-i\vec\kperp\cdot\vec\rperp} \Biggl\{e^{-i\frac{1 - \xsecquarkpfrac}{\xsecquarkpfrac}\vec\kperp\cdot\vec\sperp}\frac{1 + \xsecquarkpfrac^2}{(1 - \xsecquarkpfrac)_+}\frac{1}{\xsecquarkpfrac}\frac{\vec\sperp}{s_\perp^2}\cdot\frac{\vec\tperp}{t_\perp^2} \\
 - \delta(1 - \xsecquarkpfrac)\int_0^1\udc\xi'\frac{1 + \xi'^2}{(1 - \xi')_+}\Biggl[\frac{e^{-i(1 - \xi')\vec\kperp\cdot\vec\tperp}}{t_\perp^2} - \delta^{(2)}(\vec\tperp)\int\uddc\vec\rperp' \frac{e^{i\vec\kperp\cdot\vec{\rperp'}}}{r_\perp'^2}\Biggr]\Biggr\}
\end{multline}
where $\vec\sperp = \vec\xperp - \vec\bperp$ and $\vec\tperp = \vec\yperp - \vec\bperp$.
The first term (first and second lines) is easy: it directly translates into a plus-regulated contribution,
\begin{multline}\label{eq:h14qq:cartesian:Fs}
 \Fs14qq(\z, \xsecquarkpfrac, \vec\sperp, \vec\tperp) = -\frac{1}{4\pi^3}\frac{\alpha_s}{2\pi}\frac{N_c S_\perp}{z^2} \sum_q \xp q(\xp, \mu)D_{\Phadron/q}(z, \mu) \quadrupoleS(\vec\sperp, \vec\tperp)\\
 \times e^{-i\vec\kperp\cdot(\vec\sperp/\xsecquarkpfrac - \vec\tperp)}\frac{1 + \xsecquarkpfrac^2}{\xsecquarkpfrac}\frac{\vec\sperp}{s_\perp^2}\cdot\frac{\vec\tperp}{t_\perp^2}
\end{multline}
where I've used $\vec\rperp + \vec\sperp\frac{1 - \xsecquarkpfrac}{\xsecquarkpfrac} = \vec\rperp + \frac{\vec\sperp}{\xsecquarkpfrac} - \vec\sperp = \frac{\vec\sperp}{\xsecquarkpfrac} - \vec\tperp$ to construct the Fourier exponent.

The delta contribution is somewhat more complicated, though, because it has an additional delta function $\delta^{(2)}(\vec\tperp)$ which we need to integrate symbolically first.
We'll have to calculate this integral:
\begin{multline}
 \iint\uddc\vec\sperp\iint\uddc\vec\tperp e^{-i\vec\kperp\cdot\vec\rperp} \quadrupoleS(\vec\sperp,\vec\tperp) \\
 \times\int_0^1\udc\xi'\frac{1 + \xi'^2}{(1 - \xi')_+}\Biggl[\frac{e^{-i(1 - \xi')\vec\kperp\cdot\vec\tperp}}{t_\perp^2} - \delta^{(2)}(\vec\tperp)\int\uddc\vec\rpperp \frac{e^{i\vec\kperp\cdot\vec\rpperp}}{\rpperp^2}\Biggr]
\end{multline}
Each of the two terms in square brackets is individually divergent, so we need to regulate them, but we can't do the entire integral symbolically because of the unknown gluon distribution $\quadrupoleS$.

The trick is to distribute the $\vec\tperp$ integration and impose a lower cutoff $\epsilon$:
\begin{multline}
 \int_0^1\udc\xi'\frac{1 + \xi'^2}{(1 - \xi')_+}\Biggl[\iint\uddc\vec\sperp \iint_{\tperp>\epsilon}\uddc\vec\tperp\quadrupoleS(\vec\sperp,\vec\tperp) e^{-i\vec\kperp\cdot(\vec\sperp - \vec\tperp)} \frac{e^{-i(1 - \xi')\vec\kperp\cdot\vec\tperp}}{t_\perp^2} \\
 - \iint\uddc\vec\sperp\iint_{\rpperp>\epsilon}\uddc\vec\rpperp \quadrupoleS(\vec\sperp,\vec 0) e^{-i\vec\kperp\cdot\vec\sperp} \frac{e^{i\vec\kperp\cdot\vec\rpperp}}{\rpperp^2}\Biggr]
\end{multline}
Note that integrating with $\delta^{(2)}(\vec\tperp)$ sets $\vec\rperp\to\vec\sperp$ in the second integral.
Then we change variables $\vec\rpperp \to \xi'\vec\tperp$.
\begin{multline}\label{eq:h14qqchangedvariables}
 \iint\uddc\vec\sperp \int_0^1\udc\xi'\frac{1 + \xi'^2}{(1 - \xi')_+}\Biggl[\iint_{\tperp>\epsilon}\uddc\vec\tperp\quadrupoleS(\vec\sperp,\vec\tperp) e^{-i\vec\kperp\cdot\vec\sperp} \frac{e^{i\xi'\vec\kperp\cdot\vec\tperp}}{t_\perp^2} \\
 - \quadrupoleS(\vec\sperp,\vec 0)\iint_{\tperp>\epsilon/\xi'}\uddc\vec\tperp e^{-i\vec\kperp\cdot\vec\sperp} \frac{e^{i\xi'\vec\kperp\cdot\vec\tperp}}{\tperp^2}\Biggr]
\end{multline}
Now we split the second integral using
\begin{equation}\label{eq:h14qqintegralsplit}
 \int_{\tperp>\epsilon/\xi'} = \int_{\tperp>\epsilon} - \int_{\tperp>\epsilon}^{\tperp<\epsilon/\xi'}
\end{equation}
The first term of~\eqref{eq:h14qqintegralsplit} combines with the first term of~\eqref{eq:h14qqchangedvariables} to give
\begin{equation}\label{eq:h14qqfirstfinalterm}
 \iint\uddc\vec\sperp \iint_{\tperp>\epsilon}\uddc\vec\tperp \bigl[\quadrupoleS(\vec\sperp,\vec\tperp)- \quadrupoleS(\vec\sperp,\vec 0)\bigr] e^{-i\vec\kperp\cdot\vec\rperp} \int_0^1\udc\xi'\frac{1 + \xi'^2}{(1 - \xi')_+}\frac{e^{-i(1 - \xi')\vec\kperp\cdot\vec\tperp}}{t_\perp^2}
\end{equation}
and the second term becomes
\begin{multline}\label{eq:h14qqsecondfinalterm}
 -\iint\uddc\vec\sperp \int_0^1\udc\xi'\frac{1 + \xi'^2}{(1 - \xi')_+}\quadrupoleS(\vec\sperp,\vec 0)\iint_{\tperp>\epsilon}^{\tperp<\epsilon/\xi'}\uddc\vec\tperp \frac{e^{i\xi'\vec\kperp\cdot\vec\tperp}}{\tperp^2} \\
 = -\iint\uddc\vec\sperp \int_0^1\udc\xi'\frac{1 + \xi'^2}{(1 - \xi')_+}\quadrupoleS(\vec\sperp,\vec 0) 2\pi\ln\frac{1}{\xi'}
\end{multline}
where, in performing the $\vec\tperp$ integral, I've used the fact that $\tperp$ is small so the exponential factor can be approximated as $1$.

Finally, the integrals over $\xi'$ in~\eqref{eq:h14qqfirstfinalterm} and~\eqref{eq:h14qqsecondfinalterm} can be done symbolically, giving
\begin{multline}
 \iint\uddc\vec\sperp \iint_{\tperp>\epsilon}\uddc\vec\tperp \bigl[\quadrupoleS(\vec\sperp,\vec\tperp)- \quadrupoleS(\vec\sperp,\vec 0)\bigr] e^{-i\vec\kperp\cdot\vec\rperp} \frac{\mathcal{I}_1(\vec\kperp\cdot\vec\tperp)}{t_\perp^2} \\
 - \iint\uddc\vec\sperp \quadrupoleS(\vec\sperp,\vec 0) 2\pi\biggl(-\frac{5}{4} + \frac{\pi^2}{3}\biggr)
\end{multline}
where
\begin{equation}
 \mathcal{I}_{1}(x) = -2\gamma_E + \frac{i(2 - e^{-ix})}{x} + \frac{e^{-ix} - 1}{x^2} + 2\Ci(x) - 2i\Si(x) - 2\ln x
\end{equation}
with $\Ci$ and $\Si$ defined as
\begin{align}\label{eq:cossinintdefn}
 \Ci(z) &= -\int_z^\infty\frac{\cos t}{t}\udc t &
 \Si(z) &= \int_0^z\frac{\sin t}{t}\udc t
\end{align}
Note that the imaginary parts of $\Ci(x)$ and $\ln x$ for $x < 0$ cancel out on the principal branch of each function.

So in Cartesian position space, $\h14qq$ actually contributes both a quadrupole term,
\begin{multline}\label{eq:h14qq:cartesian:Fd:quadrupole}
 \Fd14qq(\z, \xsecquarkpfrac, \vec\sperp, \vec\tperp) = \frac{1}{4\pi^3}\frac{\alpha_s}{2\pi}\frac{\Nc \Sperp}{\z^2} \sum_q \xp q(\xp, \mu)D_{\Phadron/q}(\z, \mu) \\
 \times \Bigl[S_Y^{(4)}(\vec\sperp, \vec\tperp) - \quadrupoleS(\vec\sperp, \vec 0)\Bigr]\frac{1}{t_\perp^2}e^{-i\vec{k}_\perp\cdot\vec{r}_\perp}\mathcal{I}_{1}(\vec{k}_\perp\cdot\vec{t}_\perp)
\end{multline}
\emph{and} a dipole term
\begin{equation}\label{eq:h14qq:cartesian:Fd:dipole}
 \Fd1{2'}qq(\z, \xsecquarkpfrac, \vec\rperp) = \frac{1}{2\pi^2}\biggl(-\frac{5}{4} + \frac{\pi^2}{3}\biggr)\frac{\alpha_s}{2\pi}\frac{N_c S_\perp}{z^2} \sum_q \xp q(\xp, \mu)D_{\Phadron/q}(z, \mu)  S_Y^{(4)}(\vec\rperp, \vec 0)e^{-i\vec{k}_\perp\cdot\vec{r}_\perp}
\end{equation}
and the complete contribution to the cross section is
\begin{multline}\label{eq:h14qq:cartesian:xsec}
 \xsec14qq
 = \int_\tau^1\udc\z\int_{\tau/\z}^1\udc\xsecquarkpfrac \int\uddc\vec\sperp\uddc\vec\tperp \biggl[\frac{\Fs14qq(\z, \xsecquarkpfrac, \vec\sperp, \vec\tperp) - \Fs14qq(\z, 1, \vec\sperp, \vec\tperp)}{1 - \xsecquarkpfrac}\biggr] \\
 + \int_\tau^1\udc\z\int\uddc\vec\sperp\uddc\vec\tperp \biggl[\Fs14qq(\z, 1, \vec\sperp, \vec\tperp)\ln\biggl(1 - \frac{\tau}{\z}\biggr) + \Fd14qq(\z, 1, \vec\sperp, \vec\tperp)\biggr] \\
 + \int_\tau^1\udc\z\int\uddc\vec\rperp \Fd1{2'}qq(\z, 1, \vec\rperp)
\end{multline}

To get the momentum space formulation of the plus-regulated part, it will be easiest to start with the Cartesian space result from equations~\eqref{eq:h14qq:cartesian:Fs} as plugged into~\eqref{eq:h14qq:cartesian:xsec}. For simplicity, I'll extract only the parts that depend on $\vec\sperp$ and $\vec\tperp$:
\begin{equation}\label{eq:h14qqmomentumsingularstart}
 \int\uddc\vec\sperp\uddc\vec\tperp \quadrupoleS(\vec\sperp, \vec\tperp) e^{-i\vec\kperp\cdot(\vec\sperp/\xsecquarkpfrac - \vec\tperp)}\frac{\vec\sperp}{s_\perp^2}\cdot\frac{\vec\tperp}{\tperp^2}
\end{equation}
Although it is possible to convert this into momentum space by plugging in the inverse Fourier transform of $\quadrupoleS$, equation~\eqref{eq:quadrupoleSinvFT}, it's easier to judiciously insert some delta functions.
\begin{equation}
 \int\uddc\vec\sperp\uddc\vec\tperp \quadrupoleS(\vec\sperp, \vec\tperp) e^{-i\vec\kperp\cdot(\vec\sperp/\xsecquarkpfrac - \vec\tperp)} \int\uddc\vec{\sperp'}\uddc\vec{\tperp'} \delta^{(2)}(\vec\sperp - \vec{\sperp'}) \delta^{(2)}(\vec\tperp - \vec{\tperp'}) \frac{\vec{\sperp'}}{\sperp'^2}\cdot\frac{\vec{\tperp'}}{\tperp'^2}
\end{equation}
Note that this expression is exactly equivalent to the previous one. We can then replace the delta functions with their Fourier transforms according to equation~\eqref{eq:ident:deltaFT}
\begin{multline}
 \int\uddc\vec\sperp\uddc\vec\tperp \quadrupoleS(\vec\sperp, \vec\tperp) e^{-i\vec\kperp\cdot(\vec\sperp/\xsecquarkpfrac - \vec\tperp)} \\
 \times \int\uddc\vec{\sperp'}\uddc\vec{\tperp'} \int\frac{\uddc\vec\qiperp\uddc\vec\qiiperp}{(2\pi)^4} e^{-i\vec\qiperp\cdot(\vec\sperp - \vec{\sperp'})} e^{-i\vec\qiiperp\cdot(\vec\tperp - \vec{\tperp'})} \frac{\vec{\sperp'}}{\sperp'^2}\cdot\frac{\vec{\tperp'}}{\tperp'^2}
\end{multline}
which can be rearranged to separate the two groups of position integrals:
\begin{multline}
 \int\uddc\vec\qiperp\uddc\vec\qiiperp \int\frac{\uddc\vec\sperp\uddc\vec\tperp}{(2\pi)^4} \quadrupoleS(\vec\sperp, \vec\tperp) e^{-i(\vec\kperp/\xsecquarkpfrac + \qiperp)\cdot\vec\sperp} e^{i(\vec\kperp - \vec\qiiperp)\cdot\vec\tperp} \\
 \times \int\uddc\vec{\sperp'}\uddc\vec{\tperp'} e^{i\vec\qiperp\cdot\vec{\sperp'}} e^{i\vec\qiiperp\cdot\vec{\tperp'}} \frac{\vec{\sperp'}}{\sperp'^2}\cdot\frac{\vec{\tperp'}}{\tperp'^2}
\end{multline}
The second line works out to $-(2\pi)^2\frac{\vec\qiperp\cdot\vec\qiiperp}{\qiperp^2\qiiperp^2}$, from equation~\eqref{eq:ident:vecprod}, and the integral over $\vec\sperp$ and $\vec\tperp$ on the first line is equal to $\frac{1}{\Sperp}\quadrupoleG\bigl(\frac{\vec\kperp}{\xsecquarkpfrac} + \vec\qiperp, \vec\kperp - \vec\qiiperp\bigr)$ according to equation~\eqref{eq:quadrupoleGdefinition}. Putting this result together with the extra factors that were dropped to get equation~\eqref{eq:h14qqmomentumsingularstart}, we get
\begin{multline}\label{eq:h14qq:momentum:Fs}
 \Fs14qq(\z, \xsecquarkpfrac, \vec\qiperp, \vec\qiiperp) = \frac{1}{\pi}\frac{\alpha_s}{2\pi}\frac{N_c}{z^2} \sum_q \xp q(\xp, \mu)D_{\Phadron/q}(z, \mu) \\
 \times\frac{1 + \xsecquarkpfrac^2}{\xsecquarkpfrac} \quadrupoleG\biggl(\frac{\vec\kperp}{\xsecquarkpfrac} + \vec\qiperp, \vec\kperp - \vec\qiiperp\biggr)\frac{\vec\qiperp\cdot\vec\qiiperp}{\qiperp^2\qiiperp^2}
\end{multline}
which will be integrated over $\vec\qiperp$ and $\vec\qiiperp$ in equation~\eqref{eq:h14qq:momentum:xsec}.

For the delta function term, we can use the identity in equation (34) of~\cite{Chirilli:2012jd}, with $\bar{\vec{r}}_\perp = \vec\tperp$. Multiplying both sides of that equation by $-\frac{1}{4\pi}e^{i\vec\kperp\cdot\vec\tperp}$ and taking the complex conjugate gives
\begin{multline}
 -\frac{1}{4\pi}\iint\uddc\vec\qiperp e^{i(\vec\qiperp - \vec\kperp)\cdot\vec\tperp} \ln\frac{(\vec\qiperp - \xi'\vec\kperp)^2}{\kperp^2} \\
 = \biggl[\frac{e^{-i(1 - \xi')\vec\kperp\cdot\vec\tperp}}{\tperp^2} - \delta^{(2)}(\vec\tperp) e^{-i\vec\kperp\cdot\vec\tperp} \iint\frac{\uddc\vec\rpperp}{\rpperp^2}e^{-i\vec\kperp\cdot\vec\rpperp}\biggr]
\end{multline}
Since this is going to be integrated over $\vec\tperp$, the $e^{-i\vec\kperp\cdot\vec\tperp}$ factor in the second term will be equal to 1 anyway (because of the delta function), and taking that into account, the right side of this equation is exactly the last factor of the delta-function term in equation~\eqref{eq:xsec14qq}. Restoring the factors and the integral from the second line of~\eqref{eq:xsec14qq} gives
\begin{equation}
 -\frac{1}{4\pi}\iint\uddc\vec\qiperp \underbrace{\iint\frac{\uddc\vec\sperp\uddc\vec\tperp}{(2\pi)^4} \quadrupoleS(\vec\sperp, \vec\tperp) e^{-i\vec\kperp\cdot\vec\sperp} e^{i\vec\qiperp\cdot\vec\tperp}}_{\quadrupoleG(\vec\kperp, -\vec\qiperp)/\Sperp} \ln\frac{(\vec\qiperp - \xi'\vec\kperp)^2}{\kperp^2}
\end{equation}
Restoring the factors from the first line of~\eqref{eq:xsec14qq} gives the delta contribution. The plus prescription factor, $\frac{1}{(1 - \xi')_+}$, can be handled symbolically:
\begin{equation}
 \int_0^1 \udc\xi' \frac{1 + \xi'^2}{(1 - \xi')_+}\ln\frac{(\vec\qiperp - \xi'\vec\kperp)^2}{\kperp^2} = \int_0^1 \udc\xi' \frac{1 + \xi'^2}{1 - \xi'}\ln\frac{(\vec\qiperp - \xi'\vec\kperp)^2}{\kperp^2} - \frac{2}{1 - \xi'}\ln\frac{(\vec\qiperp - \vec\kperp)^2}{\kperp^2}
\end{equation}
making the final expression for the delta contribution\note{to get the expression in the code, have to map $\vec\qiperp\to\vec\kperp - \vec\qiperp$}
\begin{multline}\label{eq:h14qq:momentum:Fd}
 \Fd14qq = -N_c \frac{\alphas}{2\pi} \frac{1}{\z^2} \sum_q \xp q(\xp, \mu)D_{\Phadron/q}(\z, \mu) \quadrupoleG(\vec\kperp, -\vec\qiperp) \\
 \times \biggl[\frac{1 + \xi'^2}{1 - \xi'}\ln\frac{(\vec\qiperp - \xi'\vec\kperp)^2}{\kperp^2} - \frac{2}{1 - \xi'}\ln\frac{(\vec\qiperp - \vec\kperp)^2}{\kperp^2}\biggr]
\end{multline}
The complete contribution from $\h14qq$ in momentum space is then
\begin{multline}\label{eq:h14qq:momentum:xsec}
 \xsec14qq
 = \int_\tau^1\udc\z\int_{\tau/\z}^1\udc\xsecquarkpfrac \int\uddc\vec\qiperp\uddc\vec\qiiperp \biggl[\frac{\Fs14qq(\z, \xsecquarkpfrac, \vec\qiperp, \vec\qiiperp) - \Fs14qq(\z, 1, \vec\qiperp, \vec\qiiperp)}{1 - \xsecquarkpfrac}\biggr] \\
 + \int_\tau^1\udc\z\int\uddc\vec\qiperp\uddc\vec\qiiperp \Fs14qq(\z, 1, \vec\qiperp, \vec\qiiperp)\ln\biggl(1 - \frac{\tau}{\z}\biggr) \\
 + \int_\tau^1\udc\z\int\uddc\vec\qiperp\int_0^1\udc\xi' \Fd14qq(\z, 1, \vec\qiperp, \xi')
\end{multline}
with the functions $\Fs14qq$ and $\Fd14qq$ from equations~\eqref{eq:h14qq:momentum:Fs} and~\eqref{eq:h14qq:momentum:Fd} respectively.

\subsection{\texorpdfstring{$\h12gg$}{H(1)2gg}}
From equations (74) and (76) in~\cite{Chirilli:2012jd}, the contribution of the $\h12gg$ term to the cross section is
\begin{multline}\label{eq:xsec12gg}
 \xsec12gg = \Nc\Sperp\frac{\alphas}{2\pi}\int\frac{\udc\z\udc\xsecquarkpfrac}{z^2} \xp g(\xp, \mu) D_{\Phadron/g}(\z, \mu) \int\frac{\uddc\vec\rperp}{(2\pi)^2}\bigl[\dipoleS(\vec\rperp)\bigr]^2 \\
 \times \Biggl[\biggl\{2\biggl[\frac{\xsecquarkpfrac}{(1 - \xsecquarkpfrac)_+} + \frac{1 - \xsecquarkpfrac}{\xsecquarkpfrac} + \xsecquarkpfrac(1 - \xsecquarkpfrac)\biggr] + \biggl(\frac{11}{6} - \frac{2\Nf\TR}{3\Nc}\biggr)\delta(1 - \xsecquarkpfrac)\biggr\}\ln\frac{c_0^2}{\rperp^2\mu^2}\biggl(e^{-i\vec\kperp\cdot\vec\rperp} + \frac{1}{\xsecquarkpfrac^2}e^{-i\vec\kperp\cdot\vec\rperp/\xsecquarkpfrac}\biggr) \\
 - \biggl(\frac{11}{3} - \frac{4\Nf\TR}{3\Nc}\biggr)\Nc\delta(1 - \xsecquarkpfrac)e^{-i\vec\kperp\cdot\vec\rperp}\ln\frac{c_0^2}{\rperp^2\kperp^2}\Biggr]
\end{multline}
As with $\h12qq$, this consists of terms that follow the pattern of equation~\eqref{eq:xsectermbreakdown}, so it's quite straightforward to split it into the three contributions:
\begin{align}
 \Fs12gg(\z, \xsecquarkpfrac, \vec\rperp) &=
 \begin{multlined}
 \frac{2\Nc\Sperp}{(2\pi)^2}\frac{\alphas}{2\pi}\frac{\xp g(\xp, \mu) D_{\Phadron/g}(\z, \mu)}{z^2} \bigl[\dipoleS(\vec\rperp)\bigr]^2 \xsecquarkpfrac\\
 \times\ln\frac{c_0^2}{\rperp^2\mu^2}\biggl(e^{-i\vec\kperp\cdot\vec\rperp} + \frac{1}{\xsecquarkpfrac^2}e^{-i\vec\kperp\cdot\vec\rperp/\xsecquarkpfrac}\biggr)
 \end{multlined} \label{eq:h12gg:cartesian:Fs}\\
 \Fn12gg(\z, \xsecquarkpfrac, \vec\rperp) &=
 \begin{multlined}[t][.7\displaywidth]
 \frac{2\Nc\Sperp}{(2\pi)^2}\frac{\alphas}{2\pi}\frac{\xp g(\xp, \mu) D_{\Phadron/g}(\z, \mu)}{z^2} \bigl[\dipoleS(\vec\rperp)\bigr]^2 \biggl(\frac{1 - \xsecquarkpfrac}{\xsecquarkpfrac} + \xsecquarkpfrac(1 - \xsecquarkpfrac)\biggr) \\
 \times\ln\frac{c_0^2}{\rperp^2\mu^2}\biggl(e^{-i\vec\kperp\cdot\vec\rperp} + \frac{1}{\xsecquarkpfrac^2}e^{-i\vec\kperp\cdot\vec\rperp/\xsecquarkpfrac}\biggr)
 \end{multlined} \label{eq:h12gg:cartesian:Fn}\\
 \Fd12gg(\z, \xsecquarkpfrac, \vec\rperp) &=
 \begin{multlined}[t][.7\displaywidth]
 \frac{\Nc\Sperp}{(2\pi)^2}\frac{\alphas}{2\pi}\frac{\xp g(\xp, \mu) D_{\Phadron/g}(\z, \mu)}{z^2} \bigl[\dipoleS(\vec\rperp)\bigr]^2 \\
 \times \biggl[\biggl(\frac{11}{6} - \frac{2\Nf\TR}{3\Nc}\biggr) \ln\frac{c_0^2}{\rperp^2\mu^2}\biggl(e^{-i\vec\kperp\cdot\vec\rperp} + \frac{1}{\xsecquarkpfrac^2}e^{-i\vec\kperp\cdot\vec\rperp/\xsecquarkpfrac}\biggr) \\
 - \biggl(\frac{11}{3} - \frac{4\Nf\TR}{3\Nc}\biggr) e^{-i\vec\kperp\cdot\vec\rperp}\ln\frac{c_0^2}{\rperp^2\kperp^2}\biggr]
 \end{multlined}
\end{align}
though since $\Fd12gg$ will only be evaluated at $\xsecquarkpfrac = 1$, we can make that substitution (as with equation~\eqref{eq:h12qq:cartesian:Fd}) and get a simpler result:
\begin{equation}\label{eq:h12gg:cartesian:Fd}
 \Fd12gg(\z, 1, \vec\rperp) = \frac{\Nc\Sperp}{(2\pi)^2}\frac{\alphas}{2\pi}\frac{\xp g(\xp, \mu) D_{\Phadron/g}(\z, \mu)}{\z^2} \bigl[\dipoleS(\vec\rperp)\bigr]^2
 \biggl(\frac{11}{3} - \frac{4\Nf\TR}{3\Nc}\biggr)e^{-i\vec\kperp\cdot\vec\rperp} \ln\frac{\kperp^2}{\mu^2}
\end{equation}
where the contribution to the cross section is given by~\eqref{eq:dipolemasterformula} as
\begin{multline}\label{eq:h12gg:cartesian:xsec}
 \xsec12gg
 = \int_\tau^1\udc\z\int_{\tau/\z}^1\udc\xsecquarkpfrac \int\uddc\vec\rperp \biggl[\frac{\Fs12gg(\z, \xsecquarkpfrac, \vec\rperp) - \Fs12gg(\z, 1, \vec\rperp)}{1 - \xsecquarkpfrac} + \Fn12gg(\z, \xsecquarkpfrac, \vec\rperp)\biggr] \\
 + \int_\tau^1\udc\z\int\uddc\vec\rperp \biggl[\Fs12gg(\z, 1, \vec\rperp)\ln\biggl(1 - \frac{\tau}{\z}\biggr) + \Fd12gg(\z, 1, \vec\rperp)\biggr]
\end{multline}

Now using equation~\eqref{eq:maptoradial} we can obtain the radial position space formulation:
\begin{align}
 \Fs12gg(\z, \xsecquarkpfrac, \rperp) &=
 \begin{multlined}[t][.7\displaywidth]
  \frac{2\Nc\Sperp}{2\pi}\frac{\alphas}{2\pi}\frac{\xp g(\xp, \mu) D_{\Phadron/g}(\z, \mu)}{z^2} \bigl[\dipoleS(\vec\rperp)\bigr]^2 \xsecquarkpfrac \\
  \times \rperp \ln\frac{c_0^2}{\rperp^2\mu^2}\biggl[\besselJzero(\kperp\rperp) + \frac{1}{\xsecquarkpfrac^2} \besselJzero\biggl(\frac{\kperp\rperp}{\xsecquarkpfrac}\biggr)\biggr]
 \end{multlined} \label{eq:h12gg:radial:Fs}\\
 \Fn12gg(\z, \xsecquarkpfrac, \rperp) &=
 \begin{multlined}[t][.7\displaywidth]
 \frac{2\Nc\Sperp}{(2\pi)^2}\frac{\alphas}{2\pi}\frac{\xp g(\xp, \mu) D_{\Phadron/g}(\z, \mu)}{z^2} \bigl[\dipoleS(\vec\rperp)\bigr]^2 \biggl(\frac{1 - \xsecquarkpfrac}{\xsecquarkpfrac} + \xsecquarkpfrac(1 - \xsecquarkpfrac)\biggr) \\
 \times \rperp \ln\frac{c_0^2}{\rperp^2\mu^2}\biggl[\besselJzero(\kperp\rperp) + \frac{1}{\xsecquarkpfrac^2} \besselJzero\biggl(\frac{\kperp\rperp}{\xsecquarkpfrac}\biggr)\biggr]
 \end{multlined} \label{eq:h12gg:radial:Fn}\\
 \Fd12gg(\z, 1, \rperp) &=
 \begin{multlined}\frac{\Nc\Sperp}{(2\pi)^2}\frac{\alphas}{2\pi}\frac{\xp g(\xp, \mu) D_{\Phadron/g}(\z, \mu)}{\z^2} \bigl[\dipoleS(\vec\rperp)\bigr]^2
 \biggl(\frac{11}{3} - \frac{4\Nf\TR}{3\Nc}\biggr)\\
 \times\rperp\besselJzero(\kperp\rperp) \ln\frac{\kperp^2}{\mu^2}
 \end{multlined}\label{eq:h12gg:radial:Fd}
\end{align}
and
\begin{multline}\label{eq:h12gg:radial:xsec}
 \xsec12gg
 = \int_\tau^1\udc\z\int_{\tau/\z}^1\udc\xsecquarkpfrac \int_0^\infty\udc\rperp \biggl[\frac{\Fs12gg(\z, \xsecquarkpfrac, \rperp) - \Fs12gg(\z, 1, \rperp)}{1 - \xsecquarkpfrac} + \Fn12gg(\z, \xsecquarkpfrac, \rperp)\biggr] \\
 + \int_\tau^1\udc\z \int_0^\infty\udc\rperp \biggl[\Fs12gg(\z, 1, \rperp)\ln\biggl(1 - \frac{\tau}{\z}\biggr) + \Fd12gg(\z, 1, \rperp)\biggr]
\end{multline}

\subsection{\texorpdfstring{$\h12q{\bar{q}}$}{H(1)2qqbar}}
From equations (74) and (77) of~\cite{Chirilli:2012jd}, this contribution to the cross section is given as
\begin{multline}\label{eq:xsec12qqbar}
 \xsec12q{\bar q} = \Sperp\frac{\alphas}{2\pi}\int\frac{\udc\z\udc\xsecquarkpfrac}{z^2} \xp g(\xp, \mu) D_{\Phadron/g}(\z, \mu) \int\frac{\uddc\vec\sperp\uddc\vec\tperp}{(2\pi)^4}\dipoleS(\vec\sperp)\dipoleS(\vec\tperp) \\
 \times 8\pi N_f T_R e^{-i\vec\kperp\cdot\vec\tperp}\delta(1 - \xsecquarkpfrac)\int_0^1\udc\xi'\bigl[\xi'^2 + (1 - \xi')^2\bigr]\biggl[\frac{e^{-i\xi'\vec\kperp\cdot\vec\rperp}}{r_\perp^2} - \delta^{(2)}(\vec\rperp)\int\uddc\vec\rperp'\frac{e^{i\vec\kperp\cdot\vec\rperp'}}{r_\perp'^2}\biggr]
\end{multline}
Despite the notation $2q{\bar q}$, this is actually a quadrupole-type term. It's proportional to $\delta(1 - \xsecquarkpfrac)$, so only $\Fd*$ will be nonzero.

To put $\Fd12q{\bar q}$ in a a form suitable for numerical integration, we'll have to regularize the terms in brackets. The procedure is basically similar to the one used with $\h14qq$, with some slight differences because the formula is different. We start by extracting the position-dependent factors,
\begin{equation}
 \int\frac{\uddc\vec\sperp\uddc\vec\tperp}{(2\pi)^4}\dipoleS(\vec\sperp)\dipoleS(\vec\tperp) e^{-i\vec\kperp\cdot\vec\tperp}\biggl[\frac{e^{-i\xi'\vec\kperp\cdot\vec\rperp}}{r_\perp^2} - \delta^{(2)}(\vec\rperp)\int\uddc\vec\rpperp\frac{e^{i\vec\kperp\cdot\vec\rperp'}}{r_\perp'^2}\biggr]
\end{equation}
After changing variables $\vec\sperp\to\vec\rperp$ (with Jacobian $1$) and excising a region of radius $\epsilon$ around the poles in $\vec\rperp$ and $\vec\rpperp$, then changing variables $\vec\rpperp \to -\xi'\vec\rperp$ in the second term, we get
\begin{multline}
 \int\frac{\uddc\vec\tperp}{(2\pi)^2} \dipoleS(\vec\tperp) e^{-i\vec\kperp\cdot\vec\tperp} \int_{\rperp>\epsilon} \frac{\uddc\vec\rperp}{(2\pi)^2} \dipoleS(\vec\tperp - \vec\rperp) \frac{e^{-i\xi'\vec\kperp\cdot\vec\rperp}}{r_\perp^2} \\
 - \int\frac{\uddc\vec\tperp}{(2\pi)^2}\bigl[\dipoleS(\vec\tperp)\bigr]^2 e^{-i\vec\kperp\cdot\vec\tperp} \int_{\rperp > \epsilon/\xi'} \frac{\uddc\vec\rperp}{(2\pi)^2} \frac{e^{-i\xi'\vec\kperp\cdot\vec\rperp}}{\rperp^2}
\end{multline}
We split the second integral into two pieces, $\int_{\rperp > \epsilon/\xi'} = \int_{\rperp > \epsilon} - \int_{\rperp > \epsilon}^{\rperp < \epsilon/\xi'}$, so the total integral turns into
\begin{multline}
 \int\frac{\uddc\vec\tperp}{(2\pi)^2} \dipoleS(\vec\tperp) e^{-i\vec\kperp\cdot\vec\tperp} \int_{\rperp>\epsilon} \frac{\uddc\vec\rperp}{(2\pi)^2} \bigl[\dipoleS(\vec\tperp - \vec\rperp) - \dipoleS(\vec\tperp)\bigr] \frac{e^{-i\xi'\vec\kperp\cdot\vec\rperp}}{r_\perp^2} \\
 + \int\frac{\uddc\vec\tperp}{(2\pi)^2}\bigl[\dipoleS(\vec\tperp)\bigr]^2 e^{-i\vec\kperp\cdot\vec\tperp} \underbrace{\int_{\rperp > \epsilon}^{\rperp < \epsilon/\xi'} \frac{\uddc\vec\rperp}{(2\pi)^2} \frac{e^{-i\xi'\vec\kperp\cdot\vec\rperp}}{\rperp^2}}_{\ln(1/\xi')/2\pi}
\end{multline}
This gets multiplied by $\xi'^2 + (1 - \xi')^2$ and integrated, so we will need the integrals
\begin{align}
 \int_0^1\udc\xi'[\xi'^2 + (1 - \xi')^2] e^{-i\xi'x}
 &= \biggl(\frac{4i}{x^3} - \frac{i}{x}\biggr)(1 - e^{-ix}) + \frac{2}{x^2}(1 + e^{-ix}) 
 \defn \mathcal{I}_2(x) \\
 \int_0^1\udc\xi'[\xi'^2 + (1 - \xi')^2]\ln\frac{1}{\xi'}
 &= \frac{13}{18}
\end{align}
Putting everything together, the final Cartesian position space formula is
\begin{equation}\label{eq:h12qqbar:cartesian:xsec}
 \xsec12q{\bar q}
 = \int_\tau^1\udc\z\int\uddc\vec\sperp\uddc\vec\tperp \Fd12q{\bar q}(\z, 1, \vec\sperp, \vec\tperp)
 + \int_\tau^1\udc\z\int\uddc\vec\rperp \Fd1{2'}q{\bar q}(\z, 1, \vec\rperp)
\end{equation}
with 
\begin{multline}\label{eq:h12qqbar:cartesian:Fd}
 \Fd12q{\bar q}(\z, 1, \vec\sperp, \vec\tperp) = \\
 \frac{1}{2\pi^3}\frac{\alphas}{2\pi}\frac{\Nf \TR \Sperp}{\z^2} \xp g(\xp, \mu)D_{\Phadron/g}(\z, \mu) \bigl[\dipoleS(\vec\sperp) - \dipoleS(\vec\tperp)\bigr]\dipoleS(\vec\tperp) \frac{1}{\rperp^2}e^{-i\vec\kperp\cdot\vec\tperp}\mathcal{I}_2(\vec\kperp\cdot\vec\rperp)
\end{multline}
and
\begin{equation}\label{eq:h12'qqbar:cartesian:Fd}
 \Fd1{2'}q{\bar q}(\z, \xsecquarkpfrac, \vec\rperp) = \frac{13}{18\pi^2}\frac{\alphas}{2\pi}\frac{\Nf \TR \Sperp}{\z^2} \xp g(\xp, \mu)D_{\Phadron/g}(\z, \mu) \bigl[\dipoleS(\vec\rperp)\bigr]^2 e^{-i\vec\kperp\cdot\vec\rperp}
\end{equation}

To get the momentum space expression, we use equation~\eqref{eq:ident:reglog} to rewrite~\eqref{eq:xsec12qqbar} as
\begin{multline}
 -\Sperp\frac{\alphas}{2\pi}\int\frac{\udc\z\udc\xsecquarkpfrac}{z^2} \xp g(\xp, \mu) D_{\Phadron/g}(\z, \mu) \int\frac{\uddc\vec\sperp\uddc\vec\tperp}{(2\pi)^4}\dipoleS(\vec\sperp)\dipoleS(\vec\tperp) \\
 \times 8\pi N_f T_R e^{-i\vec\kperp\cdot\vec\tperp}\delta(1 - \xsecquarkpfrac)\int_0^1\udc\xi'\bigl[\xi'^2 + (1 - \xi')^2\bigr] \frac{1}{4\pi}\int\uddc\vec\qiperp e^{-i\vec\qiperp\cdot\vec\rperp} \ln\frac{(\vec\qiperp - \xi'\vec\kperp)^2}{\kperp^2}
\end{multline}
We can do the integrals over $\vec\sperp$ and $\vec\tperp$, resulting in $\frac{1}{\Sperp^2}\dipoleF(\vec\qiperp)\dipoleF(\vec\kperp - \vec\qiperp)$.
After that it's straightforward to extract the function
\begin{multline}\label{eq:h12qqbar:momentum:Fd}
 \Fd12q{\bar q}(\z, \xsecquarkpfrac, \vec\qiperp, \xi') = 
 -\frac{1}{\Sperp}\frac{\alphas}{2\pi}\frac{\xp g(\xp, \mu) D_{\Phadron/g}(\z, \mu)}{z^2}  \\
 \times 2 \Nf \TR \bigl[\xi'^2 + (1 - \xi')^2\bigr] \dipoleF(\vec\qiperp)\dipoleF(\vec\kperp - \vec\qiperp) \ln\frac{(\vec\qiperp - \xi'\vec\kperp)^2}{\kperp^2}
\end{multline}
which goes into~\eqref{eq:quadrupolemasterformulamomentum} in the form of
\begin{equation}\label{eq:h12qqbar:momentum:xsec}
 \xsec12q{\bar q} = \int_\tau^1\udc\z\int\uddc\vec\qiperp\int_0^1\udc\xi' \Fd12q{\bar q}(\z, 1, \vec\qiperp, \xi')
\end{equation}

\subsection{\texorpdfstring{$\h16gg$}{H(1)6gg}}
From equations (74) and (78) of~\cite{Chirilli:2012jd} we get this term to be
\begin{multline}\label{eq:xsec16gg}
 \xsec16gg = -\Sperp\int\frac{\udc\z\udc\xsecquarkpfrac}{\z^2}\xp g(\xp, \mu)D_{\Phadron/g}(\xp, \mu)\int\frac{\uddc\vec\sperp\uddc\vec\tperp}{(2\pi)^4} \dipoleS(\vec\sperp)\dipoleS(\vec\tperp)\dipoleS(\vec\rperp)\frac{\alphas}{2\pi} \\
 \times 16\pi N_c e^{-i\vec\kperp\cdot\vec\rperp}\Biggl\{-e^{-i\frac{\vec\kperp}{\xsecquarkpfrac}\cdot\vec\tperp}\frac{[1 - \xsecquarkpfrac(1-\xsecquarkpfrac)]^2}{(1 - \xsecquarkpfrac)_+}\frac{1}{\xsecquarkpfrac^2}\frac{\vec\rperp}{r_\perp^2}\cdot\frac{\vec\tperp}{t_\perp^2} \\
 -\delta(1 - \xsecquarkpfrac)\int_0^1\udc\xi'\biggl[\frac{\xi'}{(1-\xi')_+} + \frac{1}{2}\xi'(1 - \xi')\biggr]\Biggl[\frac{e^{-i\xi'\vec\kperp\cdot\vec\tperp}}{t_\perp^2} - \delta^{(2)}(\vec\tperp)\int\uddc\vec\rperp'\frac{e^{i\vec\kperp\cdot\vec\rperp'}}{r_\perp'^2}\Biggr]\Biggr\}
\end{multline}
This is a quadrupole-type term with $\mathcal{S}(\vec\sperp, \vec\tperp) = \dipoleS(\vec\sperp)\dipoleS(\vec\tperp)\dipoleS(\vec\rperp)$. (Recall that $\vec\rperp = \vec\sperp - \vec\tperp$.)

The first and second lines lead to a plus-regulated contribution which is straightforward to extract in Cartesian position space:
\begin{multline}\label{eq:h16gg:cartesian:Fs}
 \Fs16gg(\z, \xsecquarkpfrac, \vec\sperp, \vec\tperp) = \frac{1}{\pi^3}\frac{\alpha_s}{2\pi}\frac{N_c S_\perp}{z^2} \xp g(\xp, \mu)D_{\Phadron/g}(z, \mu)\dipoleS(\vec\sperp)\dipoleS(\vec\tperp)\dipoleS(\vec\rperp)\\ \times e^{-i\vec\kperp\cdot(\vec\rperp + \vec\tperp/\xsecquarkpfrac)}\frac{[1 - \xsecquarkpfrac(1-\xsecquarkpfrac)]^2}{\xsecquarkpfrac^2}\frac{\vec\rperp}{r_\perp^2}\cdot\frac{\vec\tperp}{t_\perp^2}
\end{multline}

For the delta contribution, which comes from the first and third lines, we again need to use cutoff regularization. The position-dependent factors are
\begin{equation}
 \int\frac{\uddc\vec\sperp\uddc\vec\tperp}{(2\pi)^4} \dipoleS(\vec\sperp)\dipoleS(\vec\tperp)\dipoleS(\vec\rperp)e^{-i\vec\kperp\cdot\vec\rperp}\Biggl[\frac{e^{-i\xi'\vec\kperp\cdot\vec\tperp}}{t_\perp^2} - \delta^{(2)}(\vec\tperp)\int\uddc\vec\rperp'\frac{e^{i\vec\kperp\cdot\vec\rpperp}}{\rpperp^2}\Biggr]
\end{equation}
We'll first need to change variables $\vec\sperp\to\vec\rperp$ so that the Fourier factor $e^{-i\vec\kperp\cdot\vec\rperp}$ doesn't change when we apply the delta function of $\vec\tperp$, but otherwise, the procedure is the same as in section~\ref{sec:h14qq}. It gives two terms,
\begin{multline}
 \int\frac{\uddc\vec\rperp\uddc\vec\tperp}{(2\pi)^4} \dipoleS(\vec\rperp)\bigl[\dipoleS(\vec\sperp)\dipoleS(\vec\tperp) - \dipoleS(\vec\rperp)\dipoleS(\vec 0)\bigr]e^{-i\vec\kperp\cdot\vec\rperp}\frac{e^{-i\xi'\vec\kperp\cdot\vec\tperp}}{t_\perp^2} \\
 - \int\frac{\uddc\vec\rperp}{(2\pi)^2} \bigl[\dipoleS(\vec\rperp)\bigr]^2\dipoleS(\vec 0)e^{-i\vec\kperp\cdot\vec\rperp}\frac{1}{2\pi}\ln\frac{1}{\xi'}
\end{multline}
Then we need the integrals
\begin{multline}
 \int_0^1 \udc\xi'\biggl[\frac{\xi'}{(1 - \xi')_+} + \frac{1}{2}\xi'(1 - \xi')\biggr]e^{-i\xi'x} \\
 = i(e^{-ix} - 1)\biggl(\frac{1}{x^3} - \frac{1}{x}\biggr) - \frac{e^{-ix} + 1}{2x^2} + e^{-ix}[-\gamma_E + \Ci(x) - \ln(x) + i\Si(x)]
     \defn \mathcal{I}_3(x)
\end{multline}
and
\begin{equation}
 \int_0^1 \udc\xi'\biggl[\frac{\xi'}{(1 - \xi')_+} + \frac{1}{2}\xi'(1 - \xi')\biggr]\ln\frac{1}{\xi'}
 = -\frac{67}{72} + \frac{\pi^2}{6}
\end{equation}
Putting everything together, we wind up with the quadrupole-type contribution\note{may be a discrepancy between this and the code, in the arguments of $\dipoleS$}
\begin{multline}\label{eq:h16gg:cartesian:Fd:quadrupole}
 \Fd16gg(\z, 1, \vec\sperp, \vec\tperp) = \frac{1}{\pi^3}\frac{\alpha_s}{2\pi}\frac{N_c S_\perp}{z^2} \xp g(\xp, \mu)D_{\Phadron/g}(z, \mu) \\
 \times \dipoleS(\vec\rperp)\Bigl[\dipoleS(\vec\sperp)\dipoleS(\vec\tperp) - \dipoleS(\vec\rperp) \dipoleS(\vec 0)\Bigr]\frac{1}{t_\perp^2}e^{-i\vec\kperp\cdot\vec\rperp}\mathcal{I}_{3}(\vec\kperp\cdot\vec\tperp)
\end{multline}
and the dipole-type contribution
\begin{equation}\label{eq:h16gg:cartesian:Fd:dipole}
 \Fd1{2''}gg(\z, 1, \vec\rperp) = \biggl(-\frac{67}{36\pi^2} + \frac{1}{3}\biggr)\frac{\alpha_s}{2\pi}\frac{N_c S_\perp}{z^2} \xp g(\xp, \mu)D_{\Phadron/g}(z, \mu) \bigl[\dipoleS(\vec\rperp)\bigr]^2 \dipoleS(\vec 0) e^{-i\vec\kperp\cdot\vec\rperp} 
\end{equation}
which both go into
\begin{multline}\label{eq:h16gg:cartesian:xsec}
 \xsec16gg
 = \int_\tau^1\udc\z\int_{\tau/\z}^1\udc\xsecquarkpfrac \int\uddc\vec\sperp\uddc\vec\tperp \frac{\Fs16gg(\z, \xsecquarkpfrac, \vec\sperp, \vec\tperp) - \Fs16gg(\z, 1, \vec\sperp, \vec\tperp)}{1 - \xsecquarkpfrac} \\
 + \int_\tau^1\udc\z\int\uddc\vec\sperp\uddc\vec\tperp \biggl[\Fs16gg(\z, 1, \vec\sperp, \vec\tperp)\ln\biggl(1 - \frac{\tau}{\z}\biggr) + \Fd16gg(\z, 1, \vec\sperp, \vec\tperp)\biggr] \\
 + \int_\tau^1\udc\z\int\uddc\vec\rperp \Fd1{2''}gg(\z, 1, \vec\rperp)
\end{multline}

To get the plus-regulated contribution in momentum space, we start with the expression from the first two lines of~\eqref{eq:xsec16gg}, which can be rewritten
\begin{multline}
 16\pi\Nc\Sperp\frac{\alphas}{2\pi} \int\frac{\udc\z\udc\xsecquarkpfrac}{\z^2}\xp g(\xp, \mu)D_{\Phadron/g}(\z, \mu)\frac{[1 - \xsecquarkpfrac(1 - \xsecquarkpfrac)]^2}{(1 - \xsecquarkpfrac)_+}\frac{1}{\xsecquarkpfrac^2} \\
 \times \int\frac{\uddc\vec\rperp\uddc\vec\tperp}{(2\pi)^4} \dipoleS(\vec\sperp)\dipoleS(\vec\tperp)\dipoleS(\vec\rperp) e^{-i\vec\kperp\cdot\vec\rperp} e^{-i\vec\kperp\cdot\vec\tperp/\xsecquarkpfrac} \frac{\vec\rperp\cdot\vec\tperp}{\rperp^2\tperp^2}
\end{multline}
We take the expression in the second line and insert delta functions as follows:
\begin{multline}
 \int\frac{\uddc\vec\rperp\uddc\vec\tperp}{(2\pi)^4}\int\uddc\vec\sperp\int\uddc\vec\rpperp\uddc\vec\tpperp \delta^{(2)}(\vec\rperp - \vec\rpperp)\delta^{(2)}(\vec\tperp - \vec\tpperp)\delta^{(2)}(\vec\sperp - \vec\rperp - \vec\tperp) \\
 \times\dipoleS(\vec\sperp)\dipoleS(\vec\tperp)\dipoleS(\vec\rperp) e^{-i\vec\kperp\cdot\vec\rperp} e^{-i\vec\kperp\cdot\vec\tperp/\xsecquarkpfrac} \frac{\vec\rpperp\cdot\vec\tpperp}{\rpperp^2\tpperp^2}
\end{multline}
This ``elevates'' $\vec\sperp$ to an independent integration variable, rather than simply a shorthand for $\vec\rperp + \vec\tperp$. Next, we replace each of these delta functions with its Fourier transform according to equation~\eqref{eq:ident:deltaFT}, using $\vec\qiperp$, $\vec\qiiperp$, and $\vec\qiiiperp$ as the momentum variables, respectively. The resulting expression can be rearranged into
\begin{multline}
 (2\pi)^2\int \frac{\uddc\vec\qiperp}{(2\pi)^2} \frac{\uddc\vec\qiiperp}{(2\pi)^2} \frac{\uddc\vec\qiiiperp}{(2\pi)^2}
 \Biggl(\int\frac{\uddc\vec\rperp}{(2\pi)^2}\dipoleS(\vec\rperp)e^{-i(\vec\kperp + \vec\qiiiperp - \vec\qiperp)\cdot\vec\rperp}\Biggr)
 \Biggl(\int\frac{\uddc\vec\sperp}{(2\pi)^2}\dipoleS(\vec\sperp)e^{i\vec\qiiiperp\cdot\vec\sperp}\Biggr) \\
 \times
 \Biggl(\int\frac{\uddc\vec\tperp}{(2\pi)^2}\dipoleS(\vec\tperp)e^{-i(\vec\kperp/\xsecquarkpfrac + \vec\qiiiperp - \vec\qiiperp)\cdot\vec\tperp}\Biggr)
 \Biggl(\int\uddc\vec\rpperp\uddc\vec\tpperp e^{-i\vec\qiperp\cdot\vec\rpperp} e^{-i\vec\qiiperp\cdot\vec\tpperp} \frac{\vec\rpperp\cdot\vec\tpperp}{\rpperp^2\tpperp^2}\Biggr)
\end{multline}
The first three integrals in parentheses give factors of $\dipoleF/\Sperp$ according to~\eqref{eq:dipoleFdefinition}, and the last is an application of~\eqref{eq:ident:vecprod}, so we wind up with
\begin{equation}
 -\frac{1}{(2\pi)^2}\frac{1}{\Sperp^3}\int \uddc\vec\qiperp \uddc\vec\qiiperp \uddc\vec\qiiiperp
 \dipoleF(\vec\kperp + \vec\qiiiperp - \vec\qiperp)
 \dipoleF(\vec\qiiiperp)
 \dipoleF\biggl(\frac{\vec\kperp}{\xsecquarkpfrac} + \vec\qiiiperp - \vec\qiiperp\biggr)
 \frac{\vec\qiperp\cdot\vec\qiiperp}{\qiperp^2\qiiperp^2}
\end{equation}
Restoring the prefactors gives
\begin{multline}\label{eq:h16gg:momentum:Fs}
 \Fs16gg(\z, \xsecquarkpfrac, \vec\qiperp, \vec\qiiperp, \vec\qiiiperp) =
 -\frac{4}{\pi}\frac{\Nc}{\Sperp^2}\frac{\alphas}{2\pi} \frac{\xp g(\xp, \mu)D_{\Phadron/g}(\z, \mu)}{\z^2}\frac{[1 - \xsecquarkpfrac(1 - \xsecquarkpfrac)]^2}{\xsecquarkpfrac^2} \\
 \times\dipoleF(\vec\kperp + \vec\qiiiperp - \vec\qiperp)
 \dipoleF(\vec\qiiiperp)
 \dipoleF\biggl(\frac{\vec\kperp}{\xsecquarkpfrac} + \vec\qiiiperp - \vec\qiiperp\biggr)
 \frac{\vec\qiperp\cdot\vec\qiiperp}{\qiperp^2\qiiperp^2}
\end{multline}

For the rest of the term, we use equation~\eqref{eq:ident:reglog} to turn the third line of~\eqref{eq:xsec16gg} into
\begin{multline}
 \delta(1 - \xsecquarkpfrac)\int_0^1\udc\xi'\biggl[\frac{\xi'}{(1 - \xi')_+} + \frac{1}{2}\xi'(1 - \xi')\biggr]\frac{1}{4\pi}\int\uddc\vec\qiiperp e^{-i\vec\qiiperp\cdot\vec\tperp}\ln\frac{(\vec\qiiperp - \xi'\vec\kperp)^2}{\kperp^2} \\
 = \frac{\delta(1 - \xsecquarkpfrac)}{4\pi}\int_0^1\udc\xi' \int\uddc\vec\qiiperp e^{-i\vec\qiiperp\cdot\vec\tperp} \biggl\{
 \biggl[\frac{\xi'}{1 - \xi'} + \frac{1}{2}\xi'(1 - \xi')\biggr] \ln\frac{(\vec\qiiperp - \xi'\vec\kperp)^2}{\kperp^2} \\
 - \frac{1}{1 - \xi'}\ln\frac{(\vec\qiiperp - \vec\kperp)^2}{\kperp^2}\biggr\}
\end{multline}
This basically takes the form $\iint\uddc\vec\qiiperp e^{-i\vec\qiiperp\cdot\vec\tperp}(\cdots)$. If we combine that with the other position-dependent factors from~\eqref{eq:xsec16gg}, we get
\begin{equation}
 \int\frac{\uddc\vec\sperp\uddc\vec\tperp}{(2\pi)^4} \dipoleS(\vec\sperp)\dipoleS(\vec\tperp)\dipoleS(\vec\rperp) e^{-i\vec\kperp\cdot\vec\rperp} \iint\uddc\vec\qiiperp e^{-i\vec\qiiperp\cdot\vec\tperp}(\cdots)
\end{equation}
Inserting $\iint\uddc\vec\rperp \delta^{(2)}(\vec\rperp - \vec\sperp + \vec\tperp)$ and expanding it with~\eqref{eq:ident:deltaFT}, with $\vec\qiperp$ as the momentum variable, turns this into
\begin{equation}
 \int\uddc\vec\qiperp\uddc\vec\qiiperp \dipoleF(\vec\qiperp)\dipoleF(\vec\kperp - \vec\qiiperp)\dipoleF(\vec\qiperp + \vec\qiiperp - \vec\kperp)(\cdots)
\end{equation}
All in all, the delta contribution is\note{need to map $\vec\qiperp\to\vec\qiperp - \vec\kperp$ to match the code}
\begin{multline}\label{eq:h16gg:momentum:Fd}
 \Fd16gg(\z, \xsecquarkpfrac, \vec\qiperp, \vec\qiiperp, \xi') = -\frac{4 N_c}{\Sperp^2} \frac{\alphas}{2\pi} \frac{\xp g(\xp, \mu)D_{\Phadron/g}(\xp, \mu)}{\z^2}
 \dipoleF(\vec\qiperp)\dipoleF(\vec\kperp - \vec\qiiperp) \\
 \times \dipoleF(\vec\qiperp + \vec\qiiperp - \vec\kperp)\biggl[
 \biggl(\frac{\xi'}{1 - \xi'} + \frac{1}{2}\xi'(1 - \xi')\biggr) \ln\frac{(\vec\qiiperp - \xi'\vec\kperp)^2}{\kperp^2}
 - \frac{1}{1 - \xi'}\ln\frac{(\vec\qiiperp - \vec\kperp)^2}{\kperp^2}\biggr]
\end{multline}
This and equation~\eqref{eq:h16gg:momentum:Fs} go into~\eqref{eq:quadrupolemasterformulamomentum} in the form of
\begin{multline}\label{eq:h16gg:momentum:xsec}
  \xsec16gg = \\
 \int_\tau^1\udc\z\int_{\tau/\z}^1\udc\xsecquarkpfrac \int\uddc\vec\qiperp\uddc\vec\qiiperp\uddc\vec\qiiiperp \biggl[\frac{\Fs16gg(\z, \xsecquarkpfrac, \vec\qiperp, \vec\qiiperp, \vec\qiiiperp) - \Fs16gg(\z, 1, \vec\qiperp, \vec\qiiperp, \vec\qiiiperp)}{1 - \xsecquarkpfrac}\biggr] \\
 + \int_\tau^1\udc\z\int\uddc\vec\qiperp\uddc\vec\qiiperp\uddc\vec\qiiiperp \Fs16gg(\z, 1, \vec\qiperp, \vec\qiiperp, \vec\qiiiperp)\ln\biggl(1 - \frac{\tau}{\z}\biggr) \\
 + \int_\tau^1\udc\z\int\uddc\vec\qiperp\uddc\vec\qiiperp \int_0^1\udc\xi' \Fd14qq(\z, 1, \vec\qiperp, \vec\qiiperp, \xi')
\end{multline}

\subsection{\texorpdfstring{$\h{1,1}2gq$}{H(1,1)2gq}}
From equations (82) and (83) in~\cite{Chirilli:2012jd}, we get this term to be
\begin{multline}\label{eq:xsec112gq}
 \xsec{1,1}2gq = \frac{\alphas}{2\pi}\frac{\Sperp\Nc}{2}\int\frac{\udc\z\udc\xsecquarkpfrac}{\z^2}\int\frac{\uddc\vec\rperp}{(2\pi)^2} \sum_q \xp q(\xp, \mu)D_{\Phadron/g}(\z, \mu) \dipoleS(\vec\rperp) \\
 \times\frac{1}{\xsecquarkpfrac^2}e^{-i\vec\kperp\cdot\vec\rperp/\xsecquarkpfrac}\frac{1}{\xsecquarkpfrac}\bigl[1 + (1 - \xsecquarkpfrac)^2\bigr]\ln\frac{c_0^2}{r_\perp^2 \mu^2}
\end{multline}
This has no plus-regulated or delta-function contribution, so the Cartesian position space representation is fairly trivial to find: we just compare the form of the term to equation~\eqref{eq:masterformula2} and find
\begin{equation}\label{eq:h112gq:cartesian:Fn}
 \Fn{1,1}2gq(\z,\xsecquarkpfrac,\vec\rperp) = \frac{1}{8\pi^2}\frac{\alpha_s}{2\pi}\frac{N_c S_\perp}{z^2} \sum_q \xp q(\xp, \mu)D_{\Phadron/g}(z, \mu)\dipoleS(\vec\rperp)e^{-i\vec\kperp\cdot\vec\rperp/\xsecquarkpfrac}\frac{1 + (1 - \xsecquarkpfrac)^2}{\xsecquarkpfrac^3}\ln\frac{c_0^2}{\rperp^2 \mu^2}
\end{equation}
which goes into
\begin{equation}\label{eq:h112gq:cartesian:xsec}
 \xsec112gq = \int_\tau^1 \udc\z \int_{\tau/z}^1 \udc\xsecquarkpfrac \int\uddc\vec\rperp \Fn{1,1}2gq(\z, \xsecquarkpfrac, \vec\rperp)
\end{equation}
to produce the contribution to the cross section.

Because of the logarithmic factor, this can't be transformed into momentum space, but we can get the radial representation using equation~\eqref{eq:maptoradial}:
\begin{equation}\label{eq:h112gq:radial:Fn}
 \Fn{1,1}2gq(\z,\xsecquarkpfrac,\rperp) = \frac{1}{4\pi}\frac{\alpha_s}{2\pi}\frac{N_c S_\perp}{z^2} \sum_q \xp q(\xp, \mu)D_{\Phadron/g}(z, \mu)\dipoleS(\vec\rperp) \besselJzero\biggl(\frac{\kperp\rperp}{\xsecquarkpfrac}\biggr) \frac{1 + (1 - \xsecquarkpfrac)^2}{\xsecquarkpfrac^3}\ln\frac{c_0^2}{\rperp^2 \mu^2}
\end{equation}
which goes into
\begin{equation}\label{eq:h112gq:radial:xsec}
 \xsec112gq = \int_\tau^1 \udc\z \int_{\tau/z}^1 \udc\xsecquarkpfrac \int\udc\rperp \Fn{1,1}2gq(\z, \xsecquarkpfrac, \rperp)
\end{equation}

\subsection{\texorpdfstring{$\h{1,2}2gq$}{H(1,2)2gq}}
Again from equations (82) and (83) in~\cite{Chirilli:2012jd}
\begin{multline}\label{eq:xsec122gq}
 \xsec{1,2}2gq = \frac{\alphas}{2\pi}\frac{\Sperp\Nc}{2}\int\frac{\udc\z\udc\xsecquarkpfrac}{\z^2}\int\frac{\uddc\vec\rperp}{(2\pi)^2} \sum_q \xp q(\xp, \mu)D_{\Phadron/g}(\z, \mu) \bigl[\dipoleS(\vec\rperp)\bigr]^2 e^{-i\vec\kperp\cdot\vec\rperp}\\
 \times\frac{1}{\xsecquarkpfrac}\bigl[1 + (1 - \xsecquarkpfrac)^2\bigr]\ln\frac{c_0^2}{r_\perp^2 \mu^2}
\end{multline}
And like $\h{1,1}2gq$, this has only a normal contribution (no plus-regulated or delta-function terms), so again we can compare it to~\eqref{eq:masterformula2} to find
\begin{equation}\label{eq:h122gq:cartesian:Fn}
 \Fn{(1,2)}2gq(\z,\xsecquarkpfrac,\vec\rperp) = \frac{1}{8\pi^2}\frac{\alpha_s}{2\pi}\frac{\Nc \Sperp}{\z^2} \sum_q \xp q(\xp, \mu)D_{\Phadron/g}(z, \mu)\dipoleS(\vec\rperp)e^{-i\vec\kperp\cdot\vec\rperp}\frac{1 + (1 - \xsecquarkpfrac)^2}{\xsecquarkpfrac}\ln\frac{c_0^2}{r_\perp^2 \mu^2}
\end{equation}
which goes into
\begin{equation}\label{eq:h122gq:cartesian:xsec}
 \xsec122gq = \int_\tau^1 \udc\z \int_{\tau/z}^1 \udc\xsecquarkpfrac \int\uddc\vec\rperp \Fn{1,2}2gq(\z, \xsecquarkpfrac, \vec\rperp)
\end{equation}

The radial representation is easily obtained using~\eqref{eq:maptoradial}:
\begin{equation}\label{eq:h122gq:radial:Fn}
 \Fn{(1,2)}2qg(\z,\xsecquarkpfrac,\vec\rperp) = \frac{1}{4\pi}\frac{\alpha_s}{2\pi}\frac{\Nc \Sperp}{\z^2} \sum_q \xp q(\xp, \mu)D_{\Phadron/g}(z, \mu)\dipoleS(\vec\rperp) \besselJzero(\kperp\rperp) \frac{1 + (1 - \xsecquarkpfrac)^2}{\xsecquarkpfrac}\ln\frac{c_0^2}{r_\perp^2 \mu^2}
\end{equation}
which goes into
\begin{equation}\label{eq:h122gq:radial:xsec}
 \xsec122gq = \int_\tau^1 \udc\z \int_{\tau/z}^1 \udc\xsecquarkpfrac \int\udc\rperp \Fn{1,2}2gq(\z, \xsecquarkpfrac, \rperp)
\end{equation}
Again, because of the logarithmic factor, there is no momentum-space representation.

\subsection{\texorpdfstring{$\h14gq$}{H(1)4gq}}\label{sec:h14gq}
Equations (82) and (83) in~\cite{Chirilli:2012jd} give us
\begin{multline}\label{eq:xsec14gq}
 \xsec14gq = 4\pi\Nc\Sperp\frac{\alphas}{2\pi}\int\frac{\udc\z\udc\xsecquarkpfrac}{\z^2} \sum_q \xp q(\xp, \mu)D_{\Phadron/g}(\z, \mu) \\
 \times \int\frac{\uddc\vec\sperp\uddc\vec\tperp}{(2\pi)^4} \quadrupoleS(\vec\sperp, \vec\tperp) e^{-i\vec\kperp\cdot\vec\rperp/\xsecquarkpfrac} e^{-i\vec\kperp\cdot\vec\tperp}
 \frac{1 + (1 - \xsecquarkpfrac)^2}{\xsecquarkpfrac^2}\frac{\vec\rperp\cdot\vec\tperp}{\rperp^2\tperp^2}
\end{multline}
Again there is only a normal contribution, no plus-regulated or delta-function term, so the Cartesian position space representation of this term can be directly extracted from the last expression:
\begin{multline}\label{eq:h14gq:cartesian:Fn}
 \Fn14gq(\z, \xsecquarkpfrac, \vec\sperp, \vec\tperp) = \frac{1}{4\pi^3}\Nc\Sperp\frac{\alphas}{2\pi} \sum_q \frac{\xp q(\xp, \mu)D_{\Phadron/g}(\z, \mu)}{\z^2} \\
 \times \quadrupoleS(\vec\sperp, \vec\tperp) e^{-i\vec\kperp\cdot\vec\rperp/\xsecquarkpfrac} e^{-i\vec\kperp\cdot\vec\tperp}
 \frac{1 + (1 - \xsecquarkpfrac)^2}{\xsecquarkpfrac^2}\frac{\vec\rperp\cdot\vec\tperp}{\rperp^2\tperp^2}
\end{multline}
and that goes into
\begin{equation}\label{eq:h14gq:cartesian:xsec}
 \xsec14gq = \int_\tau^1 \udc\z \int_{\tau/z}^1 \udc\xsecquarkpfrac \int\uddc\vec\sperp\uddc\vec\tperp \Fn14gq(\z, \xsecquarkpfrac, \vec\sperp, \vec\tperp)
\end{equation}

To convert this into momentum space, we rewrite the position-dependent parts of~\eqref{eq:xsec14gq} as
\begin{equation}
 \int\frac{\uddc\vec\sperp\uddc\vec\tperp}{(2\pi)^4}\quadrupoleS(\vec\sperp, \vec\tperp) e^{-i\vec\kperp\cdot\vec\rperp/\xsecquarkpfrac} e^{-i\vec\kperp\cdot\vec\tperp}\frac{\vec\rperp\cdot\vec\tperp}{\rperp^2\tperp^2}
\end{equation}
and insert delta functions, in the form of~\eqref{eq:ident:deltaFT}, to get
\begin{multline}
 \int\frac{\uddc\vec\sperp\uddc\vec\tperp}{(2\pi)^4}\quadrupoleS(\vec\sperp, \vec\tperp) e^{-i\vec\kperp\cdot\vec\rperp/\xsecquarkpfrac} e^{-i\vec\kperp\cdot\vec\tperp} \\
 \times \int\uddc\vec\rpperp\uddc\vec\tpperp \int\frac{\uddc\vec\qiperp}{(2\pi)^2} e^{-i\vec\qiperp\cdot(\vec\sperp - \vec\tperp - \vec\rpperp)} \int\frac{\uddc\vec\qiiperp}{(2\pi)^2} e^{-i\vec\qiiperp\cdot(\vec\tperp - \vec\tpperp)} \frac{\vec\rpperp\cdot\vec\tpperp}{\rpperp^2\tpperp^2}
\end{multline}
which simplifies to
\begin{equation}
 -\frac{1}{\Sperp}\int\frac{\uddc\vec\qiperp\uddc\vec\qiiperp}{(2\pi)^2} \quadrupoleG\biggl(\frac{\vec\kperp}{\xsecquarkpfrac} + \vec\qiperp, \frac{\vec\kperp(\xsecquarkpfrac - 1)}{\xsecquarkpfrac} - \vec\qiperp + \vec\qiiperp\biggr)\frac{\vec\qiperp\cdot\vec\qiiperp}{\qiperp^2\qiiperp^2}
\end{equation}
The complete contribution is then
\begin{equation}\label{eq:h14gq:momentum:xsec}
 \xsec14gq = \int_\tau^1 \udc\z \int_{\tau/z}^1 \udc\xsecquarkpfrac \int\uddc\vec\qiperp\uddc\vec\qiiperp \Fn14gq(\z, \xsecquarkpfrac, \vec\qiperp, \vec\qiiperp)
\end{equation}
with
\begin{multline}\label{eq:h14gq:momentum:Fn}
 \Fn14gq(\z, \xsecquarkpfrac, \vec\qiperp, \vec\qiiperp) =
 -\frac{1}{\pi}\Nc\frac{\alphas}{2\pi} \sum_q \frac{\xp q(\xp, \mu)D_{\Phadron/g}(\z, \mu)}{\z^2} \frac{1 + (1 - \xsecquarkpfrac)^2}{\xsecquarkpfrac^2}\\
 \times \quadrupoleG\biggl(\frac{\vec\kperp}{\xsecquarkpfrac} + \vec\qiperp, \frac{\vec\kperp(\xsecquarkpfrac - 1)}{\xsecquarkpfrac} - \vec\qiperp + \vec\qiiperp\biggr)\frac{\vec\qiperp\cdot\vec\qiiperp}{\qiperp^2\qiiperp^2}
\end{multline}

\subsection{\texorpdfstring{$\h{1,1}2qg$}{H(1,1)2qg}}
The $qg$ channel is very similar to the $gq$ channel. From equations (87) and (88) of~\cite{Chirilli:2012jd} we get
\begin{multline}\label{eq:xsec112qg}
 \xsec{1,1}2qg = \frac{\alphas}{2\pi}\frac{\Sperp}{2}\int\frac{\udc\z\udc\xsecquarkpfrac}{\z^2}\int\frac{\uddc\vec\rperp}{(2\pi)^2} \sum_q \xp g(\xp, \mu)D_{\Phadron/q}(\z, \mu) \bigl[\dipoleS(\vec\rperp)\bigr]^2 e^{-i\vec\kperp\cdot\vec\rperp}\\
 \times\bigl[(1 - \xsecquarkpfrac)^2 + \xsecquarkpfrac^2\bigr]\biggl(\ln\frac{c_0^2}{r_\perp^2 \mu^2} - 1\biggr)
\end{multline}
Again we compare it to~\eqref{eq:masterformula2} and find only a normal term:
\begin{multline}\label{eq:h112qg:cartesian:Fn}
 \Fn{(1,1)}2qg(\z,\xsecquarkpfrac,\vec\rperp) = \frac{1}{8\pi^2}\frac{\alpha_s}{2\pi}\frac{S_\perp}{z^2} \sum_i \xp g(\xp, \mu)D_{\Phadron/q}(z, \mu)\\
 \times\dipoleS(\vec\rperp)e^{-i\vec\kperp\cdot\vec\rperp}\bigl[(1 - \xsecquarkpfrac)^2 + \xsecquarkpfrac^2\bigr]\ln\frac{c_0^2}{r_\perp^2 \mu^2}
\end{multline}
in the Cartesian representation, which gets plugged into
\begin{equation}\label{eq:h112qg:cartesian:xsec}
 \xsec{1,1}2qg = \int_\tau^1 \udc\z \int_{\tau/z}^1 \udc\xsecquarkpfrac \int\uddc\vec\rperp \Fn{1,1}2qg(\z, \xsecquarkpfrac, \vec\rperp)
\end{equation}
or
\begin{multline}\label{eq:h112qg:radial:Fn}
 \Fn{(1,1)}2qg(\z,\xsecquarkpfrac,\rperp) = \frac{1}{4\pi}\frac{\alpha_s}{2\pi}\frac{S_\perp}{z^2} \sum_i \xp g(\xp, \mu)D_{\Phadron/q}(z, \mu)\dipoleS(\vec\rperp) \\
 \times\besselJzero(\kperp\rperp) \bigl[(1 - \xsecquarkpfrac)^2 + \xsecquarkpfrac^2\bigr]\ln\frac{c_0^2}{r_\perp^2 \mu^2}
\end{multline}
which goes into
\begin{equation}\label{eq:h112qg:radial:xsec}
 \xsec{1,1}2qg = \int_\tau^1 \udc\z \int_{\tau/z}^1 \udc\xsecquarkpfrac \int\udc\rperp \Fn{1,1}2qg(\z, \xsecquarkpfrac, \rperp)
\end{equation}
in the radial representation. Again, the logarithmic factor precludes a momentum-space formula.

\subsection{\texorpdfstring{$\h{1,2}2qg$}{H(1,2)2qg}}
From equations (87) and (88) of~\cite{Chirilli:2012jd} we get
\begin{multline}\label{eq:xsec122qg}
 \xsec{1,2}2qg = \frac{\alphas}{2\pi}\frac{\Sperp}{2}\int\frac{\udc\z\udc\xsecquarkpfrac}{\z^2}\int\frac{\uddc\vec\rperp}{(2\pi)^2} \sum_q \xp g(\xp, \mu)D_{\Phadron/q}(\z, \mu) \\
 \times \bigl[\dipoleS(\vec\rperp)\bigr]^2 e^{-i\vec\kperp\cdot\vec\rperp/\xsecquarkpfrac}
 \frac{\bigl[(1 - \xsecquarkpfrac)^2 + \xsecquarkpfrac^2\bigr]}{\xsecquarkpfrac^2}\biggl(\ln\frac{c_0^2}{r_\perp^2 \mu^2} - 1\biggr)
\end{multline}
which again, has no plus-regulated or delta contribution. Comparing to~\eqref{eq:masterformula2} we find
\begin{multline}\label{eq:h122qg:cartesian:Fn}
 \Fn{(1,2)}2qg(\z,\xsecquarkpfrac,\vec\rperp) = \frac{1}{8\pi^2}\frac{\alpha_s}{2\pi}\frac{S_\perp}{z^2} \sum_i \xp g(\xp, \mu)D_{\Phadron/q}(z, \mu)\\
 \times\Bigl(\dipoleS(\vec\rperp)\Bigr)^2 e^{-i\vec\kperp\cdot\vec\rperp/\xsecquarkpfrac}\biggl(\frac{(1 - \xsecquarkpfrac)^2}{\xsecquarkpfrac^2} + 1\biggr)\ln\frac{c_0^2}{r_\perp^2 \mu^2}
\end{multline}
which goes into
\begin{equation}\label{eq:h122qg:cartesian:xsec}
 \xsec122qg = \int_\tau^1 \udc\z \int_{\tau/z}^1 \udc\xsecquarkpfrac \int\udc\rperp \Fn{1,2}2qg(\z, \xsecquarkpfrac, \rperp)
\end{equation}
for Cartesian position space and
\begin{multline}\label{eq:h122qg:radial:Fn}
 \Fn{(1,2)}2qg(\z,\xsecquarkpfrac,\rperp) = \frac{1}{4\pi}\frac{\alpha_s}{2\pi}\frac{S_\perp}{z^2} \sum_i \xp g(\xp, \mu)D_{\Phadron/q}(z, \mu) \\
 \times\Bigl(\dipoleS(\vec\rperp)\Bigr)^2 \besselJzero\biggl(\frac{\kperp\rperp}{\xsecquarkpfrac}\biggr) \biggl(\frac{(1 - \xsecquarkpfrac)^2}{\xsecquarkpfrac^2} + 1\biggr)\ln\frac{c_0^2}{r_\perp^2 \mu^2}
\end{multline}
which goes into
\begin{equation}\label{eq:h122qg:radial:xsec}
 \xsec122qg = \int_\tau^1 \udc\z \int_{\tau/z}^1 \udc\xsecquarkpfrac \int\udc\rperp \Fn{1,2}2qg(\z, \xsecquarkpfrac, \rperp)
\end{equation}
for radial position space.

\subsection{\texorpdfstring{$\h14qg$}{H(1)4qg}}
Equations (87) and (88) of~\cite{Chirilli:2012jd} give us
\begin{multline}\label{eq:xsec14qg}
 \xsec14qg = 4\pi\Sperp\frac{\alphas}{2\pi} \int\frac{\udc\z\udc\xsecquarkpfrac}{\z^2} \sum_q \xp g(\xp, \mu)D_{\Phadron/q}(\z, \mu) \\
 \times \int\frac{\uddc\vec\sperp\uddc\vec\tperp}{(2\pi)^4} \quadrupoleS(\vec\sperp, \vec\tperp) e^{-i\vec\kperp\cdot\vec\rperp} e^{-i\vec\kperp\cdot\vec\tperp/\xsecquarkpfrac}
 \frac{(1 - \xsecquarkpfrac)^2 + \xsecquarkpfrac^2}{\xsecquarkpfrac}\frac{\vec\rperp\cdot\vec\tperp}{\rperp^2\tperp^2}
\end{multline}
This is almost identical to $\h14gq$, so the same procedure used in section~\ref{sec:h14gq} is applicable here. In Cartesian space, we get
\begin{multline}\label{eq:h14qg:cartesian:Fn}
 \Fn14qg(\z, \xsecquarkpfrac, \vec\sperp, \vec\tperp) = \frac{1}{4\pi^3}\Sperp\frac{\alphas}{2\pi} \sum_q \frac{\xp g(\xp, \mu)D_{\Phadron/q}(\z, \mu)}{\z^2} \\
 \times \quadrupoleS(\vec\sperp, \vec\tperp) e^{-i\vec\kperp\cdot\vec\rperp} e^{-i\vec\kperp\cdot\vec\tperp/\xsecquarkpfrac}
 \frac{(1 - \xsecquarkpfrac)^2 + \xsecquarkpfrac^2}{\xsecquarkpfrac}\frac{\vec\rperp\cdot\vec\tperp}{\rperp^2\tperp^2}
\end{multline}
which goes into
\begin{equation}\label{eq:h14qg:cartesian:xsec}
 \xsec14qg = \int_\tau^1 \udc\z \int_{\tau/z}^1 \udc\xsecquarkpfrac \int\uddc\vec\sperp\uddc\vec\tperp \Fn14qg(\z, \xsecquarkpfrac, \vec\sperp, \vec\tperp)
\end{equation}

In momentum space, we get\note{need to switch signs of both $\vec\qiperp$ and $\vec\qiiperp$ to match code}
\begin{multline}\label{eq:h14qg:momentum:Fn}
 \Fn14qg(\z, \xsecquarkpfrac, \vec\qiperp, \vec\qiiperp) =
 -\frac{1}{\pi}\frac{\alphas}{2\pi} \sum_q \frac{\xp g(\xp, \mu)D_{\Phadron/q}(\z, \mu)}{\z^2} \frac{(1 - \xsecquarkpfrac)^2 + \xsecquarkpfrac^2}{\xsecquarkpfrac}\\
 \times \quadrupoleG\biggl(\vec\kperp + \vec\qiperp, \frac{\vec\kperp(1 - \xsecquarkpfrac)}{\xsecquarkpfrac} - \vec\qiperp + \vec\qiiperp\biggr)\frac{\vec\qiperp\cdot\vec\qiiperp}{\qiperp^2\qiiperp^2}
\end{multline}
which goes into
\begin{equation}\label{eq:h14qg:momentum:xsec}
 \xsec14qg = \int_\tau^1 \udc\z \int_{\tau/z}^1 \udc\xsecquarkpfrac \int\uddc\vec\qiperp\uddc\vec\qiiperp \Fn14qg(\z, \xsecquarkpfrac, \vec\qiperp, \vec\qiiperp)
\end{equation}

\section{Integration Methods}\label{sec:integrationmethods}

The analytic formulas derived in the previous section, indexed in table~\ref{tbl:hardfactorformulas}, are numerically well-behaved and thus straightforward to implement in a computer program.
In a full calculation, SOLO integrates over anywhere from one variable, as in equation~\eqref{eq:h02qq:momentum:xsec}, to eight, in~\eqref{eq:h16gg:momentum:xsec}.
We'll need to use different integration algorithms to handle different kinds of integrals.

\subsection{Cubature}

The most straightforward method of integration is called multidimensional quadrature, or cubature.
There are a tremendous number of variations on this method, and a full overview would be beyond the scope of this dissertation,\footnote{An overview of cubature can be found in reference~\cite[ch. 4]{NumericalRecipes} or any other textbook on numerical methods.}
but the core of the method is as follows.
Given a one-dimensional function $f(x)$ to be integrated over an interval $[a,b]$, we choose $N$ points $x_i\in[a,b]$, called \iterm{abcissas}, evaluate the function at these points, and approximate the integral as
\begin{equation}
 \int_a^b f(x)\udc x \approx \sum_i w_i f(x_i)
\end{equation}
where $w_i$ are a set of weights chosen to correspond to the abcissas.
Simple examples of this method include the trapezoidal rule,
\begin{align}
 x_i &= a + \frac{i(b - a)}{N} &
 w_i &=
 \begin{cases}
  \frac{1}{2}, & i = 0,N-1 \\
  1,           & 0 < i < N-1
 \end{cases}
\end{align}
and Simpson's rule (valid for $N$ odd),
\begin{align}
 x_i &= a + \frac{i(b - a)}{N} &
 w_i &=
 \begin{cases}
  \frac{1}{3}, & i = 0,N-1 \\
  \frac{2}{3}, & 0 < i < N-1, i\text{ even} \\
  \frac{4}{3}, & 0 < i < N-1, i\text{ odd}
 \end{cases}
\end{align}
These methods fall into a class that one might call polynomial-based integration rules, because they are derived by fitting polynomials (linear for the trapezoidal rule, quadratic for Simpson's rule) to successive sample points.
For reasonably well-behaved functions, one can estimate that the error in these approximations scales as $N^{-k}$ for some $k$, e.g. $k = 2$ for the trapezoidal rule or $k = 4$ for Simpson's rule.

There are also a variety of more complex algorithms, in particular the several variants of Gaussian quadrature, which use nonuniformly spaced abcissas to obtain exponential convergence --- that is, the error estimate scales as $a^{-N}$.
Reference~\cite{NumericalRecipes} contains a good overview of these methods.

Cubature methods can be applied to multidimensional integration by simply performing each integral in succession.
\begin{equation}
 \idotsint f(\vec{x}) \udc^d\vec{x} = 
 \int_{a_0}^{b_0} \Biggl(\int_{a_1(x_0)}^{b_1(x_0)} \cdots \Biggl(\int_{a_d(x_0,\ldots,x_{d-1})}^{b_d(x_0,\ldots,x_{d-1})} f(\vec{x}) \udc x_d\Biggr) \cdots \udc x_1\Biggr) \udc x_0
\end{equation}
For SOLO we use an adaptive multidimensional cubature implementation~\cite{adaptint} based on the algorithms described in references~\cite{VanDooren1976,Berntsen1991}.

\subsection{Monte Carlo}

Since multidimensional cubature performs an inner integral at every point needed to evaluate the outer integral, its runtime scales exponentially in the number of dimensions, which makes it unsuitable for anything beyond a 2D or maybe 3D integration region.
For higher-dimensional integrals, we turn to Monte Carlo integration.
This takes a random sampling of points throughout a $d$-dimensional integration region and uses them to estimate the value of the integral.

Consider a function of one variable.
There are two ways to think of Monte Carlo integration.
One way is to choose a rectangle $[a,b]\times[0,f_\text{max}]$, pick $N$ uniformly distributed random points $(x,y)$ in this rectangle, and count how many fall under the function.
The integral can then be approximated as the fraction of these points which fall below the function, times the area of the rectangle.\iftime{include a mathematical demonstration that this works}
\begin{equation}
 \int_a^b f(x)\udc x \approx \frac{(b - a)f_\text{max}}{N}\sum_i \begin{cases}1, & y \leq f(x_i) \\ 0, & y > f(x_i)\end{cases}
\end{equation}
But a more efficient way is to pick $N$ random values in $[a,b]$, evaluate the function at each, and compute the average.
We can assume the average of the $N$ values approximates the average of the function.
\begin{equation}
 \int_a^b f(x)\udc x \approx \frac{b - a}{N}\sum_i f(x_i) = (b - a)\Braket{f(x)}
\end{equation}
where the angle brackets here indicate the average of $f(x)$ over all sample points $x_i$.
The theoretical standard deviation of this estimate is
\begin{equation}
 (b - a)\sqrt{\frac{\Braket{[f(x)]^2} - \Braket{f(x)}^2}{N}}
\end{equation}
This generalizes to multidimensional integrals easily:
\begin{equation}
 \int_V f(\vec{x})\udc^d\vec{x} \approx \frac{V}{N}\sum f(\vec{x}_i) = V\Braket{f(\vec{x})}
\end{equation}
with theoretical standard deviation
\begin{equation}
 V\sqrt{\frac{\Braket{[f(x)]^2} - \Braket{f(x)}^2}{N}}
\end{equation}
The appeal of Monte Carlo integration is that the standard deviation is proportional to $1/\sqrt{N}$ regardless of the number of dimensions, as opposed to cubature, where the standard deviation scales as $N^{-d}$.
For high-dimensional integrals, Monte Carlo generally yields better results for a given number of sample points.

In most applications of Monte Carlo integration, including SOLO, the function being integrated ($f$) has a large magnitude in some parts of the integration region and is relatively close to zero elsewhere.
Clearly, most of the value of the integral comes from the regions where the integrand's magnitude is large.
If, instead of choosing sample points from a uniform random distribution, we can sample adaptively --- that is, choose more samples from these regions where the integrand is large --- it will improve the accuracy of the final integral.
A mathematical derivation~\cite[sec. 7.9.1]{NumericalRecipes} shows that the optimal choice of the sampling distribution is
\begin{equation}\label{eq:importancesamplingprobability}
 p(\vec{x}) = \frac{\abs{f(\vec{x}}}{\int \abs{f(\vec{x})} \udc^d\vec{x}}
\end{equation}
Choosing sample points with this probability distribution 
This seems circular, because knowing $p$ requires knowing the integral $\int \abs{f(\vec{x})} \udc^d\vec{x}$ which we are trying to compute, but what we can do in practice is perform a quick Monte Carlo integration with uniform sampling to get an estimate of the integral, and then use~\eqref{eq:importancesamplingprobability} to choose more samples adaptively.
This technique is called \iterm{importance sampling}, and it forms the basis of the VEGAS adaptive Monte Carlo algorithm~\cite{Lepage:1977sw,Lepage:1980dq}.
VEGAS was developed for applications in particle physics, so it is not surprising that it is particularly well suited for use in SOLO.
Experimentation has shown that the VEGAS algorithm gives by far the \iftime{include some plots or data to show this}best accuracy for the computations.

\subsection{Fixing the Domain of Integration}
The integration region over $\xsecquarkpfrac$ (or $\xp$) and $\z$ is bounded by kinematics: $\frac{\tau}{\z} \le \xsecquarkpfrac,\xp \le 1$ and $\tau \le \z \le 1$, resulting in the shape shown in figure~\ref{fig:integrationdomain}. However, the C++ implementation uses integration routines from the GNU Scientific Library (GSL)~\cite{GSL}, either Monte Carlo or cubature, which can only handle simple rectangular integration domains with constants as limits.

Because all terms in equation~\eqref{eq:contributions} involve an integration over $z$ with constant limits $[\tau,1]$, it simplifies the code to use that as one of the variables of integration. The other variable needs to be chosen so that the integration region is a rectangle. The most straightforward way to do that is with a simple rescaling:
\begin{equation}
 \frac{\inty - \inty_\text{min}}{\inty_\text{max} - \inty_\text{min}} = \frac{\xsecquarkpfrac - \xsecquarkpfrac_\text{min}}{\xsecquarkpfrac_\text{max} - \xsecquarkpfrac_\text{min}}
\end{equation}
In this case, setting $\inty_\text{min} = \tau$, $\xsecquarkpfrac_\text{min} = \frac{\tau}{z}$, and $\inty_\text{max} = \xsecquarkpfrac_\text{max} = 1$, we get
\begin{equation}
 \inty = \frac{\z\xsecquarkpfrac(1 - \tau) + \tau(\z - 1)}{\z - \tau}
 \qquad\text{and its inverse}\qquad
 \xsecquarkpfrac = \frac{y(\z - \tau) - \tau(\z - 1)}{\z(1 - \tau)}
\end{equation}
The accompanying Jacobian is
\begin{equation}
 \udc z\udc\xsecquarkpfrac = \begin{vmatrix}\pd{z}{z} & \pd{z}{y} \\ \pd{\xsecquarkpfrac}{z} & \pd{\xsecquarkpfrac}{y}\end{vmatrix}\udc z\udc y = \begin{vmatrix}1 & \frac{(z - \tau)^2}{(1 - \xsecquarkpfrac)(1 - \tau)\tau} \\ 0 & \frac{z - \tau}{z(1 - \tau)}\end{vmatrix}\udc z\udc y = \frac{z - \tau}{z(1 - \tau)}\udc z\udc y
\end{equation}
giving an overall transformation of
\begin{equation}
 \int_\tau^1\int_{\frac{\tau}{z}}^1 \frac{\udc\z}{\z^2}\frac{\udc\xp}{\xp}\xsecquarkpfrac(\ldots) = \int_\tau^1\int_\tau^1\udc\z\udc\inty\frac{\z - \tau}{\z^3(1 - \tau)}(\cdots)
\end{equation}
This extra factor in the Jacobian is the last addition we need to produce the numerical results that will be shown in the next chapter.

%% file: xsecanalysis.tex
\chapter{Computational Results and Analysis}\label{ch:results}

After the algebraic manipulations of chapter~\ref{ch:crosssection}, we have an expression for the $\pA\to hX$ cross section, complete up to next-to-leading order, and free of both divergences and analytic singularities. We've developed a program called SOLO, Saturation at One-Loop Order~\cite{SOLO}, to implement the calculation.

\begin{figure}
 \setplotshift{brahms22}{1}
 \setplotshift{brahms32}{1}
 \tikzsetnextfilename{brahmsresults}
 \begin{tikzpicture}
  \begin{groupplot}[result axis,ymin=1e-8,ymode=log,group style={group size=2 by 1,ylabels at=edge left}]
  \nextgroupplot[title={BRAHMS $\eta=2.2$}]
   \resultplot{rcBK GBW LO}{\thisrow{pT}}{\hadronconversionfactor*\thisrow{lo-mean}}{brahmsrcBKYA}
   \resultplot{rcBK GBW NLO}{\thisrow{pT}}{\hadronconversionfactor*(\thisrow{lo-mean}+\coupling{\thisrow{mu2}}*\thisrow{nlo-mean})}{brahmsrcBKYA}
   \addplot[data plot] table[x=pt,y expr={\thisrow{yield}},y error expr={(\thisrow{staterr} + \thisrow{syserr})}] {\brahmsdAuloY};
  \nextgroupplot[title={BRAHMS $\eta=3.2$}]
   \resultplot{rcBK GBW LO}{\thisrow{pT}}{\hadronconversionfactor*\thisrow{lo-mean}}{brahmsrcBKYB}
   \resultplot{rcBK GBW NLO}{\thisrow{pT}}{\hadronconversionfactor*(\thisrow{lo-mean}+\coupling{\thisrow{mu2}}*\thisrow{nlo-mean})}{brahmsrcBKYB}
   \addplot[data plot] table[x=pt,y expr={\thisrow{yield}},y error expr={(\thisrow{staterr} + \thisrow{syserr})}] {\brahmsdAuhiY};
   \legend{LO,LO+NLO,data}
  \end{groupplot}
 \end{tikzpicture}
 \caption[Calculation results and data for \HepProcess{\Pdeuteron\Pnucleus \HepTo \Phadron\Panything} cross section at BRAHMS]{Comparisons of BRAHMS \cite{Arsene:2004ux} ($h^-$) yields in \dAu{} collisions to results of the numerical calculation with the rcBK gluon distribution, both at leading order (tree level) and with NLO corrections included. The edges of the solid bands were computed using $\mu^2 = \text{\SIrange{10}{50}{GeV^2}}$. This is an updated version of figure 1 of~\cite{Stasto:2013cha} with reduced statistical errors.
 }
 \label{fig:brahmsresults}
\end{figure}
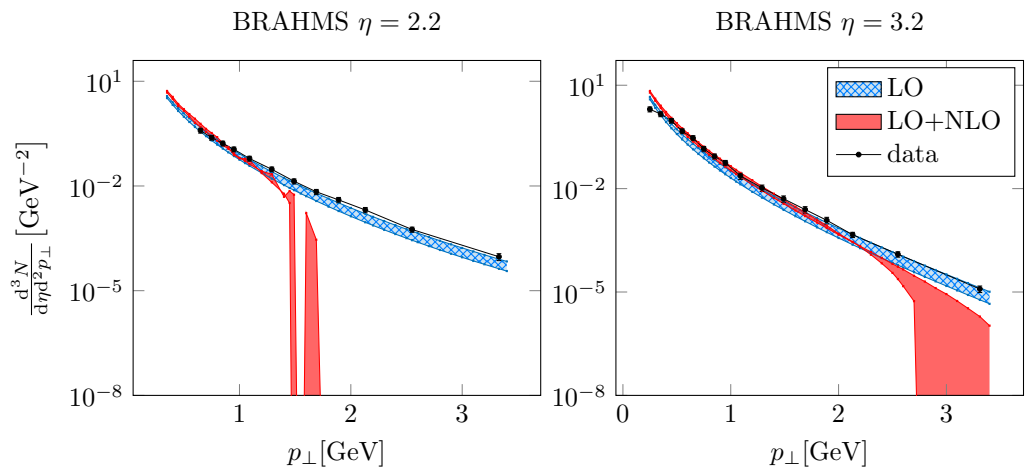

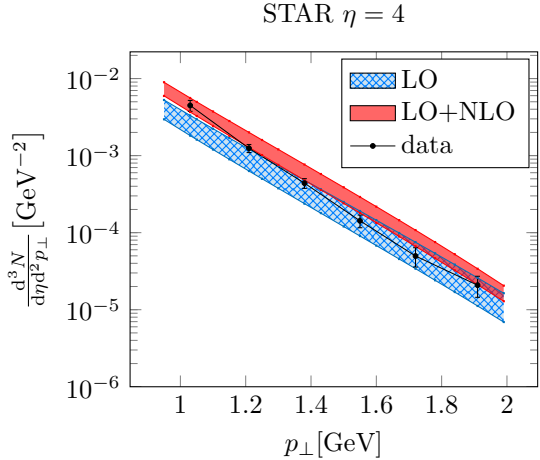
\begin{figure}
 \tikzsetnextfilename{starresults}
 \begin{tikzpicture}
  \begin{axis}[result axis,ymin=1e-6,ymode=log,title={STAR $\eta=4$}]
   \resultplot{rcBK GBW LO}{\thisrow{pT}}{\thisrow{lo-mean}}{starrcBK}
   \resultplot{rcBK GBW NLO}{\thisrow{pT}}{\thisrow{lo-mean}+\coupling{\thisrow{mu2}}*\thisrow{nlo-mean}}{starrcBK}
   \addplot[data plot] table[x=pt,y expr={\thisrow{xsec}/\stardAusigmainel},y error expr={(\thisrow{staterr} + \thisrow{syserr})/\stardAusigmainel}] {\stardAu};
   \legend{LO,LO+NLO,data}
  \end{axis}
 \end{tikzpicture}
 \caption[Calculation results and data for \HepProcess{\Pdeuteron\Pnucleus \HepTo \Phadron\Panything} cross section at STAR]{Comparisons of STAR \cite{Adams:2006uz} ($\pi^0$) yields in \dAu{} collisions to results of the numerical calculation with the rcBK gluon distribution, both at leading order (tree level) and with NLO corrections included. The edges of the solid bands were computed using $\mu^2 = \text{\SIrange{10}{50}{GeV^2}}$. This is an updated version of figure 1 of~\cite{Stasto:2013cha} with reduced statistical errors.
 }
 \label{fig:starresults}
\end{figure}

Our first results from SOLO, published in reference~\cite{Stasto:2013cha}, correspond to the experimental data collected at RHIC by the BRAHMS~\cite{Arsene:2004ux} and STAR~\cite{Adams:2006uz} experiments.
Updated versions of these results are shown in figures~\ref{fig:brahmsresults} and~\ref{fig:starresults}.
It's immediately evident from these plots that something interesting is going on: at high momenta, the LO+NLO cross section becomes negative!

Needless to say, this was quite a surprising result.
Physically, of course, it's impossible for a cross section to be negative, so at first glance the output from SOLO appears completely nonsensical.
But it's actually not that unreasonable for a fixed-order perturbative result such as this one to be negative.
The true cross section is a sum of terms at all orders in $\alphas$,
\begin{equation}
 \xsec* = (\text{LO}) + \alphas(\text{NLO}) + \alphas^2(\text{NNLO}) + \cdots
\end{equation}
and it's only the final sum that corresponds to the physical cross section and thus has to be positive.
Negative contributions from the NLO term could be compensated by large positive contributions from further terms in the series.

In fact, previous work by Albacete et al.~\cite[fig. 9]{Albacete:2012xq} has already shown a trend toward negative results at high $\pperp$ when some of the next-to-leading order contributions are included.
With two independent implementations both showing the same feature, we can be reasonably confident that we're looking at a characteristic of the NLO formulas, not a quirk of the program.\footnote{Reference~\cite{Kang:2014lha} claims that a different subtraction scheme should be used to perform the rapidity subtraction of section~\ref{sec:divergence}, and under this alternate scheme, the LO+NLO result becomes positive and more closely matches the data. However, it has been argued~\cite{Xiao:2014uba} that this alternate scheme is not correct. This issue is the focus of ongoing work.}

At this point, there are several questions:
\begin{itemize}
 \item Does the negativity imply anything about the validity of the small-$x$ formalism?
 \item What is the significance of the cutoff momentum at which the cross section turns negative?
 \item What benefit, if any, do we get by incorporating the NLO correction?
 \item Does the LO+NLO formula give an accurate result for any kinematic conditions?
 \item And the largest question: what can we meaningfully learn about the gluon distribution from the predictions of SOLO?
\end{itemize}

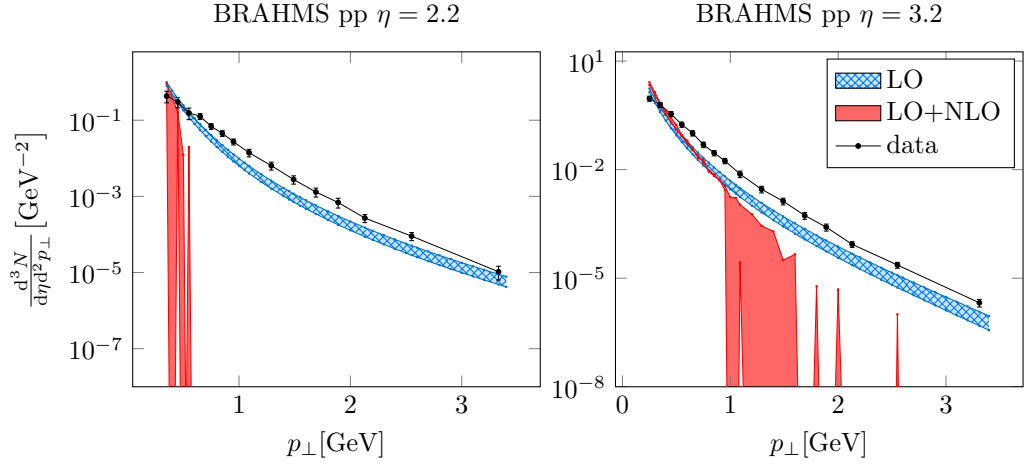
\begin{figure}
 \tikzsetnextfilename{brahmsppresults}
 \begin{tikzpicture}
  \begin{groupplot}[result axis,ymin=1e-8,ymode=log,group style={group size=2 by 1,ylabels at=edge left}]
  \nextgroupplot[title={BRAHMS $\pp$ $\eta=2.2$}]
   \resultplot{rcBK GBW LO}{\thisrow{pT}}{\hadronconversionfactor*\thisrow{lo-mean}}{brahmspprcBKYA}
   \resultplot{rcBK GBW NLO}{\thisrow{pT}}{\hadronconversionfactor*(\thisrow{lo-mean}+\coupling{\thisrow{mu2}}*\thisrow{nlo-mean})}{brahmspprcBKYA}
   \addplot[data plot] table[x=pt,y expr={\thisrow{yield}},y error expr={(\thisrow{staterr} + \thisrow{syserr})}] {\brahmspploY};
  \nextgroupplot[title={BRAHMS $\pp$ $\eta=3.2$}]
   \resultplot{rcBK GBW LO}{\thisrow{pT}}{\hadronconversionfactor*\thisrow{lo-mean}}{brahmspprcBKYB}
   \resultplot{rcBK GBW NLO}{\thisrow{pT}}{\hadronconversionfactor*(\thisrow{lo-mean}+\coupling{\thisrow{mu2}}*\thisrow{nlo-mean})}{brahmspprcBKYB}
   \addplot[data plot] table[x=pt,y expr={\thisrow{yield}},y error expr={(\thisrow{staterr} + \thisrow{syserr})}] {\brahmspphiY};
   \legend{LO,LO+NLO,data}
  \end{groupplot}
 \end{tikzpicture}
 \caption[Calculation results and data for \HepProcess{\Pproton\Pproton \HepTo \Phadron\Panything} cross section at BRAHMS]{Comparisons of BRAHMS \cite{Arsene:2004ux} ($h^-$) yields in \pp{} collisions to results of the numerical calculation with the rcBK gluon distribution, both at leading order (tree level) and with NLO corrections included. The edges of the solid bands were computed using $\mu^2 = \text{\SIrange{10}{50}{GeV^2}}$. Here the calculation completely breaks down at $\pseudorapidity = 2.2$. This is not \emph{entirely} surprising, though, because a proton does not have the extremely high gluon densities of a large atomic nucleus, and thus the CGC model cannot be expected to perform as well.}
 \label{fig:brahmsppresults}
\end{figure}

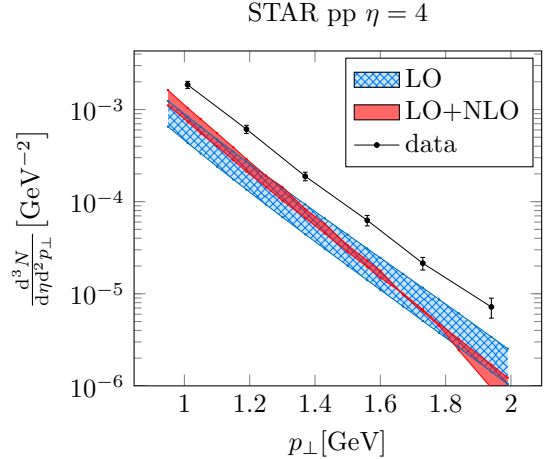
\begin{figure}
 \tikzsetnextfilename{starppresults}
 \begin{tikzpicture}
  \begin{axis}[result axis,ymin=1e-6,ymode=log,title={STAR $\pp$ $\eta=4$}]
   \resultplot{rcBK GBW LO}{\thisrow{pT}}{\thisrow{lo-mean}}{starpprcBK}
   \resultplot{rcBK GBW NLO}{\thisrow{pT}}{\thisrow{lo-mean}+\coupling{\thisrow{mu2}}*\thisrow{nlo-mean}}{starpprcBK}
   \addplot[data plot] table[x=pt,y expr={\thisrow{xsec}/\starppsigmainel},y error expr={(\thisrow{staterr} + \thisrow{syserr})/\starppsigmainel}] {\starppC};
   \legend{LO,LO+NLO,data}
  \end{axis}
 \end{tikzpicture}
 \caption[Calculation results and data for \HepProcess{\Pproton\Pproton \HepTo \Phadron\Panything} cross section at STAR]{Comparisons of STAR \cite{Adams:2006uz} ($\pi^0$) yields in \pp{} collisions to results of the numerical calculation with the rcBK gluon distribution, both at leading order (tree level) and with NLO corrections included. The edges of the solid bands were computed using $\mu^2 = \text{\SIrange{10}{50}{GeV^2}}$.
 }
 \label{fig:starppresults}
\end{figure}

\begin{figure}
 \tikzsetnextfilename{brahmsscaledppresults}
 \begin{tikzpicture}
  \begin{groupplot}[result axis,ymin=1e-8,ymode=log,group style={group size=2 by 1,ylabels at=edge left}]
  \nextgroupplot[title={BRAHMS $\pp$ $\eta=2.2$, increased $\Qs$}]
   \resultplot{rcBK GBW LO}{\thisrow{pT}}{\hadronconversionfactor*\thisrow{lo-mean}}{brahmsscaledpprcBKYA}
   \resultplot{rcBK GBW NLO}{\thisrow{pT}}{\hadronconversionfactor*(\thisrow{lo-mean}+\coupling{\thisrow{mu2}}*\thisrow{nlo-mean})}{brahmsscaledpprcBKYA}
   \addplot[data plot] table[x=pt,y expr={\thisrow{yield}},y error expr={(\thisrow{staterr} + \thisrow{syserr})}] {\brahmspploY};
  \nextgroupplot[title={BRAHMS $\pp$ $\eta=3.2$, increased $\Qs$}]
   \resultplot{rcBK GBW LO}{\thisrow{pT}}{\hadronconversionfactor*\thisrow{lo-mean}}{brahmsscaledpprcBKYB}
   \resultplot{rcBK GBW NLO}{\thisrow{pT}}{\hadronconversionfactor*(\thisrow{lo-mean}+\coupling{\thisrow{mu2}}*\thisrow{nlo-mean})}{brahmsscaledpprcBKYB}
   \addplot[data plot] table[x=pt,y expr={\thisrow{yield}},y error expr={(\thisrow{staterr} + \thisrow{syserr})}] {\brahmspphiY};
   \legend{LO,LO+NLO,data}
  \end{groupplot}
 \end{tikzpicture}
 \caption[Calculation results and data for \HepProcess{\Pproton\Pproton \HepTo \Phadron\Panything} cross section at BRAHMS, with increased saturation scale]{Comparisons of BRAHMS \cite{Arsene:2004ux} ($h^-$) yields in \pp{} collisions to results of the numerical calculation with the rcBK gluon distribution with $\Qs^2$ in the initial condition increased by a factor of $6$, both at leading order (tree level) and with NLO corrections included. The edges of the solid bands were computed using $\mu^2 = \text{\SIrange{10}{50}{GeV^2}}$.}
 \label{fig:brahmsscaledppresults}
\end{figure}

\begin{figure}
 \tikzsetnextfilename{starscaledppresults}
 \begin{tikzpicture}
  \begin{axis}[result axis,ymin=1e-6,ymode=log,title={STAR $\pp$ $\eta=4$, increased $\Qs$}]
   \resultplot{rcBK GBW LO}{\thisrow{pT}}{\thisrow{lo-mean}}{starscaledpprcBK}
   \resultplot{rcBK GBW NLO}{\thisrow{pT}}{\thisrow{lo-mean}+\coupling{\thisrow{mu2}}*\thisrow{nlo-mean}}{starscaledpprcBK}
   \addplot[data plot] table[x=pt,y expr={\thisrow{xsec}/\starppsigmainel},y error expr={(\thisrow{staterr} + \thisrow{syserr})/\starppsigmainel}] {\starppC};
   \legend{LO,LO+NLO,data}
  \end{axis}
 \end{tikzpicture}
 \caption[Calculation results and data for \HepProcess{\Pproton\Pproton \HepTo \Phadron\Panything} cross section at STAR, with increased saturation scale]{Comparisons of STAR \cite{Adams:2006uz} ($\pi^0$) yields in \pp{} collisions to results of the numerical calculation with the rcBK gluon distribution with $\Qs^2$ in the initial condition increased by a factor of $6$, both at leading order (tree level) and with NLO corrections included. The edges of the solid bands were computed using $\mu^2 = \text{\SIrange{10}{50}{GeV^2}}$.
 }
 \label{fig:starscaledppresults}
\end{figure}

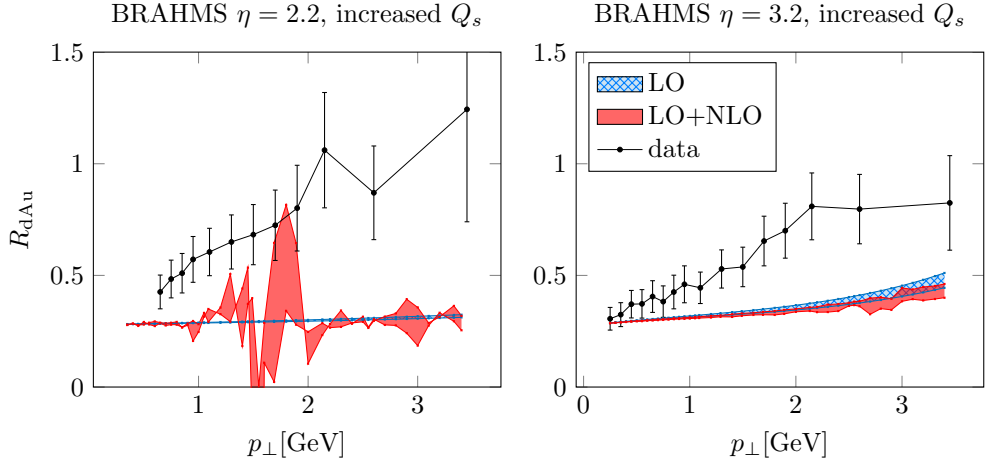
\begin{figure}
 \tikzsetnextfilename{brahmsRdAu}
 \begin{tikzpicture}
  \begin{groupplot}[result axis,ylabel={$R_{\dAu}$},ymin=0,ymax=1.5,group style={group size=2 by 1,ylabels at=edge left}]
  \nextgroupplot[title={BRAHMS $\eta=2.2$, increased $\Qs$}]
   \resultplot{rcBK GBW LO}{\thisrow{pT}}{\thisrow{lo-mean}/\thisrow{pp-lo-mean}/\brahmsNcoll}{brahmsrcBKYA}
   \resultplot{rcBK GBW NLO}{\thisrow{pT}}{(\thisrow{lo-mean}+\coupling{\thisrow{mu2}}*\thisrow{nlo-mean})/(\thisrow{pp-lo-mean}+\coupling{\thisrow{mu2}}*\thisrow{pp-nlo-mean})/\brahmsNcoll}{brahmsrcBKYA}
   \addplot[data plot] table[x=pt,y expr={\thisrow{RdAu}},y error expr={(\thisrow{staterr} + \thisrow{syserr})}] {\brahmsRdAuloY};
  \nextgroupplot[title={BRAHMS $\eta=3.2$, increased $\Qs$},legend pos=north west]
   \resultplot{rcBK GBW LO}{\thisrow{pT}}{\thisrow{lo-mean}/\thisrow{pp-lo-mean}/\brahmsNcoll}{brahmsrcBKYB}
   \resultplot{rcBK GBW NLO}{\thisrow{pT}}{(\thisrow{lo-mean}+\coupling{\thisrow{mu2}}*\thisrow{nlo-mean})/(\thisrow{pp-lo-mean}+\coupling{\thisrow{mu2}}*\thisrow{pp-nlo-mean})/\brahmsNcoll}{brahmsrcBKYB}
   \addplot[data plot] table[x=pt,y expr={\thisrow{RdAu}},y error expr={(\thisrow{staterr} + \thisrow{syserr})}] {\brahmsRdAuhiY};
   \legend{LO,LO+NLO,data}
  \end{groupplot}
 \end{tikzpicture}
 \caption[Calculation and data for \HepProcess{\Pdeuteron\Pnucleus \HepTo \Phadron\Panything} nuclear modification factor at BRAHMS]{Nuclear modification factors measured by BRAHMS compared to the equivalent results computed using SOLO.}
 \label{fig:brahmsRdAu}
\end{figure}

\begin{figure}
 \tikzsetnextfilename{starRdAu}
 \begin{tikzpicture}
  \begin{axis}[result axis,ylabel={$R_{\dAu}$},ymin=0,ymax=1.5,title={STAR $\pseudorapidity=4$, increased $\Qs$},legend pos=north west]
   \resultplot{rcBK GBW LO}{\thisrow{pT}}{\thisrow{lo-mean}/\thisrow{pp-lo-mean}/\starNcoll}{starrcBK}
   \resultplot{rcBK GBW NLO}{\thisrow{pT}}{(\thisrow{lo-mean}+\coupling{\thisrow{mu2}}*\thisrow{nlo-mean})/(\thisrow{pp-lo-mean}+\coupling{\thisrow{mu2}}*\thisrow{pp-nlo-mean})/\starNcoll}{starrcBK}
   \addplot[data plot] table[x=pt,y expr={\thisrow{RdAu}},y error expr={(\thisrow{staterr} + \thisrow{syserr})}] {\starRdAu};
   \legend{LO,LO+NLO,data}
  \end{axis}
 \end{tikzpicture}
 \caption[Calculation and data for \HepProcess{\Pdeuteron\Pnucleus \HepTo \Phadron\Panything} nuclear modification factor at STAR]{Nuclear modification factors measured by STAR compared to the equivalent results computed using SOLO. In this plot the NLO correction is small enough that the two theoretical result plots (LO, and LO+NLO) are visually indistinguishable.}
 \label{fig:starRdAu}
\end{figure}
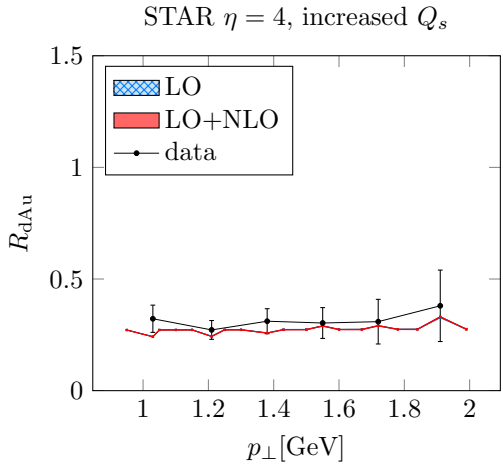

\section{Examining Negativity}

\subsection{Validity of the Results}

In light of the negative results, it's natural to wonder whether the results from SOLO are of any use for predicting cross sections.
Looking at the results in figure~\ref{fig:brahmsresults}, there seem to be three distinct regions.
\begin{itemize}
 \item
 At \emph{very} low $\pperp$, the computational results overpredict the measurements by BRAHMS.
 This is expected, though, because the hadron transverse momentum --- or, really, $\xseccgcpperp$ --- is related to the characteristic energy scale for the strong interaction, the scale at which $\alphas$ is evaluated.
 When $\pperp$ gets very small, the kinematic regions where $\kperp \lesssim \SI{1}{GeV}$ make a significant contribution to the integral.
 But nonperturbative effects start to influence strong interactions below roughly $\SI{1}{GeV}$, and accordingly we shouldn't expect our perturbative calculations to have any predictive value at the lowest $\pperp$.
 \item
 At low to moderate $\pperp$, the data and the measurements match fairly well. The exact boundaries of this region depend on the kinematic conditions.
 \item
 At high $\pperp$, above some cutoff momentum, the cross section predicted by SOLO is negative.
 \note{why does the source of the negativity not affect lower $\pperp$?}We can trace this to the plus prescription that comes from the virtual diagrams and the subtractions in the unintegrated gluon distribution, as we'll see in section~\ref{sec:sources}.
\end{itemize}

The situation is rather different for proton-proton collisions, shown in figures~\ref{fig:brahmsppresults} and~\ref{fig:starppresults}.
Statistical errors on the NLO corrections are very large, and at $\pseudorapidity = 2.2$ the middle region in which the calculated results match the data is entirely nonexistent, while at the higher pseudorapidity the calculation underpredicts the data for all $\xsechadronpperp$.

To resolve this discrepancy, we can consider adjusting the parameters of the gluon distribution to fit the curve to the BRAHMS data.
It reduces the predictive power of the model, but then again it is reasonable that the parameters of the distribution may differ for protons and nuclei in some nontrivial way --- this would be a sign of interesting nuclear effects.
As a first step, figures~\ref{fig:brahmsscaledppresults} and~\ref{fig:starscaledppresults} show that if we naively use the gluon distribution from the nucleus, in which $\Qs^2$ in the initial condition of the BK evolution is larger by a factor of $\massnumber^{1/3} \approx 6$, we get a decent match to the proton-proton data.

Continuing with that result, something interesting appears when we use SOLO to calculate the nuclear modification factor $R_{\dAu}$, defined as the ratio of the cross section (or yield) for $\dAu$ collisions to the cross section for $\pp$ collisions scaled up by the number of individual nucleon-nucleon collisions involved, $\Ncoll$.
\begin{equation}\label{eq:RdAudefn}
 R_{\dAu} \defn \frac{1}{\Ncoll}\frac{\displaystyle\biggl(\xsec*\biggr)_{\dAu}}{\displaystyle\biggl(\xsec*\biggr)_{\pp}}
\end{equation}
Figures~\ref{fig:brahmsRdAu} and~\ref{fig:starRdAu} show that $R_{\dAu}$ at next-to-leading order is fairly close to the leading-order result both above and below the cutoff momentum (albeit quite far from the BRAHMS experimental measurements).
And even in the vicinity of the cutoff momentum, where we get large fluctuations because the denominator of equation~\eqref{eq:RdAudefn} approaches zero, the ratio is still centered on the leading-order result.

As a purely phenomenological observation, the (very rough) continuity of $R_{\dAu}$ suggests that there may be factors common to the LO and NLO terms which are positive for all $\pperp$, and that the negativity arises from other factors which appear only in the NLO corrections.
In other words, consider the following reasoning: if we let $D\sigma$ represent the cross section, then what we see from the plot is
\begin{equation}\label{eq:RdAuatorders}
 \frac{D\sigma_{\dAu}^{\text{LO}}}{D\sigma_{\pp}^{\text{LO}}} \approx \frac{D\sigma_{\dAu}^{\text{LO+NLO}}}{D\sigma_{\pp}^{\text{LO+NLO}}}
 \qquad\text{which implies that}\qquad
 \frac{D\sigma_{\dAu}^{\text{NLO}}}{D\sigma_{\dAu}^{\text{LO}}} \approx \frac{D\sigma_{\pp}^{\text{NLO}}}{D\sigma_{\pp}^{\text{LO}}}
\end{equation}
which would be satisfied if the cross section decomposes as
\begin{equation}\label{eq:xsecdecomposition}
 D\sigma_{\text{collision}}^{\text{order}} \sim (\underbrace{\text{collision-specific factor}}_\text{positive})\times(\underbrace{\text{order-specific factor}}_{\substack{\text{source of negativity}\\\text{at NLO}}})
\end{equation}
Of course, there is no guarantee that~\eqref{eq:xsecdecomposition} follows from~\eqref{eq:RdAuatorders}, but the plots in figure~\ref{fig:brahmsRdAu} are fairly suggestive.
If the cross section does follow equation~\eqref{eq:xsecdecomposition} at leading and next-to-leading order, the same pattern might extend to higher orders, which would make it possible to extract meaningful information about the all-order cross section from finite-order approximations to $R_{\dAu}$.

\subsection{Sources of Negative Contributions}\label{sec:sources}
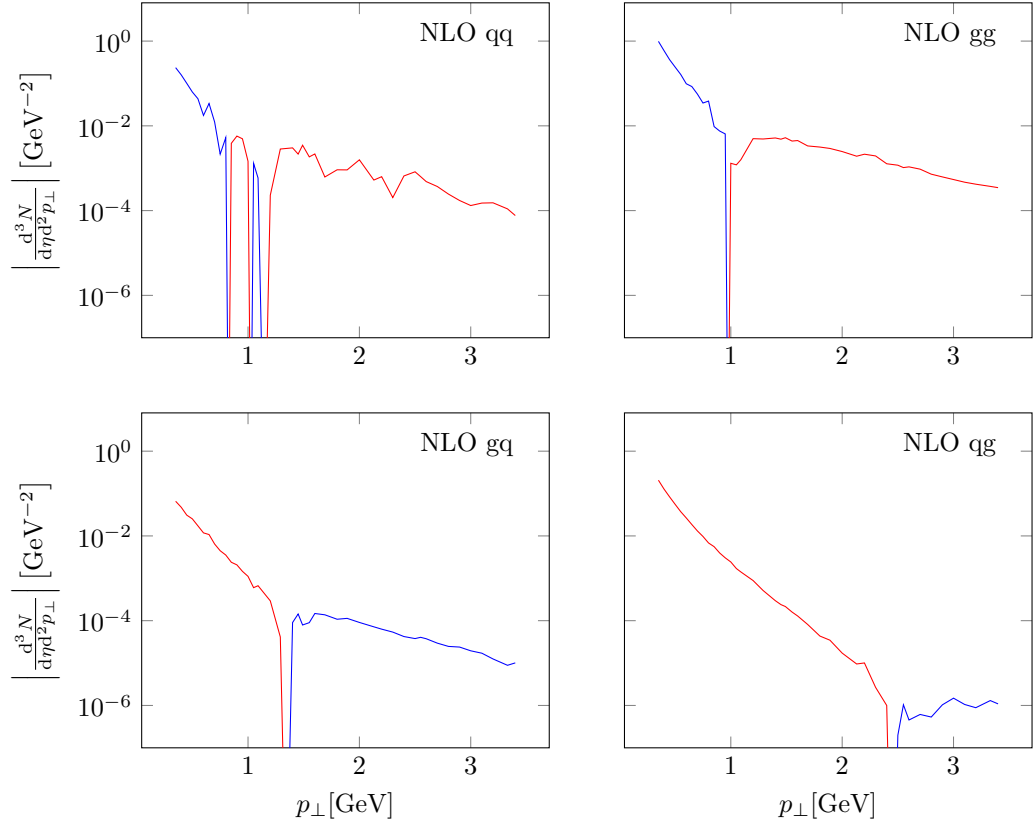
\begin{figure}
 \tikzset{
  faded/.style={opacity=0.3},
  normal/.style={opacity=1},
  qq lo/.style={normal},
  gg lo/.style={normal},
  qq nlo/.style={normal},
  gg nlo/.style={normal},
  gq nlo/.style={normal},
  qg nlo/.style={normal}
 }
 \tikzsetnextfilename{channelbreakdown}
 \begin{tikzpicture}
  \begin{groupplot}[
   result axis,
   ylabel={$\abs{\frac{\udddc N}{\udc \eta\uddc p_\perp}} \bigl[\si{GeV^{-2}}\bigr]$},
   ymode=log,
   every plot/.style={mark=*,mark size={0.5pt},width=7.4cm,height=6.5cm},
   pos/.style=blue,neg/.style=red,
   title style={at={(1,1)},below left=5mm},
   ymax=8,ymin=1e-7,
   group style={group size=2 by 2,ylabels at=edge left,xlabels at=edge bottom,yticklabels at=edge left}
  ]
  \nextgroupplot[title={NLO qq}]
   \addplot[pos,qq nlo] table[x expr={\thisrow{pT}},y expr={\thisrow{r:h12qq-mean}+\thisrow{m:h14qq-mean}}] {\brahmslomurcBKloYsep};
   \addplot[neg,qq nlo] table[x expr={\thisrow{pT}},y expr={-\thisrow{r:h12qq-mean}-\thisrow{m:h14qq-mean}}] {\brahmslomurcBKloYsep};
  \nextgroupplot[title={NLO gg}]
   \addplot[pos,gg nlo] table[x expr={\thisrow{pT}},y expr={\thisrow{r:h12gg-mean}+\thisrow{m:h12qqbar-mean}+\thisrow{m:h16gg-mean}}] {\brahmslomurcBKloYsep};
   \addplot[neg,gg nlo] table[x expr={\thisrow{pT}},y expr={-\thisrow{r:h12gg-mean}-\thisrow{m:h12qqbar-mean}-\thisrow{m:h16gg-mean}}] {\brahmslomurcBKloYsep};
  \nextgroupplot[title={NLO gq}]
   \addplot[pos,gq nlo] table[x expr={\thisrow{pT}},y expr={\thisrow{r:h112gq-mean}+\thisrow{r:h122gq-mean}+\thisrow{m:h14gq-mean}}] {\brahmslomurcBKloYsep};
   \addplot[neg,gq nlo] table[x expr={\thisrow{pT}},y expr={-\thisrow{r:h112gq-mean}-\thisrow{r:h122gq-mean}-\thisrow{m:h14gq-mean}}] {\brahmslomurcBKloYsep};
  \nextgroupplot[title={NLO qg}]
   \addplot[pos,qg nlo] table[x expr={\thisrow{pT}},y expr={\thisrow{r:h112qg-mean}+\thisrow{r:h122qg-mean}+\thisrow{m:h14qg-mean}}] {\brahmslomurcBKloYsep};
   \addplot[neg,qg nlo] table[x expr={\thisrow{pT}},y expr={-\thisrow{r:h112qg-mean}-\thisrow{r:h122qg-mean}-\thisrow{m:h14qg-mean}}] {\brahmslomurcBKloYsep};
  \end{groupplot}
 \end{tikzpicture}
 \caption[Breakdown of NLO contributions by channel]{Breakdown by channel of the NLO contributions to the BRAHMS $h^-$ cross section for $\pseudorapidity = 2.2$ (the one shown in figure~\ref{fig:brahmsresults}). These plots display the magnitude of the contribution, and blue indicates where the contribution is positive while red indicates where it's negative.}
 \label{fig:channelbreakdown}
\end{figure}

\begin{figure}
 \tikzset{
  faded/.style={opacity=0.3},
  normal/.style={opacity=1},
  qq lo/.style={normal},
  gg lo/.style={normal},
  qq nlo/.style={normal},
  gg nlo/.style={normal},
  gq nlo/.style={normal},
  qg nlo/.style={normal}
 }
 \tikzsetnextfilename{termbreakdown}
 \begin{tikzpicture}
  \begin{groupplot}[
   result axis,
   ylabel={$\abs{\frac{\udddc N}{\udc \eta\uddc p_\perp}} \bigl[\si{GeV^{-2}}\bigr]$},
   ymode=log,
   every plot/.style={mark=*,mark size={0.5pt},width=7.4cm,height=4cm},
   pos/.style=blue,neg/.style=red,
   title style={at={(1,1)},below left=5mm},
   group style={group size=2 by 1,ylabels at=edge left,xlabels at=edge bottom}
  ]
  \def\ptexp{3}
  \nextgroupplot[
   ymin=5e-2,ymax=2e-1,
   yticklabel style={/pgf/number format/fixed},
   exp yticklabels off
  ]
   \addplot[neg,qq nlo] table[x expr={\thisrow{pT}},y expr={-\thisrow{pT}^\ptexp*\thisrow{r:h12qq-mean}}] {\brahmslomurcBKloYsep} node[above,pos=0.5] {$\h12qq$};
   \addplot[pos,qq nlo] table[x expr={\thisrow{pT}},y expr={\thisrow{pT}^\ptexp*\thisrow{m:h14qq-mean}}] {\brahmslomurcBKloYsep} node[below,pos=0.5] {$\h14qq$};
  \nextgroupplot[ymin=1e-4,ymax=4e-1]
   \addplot[neg,gg nlo] table[x expr={\thisrow{pT}},y expr={-\thisrow{pT}^\ptexp*\thisrow{r:h12gg-mean}}] {\brahmslomurcBKloYsep} node[above,pos=0.6] {$\h12gg$};
   \addplot[pos,gg nlo] table[x expr={\thisrow{pT}},y expr={\thisrow{pT}^\ptexp*\thisrow{m:h12qqbar-mean}}] {\brahmslomurcBKloYsep} node[above,pos=0.5] {$\h12q{\bar q}$};
   \addplot[pos,gg nlo] table[x expr={\thisrow{pT}},y expr={\thisrow{pT}^\ptexp*\thisrow{m:h16gg-mean}}] {\brahmslomurcBKloYsep} node[below,pos=0.5] {$\h16gg$};
  \end{groupplot}
 \end{tikzpicture}
 \caption[Breakdown of NLO diagonal-channel contributions by term]{A further breakdown of the NLO diagonal channels from figure~\ref{fig:channelbreakdown} shows that the negativity comes from $\h12qq$ and $\h12gg$. As in figure~\ref{fig:channelbreakdown}, the plot shows the magnitude of the corresponding term and the color indicates the sign: blue for positive, red for negative.}
 \label{fig:termbreakdown}
\end{figure}
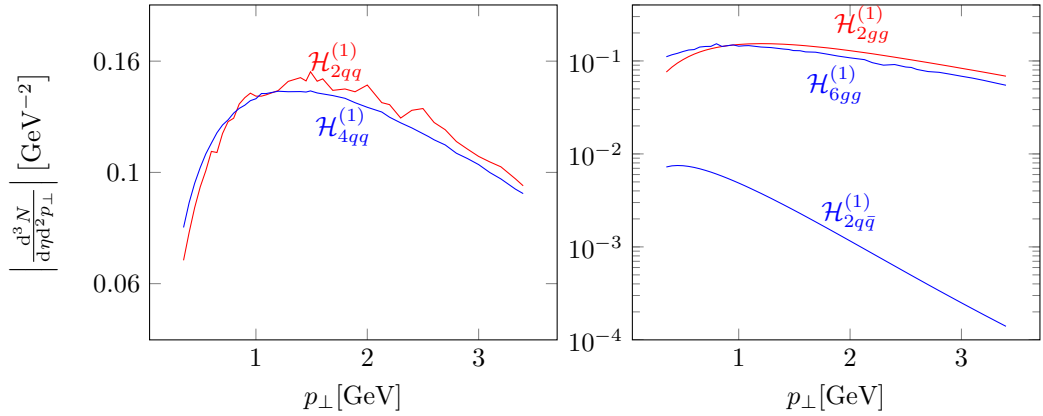

To investigate possible patterns of the form~\eqref{eq:xsecdecomposition}, we'll need to understand where the negativity comes from.
A logical place to start is breaking down the contributions term by term to isolate the origin of the negative results.
Figure~\ref{fig:channelbreakdown} shows the magnitudes of the NLO contributions from the four channels, with red lines representing negative contributions and blue lines positive contributions.
At high $\pperp$, the lines for the diagonal $qq$ and $gg$ channels are negative, indicating that those terms are the sources of the negativity.
We can do the same thing for individual terms within those channels, as in figure~\ref{fig:termbreakdown}, and thereby narrow down the negative contributions to the dipole terms $\h12qq$ and $\h12gg$.

With this in mind, let's look back at the forms of these two terms, equations~\eqref{eq:h12qq:radial:xsec} and~\eqref{eq:h12gg:radial:xsec} and the functions that contribute to them.
Roughly, the normal and delta-function contributions~\eqref{eq:h12qq:radial:Fd}, \eqref{eq:h12gg:radial:Fn}, and~\eqref{eq:h12gg:radial:Fd} take the form
\begin{equation}
 (\text{positive factors})\times \bigl[\dipoleS(\vec\rperp)\bigr]^{(\text{$1$ or $2$})} \rperp \ln\frac{C}{\rperp^2} \besselJzero(\kperp\rperp)
\end{equation}
with the logarithm absent from the delta-function contributions.
Knowing that $\dipoleS\approx 1$ for $\rperp\lesssim 1/\Qs$ but decays as $\rperp^{-4}$ or faster for large $\rperp$ is enough to determine that this expression is positive.
The plus-regulated contributions~\eqref{eq:h12qq:radial:Fs} and~\eqref{eq:h12gg:radial:Fs} are similar except for being integrated over the plus-regulated term:
\begin{equation}
 \int_{\tau/\z}^1 \udc\xisym\frac{f(\xisym)}{(1 - \xisym)_+} = \int_{\tau/\z}^1 \udc\xisym\frac{f(\xisym) - f(1)}{1 - \xisym} + f(1)\ln\biggl(1 - \frac{\tau}{\z}\biggr)
\end{equation}
Since $\tau/\z \leq 1$, $\ln\bigl(1 - \frac{\tau}{\z}\bigr)$ is negative (or zero), and $f(\xisym) < f(1)$ for $\xisym < 1$, so the integral is also negative.
Clearly, the blame for the negative results falls squarely on the plus-prescription regulator.
Note that a similar structure occurs in $\h14qq$ and $\h16gg$, equations~\eqref{eq:h14qq:momentum:Fs} and~\eqref{eq:h16gg:momentum:Fs} respectively, but those terms carry an overall minus sign which makes their final contributions positive.

Physically, we can trace these plus-regulated contributions back to the subtractions performed to remove the divergences.
\iftime{add some more detail?}
This suggests that perhaps the subtraction procedure could be modified to make the overall result positive, though doing so is not a trivial change because it's necessary to ensure consistency with the renormalization scheme used in the fragmentation functions and parton distributions.
Whether this approach can address the negativity is a question for future work to address (though see chapter~\ref{ch:beyondnlo} for some initial attempts).

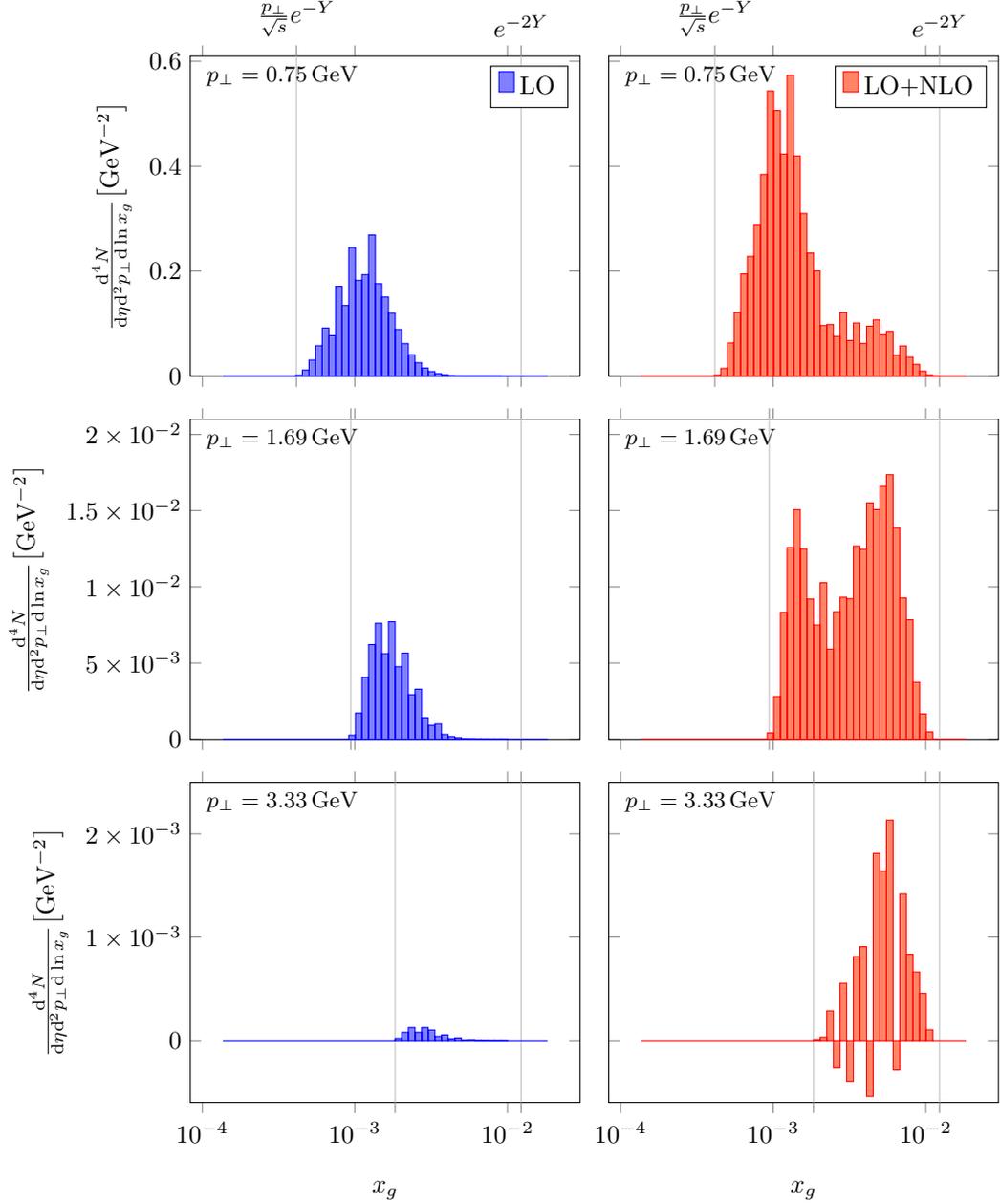
\begin{figure}
 \tikzsetnextfilename{xgdist}
 \begin{tikzpicture}
  \begin{groupplot}[
   group style={
    group size=2 by 3,
    xlabels at=edge bottom,
    xticklabels at=edge bottom,
    ylabels at=edge left,
    yticklabels at=edge left,
    vertical sep=6mm,
    horizontal sep=4mm
   },
   ybar,
   xtick={-4.605,-6.908,-9.210}, 
   xticklabels={\num{e-2},\num{e-3},\num{e-4}},
   xlabel={$\xg$},
   ylabel={$\frac{\udc^4 N}{\udc \eta\uddc p_\perp\udc\ln\xg} \bigl[\si{GeV^{-2}}\bigr]$},
   every axis plot/.append style={
    ybar interval,
    table/x expr={-\thisrow{ygmin}}
   },
   lo/.style={
    draw=blue,
    fill=blue!50,
    table/y expr={10*\hadronconversionfactor*(\thisrow{lo})}
   },
   nlo/.style={
    draw=red,
    fill=red!80!yellow!60,
    table/y expr={10*\hadronconversionfactor*(\thisrow{lo}+\coupling{10}*\thisrow{nlo})} 
   },
   scaled ticks=false,
   extra x tick labels={},
   extra x tick style={grid=major,xticklabel pos=upper,anchor=north},
   label extra ticks/.style={extra x tick labels={$e^{-2\rapidity}$,$\frac{\xsechadronpperp}{\sqs}e^{-\rapidity}$}},
   every axis title/.append style={
    at={(0.02,0.95)},
    below right,
    font=\small
   },
   legend pos=north east,
   pT 075/.style={
    ymin=0,
    ymax=6.1e-1,
    extra x ticks={-4.4,-7.79}, 
    title={$\xsechadronpperp = \SI{0.75}{GeV}$}
   },
   pT 169/.style={
    ymin=0,
    ymax=2.1e-2,
    extra x ticks={-4.4,-6.97},
    title={$\xsechadronpperp = \SI{1.69}{GeV}$}
   },
   pT 333/.style={
    ymin=-6e-4,
    ymax=2.5e-3,
    extra x ticks={-4.4,-6.30},
    title={$\xsechadronpperp = \SI{3.33}{GeV}$}
   }
  ]
  \nextgroupplot[label extra ticks,pT 075]
   \addplot+[lo] table {datafiles/xgdist/results0.75.dat};
   \legend{LO}
  \nextgroupplot[label extra ticks,pT 075]
   \addplot+[nlo] table {datafiles/xgdist/results0.75.dat};
   \legend{LO+NLO}
  \nextgroupplot[pT 169]
   \addplot+[lo] table {datafiles/xgdist/results1.69.dat};
  \nextgroupplot[pT 169]
   \addplot+[nlo] table {datafiles/xgdist/results1.69.dat};
  \nextgroupplot[pT 333]
   \addplot+[lo] table {datafiles/xgdist/results3.33.dat};
  \nextgroupplot[pT 333]
   \addplot+[nlo] table {datafiles/xgdist/results3.33.dat};
  \end{groupplot}
 \end{tikzpicture}
 \caption[Distribution of the cross section across values of $\xg$]{The distribution of contributions to the cross section across values of $\xg$, for three different values of $\xsechadronpperp$, taking into account only the leading order calculation on the left and the LO+NLO calculation on the right. Gray vertical lines show the theoretical limits of $\xg$. These data were generated from the BRAHMS $\pseudorapidity = 2.2$ results at $\mu^2 = \SI{10}{GeV^2}$.}
 \label{fig:xgdist}
\end{figure}

\begin{figure}
 \tikzsetnextfilename{satscalebrahms}
 \begin{tikzpicture}
  \begin{groupplot}[
   group style={group size=2 by 1,ylabels at=edge left},
   no markers,
   xlabel={$\pperp \bigl[\si{GeV}\bigr]$},
   ylabel={$Q_s \bigl[\si{GeV}\bigr]$}
  ]
  \nextgroupplot[title={BRAHMS $\pseudorapidity = 2.2$}]
   \addplot+[blue,name path=upper bound] table[x=pT,y=Qs] {\kovrsatscaleA};
   \addplot+[blue,name path=lower bound] table[x=pT,y expr={0.7744232289912751}] {\kovrsatscaleA};
   \addplot+[blue!10!white] fill between[of=upper bound and lower bound];
  \nextgroupplot[title={BRAHMS $\pseudorapidity = 3.2$}]
   \addplot+[blue,name path=upper bound] table[x=pT,y=Qs] {\kovrsatscaleB};
   \addplot+[blue,name path=lower bound] table[x=pT,y expr={0.8809591482553149}] {\kovrsatscaleB};
   \addplot+[blue!10!white] fill between[of=upper bound and lower bound];
  \end{groupplot}
 \end{tikzpicture}
 \caption[Range of saturation scales probed by the calculation]{The range of saturation scales probed by the calculation using BRAHMS parameters as a function of $\xsechadronpperp$. As in figure~\ref{fig:brahmsresults}, the plot on the left is for $\pseudorapidity = 2.2$ and the one on the right for $\pseudorapidity = 3.2$.}
 \label{fig:satscalebrahms}
\end{figure}

\subsection{The Cutoff Momentum and the Saturation Scale}\label{sec:cutoff}

An interesting phenomenological observation is that the cutoff momentum at which the cross section turns negative grows with rapidity. Now, the maximum $\pperp$ that is kinematically allowed is given by $\tau = 1$, or $(\pperp)_\text{max} = \sqs e^{-Y}$. For RHIC at $\pseudorapidity = 4$, as in figure~\ref{fig:starresults}, this is $(\pperp)_\text{max} = \SI{3.7}{GeV}$, and we can see by extrapolating from figure~\ref{fig:brahmsresults} that the cutoff momentum exceeds $(\pperp)_\text{max}$ at that pseudorapidity, which explains why we don't see negativity in the STAR results.

Let's consider the BRAHMS $\pseudorapidity=2.2$ and $\pseudorapidity=3.2$ results shown in figure~\ref{fig:brahmsresults}.
Each point on the graph, at a given value of $\pperp$, incorporates evaluations of the gluon distribution at values of $\xg$ in the range $\frac{\pperp}{\sqs}e^{-Y} \leq \xg \leq e^{-2Y}$. The distribution among values of $\xg$ is shown for selected values of $\xsechadronpperp$ in figure~\ref{fig:xgdist}.
We can see from the figure that the distribution is not highly concentrated toward one end or the other, so a value of $\xg$ in the middle of the theoretically allowed range is a reasonable approximation to the average.

Now, we can look at the values of the saturation scale corresponding to the theoretical maximum and minimum $\xg$, using the relationship shown in figure~\ref{fig:satscale} to relate $\xg$ to $\Qs$.
From this we get the graphs shown in figure~\ref{fig:satscalebrahms}.
The saturation scales encompass a fairly narrow range around $\SI{0.9}{GeV}$ at $\pseudorapidity = 2.2$, and $\SI{1}{GeV}$ at $\pseudorapidity = 3.2$. In the former case, the cutoff at which the cross section turns negative is about $1.7\Qs$, and in the latter case around $3.1\Qs$, so clearly the relationship between these two momentum scales is rapidity-dependent in some nontrivial way.


\subsection{Benefits of NLO}

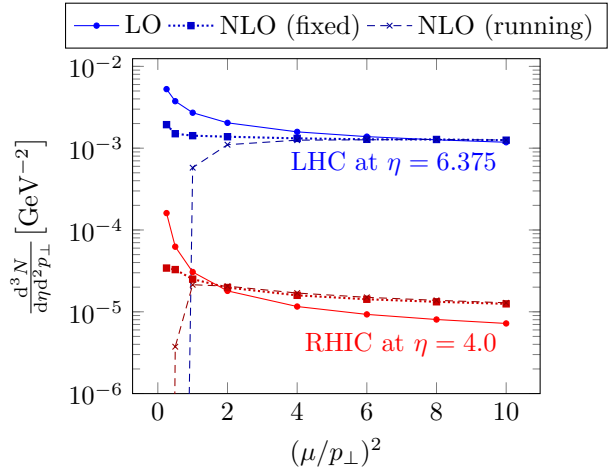
\begin{figure}
 \tikzsetnextfilename{mudependence}
 \begin{tikzpicture}
  \begin{axis}[result axis,ymode=log,xlabel=$(\mu/p_\perp)^2$,legend columns=4,every axis legend/.append style={at={(axis description cs:0.5,1.03)},anchor=south,cells={anchor=west}}]
   \addplot[blue,mark=*,mark size=1pt] table[x index=2,y expr=\thisrow{lomean}] {\lhcvsmurcBKhiY} node[pos=1,below left] {LHC at $\eta = 6.375$};
   \addplot[blue!80!black,mark=square*,mark options={solid},mark size=1pt,densely dotted,thick] table[x index=2,y expr={\thisrow{lomean}+\fixedcoupling{\thisrow{pT}^2*\thisrow{mu2factor}}*\thisrow{nlomean}}] {\lhcvsmurcBKhiY};
   \addplot[blue!60!black,mark=x,mark size=1.5pt,densely dashed] table[x index=2,y expr={\thisrow{lomean}+\runningcoupling{\thisrow{pT}^2*\thisrow{mu2factor}}*\thisrow{nlomean}}] {\lhcvsmurcBKhiY};

   \addplot[red,mark=*,mark size=1pt] table[x index=2,y expr=\thisrow{lomean}] {\rhicvsmurcBK} node[pos=1,below left] {RHIC at $\eta = 4.0$};
   \addplot[red!80!black,mark=square*,mark options={solid},mark size=1pt,densely dotted,thick] table[x index=2,y expr={\thisrow{lomean}+\fixedcoupling{\thisrow{pT}^2*\thisrow{mu2factor}}*\thisrow{nlomean}}] {\rhicvsmurcBK};
   \addplot[red!60!black,mark=x,mark size=1.5pt,densely dashed] table[x index=2,y expr={\thisrow{lomean}+\runningcoupling{\thisrow{pT}^2*\thisrow{mu2factor}}*\thisrow{nlomean}}] {\rhicvsmurcBK};
   
   \legend{LO,NLO (fixed),NLO (running)};
  \end{axis}
 \end{tikzpicture}
 \caption[Factorization scale dependence of the cross section]{Factorization scale dependence of the cross section at $\pperp = \SI{2}{GeV}$, as a representative result. The NLO results for fixed coupling ($\alphas = 0.2$) and one-loop running coupling are both presented. The dramatic drop of the NLO curve with the running coupling at low $\mu^2$ is simply due to the breakdown of perturbative calculations at large $\alphas(\mu)$. Nevertheless, these two NLO curves almost overlap with each other at large values of $\mu^2$, showing that the nature of the coupling does not significantly affect the qualitative result that the NLO corrections reduce the factorization scale dependence. Figure from reference~\cite{Stasto:2013cha}.
 }
 \label{fig:mudependence}
\end{figure}
We would expect that at each order, the theoretical uncertainty due to the factorization scale decreases. And this is indeed what we see; figure~\ref{fig:mudependence} shows the variation of the width of the cross section with $\mu$, and the curve is flatter with the inclusion of the NLO corrections.

Including the NLO corrections is also useful because it gives us an indication of the behavior of further corrections: if the NLO corrections are small relative to the leading order result, perhaps we can expect the NNLO corrections to be smaller still. Conversely, the negative results we're getting may indicate an alternating series of corrections at progressively higher orders. Of course computing the NNLO or higher corrections would be a massive undertaking, not only because of the sheer number of Feynman diagrams involved but also because several approximations made in deriving the NLO results would no longer apply. In the absence of the actual result, the NLO corrections give us an indication of circumstances under which we could expect the series to be well-behaved as more and more terms are incorporated.

\section{Uncertainties}\label{sec:uncertainty}
Computation with SOLO incorporates several sources of computational uncertainty, in addition to the theoretical uncertainty associated with the factorization scale.
These uncertainties limit our ability to relate a measured cross section to an underlying gluon distribution.

\subsection{Monte Carlo Statistical Uncertainty}
By far the \note{is it?}largest contribution to the uncertainty in the results from SOLO is the Monte Carlo statistical uncertainty, described in section~\ref{sec:integrationmethods}.
If our results are to tell us anything meaningful about the gluon distribution, we have to ensure that this uncertainty doesn't overwhelm the difference between the results for different gluon distributions.

Figure~\ref{fig:errorbands} shows the uncertainty in SOLO output for a sample result, namely the BRAHMS result at $\pseudorapidity = 2.2$ and with $\mu = 10$ (one of the same curves plotted in figure~\ref{fig:brahmsresults}). From the upper plots, it's clear that the leading order result is quite precise; the uncertainty is too small to appear on the plot, which makes sense for two reasons: first, because the leading order integrands, $\Fd02qq$ and $\Fd02gg$, are particularly simple, and also because $\Fd02qq$ only needs to be integrated over two variables ($\z$ and $\xisym$). We can use two-dimensional cubature, rather than Monte Carlo integration, which decreases the uncertainty by several orders of magnitude.

The uncertainty in the next-to-leading order result is clearly much more significant. However, it's worth noting that the upper bound of the error band is still negative (or at least less than $\SI{1e-6}{GeV^{-2}}$), for the most part, above the cutoff momentum, and the error allows a range of only $\pm\SI{.5}{GeV}$ or so for that cutoff. So the qualitative conclusion, that the LO+NLO result turns negative at high $\pperp$, is fairly robust.
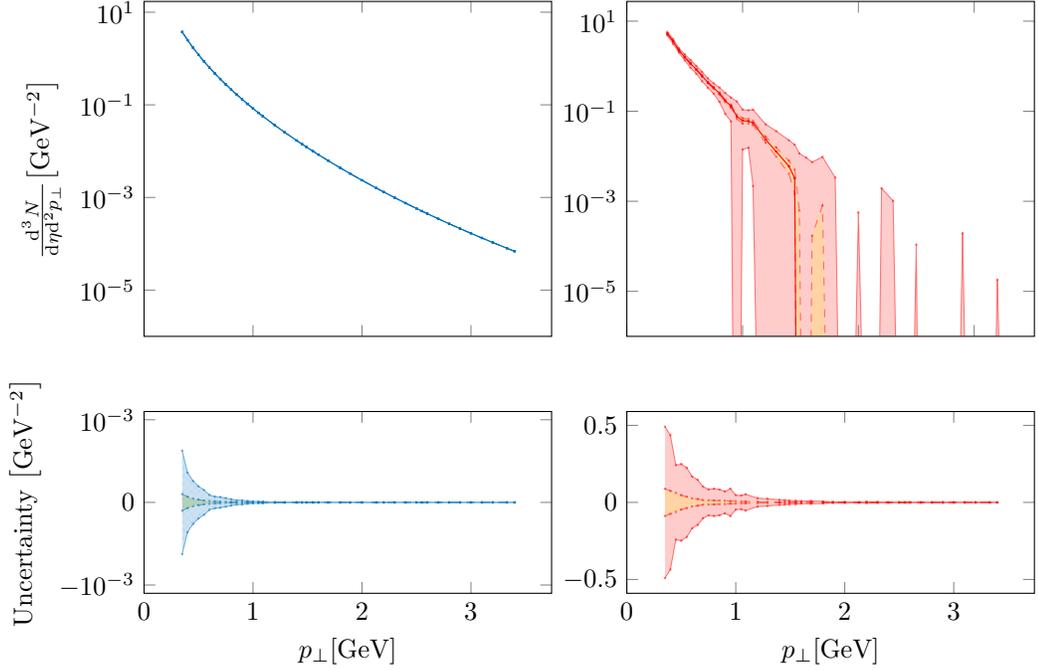
\begin{figure}
 \tikzsetnextfilename{errorbands}
 \begin{tikzpicture}
  \begin{groupplot}[scaled ticks=false,group style={group size=2 by 2,ylabels at=edge left,xlabels at=edge bottom,xticklabels at=edge bottom},xmin=0]
  \nextgroupplot[result axis,ymode=log,xlabel={}]
   \addplot[rcBK GBW LO,result plot line,xsec LO=\hadronconversionfactor] table {\brahmsrcBKYAlomu};
   \addplot[rcBK GBW LO,result plot line,opacity=0.5,name path=LO plus sigma,xsec LO plus sigma=\hadronconversionfactor] table {\brahmsrcBKYAlomu};
   \addplot[rcBK GBW LO,result plot line,opacity=0.5,name path=LO minus sigma,xsec LO minus sigma=\hadronconversionfactor] table {\brahmsrcBKYAlomu};
   \addplot[rcBK GBW LO,result plot line,dashed,opacity=0.5,name path=LO upper errbound,xsec LO upper error estimate=\hadronconversionfactor] table {\brahmsrcBKYAlomu};
   \addplot[rcBK GBW LO,result plot line,dashed,opacity=0.5,name path=LO lower errbound,xsec LO lower error estimate=\hadronconversionfactor] table {\brahmsrcBKYAlomu};
   \addplot[rcBK GBW LO,result plot fill,opacity=0.2] fill between[of=LO plus sigma and LO minus sigma];
   \addplot[rcBK GBW LO,result plot fill,fill=yellow,opacity=0.2] fill between[of=LO upper errbound and LO lower errbound];
  \nextgroupplot[result axis,ymode=log,xlabel={},ylabel={}] 
   \addplot[rcBK GBW NLO,result plot line,xsec NLO=\hadronconversionfactor] table {\brahmsrcBKYAlomu};
   \addplot[rcBK GBW NLO,result plot line,opacity=0.5,name path=NLO plus sigma,xsec NLO plus sigma=\hadronconversionfactor] table {\brahmsrcBKYAlomu};
   \addplot[rcBK GBW NLO,result plot line,opacity=0.5,name path=NLO minus sigma,xsec NLO minus sigma=\hadronconversionfactor] table {\brahmsrcBKYAlomu};
   \addplot[rcBK GBW NLO,result plot line,dashed,opacity=0.5,name path=NLO upper errbound,xsec NLO upper error estimate=\hadronconversionfactor] table {\brahmsrcBKYAlomu};
   \addplot[rcBK GBW NLO,result plot line,dashed,opacity=0.5,name path=NLO lower errbound,xsec NLO lower error estimate=\hadronconversionfactor] table {\brahmsrcBKYAlomu};
   \addplot[rcBK GBW NLO,result plot fill,opacity=0.2] fill between[of=NLO plus sigma and NLO minus sigma];
   \addplot[rcBK GBW NLO,result plot fill,fill=yellow,opacity=0.2] fill between[of=NLO upper errbound and NLO lower errbound];
  \nextgroupplot[height=4cm,xlabel={$p_\perp [\si{GeV}]$},ylabel={Uncertainty $\bigl[\si{GeV^{-2}}\bigr]$},ymin={-1.1e-3},ymax={1.1e-3},ytick={-1e-3,0,1e-3},yticklabels={$-10^{-3}$,$0$,$10^{-3}$}]
   \addplot[rcBK GBW LO,result plot line,opacity=0.5,name path=LO plus sigma] table[x=pT,y expr={\hadronconversionfactor*\thisrow{lo-stddev}}] {\brahmsrcBKYAlomu};
   \addplot[rcBK GBW LO,result plot line,opacity=0.5,name path=LO minus sigma] table[x=pT,y expr={-\hadronconversionfactor*\thisrow{lo-stddev}}] {\brahmsrcBKYAlomu};
   \addplot[rcBK GBW LO,result plot line,dashed,opacity=0.5,name path=LO upper errbound] table[x=pT,y expr={\hadronconversionfactor*\thisrow{lo-errbound}}] {\brahmsrcBKYAlomu};
   \addplot[rcBK GBW LO,result plot line,dashed,opacity=0.5,name path=LO lower errbound] table[x=pT,y expr={-\hadronconversionfactor*\thisrow{lo-errbound}}] {\brahmsrcBKYAlomu};
   \addplot[rcBK GBW LO,result plot fill,opacity=0.2] fill between[of=LO plus sigma and LO minus sigma];
   \addplot[rcBK GBW LO,result plot fill,fill=yellow,opacity=0.2] fill between[of=LO upper errbound and LO lower errbound];
  \nextgroupplot[height=4cm,xlabel={$p_\perp [\si{GeV}]$},ylabel={}]
   \addplot[rcBK GBW NLO,result plot line,opacity=0.5,name path=NLO plus sigma] table[x=pT,y expr={\hadronconversionfactor*(\thisrow{lo-stddev}+\coupling{\thisrow{mu2}}*\thisrow{nlo-stddev})}] {\brahmsrcBKYAlomu};
   \addplot[rcBK GBW NLO,result plot line,opacity=0.5,name path=NLO minus sigma] table[x=pT,y expr={-\hadronconversionfactor*(\thisrow{lo-stddev}+\coupling{\thisrow{mu2}}*\thisrow{nlo-stddev})}] {\brahmsrcBKYAlomu};
   \addplot[rcBK GBW NLO,result plot line,dashed,opacity=0.5,name path=NLO upper errbound] table[x=pT,y expr={\hadronconversionfactor*(\thisrow{lo-errbound}+\coupling{\thisrow{mu2}}*\thisrow{nlo-errbound})}] {\brahmsrcBKYAlomu};
   \addplot[rcBK GBW NLO,result plot line,dashed,opacity=0.5,name path=NLO lower errbound] table[x=pT,y expr={-\hadronconversionfactor*(\thisrow{lo-errbound}+\coupling{\thisrow{mu2}}*\thisrow{nlo-errbound})}] {\brahmsrcBKYAlomu};
   \addplot[rcBK GBW NLO,result plot fill,opacity=0.2] fill between[of=NLO plus sigma and NLO minus sigma];
   \addplot[rcBK GBW NLO,result plot fill,fill=yellow,opacity=0.2] fill between[of=NLO upper errbound and NLO lower errbound];
  \end{groupplot}
 \end{tikzpicture}
 \caption[Monte Carlo uncertainty in the cross section]{Quantifying the uncertainty in a sample output of SOLO, using BRAHMS at $\pseudorapidity=2.2$ and $\mu=10$. The shaded region shows the range $\pm1\sigma$ with $\sigma$ being the standard deviation of the mean of the results from different random seeds, and the yellow highlight in the middle shows the extent of the error estimate from the Monte Carlo algorithm. The top row shows the cross section with an error band, and the bottom row shows only the uncertainty. In the leading order plot, the uncertainty is negligible, but when the next-to-leading order correction is included it becomes potentially significant. We can see that the absolute uncertainty using either metric shrinks as we go to higher $\pperp$, but its relative effect on the cross section increases.}
 \label{fig:errorbands}
\end{figure}

\subsection{Integral Truncation Error}
Another source of error comes from the procedure SOLO uses to simulate infinite integrals like $\iint_{\mathbb{R}^2}\uddc\vec\qperp$. The quantities being integrated ($\Fs*$, $\Fn*$, $\Fd*$) fall off fairly quickly with $\qperp$ or $\rperp$ or so on, outside of a central region in which the integrand is large. This is primarily due to the gluon distribution $\dipoleF$ having a power law dependence.

There are more accurate methods for handling infinite integrals using variable transformations, but we believe that, for the results presented in this work, the error introduced by a simple cutoff is small enough to be safely neglected.\iftime{justify this with some plots}

\section{Relating the Cross Section to the Gluon Distribution}
\iftime{Consider trying to evaluate this with exponential or polynomial splines as the gluon distribution}

As the purpose of this project is to use the cross section to constrain the form of the unintegrated gluon distribution, it will be useful to see how varying the shape of the gluon distribution affects the cross section.
From physical arguments, we know $\dipoleF(\vec\kperp)$ has to peak around $\vec\kperp = \vec0$ and fall off at large $\kperp$, but there's considerable flexibility in how that can be achieved: the characteristic size of the peak, $\Qs$, and the form at which it falls off are both relatively \note{not really ``free''}free and can have a considerable effect on the cross section.

\begin{figure}
 \tikzsetnextfilename{gdistcomparison}
 \begin{tikzpicture}
  \begin{groupplot}[result axis,ymode=log,ymin=8e-8,ymax=5e1,xmin=0,xmax=3.5,group style={group size=2 by 2,horizontal sep=0pt,vertical sep=0pt,x descriptions at=edge bottom,y descriptions at=edge left}]
  \nextgroupplot
   \resultplot{GBW LO}{\thisrow{pT}}{\hadronconversionfactor*\thisrow{lomean}}{brahmsGBW}
   \resultplot{GBW NLO}{\thisrow{pT}}{\hadronconversionfactor*(\thisrow{lomean}+\coupling{\thisrow{mu2}}*\thisrow{nlomean})}{brahmsGBW}
   \addplot[data plot,mark size=0.6pt] table[x index=0,y expr={\thisrow{yield}},y error expr={(\thisrow{staterr} + \thisrow{syserr})}] {\brahmsdAuhiY};
   \node[anchor=north east] (gbwlabel) at (axis description cs:0.97,0.97) {GBW};
   
  \nextgroupplot
   \resultplot{MV LO}{\thisrow{pT}}{\hadronconversionfactor *\thisrow{lomean}}{brahmsMV}
   \resultplot{MV NLO}{\thisrow{pT}}{\hadronconversionfactor *(\thisrow{lomean}+\coupling{\thisrow{mu2}}*\thisrow{nlomean})}{brahmsMV}
   \addplot[data plot,mark size=0.6pt] table[x index=0,y expr={\thisrow{yield}},y error expr={(\thisrow{staterr} + \thisrow{syserr})}] {\brahmsdAuhiY};
   \node[anchor=north east] (mvlabel) at (axis description cs:0.97,0.97) {MV};
   
   \nextgroupplot
   \resultplot{BK LO}{\thisrow{pT}}{\hadronconversionfactor *\thisrow{lomean}}{brahmsBK}
   \resultplot{BK NLO}{\thisrow{pT}}{\hadronconversionfactor *(\thisrow{lomean}+\coupling{\thisrow{mu2}}*\thisrow{nlomean})}{brahmsBK}
   \addplot[data plot,mark size=0.6pt] table[x index=0,y expr={\thisrow{yield}},y error expr={(\thisrow{staterr} + \thisrow{syserr})}] {\brahmsdAuhiY};
   \node[anchor=north east] (bklabel) at (axis description cs:0.97,0.97) {BK};
   
   \nextgroupplot
   \resultplot{rcBK GBW LO}{\thisrow{pT}}{\hadronconversionfactor *\thisrow{lo-mean}}{brahmsrcBKYB}
   \resultplot{rcBK GBW NLO}{\thisrow{pT}}{\hadronconversionfactor *(\thisrow{lo-mean}+\coupling{\thisrow{mu2}}*\thisrow{nlo-mean})}{brahmsrcBKYB}
   \addplot[data plot,mark size=0.6pt] table[x index=0,y expr={\thisrow{yield}},y error expr={(\thisrow{staterr} + \thisrow{syserr})}] {\brahmsdAuhiY};
   \node[anchor=north east] (rcbklabel) at (axis description cs:0.97,0.97) {rcBK};
  \end{groupplot}
 \end{tikzpicture}
 \caption[Calculation results and data for \HepProcess{\Pdeuteron\Pnucleus \HepTo \Phadron\Panything} cross section at BRAHMS with varying gluon distributions]{Comparisons of BRAHMS data~\cite{Arsene:2004ux} at $\pseudorapidity=3.2$ with the theoretical results for four choices of gluon distribution: GBW~\eqref{eq:GBWdipoleS}, MV~\eqref{eq:MVdipoleS} with $\Lambda_\text{MV}=\SI{0.24}{GeV}$, BK solution with fixed coupling at $\alphas = 0.1$, and rcBK with $\Lambda_\textrm{QCD} = \SI{0.1}{GeV}$. The edges of the solid bands show results for $\mu^2=\text{\SIrange{10}{50}{GeV^2}}$. As in other figures, the crosshatch fill shows LO results and the solid fill shows LO+NLO results.}
 \label{fig:gdistcomparison}
\end{figure}

Figure~\ref{fig:gdistcomparison} shows the cross section resulting from four different gluon distributions: the GBW~\eqref{eq:GBWdipoleS} and MV~\eqref{eq:MVdipoleS} models, which have analytic expressions in coordinate space, and the BK equation with fixed and running coupling.
We can look back at figures~\ref{fig:dipoleFGBWMV} and~\ref{fig:dipoleNmomentumspace} to get a sense of how these gluon distributions differ from each other, and relate those differences to the features appearing in the plots of figure~\ref{fig:gdistcomparison}.

First of all, the distinguishing feature of the GBW gluon distribution is its exponential falloff at high $\xseccgcpperp$.
Looking at figure~\ref{fig:gdistcomparison}, we can see that the GBW plot exhibits a unique downward bend in the leading order cross section at high $\xsechadronpperp$.
This shows that the cross section at high transverse momentum preferentially receives contributions from the high-momentum region of the gluon distribution, as expected based on the formulas of chapter~\ref{ch:crosssection}.

We can see an odd feature in the MV gluon distribution results in the top right plot of figure~\ref{fig:gdistcomparison}: the difference between the leading and next to leading order contributions is larger than in any other plot.
Looking at figures~\ref{fig:dipoleFGBWMV} and~\ref{fig:dipoleNmomentumspace}, the MV distribution appears to have a slower falloff at large $\xseccgcpperp$, perhaps suggesting that the NLO corrections at medium $\xsechadronpperp\approx\SI{2}{GeV}$ are particularly sensitive to the high-$\xseccgcpperp$ region of the gluon distribution.
This may be worth investigating further.

The difference between the BK and rcBK results is an overall suppression of the cross section at high $\xsechadronpperp$ in the fixed coupling case.
This not only pushes the leading order prediction with fixed coupling below the data, but it also causes the cutoff momentum where the LO+NLO cross section turns negative to be lower than in the rcBK plot.
Since one of the main differences between the BK and rcBK gluon distributions is the reduced evolution speed of the saturation scale with running coupling, we can conclude that slower evolution of the saturation scale causes the cross section at high transverse momentum to be suppressed.
This raises the question of whether the cutoff momentum is directly related to the saturation scale, as one would naively expect, since the slower evolution of the rcBK equation reduces the saturation scale at high momenta and we might expect to see a lower cutoff in the rcBK results if the relationship is direct.

\begin{figure}
 \begin{subfigure}{\linewidth}
  \tikzsetnextfilename{xsecvariationwithc}
  \begin{tikzpicture}
   \begin{groupplot}[
    result axis,
    ymode=log,
    group style={group size=2 by 1,ylabels at={edge left}},
    every axis plot/.append style={opacity=0.7,result plot marks},
    cycle list name=rainbow
   ]
   \nextgroupplot[title={BRAHMS $\pseudorapidity=3.2$ LO},every axis plot/.append style={smooth,brahms xsec LO}]
    \addplot table {\brahmsGBWcA} node[pos=0.2,below left] {$c = \num{0.07}$};
    \addplot table {\brahmsGBWcB};
    \addplot table {\brahmsGBWcC};
    \addplot table {\brahmsGBWcD};
    \addplot table {\brahmsGBWcE} node[pos=0.7,above right] {$c = \num{0.7}$};
   \nextgroupplot[title={BRAHMS $\pseudorapidity=3.2$ LO+NLO},every axis plot/.append style={brahms xsec NLO}]
    \addplot table {\brahmsGBWcA} node[pos=0.3,below left,rotate=-90] {$c = \num{0.07}$};
    \addplot table {\brahmsGBWcB};
    \addplot table {\brahmsGBWcC};
    \addplot table {\brahmsGBWcD};
    \addplot table {\brahmsGBWcE} node[pos=0.2,above right] {$c = \num{0.7}$};
   \end{groupplot}
  \end{tikzpicture}
  \caption{Response in the leading and next-to-leading order cross sections as $c$ in the GBW model is varied between $\num{0.07}$ and $\num{0.7}$. As the gluon distribution pushes out to higher $\kperp$, the large-$\pperp$ tail of the cross section increases, but the low-$\pperp$ part of the cross section remains more or less unchanged.}
  \label{fig:xsecvariationwithc}
 \end{subfigure}
 \begin{subfigure}{\linewidth}
  \tikzsetnextfilename{gdistvariationwithc}
  \begin{tikzpicture}[
   declare function={
    Qsq(\c,\x)=\c*(197)^(1/3)*(0.000304/\x)^0.288;
    FGBW(\c,\x,\ksq)=exp(-\ksq/Qsq(\c,\x))/(pi*Qsq(\c,\x));
   }
  ]]
   \begin{groupplot}[
    group style={group size=1 by 2,xlabels at=edge bottom},
    domain=1e-2:1e2,
    samples=50,
    xmode=log,ymode=log,
    xmin=1e-2,xmax=1e2,ymin=1e-8,
    xlabel={$\kperp^2 \bigl[\si{GeV^2}\bigr]$},
    ylabel={$\dipoleF(\kperp) \bigl[\si{GeV^{-2}}\bigr]$},
    every axis plot/.append style={opacity=0.7,smooth},
    width=\linewidth-1cm,height=4cm
   ]
   \nextgroupplot[cycle list={red},every node/.append style={color=red}]
    \addplot+[thick,inline label at y={1e-5}{$0.01$}] {FGBW(0.07, 0.01, x)};
    \addplot+[inline label at y={1e-5}{$0.001$}] {FGBW(0.07, 0.001, x)};
    \addplot+[inline label at y={1e-5}{$0.0001$}] {FGBW(0.07, 0.0001, x)};
    \node[below left] at (rel axis cs:0.98,0.98) {$\centrality = 0.07$};
   \nextgroupplot[cycle list={cyan},every node/.append style={color=cyan}]
    \addplot+[thick,inline label at y={1e-5}{$0.01$}] {FGBW(0.7, 0.01, x)};
    \addplot+[inline label at y={1e-5}{$0.001$}] {FGBW(0.7, 0.001, x)};
    \addplot+[inline label at y={1e-5}{$0.0001$}] {FGBW(0.7, 0.0001, x)};
    \node[below left] at (rel axis cs:0.98,0.98) {$\centrality = 0.7$};
   \end{groupplot}
  \end{tikzpicture}
  \caption{The GBW gluon distributions corresponding to the lowest and highest cross section curves in part~\subref{fig:xsecvariationwithc}. The upper plot shows $\centrality = 0.07$, corresponding to the smallest cross section in part~\subref{fig:xsecvariationwithc}, and the lower plot shows $\centrality = 0.7$ which corresponds to the largest cross section. Within each plot, different lines correspond to different values of $\xg$, as labeled, showing the evolution of the gluon distribution toward larger momenta as $\centrality$ (and the saturation scale) increases.}
  \label{fig:gdistvariationwithc}
 \end{subfigure}
 \caption[Effect on the BRAHMS cross section of varying the overall normalization of the saturation scale]{Effect on the BRAHMS cross section at $\mu^2 = \SI{10}{GeV^2}$ of varying $c$ in the saturation scale of the GBW model. Comparing parts~\subref{fig:xsecvariationwithc} and~\subref{fig:gdistvariationwithc} reveals that an overall increase in the saturation scale leads to an increase in the cross section at large $\xsechadronpperp$, and a corresponding shift of the cutoff momentum toward larger $\xsechadronpperp$, but little change at the smallest $\xsechadronpperp$.}
 \label{fig:variationwithc}
\end{figure}

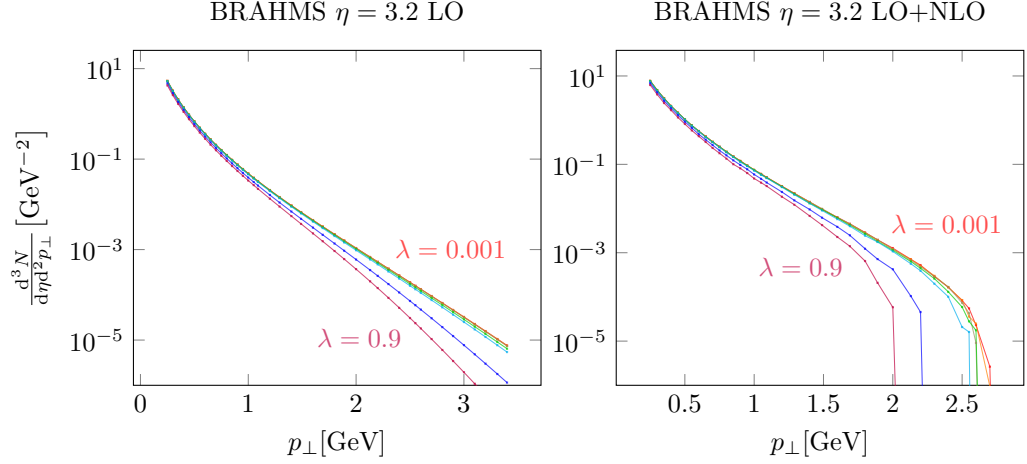
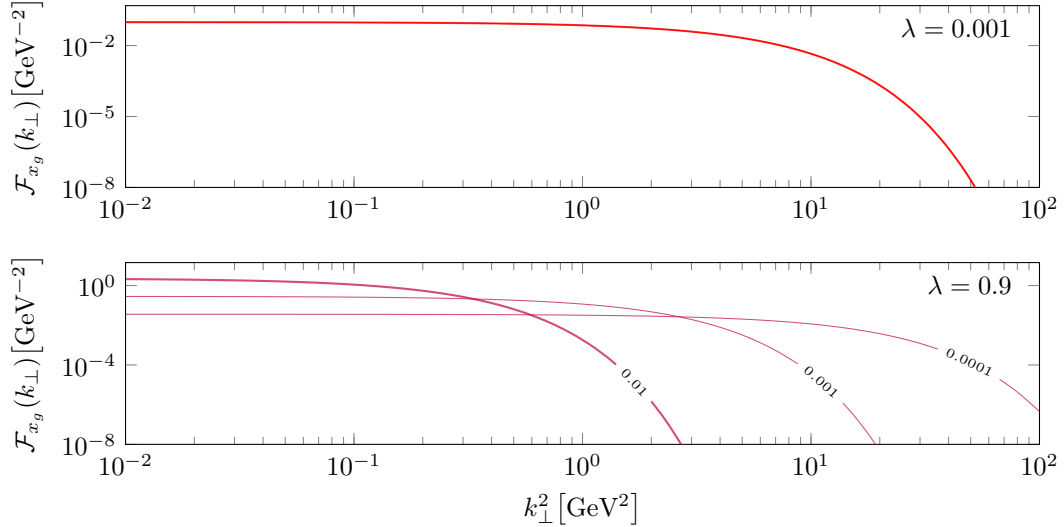
\begin{figure}
 \begin{subfigure}{\linewidth}
  \tikzsetnextfilename{xsecvariationwithlambda}
  \begin{tikzpicture}
   \begin{groupplot}[
    result axis,
    ymode=log,
    group style={group size=2 by 1,ylabels at={edge left}},
    every axis plot/.append style={opacity=0.7,result plot marks},
    cycle list name=rainbow
   ]
   \nextgroupplot[title={BRAHMS $\pseudorapidity=3.2$ LO},every axis plot/.append style={smooth,brahms xsec LO}]
    \addplot table {\brahmsGBWlA} node[pos=0.7,above right] {$\lambda = \num{0.001}$};
    \addplot table {\brahmsGBWlB};
    \addplot table {\brahmsGBWlC};
    \addplot table {\brahmsGBWlD};
    \addplot table {\brahmsGBWlE};
    \addplot table {\brahmsGBWlF};
    \addplot table {\brahmsGBWlG} node[pos=0.7,below left] {$\lambda = 0.9$};
   \nextgroupplot[title={BRAHMS $\pseudorapidity=3.2$ LO+NLO},every axis plot/.append style={brahms xsec NLO}]
    \addplot table {\brahmsGBWlA} node[pos=0.2,above right] {$\lambda = \num{0.001}$};
    \addplot table {\brahmsGBWlB};
    \addplot table {\brahmsGBWlC};
    \addplot table {\brahmsGBWlD};
    \addplot table {\brahmsGBWlE};
    \addplot table {\brahmsGBWlF};
    \addplot table {\brahmsGBWlG} node[pos=0.2,below left] {$\lambda = 0.9$};
   \end{groupplot}
  \end{tikzpicture}
  \caption{Response in the leading and next-to-leading order cross sections as $\lambda$ in the GBW model is varied between $\num{0.001}$ and $\num{0.9}$. As the gluon distribution pushes out to higher $\kperp$, the large-$\pperp$ tail of the cross section increases, but the low-$\pperp$ part of the cross section remains more or less unchanged.}
  \label{fig:xsecvariationwithlambda}
 \end{subfigure}
 \begin{subfigure}{\linewidth}
  \tikzsetnextfilename{gdistvariationwithlambda}
  \begin{tikzpicture}[
   declare function={
    Qsq(\l,\x)=0.56*(197)^(1/3)*(0.000304/\x)^\l;
    FGBW(\l,\x,\ksq)=exp(-\ksq/Qsq(\l,\x))/(pi*Qsq(\l,\x));
   }
  ]
   \begin{groupplot}[
    group style={group size=1 by 2,xlabels at={edge bottom}},
    domain=1e-2:1e2,
    samples=50,
    xmode=log,ymode=log,
    xmin=1e-2,xmax=1e2,ymin=1e-8,
    xlabel={$\kperp^2 \bigl[\si{GeV^2}\bigr]$},ylabel={$\dipoleF(\kperp) \bigl[\si{GeV^{-2}}\bigr]$},
    width=\linewidth-1cm,
    height=4cm,
    every axis plot/.append style={opacity=0.7,smooth,no marks}
   ]
   \nextgroupplot[cycle list={red}]
    \addplot+[thick] {FGBW(0.001, 0.01, x)};
    \addplot+[] {FGBW(0.001, 0.001, x)};
    \addplot+[] {FGBW(0.001, 0.0001, x)};
    \node[below left] at (rel axis cs:0.98,0.98) {$\lambda= 0.001$};
   \nextgroupplot[cycle list={purple}]
    \addplot+[thick,inline label at y={1e-5}{$0.01$}] {FGBW(0.900, 0.01, x)};
    \addplot+[inline label at y={1e-5}{$0.001$}] {FGBW(0.900, 0.001, x)};
    \addplot+[inline label at y={1e-4}{$0.0001$}] {FGBW(0.900, 0.0001, x)};
    \node[below left] at (rel axis cs:0.98,0.98) {$\lambda= 0.9$};
   \end{groupplot}
  \end{tikzpicture}
  \caption{The GBW gluon distributions corresponding to the lowest and highest cross section curves in part~\subref{fig:xsecvariationwithlambda}. The upper plot shows $\lambda = 0.001$, corresponding to the smallest cross section in part~\subref{fig:xsecvariationwithlambda}, and the lower plot shows $\lambda = 0.7$ which corresponds to the largest cross section. Within each plot, different lines correspond to different values of $\xg$, as labeled. The speed of the evolution is set by $\lambda$; in the upper plot, the evolution is slow enough that the curves at the different values of $\xg$ shown --- $0.01$, $0.001$, and $0.0001$ --- are visually indistinguishable.}
  \label{fig:gdistvariationwithlambda}
 \end{subfigure}
 \caption[Effect on the BRAHMS cross section of varying the rate of evolution of the saturation scale]{Effect on the cross section of varying $\lambda$, the rate of increase of the saturation scale with $\ln\frac{1}{\xg}$, in the GBW model. Increasing the rate of evolution suppresses the cross section slightly at high $\xsechadronpperp$, but in general the cross section seems not to be particularly sensitive to $\lambda$ until it becomes rather large (close to $1$).}
 \label{fig:variationwithlambda}
\end{figure}

We can explore the connection between the gluon distribution and the cross section further in a more controlled environment by varying the parameters of the simple GBW gluon distribution~\eqref{eq:GBWsatscale} and looking at how it affects the results.
Figures~\ref{fig:variationwithc} and~\ref{fig:variationwithlambda} show the effect of changing $\centrality$ and $\lambda$ in the saturation scale for the GBW distribution, respectively.
Essentially, $\centrality$ controls the overall normalization of the saturation scale, while $\lambda$ controls the speed at which it changes with decreases in $\xg$.

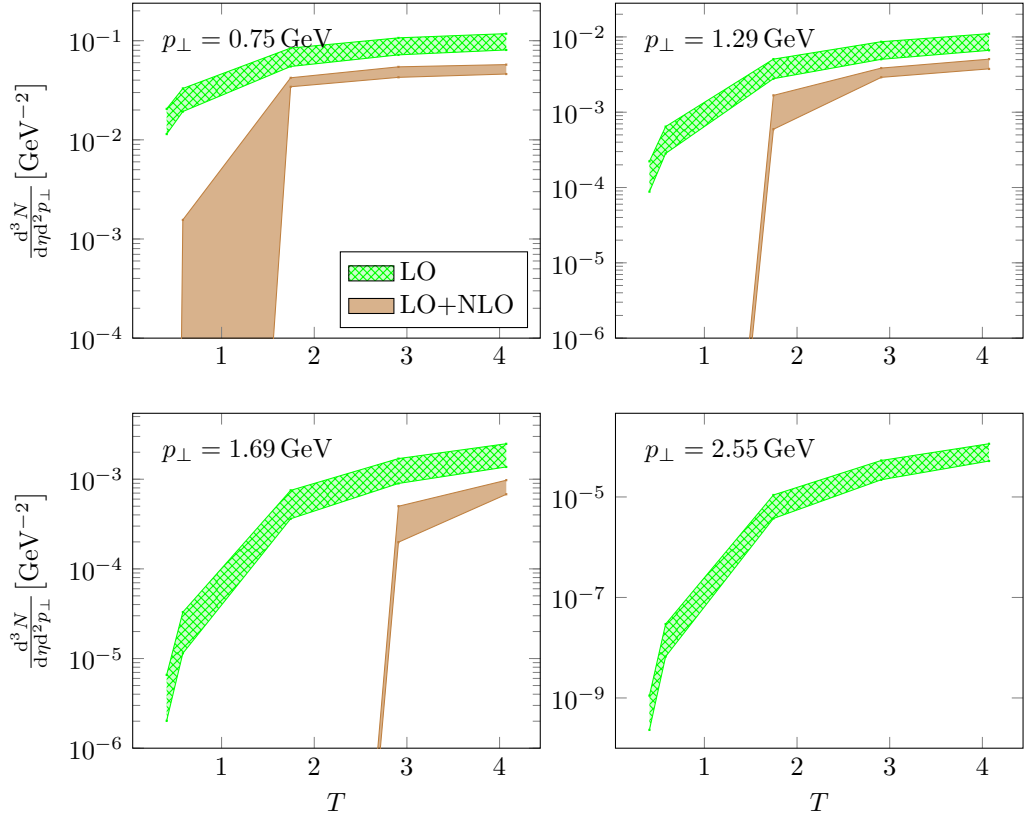
\begin{figure}
\tikzsetnextfilename{gbwthickness}
 \begin{tikzpicture}
  \begin{groupplot}[
   group style={group size=2 by 2,xlabels at=edge bottom,ylabels at=edge left},
   ymode=log,
   result axis,
   xlabel={$T$}
  ]
  \nextgroupplot[ymin=1e-4,legend pos=south east]
   \resultplot{GBW LO}{{\thisrow{c}*pow(197,1./3.)}}{\thisrow{lo-mean}}{brahmsGBWpT075}
   \resultplot{GBW NLO}{{\thisrow{c}*pow(197,1./3.)}}{\thisrow{nlo-mean}}{brahmsGBWpT075}
   \node[below right] at (axis description cs:0.05,0.95) {$\xsechadronpperp = \SI{0.75}{GeV}$};
   \legend{LO,LO+NLO}
  \nextgroupplot
   \resultplot{GBW LO}{{\thisrow{c}*pow(197,1./3.)}}{\thisrow{lo-mean}}{brahmsGBWpT129}
   \resultplot{GBW NLO}{{\thisrow{c}*pow(197,1./3.)}}{\thisrow{nlo-mean}}{brahmsGBWpT129}
   \node[below right] at (axis description cs:0.05,0.95) {$\xsechadronpperp = \SI{1.29}{GeV}$};
  \nextgroupplot
   \resultplot{GBW LO}{{\thisrow{c}*pow(197,1./3.)}}{\thisrow{lo-mean}}{brahmsGBWpT169}
   \resultplot{GBW NLO}{{\thisrow{c}*pow(197,1./3.)}}{\thisrow{nlo-mean}}{brahmsGBWpT169}
   \node[below right] at (axis description cs:0.05,0.95) {$\xsechadronpperp = \SI{1.69}{GeV}$};
  \nextgroupplot[ymin=1e-10]
   \resultplot{GBW LO}{{\thisrow{c}*pow(197,1./3.)}}{\thisrow{lo-mean}}{brahmsGBWpT255}
   \resultplot{GBW NLO}{{\thisrow{c}*pow(197,1./3.)}}{\thisrow{nlo-mean}}{brahmsGBWpT255}
   \node[below right] at (axis description cs:0.05,0.95) {$\xsechadronpperp = \SI{2.55}{GeV}$};
  \end{groupplot}
 \end{tikzpicture}
 \caption[Dependence of the cross section on thickness]{The BRAHMS cross section computed using the GBW gluon distribution as a function of the thickness parameter $T = \centrality \massnumber^{1/3}$. As with other BRAHMS results, the band shows the range from $\mu^2 = \SI{10}{GeV^2}$ to $\mu^2 = \SI{50}{GeV^2}$. The former curve is computed from the same data as in figure~\ref{fig:xsecvariationwithc}. In the plot for $\xsechadronpperp = \SI{2.55}{GeV}$, the LO+NLO result is entirely negative so is not shown.}
 \label{fig:gbwthickness}
\end{figure}

From figure~\ref{fig:variationwithc}, it seems that an increase in the overall normalization of the saturation scale leads to an increase in the cross section and an increase in the cutoff momentum.
To see this more directly, we can plot the results as a function of centrality.
Figure~\ref{fig:gbwthickness} shows this visualization of the same data from figure~\ref{fig:xsecvariationwithc} along with its counterpart at $\mu^2 = \SI{50}{GeV^2}$ as a function of the thickness parameter $T = \centrality \massnumber^{1/3}$, which for large $\massnumber$ is roughly proportional to the number of nucleons impacted by the projectile, also known as the number of binary collisions $N_\text{coll}$.

Moving on to figure~\ref{fig:variationwithlambda}, these results appear to support the observation made earlier that an increase in the speed of evolution leads to a decrease in the cross section at high $\xsechadronpperp$ at both leading and next-to-leading order, as well as a decrease in the cutoff momentum at which the LO+NLO result goes negative.
However, plot~\ref{fig:gdistvariationwithlambda} reveals that the increase in $\lambda$ is also accompanied by a large decrease in the \emph{value} of the saturation scale.
In figure~\ref{fig:variationwithc}, we are able to isolate the effect of the overall normalization of $\Qs$ and see that increasing $\Qs$ does indeed increase the cross section at high $\xsechadronpperp$.
Both figures also show that the cross section at low $\xsechadronpperp$ is not strongly dependent on the saturation scale.

Importantly, the negativity in the LO+NLO result persists regardless of the shape of the gluon distribution, at least within the examples shown.
While this is not an exhaustive search of the parameter space of the gluon distribution, it does suggest that the negativity is an essential feature of the calculation, and not an artifact of our choice of gluon distribution.
Since the publication of the results~\cite{Stasto:2013cha}, the question has been raised whether a suitable gluon distribution model could be chosen to eliminate the negativity, but in light of these results that seems unlikely.

\section{Conclusions}

Let's revisit the five questions posed at the beginning of the chapter:
\begin{itemize}
 \item \textit{Does the negativity imply anything about the validity of the small-$x$ formalism?} Not in general, since the formalism appears to work for small $\xsechadronpperp$.
 \item \textit{What is the significance of the cutoff momentum at which the cross section turns negative?} This is not currently clear. One might expect it to be related to the saturation momentum, but if so, the relationship is not an obvious one.
 \item \textit{What benefit, if any, do we get by incorporating the NLO correction?} As expected, the theoretical uncertainty arising from the factorization scale is reduced with the NLO corrections, allowing for more accurate tests of gluon distributions in suitable kinematic regimes.
 \item \textit{Does the LO+NLO formula give an accurate result for any kinematic conditions?} Yes; we've seen that there is a range of moderate $\xsechadronpperp$ where the results do match the data.
 \item \textit{What can we meaningfully learn about the gluon distribution from the predictions of SOLO?} This is a tricky question to answer: more than nothing, but not as much as we might have hoped. It's the high-momentum region of the cross section that is most sensitive to the structure of the gluon distribution, but this is the same region where the LO+NLO formula ceases to be accurate. As long as we are working at this order in $\alphas$, we can only extract useful data from the moderate-momentum region, which can tell us something about the normalization of the saturation scale, but less so about its rate of evolution.
\end{itemize}
Clearly, a better calculation of the cross section at high $\xsechadronpperp$ will be very interesting because it might allow us to probe the high-momentum structure of the gluon distribution in more detail.
In the next chapter, we'll see some first steps toward that goal.

%% file: beyondnlo.tex
\chapter{Extensions Beyond NLO}\label{ch:beyondnlo}

With the results of the last chapter in hand, it's clear that there are kinematic regions where the next-to-leading order corrections to the cross section are not physically reasonable.
An obvious followup question is whether we can correct this by pulling in contributions from higher orders.

\section{Resummation}


In order to resum higher order contributions, let's take a look at the NLO contributions in figures~\ref{fig:channelbreakdown}.
The largest contributions come from the gg channel, specifically from those diagrams that involve gluon loops.
Presumably, then, the largest corrections at higher orders come from diagrams with additional gluon loops.\iftime{show a figure with some such diagrams}

Resumming the dominant contributions from these diagrams can be accomplished as follows.
We start with the contribution from the quark-quark channel, given by the sum of equations~\eqref{eq:xsec02qq}, \eqref{eq:xsec12qq}, and~\eqref{eq:xsec14qq}.
In the limit of high $\xsechadronpperp$, this can be approximated as
\begin{multline}\label{eq:highpTqqchannel}
 \biggl(\xsec*\biggr)_{qq} \overset{\xsechadronpperp\to\infty}{=} \frac{\alphas\Nc\Sperp}{4\pi^2} \sum_{\text{flavors}} \int_\tau^1\frac{\udc\z}{\z^2}D_{h/q}(\z)\frac{1}{\xsecquarkpperp^4}\int_{\tau/\z}^1\udc\xisym \xp q(\xp) \\
 \times \frac{(1 + \xisym^2)^2}{(1 - \xisym)_+}\int\uddc\vec\xseccgcpperp \dipoleF(\vec\xseccgcpperp)\xseccgcpperp^2
\end{multline}
(the factorization scale dependence is left implicit).
We've seen that the negativity comes from the subtraction in the plus prescription~\eqref{eq:plusprescription}, specifically the logarithmic term in
\begin{equation}
 \int_{\tau/\z}^1\udc\xisym \xp q(\xp)\frac{(1 + \xisym^2)^2}{(1 - \xisym)_+} 
 =
 \int_{\tau/\z}^1\udc\xisym \frac{(1 + \xisym^2)^2\xp q(\xp) - 4(\tau/\z) q(\tau/\z)}{1 - \xisym} + 4\frac{\tau}{\z} q\biggl(\frac{\tau}{\z}\biggr)\ln\biggl(1 - \frac{\tau}{\z}\biggr)
\end{equation}
At large $\pperp$, $\tau$ becomes large, which squeezes the region of integration so that $\xisym$ does not vary very far from $1$.
Under these conditions, since $\xp = \tau/\xisym\z$~\eqref{eq:xpnlo}, we can approximate $\xp \approx \frac{\tau}{\z}$ in the parton distribution, which lets us factor out that distribution from the preceding integral,
\begin{multline}
 \frac{\tau}{\z} q\biggl(\frac{\tau}{\z}\biggr)\int_{\tau/\z}^1\udc\xisym \frac{(1 + \xisym^2)^2 - 4}{1 - \xisym} + 4\frac{\tau}{\z} q\biggl(\frac{\tau}{\z}\biggr)\ln\biggl(1 - \frac{\tau}{\z}\biggr) \\
 =
 \frac{\tau}{\z} q\biggl(\frac{\tau}{\z}\biggr)\biggl[\underbrace{\int_{\tau/\z}^1\udc\xisym \frac{(1 + \xisym^2)^2 - 4}{1 - \xisym} + 4\ln\biggl(1 - \frac{\tau}{\z}\biggr)}_{\mathcal{I}_q(\tau/\z)}\biggr]
\end{multline}
After doing the integral, $\mathcal{I}_q(\tau/\z)$ works out to
\begin{equation}
 \mathcal{I}_q(x) = -\frac{61}{12} + 3x + \frac{3}{2}x^2 + \frac{1}{3}x^3 + \frac{1}{4}x^4 + 4\ln(1 - x)
\end{equation}
so the high-$\xsechadronpperp$ limit of the quark-quark channel~\eqref{eq:highpTqqchannel} becomes
\begin{equation}\label{eq:firstgluonloop}
 \biggl(\xsec*\biggr)_{qq} \overset{\xsechadronpperp\to\infty}{\approx} \frac{\alphas\Nc\Sperp}{4\pi^2} \sum_{\text{flavors}} \int_\tau^1\frac{\udc\z}{\z^2}D_{h/q}(\z)\frac{1}{\xsecquarkpperp^4} \frac{\tau}{\z} q\biggl(\frac{\tau}{\z}\biggr)\mathcal{I}_q\biggl(\frac{\tau}{\z}\biggr) \int\uddc\vec\xseccgcpperp \dipoleF(\vec\xseccgcpperp)\xseccgcpperp^2
\end{equation}

To perform a simple exponential resummation, we can consider this as the first term in the Taylor series for $e^x - 1$.
So we will try adding to equation~\eqref{eq:firstgluonloop} a contribution corresponding to the remainder of the series, namely
\begin{multline}
 \frac{\Sperp}{\pi} \sum_{\text{flavors}} \int_\tau^1\frac{\udc\z}{\z^2}D_{h/q}(\z)\frac{1}{\xsecquarkpperp^4} \frac{\tau}{\z} q\biggl(\frac{\tau}{\z}\biggr) \int\uddc\vec\xseccgcpperp \dipoleF(\vec\xseccgcpperp)\xseccgcpperp^2 \\
 \times \frac{1}{c}\Biggl(\exp\biggl[c\frac{\alphas\Nc}{4\pi}\mathcal{I}_q\biggl(\frac{\tau}{\z}\biggr)\biggr] - 1 - c\frac{\alphas\Nc}{4\pi}\mathcal{I}_q\biggl(\frac{\tau}{\z}\biggr)\Biggr)
\end{multline}
The constant $c$, which can be adjusted to fix the normalization, is a practical necessity.
This resummation is intended to apply at high $\pperp$, and adjusting $c$ allows us to match the high-$\pperp$ resummed expression to the low-$\pperp$ LO+NLO expression.
However this constant does not have a rigorous theoretical justification.

Going through the same procedure for the gluon-gluon channel, we express the sum of equations~\eqref{eq:xsec02gg}, \eqref{eq:xsec12gg}, \eqref{eq:xsec12qqbar}, and~\eqref{eq:xsec16gg} in the high-$\xsechadronpperp$ limit as
\begin{equation}
 \frac{\alphas\Nc\Sperp}{4\pi^2} \int_\tau^1\frac{\udc\z}{\z^2}D_{h/g}(\z)\frac{1}{\xsecquarkpperp^4} \frac{\tau}{\z} g\biggl(\frac{\tau}{\z}\biggr) 2\mathcal{I}_g\biggl(\frac{\tau}{\z}\biggr) \int\uddc\vec\xseccgcpperp \dipoleF(\vec\xseccgcpperp)\xseccgcpperp^2
\end{equation}
where
\begin{align}
 \mathcal{I}_g(x)
 &= 2\int_x^1 \udc\xisym[1 + \xisym^2 + (1 - \xisym)^2]\biggl[\frac{\xisym}{(1 - \xisym)_+} + \frac{1 - \xisym}{\xisym} + \xisym(1 - \xisym)] \\
 &= -\frac{152}{15} + 12x - 6x^2 + \frac{16}{3}x^3 - 2x^4 + \frac{4}{5}x^5 + 4\ln(1 - x) - 4\ln x
\end{align}
Then we add the following terms as the correction:
\begin{multline}
 \frac{\Sperp}{\pi} \int_\tau^1\frac{\udc\z}{\z^2}D_{h/g}(\z)\frac{1}{\xsecquarkpperp^4} \frac{\tau}{\z} g\biggl(\frac{\tau}{\z}\biggr) \int\uddc\vec\xseccgcpperp \dipoleF(\vec\xseccgcpperp)\xseccgcpperp^2 \\
 \times \frac{1}{c}\Biggl(\exp\biggl[c\frac{\alphas\Nc}{4\pi}2\mathcal{I}_g\biggl(\frac{\tau}{\z}\biggr)\biggr] - 1 - c\frac{\alphas\Nc}{4\pi}2\mathcal{I}_g\biggl(\frac{\tau}{\z}\biggr)\Biggr)
\end{multline}

\begin{figure}
 \tikzsetnextfilename{resummation}
 \begin{tikzpicture}
  \begin{groupplot}[group style={group size=2 by 1,ylabels at=edge left},result axis,ymin=1e-8,ymode=log]
  \nextgroupplot[title={BRAHMS $\eta=2.2$}]
   \resultplot{rcBK GBW LO}{\thisrow{pT}}{\hadronconversionfactor*\thisrow{lo-mean}}{brahmsrcBKYA}
   \resultplot{rcBK GBW NLO}{\thisrow{pT}}{\hadronconversionfactor*(\thisrow{lo-mean}+\coupling{\thisrow{mu2}}*\thisrow{nlo-mean})}{brahmsrcBKYA}
   \resultplot{rcBK GBW resummed 1}{\thisrow{pT}}{\hadronconversionfactor*(\thisrow{lomean}+\coupling{\thisrow{mu2}}*(\thisrow{nlomean}+\thisrow{rqqmean}+\thisrow{rggmean}))/3}{brahmsrcBKRYA}
   \resultplot{rcBK GBW resummed 2}{\thisrow{pT}}{\hadronconversionfactor*(\thisrow{lomean}+\coupling{\thisrow{mu2}}*(\thisrow{nlomean}+\thisrow{rqqmean}+\thisrow{rggmean}))/6}{brahmsrcBKRYC}
   \resultplot{rcBK GBW resummed 3}{\thisrow{pT}}{\hadronconversionfactor*(\thisrow{lomean}+\coupling{\thisrow{mu2}}*(\thisrow{nlomean}+\thisrow{rqqmean}+\thisrow{rggmean}))/10}{brahmsrcBKRYE}
   \addplot[data plot] table[x=pt,y expr={\thisrow{yield}},y error expr={(\thisrow{staterr} + \thisrow{syserr})}] {\brahmsdAuloY};
  \nextgroupplot[title={BRAHMS $\eta=3.2$}]
   \resultplot{rcBK GBW LO}{\thisrow{pT}}{\hadronconversionfactor*\thisrow{lo-mean}}{brahmsrcBKYB}
   \resultplot{rcBK GBW NLO}{\thisrow{pT}}{\hadronconversionfactor*(\thisrow{lo-mean}+\coupling{\thisrow{mu2}}*\thisrow{nlo-mean})}{brahmsrcBKYB}
   \resultplot{rcBK GBW resummed 1}{\thisrow{pT}}{\hadronconversionfactor*(\thisrow{lomean}+\coupling{\thisrow{mu2}}*(\thisrow{nlomean}+\thisrow{rqqmean}+\thisrow{rggmean}))/3}{brahmsrcBKRYB}
   \resultplot{rcBK GBW resummed 2}{\thisrow{pT}}{\hadronconversionfactor*(\thisrow{lomean}+\coupling{\thisrow{mu2}}*(\thisrow{nlomean}+\thisrow{rqqmean}+\thisrow{rggmean}))/6}{brahmsrcBKRYD}
   \resultplot{rcBK GBW resummed 3}{\thisrow{pT}}{\hadronconversionfactor*(\thisrow{lomean}+\coupling{\thisrow{mu2}}*(\thisrow{nlomean}+\thisrow{rqqmean}+\thisrow{rggmean}))/10}{brahmsrcBKRYF}
   \addplot[data plot] table[x=pt,y expr={\thisrow{yield}},y error expr={(\thisrow{staterr} + \thisrow{syserr})}] {\brahmsdAuhiY};
   \legend{LO,NLO,$c=3$,$c=6$,$c=10$,data}
  \end{groupplot}
 \end{tikzpicture}
 \caption[Results of the calculation at BRAHMS including the resummation]{Results of the computation when the resummation correction is added to the LO+NLO results from figure~\ref{fig:brahmsresults}. These are for \dAu{} collisions where the edges of the error band are computed from $\mu^2 = \SI{10}{GeV^2}$ and $\mu^2 = \SI{50}{GeV^2}$. The erratic behavior of the plots is primarily due to large statistical variation from the Monte Carlo integration. It is clearer in the left plot that larger values of $c$ produce larger values of the resummation correction, thus bringing the cross section positive at large $\pperp$ as $c$ increases, but the sharp increase at low $\pperp$ is also evident. The same trends are present in the plot on the right, but the curves have a significant overlap so the effect of the resummation correction is harder to see.}
 \label{fig:resummation}
\end{figure}
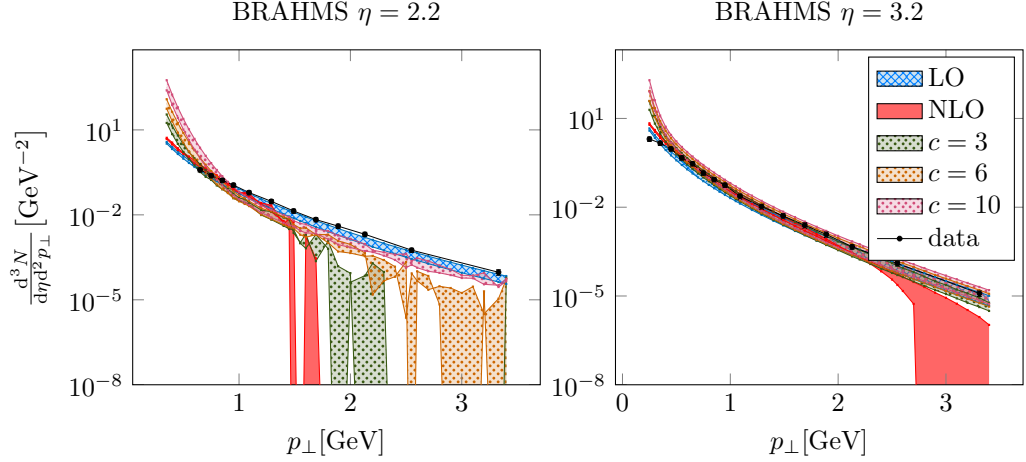

Numerical results for the complete LO+NLO+resummed cross section, for several values of $c$, are shown in figure~\ref{fig:resummation}.
As expected, the resummed expression compensates for the negativity at high $\pperp$ but overpredicts the experimental results at low $\pperp$.
This makes sense because other diagrams not included in the resummation will have a larger contribution at low momenta.

However, we find that the cross section remains negative at high $\pperp$ up to rather large values of $c\approx 10$.
Given that a true exponential resummation would have $c = 1$, this seems suspiciously large, so the results cast some doubt on the validity of the technique as a cure for the negativity.
A more detailed resummation procedure which incorporates additional higher order corrections may work well, but the simple exponential resummation appears insufficient.

\section{Matching to Collinear Factorization}

One reason the negative results of chaper~\ref{ch:results} are particularly surprising is that the cross section can also be calculated using \term{collinear factorization}, yielding an expression that is manifestly positive at high $\pperp$.
Collinear factorization describes the structures of both the projectile and target using integrated parton distributions, as opposed to the hybrid factorization used in chapter~\ref{ch:crosssection} which represents the target using the unintegrated gluon distribution of the CGC.
This suggests that we may want to investigate whether there is some fundamental difference betwen these two factorization schemes which causes their results to diverge at high transverse momentum.
Reference~\cite{Stasto:2014sea} addressed this question, and the results published there are shown below.


Let's begin by demonstrating how to calculate the cross section using collinear factorization.
For the quark-quark channel, we start with the parton-level cross sectio for the diagram shown in figure~\ref{fig:qqnlodiagram}.
\iftime{see if a separate diagram is needed}
The squared amplitude comes from applying the standard rules of Feynman perturbation theory
\begin{equation}
 \abs{\scamp}^2 = \frac{\gs^4}{\Nc}\biggl[-\CF\frac{\partonmandelstamu^2 + \partonmandelstams^2}{\partonmandelstams\partonmandelstamu} + \Nc\frac{\partonmandelstamu^2 + \partonmandelstams^2}{\partonmandelstamt^2}\biggr]
\end{equation}
where $\partonmandelstams$, $\partonmandelstamt$, and $\partonmandelstamu$ are the Mandelstam variables for the partons.
We then convolve this with the parton distributions for the projectile and target, and the fragmentation function, obtaining
\begin{equation}\label{eq:collinearxsec}
 \xsec* = \sum_i \int \frac{\udc\z}{\z^2}D_{h/i}(\z)\int \udc\xp \udc\xg \intpdf(\xp) \intg(\xg)\frac{\abs{\scamp}^2}{2\partonmandelstams}\frac{1}{2(2\pi)^3}2\pi \delta\bigl(\partonmandelstams + \partonmandelstamt + \partonmandelstamu\bigr)
\end{equation}

We can show that this is equivalent to the high-$\pperp$ limit of the hybrid factorization result as follows.
Using the definitions of the parton-level Mandelstam variables,\iftime{show where these equations come from}
\begin{align}
 \partonmandelstams
 &= (\xsecquarkp^\mu + \xsecgluonp^\mu)^2
 = \frac{\xsecquarkpperp^2}{\xisym(1 - \xisym)} &
 \partonmandelstamt
 &= (\xsecquarkp^\mu - \xseciquarkp^\mu)^2
 = -\frac{\xsecquarkpperp^2}{\xisym} &
 \partonmandelstamu
 &= (\xsecgluonp^\mu - \xseciquarkp^\mu)^2
 = -\frac{\xsecquarkpperp^2}{1 - \xisym}
\end{align}
where $\xseciquarkp^\mu$ is the momentum of the initial state quark, we can substitute into equation~\eqref{eq:collinearxsec} to obtain
\begin{equation}
 \xsec* = \frac{\alphas^2}{\Nc}\sum_i \int \frac{\udc\z}{\z^2}D_{h/i}(\z)\int \frac{\udc\xisym}{\xisym} \xp\intpdf(\xp) \intg(\xg) \frac{1 + \xisym^2}{1 - \xisym}\frac{\xisym}{\xsecquarkpperp^4}[\CF(1 - \xisym)^2 + \Nc\xisym]
\end{equation}
With the identification between integrated and unintegrated gluon distributions~\eqref{eq:dipoleFintGrelation}, this coincides with the high-$\pperp$ limit from the hybrid factorization, which comes from expanding the LO+NLO cross section as a series in $\Qs/\xsecquarkpperp$.
That expression is~\cite{Stasto:2014sea}
\begin{multline}
 \biggl(\xsec*\biggr)_{qq} = \frac{\alphas}{2\pi^2}\int_\tau^1 \frac{\udc\z}{\z^2}D_{h/\Pquark}(\z)\int_{\tau/\z}^1\udc\xisym\frac{1 + \xisym^2}{1 - \xisym}\xp q(\xp) \\
 \times \biggl(\CF\frac{(1 - \xisym)^2}{\xsecquarkpperp^4} + \Nc\frac{\xisym}{\xsecquarkpperp^4}\biggr)\int_{\mathcal{R}}\uddc\vec\xseccgcpperp \xseccgcpperp^2\dipoleF(\xseccgcpperp)
\end{multline}
where $\mathcal{R}$ is the integration region in transverse momentum space.

Because we are working at $\xsecquarkpperp \gg \Qs$, the assumption made in the calculations of chapters~\ref{ch:crosssection} and~\ref{ch:results} that $\vec\xsecquarkpperp \approx \vec\xseccgcpperp$ is no longer valid.
That means the integration region $\mathcal{R}$ is determined by the exact kinematic constraint of equation~\eqref{eq:xilimit}, or equivalently
\begin{equation}
 (\vec\xseccgcpperp - \vec\xsecquarkpperp)^2 \leq \xsecquarkpperp(\sqs e^{\rapidity} - \xsecquarkpperp)\frac{1 - \xisym}{\xisym}
\end{equation}
which selects a circle in $\vec\xseccgcpperp$ space around $\vec\xsecquarkpperp$.
Except for a single point at $\vec\xseccgcpperp = \vec\xsecquarkpperp$, this means $\xisym < 1$, which eliminates the subtracted term from the plus prescription.
Accordingly, the numerical results of this procedure should not suffer from the negativity problem we saw in the last chapter.

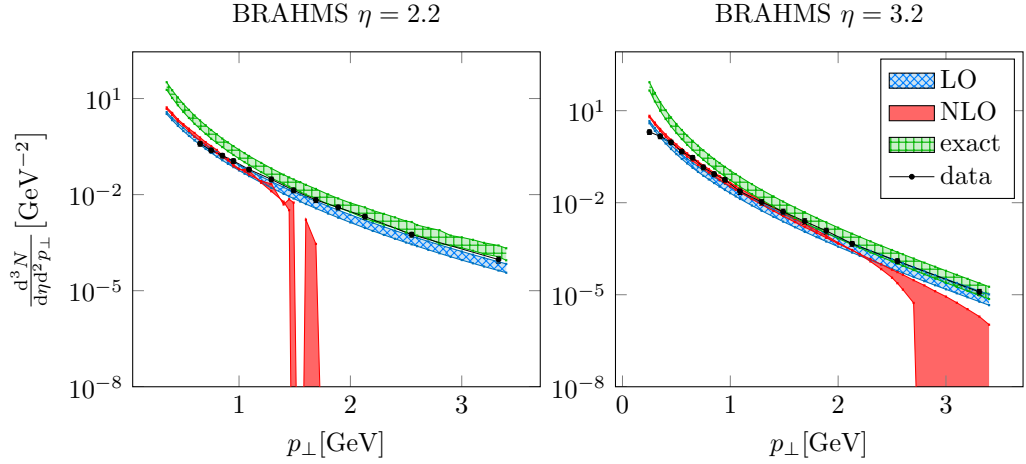
\begin{figure}
 \tikzsetnextfilename{brahmsexactresults}
 \begin{tikzpicture}
  \begin{groupplot}[result axis,ymin=1e-8,ymode=log,group style={group size=2 by 1,ylabels at=edge left}]
  \nextgroupplot[title={BRAHMS $\eta=2.2$}]
   \resultplot{rcBK GBW LO}{\thisrow{pT}}{\hadronconversionfactor*\thisrow{lo-mean}}{brahmsrcBKYA}
   \resultplot{rcBK GBW NLO}{\thisrow{pT}}{\hadronconversionfactor*(\thisrow{lo-mean}+\coupling{\thisrow{mu2}}*\thisrow{nlo-mean})}{brahmsrcBKYA}
   \resultplot{rcBK GBW exact}{\thisrow{pT}}{\hadronconversionfactor*(\thisrow{lo-mean}+\coupling{\thisrow{mu2}}*\thisrow{nlo-mean})}{brahmsrcBKeYA}
   \addplot[data plot] table[x=pt,y expr={\thisrow{yield}},y error expr={(\thisrow{staterr} + \thisrow{syserr})}] {\brahmsdAuloY};
  \nextgroupplot[title={BRAHMS $\eta=3.2$}]
   \resultplot{rcBK GBW LO}{\thisrow{pT}}{\hadronconversionfactor*\thisrow{lo-mean}}{brahmsrcBKYB}
   \resultplot{rcBK GBW NLO}{\thisrow{pT}}{\hadronconversionfactor*(\thisrow{lo-mean}+\coupling{\thisrow{mu2}}*\thisrow{nlo-mean})}{brahmsrcBKYB}
   \resultplot{rcBK GBW exact}{\thisrow{pT}}{\hadronconversionfactor*(\thisrow{lo-mean}+\coupling{\thisrow{mu2}}*\thisrow{nlo-mean})}{brahmsrcBKeYB}
   \addplot[data plot] table[x=pt,y expr={\thisrow{yield}},y error expr={(\thisrow{staterr} + \thisrow{syserr})}] {\brahmsdAuhiY};
   \legend{LO,NLO,exact,data}
  \end{groupplot}
 \end{tikzpicture}
 \caption[Results of the exact kinematic calculation at BRAHMS]{Results for BRAHMS from figure~\ref{fig:brahmsresults} compared with the corresponding results with the exact kinematics.}
\label{fig:brahmsexactresults}
\end{figure}

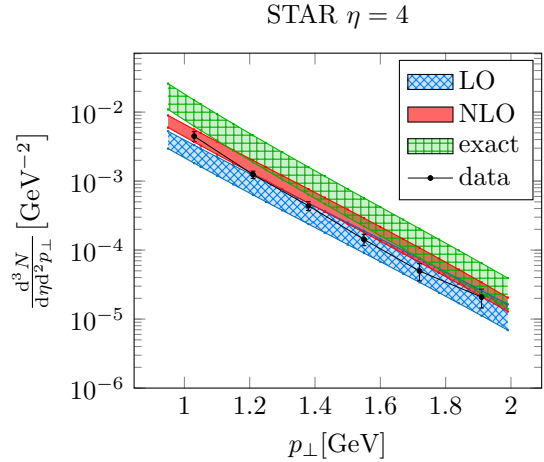
\begin{figure}
 \tikzsetnextfilename{starexactresults}
 \begin{tikzpicture}
  \begin{axis}[result axis,ymin=1e-6,ymode=log,title={STAR $\eta=4$}]
   \resultplot{rcBK GBW LO}{\thisrow{pT}}{\thisrow{lo-mean}}{starrcBK}
   \resultplot{rcBK GBW NLO}{\thisrow{pT}}{\thisrow{lo-mean}+\coupling{\thisrow{mu2}}*\thisrow{nlo-mean}}{starrcBK}
   \resultplot{rcBK GBW exact}{\thisrow{pT}}{\thisrow{lo-mean}+\coupling{\thisrow{mu2}}*(\thisrow{nlo-mean})}{starrcBKe}
   \addplot[data plot] table[x=pt,y expr={\thisrow{xsec}/\stardAusigmainel},y error expr={(\thisrow{staterr} + \thisrow{syserr})/\stardAusigmainel}] {\stardAu};
   \legend{LO,NLO,exact,data}
  \end{axis}
 \end{tikzpicture}
 \caption[Results of the exact kinematic calculation at STAR]{Results for STAR from figure~\ref{fig:starresults} compared with the corresponding results with the exact kinematics.}
\label{fig:starexactresults}
\end{figure}

Figures~\ref{fig:brahmsexactresults} and~\ref{fig:starexactresults} show that this is indeed the case.
When the exact kinematic expressions are used, the resulting cross section is positive for all $\pperp$ due to the exclusion of the negative contributions at $\xisym = 1$.
As with the resummation, the exact kinematic result overpredicts the data at small $\xsechadronpperp$, which is to be expected since we are using an expansion in $\Qs/\xsecquarkpperp$ which fails at low momenta.

An essential attribute of this matching prescription is the emergence of a ``matching momentum'' that separates the low-$\xsechadronpperp$ regime, where the small-$x$ calculation is accurate, from the high-$\xsechadronpperp$ regime, where the exact kinematic result is accurate.
Strictly speaking, though, this will be a \emph{range} of momenta, not a single value.
The lower bound of the range is approximately set by the saturation scale.
Since the exact kinematic result is a power series in $\Qs/\xsecquarkpperp$, the subleading terms which cause the result to diverge from the data should be negligible for $\xsechadronpperp \gtrsim \Qs$.
The upper bound, on the other hand, is set by the momentum at which the LO+NLO result diverges from the experimental data, which is slightly less than the cutoff momentum at which it becomes negative.
We have already seen in section~\ref{sec:cutoff} that the cutoff momentum is not straightforwardly related to the saturation scale; instead, it's somewhat larger than $\Qs$, by a factor which increases with rapidity.
This means we have a range in which both the LO+NLO formula and the exact kinematic formula are accurate descriptions of the data.

Figure~\ref{fig:brahmsexactresults} shows the matching momentum range to be roughly \SIrange{1}{1.5}{GeV} for $\pseudorapidity = 2.2$ and about \SIrange{1}{2}{GeV} for $\pseudorapidity = 3.2$.
As expected, the lower bound basically coincides with the saturation scale shown in figure~\ref{fig:satscalebrahms}, whereas the upper bound is larger, and the range of overlap grows with increasing rapidity.
We already know that the hybrid factorization used in this calculation applies only to forward rapidities, roughly $2$ and up, so the trend we seen in the results of figures~\ref{fig:brahmsexactresults} and~\ref{fig:starexactresults} suggests that the overlap region will be nonempty for all sufficiently forward rapidities.

According to these results, then, by combining the exact kinematic result at high $\pperp$ with the approximate one at low $\pperp$, we can achieve an accurate description of RHIC data over the full range of transverse momentum.

%% file: conclusion.tex
\chapter{Summary and Conclusion}\label{ch:conclusion}

In this dissertation, we've seen a variety of different phenomenological techniques for probing the unintegrated gluon distribution at small momentum fractions $\xg$.
The two calculations presented can be said to provide a representative (though not exhaustive) sample of the field of small-$x$ physics as a whole, in terms of
\begin{itemize}
 \item the quantities being calculated: partially inclusive cross sections and correlation functions
 \item the physical processes being measured: Drell-Yan lepton pair production and hadron production
 \item the orders of the phenomenological calculations involved (leading, next-to-leading, and beyond) and the theoretical difficulties inherent to each order
 \item the experimental facilities used to produce these results: BRAHMS and STAR at RHIC, and similar detectors at the LHC
\end{itemize}

As we've seen in chapter~\ref{ch:correlation}, the lepton pair-hadron correlation has several characteristics that make it useful as a probe of the gluon distribution.
In particular, the angular correlation allows us to separately probe values of $\dipoleF$ at low and high transverse momenta using, respectively, the back-to-back region $\Delta\phi\approx\pi$ and the parallel emission region $\Delta\phi\approx 0, 2\pi$).
For saturation physics, we are most interested in the region of small $\xg$, $Q < \Qs(\xg)$), which reveals itself in the shape of the central double peak.
The detailed characteristics of the peak, including its depth and width, are especially sensitive to the gluon distribution in the saturation regime.
Meanwhile the height of the peak is related to the rate of evolution of the saturation scale with $\xg$.

On the other hand, the nature of the high-momentum tail of $\dipoleF$, whether exponential or power-like, is closely correlated to the height of a parallel emission peak.
Existing experimental evidence suggests that the tail follows a power law and so, when correlation data from the LHC are collected and published, we should expect to see a large enhancement of parallel emission.
The same enhancement should be clearly evident in data from RHIC for low lepton pair masses.

The STAR experiment is considering these lepton pair-hadron correlation measurements, especially the resolution of the peak structure, as a motivation for the upcoming upgrade to their detector~\cite{STARprivate2013}.
Hopefully, then, high-quality correlation data will be available to compare to the theoretical predictions within the next several years.

Chapters~\ref{ch:crosssection} and~\ref{ch:results} describe the first numerical implementation of the complete next-to-leading order corrections (though not including NLL BK evolution) to the inclusive cross section for \HepProcess{\Pproton\Pnucleus \HepTo \Phadron\Panything}.
In chapter~\ref{ch:crosssection}, I've briefly reviewed the calculation of the NLO corrections, including the modification of the kinematics required for next-to-leading order processes.
The remainder of the chapter details the conversion of the formulas into a form suitable for numerical implementation by removing singularities and Fourier factors.

The results, presented in chapter~\ref{ch:results}, are very interesting (if not entirely surprising) due to the negativity of the LO+NLO calculated cross section at high $\xsechadronpperp$.
Evidently, incorporating the NLO corrections limits the kinematic range in which the formula gives an accurate relationship between the gluon distribution and the measured differential cross section.
The discrepancy at high tranverse momentum appears to be related to the subtractions of the divergences and the associated BK evolution.
However, at low $\xsechadronpperp$ where the relationship is accurate and the calculated results do agree with experimental data, the corrections reduce the theoretical uncertainty resulting from the factorization scale dependence.
Accordingly the LO+NLO formula allows more precise tests of saturation physics at low transverse momentum.

The observation that the calculated cross section is negative at high $\xsechadronpperp$ has significant consequences for future phenomenology: in particular, it indicates that we cannot rely on the convergence of the perturbation series at all transverse momenta, and that future work is probably best directed towards resummation of the most prominent contributions at higher orders, or other modifications to the LO+NLO cross section which were neglected in the derivation of chapter~\ref{ch:crosssection}.
Chapter~\ref{ch:beyondnlo} presents two initial forays into that effort.
First, we show an attempt at resumming the most prominent higher order contributions from multiple gluon loops.
Although the correction terms resulting from the resummation do add to the cross section at high $\xsechadronpperp$, they do not cure the negativity without a suspiciously large ``fudge factor'' being inserted in the formula.
A more promising technique is to dispense with the original calculation at high transverse momentum entirely, and use a large-$\xsechadronpperp$ series expansion that explicitly matches the result from collinear factorization, which is known not to suffer from negativity.
This gives a fairly accurate reproduction of the data at high transverse momentum.
By matching this result to the original small-$x$ calculation, we can achieve a reasonable description of the data over all transverse momenta.
This opens the door to using the high-$\xsechadronpperp$ region of the inclusive hadron cross section as a sensitive probe of the unintegrated gluon distribution.

%% file: grouptheory.tex
\Appendix{Group Theory of $\SUIII$}

Recall that a group is a set of elements together with an operation that acts on members of the set such that the inputs and outputs of the operation satisfy a particular relationship. 

Any other set of objects that satisfy the same relationships can also be considered a \term{representation}\footnotemark{} of $\SUIII$.\footnotetext{Technically, the representation is the mapping --- the group homomorphism --- from the set of abstract objects that constitute the group to another set of objects, such as matrices, that also satisfy the appropriate relationships under whatever operation is appropriate for those objects.}

The corresponding operation is matrix multiplication. For any two elements $U_1,U_2\in\SUIII$, the product $U_1 U_2 \in \SUIII$. These matrices constitute the \term{fundamental representation} of $\SUIII$.

Any matrix satisfying the three properties can be written $U = \exp(iM)$, where $M$ is a traceless hermitian $3\times 3$ matrix. The space of all $3\times 3$ traceless hermitian matrices is an eight-dimensional vector space, and thus $M$ can always be expressed as a linear combination of eight basis elements:
\begin{equation}
 M = \sum_{i=1}^{8} c_i\lambda_i
\end{equation}
We can choose these basis elements to be proportional to the \term{Gell-Mann matrices}
\begin{equation}
\begin{aligned}
 \lambda_1 &= \begin{pmatrix} 0 & 1 & 0 \\ 1 & 0 & 0 \\ 0 & 0 & 0 \end{pmatrix} &
 \lambda_2 &= \begin{pmatrix} 0 & -i & 0 \\ i & 0 & 0 \\ 0 & 0 & 0 \end{pmatrix} &
 \lambda_3 &= \begin{pmatrix} 1 & 0 & 0 \\ 0 & -1 & 0 \\ 0 & 0 & 0 \end{pmatrix} \\
 \lambda_4 &= \begin{pmatrix} 0 & 0 & 1 \\ 0 & 0 & 0 \\ 1 & 0 & 0 \end{pmatrix} &
 \lambda_5 &= \begin{pmatrix} 0 & 0 & -i \\ 0 & 0 & 0 \\ i & 0 & 0 \end{pmatrix} \\
 \lambda_6 &= \begin{pmatrix} 0 & 0 & 0 \\ 0 & 0 & 1 \\ 0 & 1 & 0 \end{pmatrix} &
 \lambda_7 &= \begin{pmatrix} 0 & 0 & 0 \\ 0 & 0 & -i \\ 0 & i & 0 \end{pmatrix} &
 \lambda_8 &= \frac{1}{\sqrt{3}} \begin{pmatrix} 1 & 0 & 0 \\ 0 & 1 & 0 \\ 0 & 0 & -2 \end{pmatrix}
\end{aligned}
\end{equation}
with the coefficient
\begin{equation}
 T^i = \frac{\lambda_i}{2}
\end{equation}
These are the generators of $\SUIII$ in the fundamental representation. They satisfy the commutation relations
\begin{equation}
 \commut{T^a}{T^b} = i f^{abc} T^c
\end{equation}
We'll also need the generators in the adjoint representation, defined as
\begin{equation}
 (\tilde{T}^c)^{ab} = f^{abc}
\end{equation}

\section{Identities}
\subsection{Completeness Identity}
Suppose we have a vector space $\mathcal{M}$ of matrices equipped with the scalar product
\begin{equation}\label{eq:traceprod:definition}
 \braket{A,B} \defn \trace\bigl(\herm{A} B\bigr)
\end{equation}
and there is a basis of matrices $X^a$ on this space which is orthonormal
\begin{equation}
 \braket{X^a,X^b} = \delta^{ab}
\end{equation}
and complete
\begin{equation}\label{eq:traceprod:expansion}
 \forall M\in\mathcal{M},\ M = M^a X^a \qquad\text{where}\qquad M^a \defn \braket{M,X^a}
\end{equation}
where the sum over the repeated index $a$ is implied.

If we let $\elementmatrix_{ij}$ be the matrix which has a $1$ at row $i$ and column $j$ and zero everywhere else,
\begin{equation}\label{eq:elementmatrix}
 (\elementmatrix_{ij})_{mn} \defn \delta_{im}\delta_{jn}
\end{equation}
then the scalar product with $\elementmatrix_{ij}$ provides a convenient way to extract one element of an arbitrary matrix
\begin{equation}
 \braket{\elementmatrix_{ij},M} = \trace\bigl(\herm{\elementmatrix_{ij}}M\bigr) = M_{ij}
\end{equation}
Now, using~\eqref{eq:traceprod:expansion}, we can write
\begin{equation}
 \elementmatrix_{ij}\elementmatrix_{kl} = \braket{\elementmatrix_{ij},X^a}X^a \braket{\elementmatrix_{kl},X^b}X^b = X^a_{ij}X^a X^b_{kl}X^b
\end{equation}
and if we take the trace of this, using
\begin{equation}
 \trace\bigl(\elementmatrix_{ij} \elementmatrix_{kl}\bigr) = (\elementmatrix_{ij})_{mn}(\elementmatrix_{kl})_{nm} = \delta_{im}\delta_{jn} \delta_{kn}\delta_{lm} = \delta_{il}\delta_{jk}
\end{equation}
and
\begin{equation}
 \trace\bigl(X^a X^b\bigr) = \trace\bigl(\herm{X}^a X^b\bigr) = \braket{X^a,X^b} = \delta^{ab}
\end{equation}
we find that
\begin{equation}\label{eq:deltaidentity}
 \delta_{il}\delta_{jk} = X^a_{ij}X^a_{kl}
\end{equation}

We can apply these conclusions to the basis consisting of the normalized Gell-Mann matrices and the identity matrix:
\begin{align}
 X^0 &= \frac{1}{\sqrt{3}}\begin{pmatrix}1 & 0 & 0 \\ 0 & 1 & 0 \\ 0 & 0 & 1\end{pmatrix} &
 X^i &= \frac{\lambda_i}{\sqrt{2}} = \sqrt{2}T_i,\quad 1\leq i \leq 8
\end{align}
In this basis, equation~\eqref{eq:deltaidentity} becomes
\begin{equation}
 \delta_{il}\delta_{jk} = X^0_{ij} X^0_{kl} + \sum_{a=1}^8 X^a_{ij}X^a_{kl} = \frac{1}{3}\delta_{ij}\delta_{kl} + \sum_{a=1}^8 X^a_{ij}X^a_{kl}
\end{equation}
which can be rearranged to
\begin{equation}
 T^a_{ij} T^a_{kl} = \frac{1}{2}\delta_{il}\delta_{jk} - \frac{1}{2\Nc}\delta_{ij}\delta_{kl}
\end{equation}
where now the implicit sum over $a$ runs from $1$ to $8$ because there are only 8 generators.

\subsection{Wilson line reduction}
This identity is useful to express a Wilson line in the adjoint representation in terms of the fundamental representation. The proof presented here comes from reference~\cite{DavidBarMoshe:122955}.
\begin{equation}\label{eq:ident:wilsonlinereduce}
 W^{ab}(\vec\xperp) = 2\trace T^a U(\vec\xperp) T^b \herm{U}(\vec\xperp)
\end{equation}

We start with the fact that tensor product of the fundamental representation and its conjugate gives the adjoint representation plus the trivial representation.
In the language of group theory, one might see this written (for $\SUIII$) as $3\otimes\bar{3} = 8\oplus 1$; in the language of Wilson lines, which are members of $\SUIII$, it becomes
\begin{equation}\label{eq:groupproduct}
 U^{cf}\herm{U}^{gd} = W^{(cd)(fg)} + \frac{1}{\Nc}\delta^{cf}\delta^{dg}
\end{equation}
Here $W^{(cd)(fg)}$ is the same object that was designated $W^{ab}$ before, but the indices $a$ and $b$, which ran from~$1$ to~$8$, have been replaced by pairs $(cd)$ and $(fg)$ where each of $c$, $d$, $f$, and $g$ runs from $1$ to $3$. 
Effectively $W^{(cd)(fg)}$ is a member of a subset of $9\times 9$ matrices that implement the adjoint representation of $\SUIII$.

In the paired index notation, the other side of the equation becomes $2\trace T^(cd) U(\vec\xperp) T^(fg) \herm{U}(\vec\xperp)$.
We can show that this is equal to~\eqref{eq:groupproduct} using the following alternate basis for the generators of $\SUIII$:
\begin{equation}
 T^{(cd)} = \frac{1}{\sqrt{2}}\biggl(\elementmatrix^{cd} - \frac{1}{\Nc}\delta^{cd}\biggr)
\end{equation}
or with the additional indices explicit,
\begin{align}
 T^{(cd)}_{ij} 
 &= \frac{1}{\sqrt{2}}\biggl(\elementmatrix^{cd}_{ij} - \frac{1}{\Nc}\delta^{cd}_{ij}\biggr) \\
 &= \frac{1}{\sqrt{2}}\biggl(\delta_i^c \delta_i^d - \frac{1}{\Nc}\delta^{cd}\delta_{ij}\biggr)
\end{align}
where I've used the definition of $\elementmatrix$ from above~\eqref{eq:elementmatrix}.
Plugging in, we get
\begin{equation}
 2\frac{1}{\sqrt{2}}\biggl(\delta_i^c \delta_i^d - \frac{1}{\Nc}\delta^{cd}\delta_{ij}\biggr) U_{jk}(\vec\xperp) \frac{1}{\sqrt{2}}\biggl(\delta_k^f \delta_l^g - \frac{1}{\Nc}\delta^{fg}\delta_{kl}\biggr) \herm{U_{li}}(\vec\xperp)
\end{equation}
Expanding this and summing over repeated indices gives
\begin{equation}
 U^{df}\herm{U}^{gc} - \frac{1}{\Nc}\delta^{cd}\underbrace{U^{if}\herm{U}^{gi}}_{\delta^{fg}} - \frac{1}{\Nc}\delta^{fg}\underbrace{U^{dk}\herm{U}^{kc}}_{\delta^{cd}} + \frac{1}{\Nc^2}\delta^{cd}\delta^{fg} \underbrace{U_{ik}\herm{U_{ki}}}_{\Nc}
 =
 U^{df}\herm{U}^{gc} - \frac{1}{\Nc}\delta^{cd}\delta^{fg}
\end{equation}
Relabeling indices $d\leftrightarrow c$ and using~\eqref{eq:groupproduct} shows that this is equal to $W^{(cd)(fg)}$, thus completing the proof.

%% file: lightcone.tex
\Appendix{Light-Cone Kinematics}\label{ap:lightcone}
In light-cone kinematics, instead of the $t$ and $z$ coordinates of a four-vector, we use $+$ and $-$ components, defined as
\begin{equation}
 x^\pm = x_t \pm x_z
\end{equation}
With these definitions, the scalar product in light-cone coordinates works out to
\begin{equation}
 a^\mu b_\mu = \frac{1}{2}(a_+ b_- + a_- b_+) - a_x b_x - a_y b_y
\end{equation}
which corresponds to a metric (in the $t$, $x$, $y$, $z$ basis) of
\begin{equation}
 \begin{pmatrix}
  0 & 0 & 0 & \frac{1}{2} \\
  0 & -1 & 0 & 0 \\
  0 & 0 & -1 & 0 \\
  \frac{1}{2} & 0 & 0 & 0
 \end{pmatrix}
\end{equation}

We most often use light-cone coordinates for momenta, in which case the definitions are
\begin{align}\label{eq:lightconemomentum}
 p^{\pm} &= E \pm p_z &
 \vec{p}_\perp &= (p_x, p_y)
\end{align}
and for the scalar product
\begin{align}
 p^\mu q_\mu &= \frac{1}{2}(p^+ q^- + p^- q^+) - \vec{p}_\perp\cdot\vec{q}_\perp = E_p E_q - p_z q_z - p_x q_x - p_y q_y \\
 p^\mu p_\mu &= p^+ p^- - p_\perp^2 = E^2 - p_z^2 - p_x^2 - p_y^2 = m^2 \label{eq:lightconerelation}
\end{align}
the latter of which can be rearranged into
\begin{equation}\label{eq:lightconereciprocalrelation}
 p^\pm = \frac{p_\perp^2 + m^2}{p^\mp}
\end{equation}

Now we take the definition of rapidity,
\begin{equation}\label{eq:rapidity}
 Y = \frac{1}{2}\ln\frac{E + p_z}{E - p_z} = \frac{1}{2}\ln\frac{p^+}{p^-}
\end{equation}
We can solve this for $p_z$ to get
\begin{equation}
 p_z = E\tanh Y
\end{equation}
or for $p^+$ or $p^-$ to get
\begin{equation}\label{eq:lightconerapidityrelation}
 p^\pm = p^\mp e^{\pm 2Y}
\end{equation}

Equation~\eqref{eq:lightconerelation} can be rearranged into
\begin{equation}\label{eq:rearrangedlightconerelation}
 p^+ p^- = p_\perp^2 + m^2
\end{equation}
From equation~\eqref{eq:lightconerapidityrelation} we get
\begin{equation}\label{eq:rearrangedlightconerapidityrelation}
 p^+ p^- = (p^\pm)^2 e^{\mp 2Y}
\end{equation}
Combining equations~\eqref{eq:rearrangedlightconerelation} and~\eqref{eq:rearrangedlightconerapidityrelation} gives
\begin{equation}\label{eq:momentumpm}
 p^\pm = \sqrt{p_\perp^2 + m^2}\ e^{\pm Y}
\end{equation}
which can also be solved for $Y$ as
\begin{equation}\label{eq:rapiditymomentumpm}
 Y = \pm\ln\frac{p^\pm}{\sqrt{p_\perp^2 + m^2}}
\end{equation}

%% file: identities.tex
\Appendix{Identities for Integration}

\section{Integration over vector product}
This identity is required to convert certain parts of quadrupole-type terms in chapter~\ref{ch:crosssection} to momentum space.
\begin{equation}\label{eq:ident:vecprod}
 \int\uddc\vec\sperp\uddc\vec\tperp e^{-i\vec\qiperp\cdot\vec\sperp} e^{-i\vec\qiiperp\cdot\vec\tperp} \frac{\vec\sperp\cdot\vec\tperp}{\sperp^2\tperp^2} = -(2\pi)^2\frac{\vec\qiperp\cdot\vec\qiiperp}{\qiperp^2\qiiperp^2}
\end{equation}
We can rewrite the left side as
\begin{equation}
 \int_0^{2\pi}\udc\theta_s\int_0^\infty\udc\sperp \int_0^{2\pi}\udc\theta_t\int_0^\infty\udc\tperp \sperp\tperp e^{-i\qiperp\sperp\cos(\theta_s - \theta_1)} e^{-i\qiiperp\tperp\cos(\theta_t - \theta_2)} \frac{\cos(\theta_s - \theta_t)}{\sperp\tperp}
\end{equation}
and substitute in the Bessel generating function
\begin{equation}
 e^{ikx\cos\theta} = \sum_n i^n J_n(kx)e^{in\theta}
\end{equation}
to get
\begin{multline}
 \int_0^{2\pi}\udc\theta_s\int_0^\infty\udc\sperp \int_0^{2\pi}\udc\theta_t\int_0^\infty\udc\tperp \sum_m (-i)^m J_m(\qiperp\sperp)e^{im(\theta_s - \theta_1)} \\
 \times \sum_n (-i)^n J_n(\qiiperp\tperp)e^{in(\theta_t - \theta_2)} \frac{1}{2}\bigl(e^{i(\theta_s - \theta_t)} + e^{i(\theta_t - \theta_s)}\bigr)
\end{multline}
which can be rearranged to
\begin{multline}
 \frac{1}{2}
 \sum_m 
 \sum_n 
 e^{-im\theta_1} e^{-in\theta_2}
 \int_0^\infty\udc\sperp (-i)^m J_m(\qiperp\sperp) \int_0^\infty\udc\tperp (-i)^n J_n(\qiiperp\tperp) \\
 \int_0^{2\pi}\udc\theta_s 
 \int_0^{2\pi}\udc\theta_t 
  \Bigl(e^{i(m + 1)\theta_s} e^{-i(n - 1)\theta_t} + e^{i(n + 1)\theta_t} e^{i(m - 1)\theta_s)}\Bigr)
\end{multline}
Performing the angular integrals turns the last line into $(2\pi)^2\bigl(\delta_{m,-1}\delta_{n,1} + \delta_{n,-1}\delta_{m,1}\bigr)$. Then we can sum over $m$ and $n$ to get
\begin{multline}
 \frac{(2\pi)^2}{2}
 e^{i\theta_1} e^{-i\theta_2}
 \int_0^\infty\udc\sperp i J_{-1}(\qiperp\sperp) \int_0^\infty\udc\tperp (-i) J_{1}(\qiiperp\tperp) \\
 +
 \frac{(2\pi)^2}{2}
 e^{-i\theta_1} e^{i\theta_2}
 \int_0^\infty\udc\sperp (-i) J_{1}(\qiperp\sperp) \int_0^\infty\udc\tperp i J_{-1}(\qiiperp\tperp)
\end{multline}
The radial integrals can be done with the identity $\int_0^\infty J_{\pm 1}(kx)\udc x = \pm\frac{1}{k}$ (valid for $k\in\realset$) which leaves us with
\begin{equation}
 -\frac{(2\pi)^2}{2\qiperp\qiiperp}\bigl(e^{i\theta_1}e^{-i\theta_2} + e^{-i\theta_1}e^{i\theta_2}\bigr)
 = -(2\pi)^2\frac{\cos(\theta_1 - \theta_2)}{\qiperp\qiiperp}
 = -(2\pi)^2\frac{\vec\qiperp\cdot\vec\qiiperp}{\qiperp^2\qiiperp^2}
\end{equation}

\section{Fourier Transforms}
The Fourier transform can be defined in various ways, corresponding to different choices of the constants $a$ and $b$ in the definition
\begin{equation}\label{eq:ident:FT}
 \tilde f(k) = \sqrt{\frac{\abs{b}}{(2\pi)^{1 - a}}}\int_{-\infty}^{\infty} f(x) e^{-ibkx} \udc x
\end{equation}
and its inverse
\begin{equation}\label{eq:ident:IFT}
 f(x) = \sqrt{\frac{\abs{b}}{(2\pi)^{1 + a}}}\int_{-\infty}^{\infty} \tilde f(k) e^{ibkx} \udc k
\end{equation}
To avoid ambiguity, in this dissertation I don't use the notation $\tilde f(k)$; all Fourier-type integrals are written out explicitly.

At several points it's useful to have the Fourier integral of the 2D Dirac delta function,
\begin{equation}\label{eq:ident:deltaFT}
 \delta^{(2)}(\vec\rperp) = \frac{1}{(2\pi)^2}\iint e^{\pm i\vec k\cdot\vec\rperp}\uddc\vec k
\end{equation}
which is easily derived from the definitions~\eqref{eq:ident:FT} and~\eqref{eq:ident:IFT} with $a = 1$ and $b = \pm 1$.
For one dimension,
\begin{equation}
 \int_{-\infty}^{\infty} \delta(x) e^{-ikx}\udc x = 1
 \qquad\implies\qquad
 \frac{1}{2\pi}\int_{-\infty}^{\infty} e^{ikx}\udc k = \delta(x)
\end{equation}
and then $\delta^{(2)}(\vec\rperp) = \delta(x)\delta(y)$.

We can use this to prove a useful identity for multiplication in Fourier integrals as follows:
\begin{align}
    \int f(x)g(x) e^{-ikx} \udc x 
 &= \iint f(x)g(y) \delta(x - y) e^{-ikx} \udc x \udc y \notag \\
 &= \iiint f(x)g(y) e^{iq(x - y)} e^{-ikx} \udc x \udc y \udc q \notag \\
    \int f(x)g(x) e^{-ikx} \udc x 
 &= \int \udc q \biggl(\int f(x) e^{-i(k - q)x} \udc x\biggr) \biggl(\int g(y) e^{-iqy} \udc y\biggr) \label{eq:ident:FTconvolution1D}
\end{align}
The derivation is essentially the same for the multidimensional version,
\begin{equation}
 \int f(\vec{x})g(\vec{x}) e^{-i\vec{k}\cdot\vec{x}} \udc^n\vec{x} = \int\udc^n\vec{q} \biggl(\int f(\vec{x}) e^{-i(\vec{k} - \vec{q})\cdot\vec{x}} \udc^n\vec{x}\biggr) \biggl(\int g(\vec{y}) e^{-i\vec{q}\cdot\vec{y}} \udc^n\vec{y}\biggr) \label{eq:ident:FTconvolutionnD}
\end{equation}

\section{Regularization to Logarithm}
This identity, equation (34) from reference~\cite{Chirilli:2012jd}, is the easiest way to convert some parts of quadrupole-type terms in chapter~\ref{ch:crosssection} to momentum space.
\begin{equation}\label{eq:ident:reglog}
 \frac{1}{4\pi}\int\uddc\vec\qperp e^{-i\vec\qperp\cdot\vec\rperp}\ln\frac{(\vec\qperp - \xi'\vec\kperp)^2}{\kperp^2} = \delta^{(2)}(\vec\rperp)\int\frac{\uddc\vec\rpperp}{\rpperp^2}e^{i\vec\kperp\cdot\vec\rpperp} - \frac{1}{\rperp^2}e^{-i\xi'\vec\kperp\cdot\vec\rperp}
\end{equation}

A proof starts with the relation
\begin{equation}
 \ln\frac{(\vec\qperp - \xi'\vec\kperp)^2}{\kperp^2} = \lim_{a\to 0}2\bigl[K_0(a\kperp) - K_0(a\norm{\vec\qperp - \xi'\vec\kperp})\bigr]
\end{equation}
which we can derive by noting that $K_0(x)$ goes to $-\ln x + \ln 2 - \eulergamma$ as $x\to 0^+$.
\begin{multline}
 \lim_{a\to 0}2\bigl[K_0(a\kperp) - K_0(a\norm{\vec\qperp - \xi'\vec\kperp})\bigr] \\
 = 2\bigl[-\ln \kperp - \ln a + \ln 2 - \eulergamma + \ln \norm{\vec\qperp - \xi'\vec\kperp} + \ln a - \ln 2 + \eulergamma\bigr]
 = \ln\frac{(\vec\qperp - \xi'\vec\kperp)^2}{\kperp^2}
\end{multline}
With this, we can express the left side of~\eqref{eq:ident:reglog} as
\begin{equation}
 \frac{2}{4\pi}\lim_{a\to 0}\int\uddc\vec\qperp e^{-i\vec\qperp\cdot\vec\rperp}\bigl[K_0(a\kperp) - K_0(a\norm{\vec\qperp - \xi'\vec\kperp})\bigr]
\end{equation}
which simplifies to
\begin{equation}
 \frac{1}{2\pi}\biggl[(2\pi)^2\delta^{(2)}(\vec\rperp)\lim_{a\to 0}K_0(a\kperp) - e^{-i\xi'\vec\kperp\cdot\vec\rperp}\lim_{a\to 0}\int\uddc\vec\qiperp e^{-i\vec\qiperp\cdot\vec\rperp}K_0(a\qiperp)\biggr]
\end{equation}
We can now use the fact that the modified Bessel function can be expressed as
\begin{equation}
 K_0(a\kperp) = \int\frac{\uddc\vec\rpperp}{2\pi} \frac{e^{i\vec\kperp\cdot\vec\rpperp}}{\rpperp^2 + a^2}
\end{equation}
Substituting this in and performing the limit, we obtain
\begin{equation}
 \delta^{(2)}(\vec\rperp)\frac{e^{i\vec\kperp\cdot\vec\rpperp}}{\rpperp^2} - e^{-i\xi'\vec\kperp\cdot\vec\rperp}\int\frac{\uddc\vec\qiperp}{2\pi} e^{-i\vec\qiperp\cdot\vec\rperp} \int\frac{\uddc\vec\rpperp}{2\pi} \frac{e^{i\vec\qiperp\cdot\vec\rpperp}}{\rpperp^2}
\end{equation}
Applying equation~\eqref{eq:ident:deltaFT} to the second term gives us the desired result~\eqref{eq:ident:reglog}.

\section{Sum of Polarizations}
This identity is often used to sum over polarizations.
\begin{equation}\label{eq:ident:polarization}
 \sum_{\lambda} \bigl(\vec{a}\cdot\epsilon_{\vec k}^{(\lambda)}\bigr) \bigl(\vec{b}\cdot\epsilon_{\vec k}^{(\lambda)}\bigr) = \vec{a}\cdot\vec{b} - (\vec{a}\cdot\unitk) (\vec{b}\cdot\unitk)
\end{equation}
Its proof follows simply from the fact that the $\epsilon_{\vec k}^{(\lambda)}$ together with $\unitk$ form a complete and orthonormal basis.
The completeness property means that
\begin{equation}
 \vec{a} = (\vec{a}\cdot\unitk)\unitk + \sum_\lambda(\vec{a}\cdot\epsilon_{\vec k}^{(\lambda)})\epsilon_{\vec k}^{(\lambda)}
\end{equation}
Plugging this decomposition and the corresponding formula for $\vec{b}$ into the product $\vec{a}\cdot\vec{b}$, and using the orthonormality of the basis vectors, gives
\begin{equation}
 \vec{a}\cdot\vec{b} = (\vec{a}\cdot\unitk)(\vec{b}\cdot\unitk) + \sum_\lambda(\vec{a}\cdot\epsilon_{\vec k}^{(\lambda)})(\vec{b}\cdot\epsilon_{\vec k}^{(\lambda)})
\end{equation}
which is trivially rearranged to give~\eqref{eq:ident:polarization}.

%% file: vita.tex

\section*{Publications}
\begin{description}
 \item[Next-to-leading and resummed BFKL evolution with saturation boundary]~\\
  Emil Avsar, Anna M. Sta\'sto, D.~N. Triantafyllopoulos, and David Zaslavsky.\\
  JHEP \textbf{2011}, 34 (2011). \texttt{arXiv:1107.1252}.
 \item[Drell-Yan lepton-pair-hadron correlation in pA collisions]~\\
  Anna M. Sta\'sto, Bo-Wen Xiao, and David Zaslavsky. \\
  Phys. Rev. D \textbf{86}, 014009 (2012). \texttt{arXiv:1204.4861}.
 \item[Towards the Test of Saturation Physics Beyond Leading Logarithm]~\\
  Anna M. Sta\'sto, Bo-Wen Xiao, and David Zaslavsky. \\
  Phys. Rev. Lett. \textbf{112}, 012302 (2014). \texttt{arXiv:1307.4057}.
 \item{\hspace*{-\labelsep}}\raggedright\textbf{Matching collinear and small $x$ factorization calculations for inclusive hadron production in pA collisions}~\\
  Anna M. Sta\'sto, Bo-Wen Xiao, Feng Yuan, and David Zaslavsky. \\
  Phys. Rev. D \textbf{90}, 014047 (2014). \texttt{arXiv:1405.6311}.
\end{description}

\section*{Education}
\begin{description}
 \item[Ph.D. Pennsylvania State University]~\\
   Ph.D. in nuclear and particle physics focusing on small-$x$ phenomenology\\
   December 2014
 \item[A.B. Princeton University]~\\
  A.B. in physics, with concentration in applications of computing\\
  June 2008
\end{description}

\section*{Awards}
\begin{description}
 \item[Peter Eklund Lecturing Competition] Winner 2014, Honorable Mention 2013
 \item[W. Donald Miller Graduate Fellowship] 2013
 \item[Duncan Graduate Fellowship] 2012
 \item[Penn State University Graduate Fellowship] 2008
\end{description}

\section*{Outreach Activities}
\begin{description}
 \item[Ellipsix Informatics] \texttt{http://www.ellipsix.net}\\
 Science Communicator and Blogger\\
 Write about scientific research for a broader audience
 \item[Physics Stack Exchange] \texttt{http://physics.stackexchange.com}\\
 Contributor and Community Moderator\\
 Provide answers to physics questions and help manage the community
\end{description}